\documentclass[12pt,german,english,titlepage,a4paper]{report}
     \usepackage{babel}
     \usepackage{longtable}
     \usepackage{graphicx}
     \usepackage[latin1]{inputenc}
\usepackage{amsmath}
\usepackage{amssymb}          % for \mathfrak{} instead of the eufrak package
\usepackage{amscd}
\usepackage{fancyhdr}
\newif\ifpdf
\ifx\pdfoutput\undefined
   \pdffalse
\else
   \pdfoutput=1
   \pdftrue
\fi
\ifpdf  %......... hyperref ..................................................
  \usepackage[pdftex=true, pdftitle={Simulations of Lattice Fermions with Chiral Symmetry in Quantum Chromodynamics} 
              ,pdfauthor={Stanislav Shcheredin}, pdfsubject={Digital thesis}
              ,pdfkeywords={chiral fermions}
              ,plainpages=false, hypertexnames=false, pdfpagelabels=true
              ,pdfpagemode=UseOutlines, bookmarks=true, bookmarksopen=true
              ,bookmarksopenlevel=0, bookmarksnumbered=true,unicode,
              ,hyperindex=true, colorlinks=true, pdfstartview={FitV}%
             ]{hyperref}
   \usepackage{color}
   \definecolor{darkred}{rgb}{0.5,0,0}
   \definecolor{darkgreen}{rgb}{0,0.5,0}
   \definecolor{darkblue}{rgb}{0,0,0.5}
   \hypersetup{colorlinks, linkcolor=black, filecolor=darkgreen, 
                           urlcolor=darkred, citecolor=darkblue}
\else  
   \usepackage{hyperref}
\fi  %........................................................................

\usepackage[numbers,sort&compress]{natbib}  % for [1,2,3] --> [1-3] because cite
\usepackage{hypernat}                       % doesn't work with hyperref package

\newcommand{\dcauthorpre}{Herr Dipl.-Phys. } 
\newcommand{\dcauthorsurname}{Shcheredin } 
\newcommand{\dcauthorname}{Stanislav } 
\newcommand{\dcauthoradd}{geboren am 17.12.1978 im Moskauer Gebiet} 
\newcommand{\dctitle}{Simulations of Lattice Fermions with Chiral Symmetry in Quantum Chromodynamics} 
\newcommand{\dcsubtitle}{~}  
\newcommand{\dcapprovala}{Prof. Dr. P. Damgaard} 
\newcommand{\dcapprovalb}{Prof. Dr. M. M\"uller-Preu\ss ker} 
\newcommand{\dcapprovalc}{Prof. Dr. P. Weisz} 

\newcommand{\dcdegree}{doctor rerum naturalium\\ (dr. rer. nat.)} 
\newcommand{\dcsubject}{Physik} 
\newcommand{\dcfaculty}{Mathematisch-Naturwissenschaftlichen Fakult\"at I}
\newcommand{\dcuniversity}{Humboldt-Universit\"at zu Berlin}
\newcommand{\dcdean}{Prof. Thomas Buckhout, Ph.D.}
\newcommand{\dcpresident}{Prof. Dr. J\"{u}rgen\ Mlynek}
\newcommand{\dcdatesubmitted}{1. November 2004} 
\newcommand{\dcdateexam}{17. Januar 2005} 

\newcommand{\Psibar}{\overline\Psi}
\newcommand{\chibar}{\overline\chi}

\newcommand{\fr}[2]{{\frac{#1}{#2}}}
\newcommand{\fm}{\,{\rm fm}}
\newcommand{\nn}{\nonumber}

\newcommand{\Z}{Z \!\!\! Z}
\newcommand{\ie}{{\frenchspacing i.\hspace{0.4mm}e.{}}}

\newcommand{\eg}{{\frenchspacing e.\hspace{0.4mm}g.{}}}

\author{von \\ \dcauthorpre  \dcauthorname  \dcauthorsurname  \\ \dcauthoradd}

\title{ \vspace{-5cm}\dctitle \\ 
\vspace{0.5cm}
\large{\dcsubtitle} \\ 
\vspace{0.5cm} {\Large{D I S S E R T A T I O N }}\\ 
\vspace{0.5cm} \large{zur Erlangung des akademischen Grades \\ 
\dcdegree\\ im Fach \dcsubject \\ 
\vspace{0.5cm} eingereicht an der \\ 
\dcfaculty \\ 
\dcuniversity \\}}
\date{\vspace{0.5cm}
\raggedright{
Pr\"asident der Humboldt-Universit\"at zu Berlin:\\
\dcpresident \vspace{-0.3cm}
}\vspace{0.5cm}\\
\raggedright{
Dekan der \dcfaculty:\\
\dcdean \vspace{-0.3cm}
}\vspace{0.5cm}\\
\raggedright{
Gutachter:
\begin{enumerate} 
\item{\dcapprovala} \vspace{-0.3cm}
\item{\dcapprovalb} \vspace{-0.3cm}
\item{\dcapprovalc} \vspace{-0.3cm}
\end{enumerate}} \vspace{0.5cm}
\raggedright{
\begin{tabular}{lll}
eingereicht am: &  &\dcdatesubmitted\\
Tag der m\"undlichen Pr\"ufung: & & \dcdateexam
\end{tabular}}\\ 
}
\begin{document}
\pagenumbering{roman}
\maketitle
%---------------------
\selectlanguage{english}
\abstract
The recent years have seen a tangible progress in the fermion field discretization with an exact chiral symmetry, which is realized on the lattice through the Ginsparg-Wilson relation. At present the simulation costs of Ginsparg-Wilson fermions are considerably higher than for the conventional formulations without the exact chiral symmetry and therefore only quenched simulations in Quantum Chromodynamics (QCD) are feasible so far. However, the chiral fermions are especially important to address the ab initio QCD calculations at small quark masses of the low energy constants of the chiral Lagrangian. This thesis is dedicated to explore the feasibility of such calculations in the $\epsilon$--regime of QCD for the extraction of physical information. We apply two formulations of the Ginsparg-Wilson fermions, namely, the Neuberger operator and the hypercube overlap operator to compute the observables of interest. Since the calculations are done in the $\epsilon$--regime, each observable has to be sampled within a fixed topological sector, which is well defined for the Ginsparg-Wilson fermions as the index of the Dirac operator. As a main result we present the comparison of the distributions of the leading individual eigenvalues of the Neuberger operator in QCD and the analytical predictions of chiral random matrix theory. We observe a good agreement as long as each side of the physical volume exceeds about $L\approx 1.12\fm$. At the same time the chiral condensate $\Sigma$ can also be estimated. It turns out that this bound for $L$ is generic and sets the size of the physical volume where the axial correlator behaves according to chiral perturbation theory. This allows us to compute a value for the pion decay constant $F_{\pi}$. The simulations also show that due to the high probability of the near-zero modes it is prohibitively difficult to sample the axial correlator in the neutral topological sector. In the higher sectors, however, we observe that the sensitivity of the analytical predictions for the axial correlator to extract $\Sigma$ is lost to a large extent. As an alternative procedure we only consider the contribution from the zero modes. Here we are able to obtain an estimate for $F_{\pi}$ and $\alpha$, where $\alpha$ is a low energy constant peculiar to quenching. We calculate the topological susceptibility, both for the Neuberger operator and for the overlap hypercube operator. It turns out that the result with the overlap hypercube operator is closer to the continuum limit. Also the locality properties are superior to those of the Neuberger fermions. As a theoretical development the L\"uscher topology conserving gauge action is investigated. This enables us to sample the observables of interest in the $\epsilon$--regime without recomputing the index.  We can report that a promising gauge action has been identified.
%\dckeywordsen
%_____________________
\selectlanguage{german}
\abstract \setcounter{page}{2}  %Gegebenfalls die Seitenzahl "andern
In den letzten Jahren gab es konkrete Fortschritte in der Diskretisierung von Fermionfeldern mit exakter chiraler Symmetrie. Diese Symmetrie wird auf dem Gitter durch die Ginsparg-Wilson Beziehung realisiert. Momentan sind alle bekann\-ten Formulierungen wesentlich aufw\"andiger als konventionelle Fermionen ohne chirale Symmetrie. Deswegen sind heutzutage nur ,,quenched'' Simulationen der Quantenchromodynamik (QCD) m\"oglich. Doch die chiralen Fermionen sind besonders wichtig f\"ur ab initio Berechnungen der Niederenergie-Konstanten  der chiralen Lagrangedichte bei kleinen Quarkmassen. Das Ziel dieser Dissertation besteht darin, die  Realisierbarkeit dieser Methode zur Gewinnung physikalischer Informationen im $\epsilon$--Regime der QCD zu erforschen. Wir haben zwei Formulierungen eingesetzt. Dies sind der Neuberger Operator und Overlap Hyperkubus Operator. Da die Rechnungen im $\epsilon$--Regime durchgef\"urt werden, muss man jede Observable in einem bestimmten topologischen Sektoren messen. Bei Ginsparg-Wilson Fermionen ist dieser durch den Index des Dirac Operators definiert. Ein Hauptergebniss dieser Arbeit ist der Vergleich der Wahrscheinlichkeits\-verteilungen einzelner Eigenwerte des Neuberger Operators in der QCD mit entsprechenden analytischen Vorhersagen der Theorie der Zufallsmatritzen. Wir beobachten eine gute \"Ubereinstimmung solange jede Seite $L$ des physikalischen Volumens gr\"o\ss er als etwa $1.12\fm$ ist. Dabei kann auch das chirale Kondensat $\Sigma$ abgesch\"atzt werden. Es ergab sich, dass diese untere Schranke von $L$ allgemein gilt und die Gr\"o\ss e des physikalischen Volumens, auf dem der Axialkorrelator den Vorhersagen des chiralen St\"orungstheorie folgt, festlegt. Damit k\"onnen wir die Pionzerfallskonstante $F_{\pi}$ bestimmen. Unsere Simulationen zeigen, dass wegen der gro\ss en Wahrscheinlichkeit niedriger Eigenwerte die Messung des Axial\-korrelators im topologischen neutralen Sektor extrem aufw\"andig ist.  Doch reicht die Empfindlichkeit der Vorhersagen der chiralen  St\"orungstheorie in h\"oheren topologischen Sektoren bei der gegebenen Statistik nicht zur Bestimmung von $\Sigma$  aus. Als alternative Methode, gehen wir dazu \"uber, allein den Beitrag der Nullmoden zu betrachten. Hier k\"onnen wir Absch\"atzungen f\"ur $F_{\pi}$ und $\alpha$ gewinnen. Dabei ist $\alpha$ eine Niederenergie-Konstante die nur in der ,,quenched'' N\"aherung erscheint. Wir berechnen die topologische Suszeptibilit\"at f\"ur den Neuberger und Overlap Hyperkubus Operator. Im letzten Fall ist der berechnete Wert n\"aher beim Kontinuumslimes. Die Lokalisierung f\"ur den Overlap Hyperkubus Operator ist auch besser als f\"ur den Neuberger Operator. Unser anderes Ziel ist die Erforschung einer topologieerhaltenden Eichwirkung. Hier k\"onnen wir berichten, dass eine vielversprechende Eichwirkung gefunden worden ist.
%\dckeywordsde
%---------------------
% Widmung ist optional (bitte ggf. auskommentieren)
%\chapter*{Widmung}
%Dieses Manual ist allen DiDi-Mitarbeitern gewidmet.
%---------------------
\selectlanguage{english}
% Danksagung ist optional (bitte ggf. auskommentieren)
\chapter*{Acknowledgments}
%Vielen Dank an alle DiDi-Mitarbeiter! Ihr habt mir sehr geholfen!
First of all I would like to express my gratitude to Prof. Michael M\"uller-Preu\ss ker for having made it possible for me to make my Ph.D. in his theoretical group. In particular I received a great scientific supervision as well as the all possible support. Especially I am indebted to Dr. Wolfgang Bietenholz who became my direct supervisor during my Ph.D. His careful and kind supervision as well as the fascinating ideas filled me all these years and contributed a great deal to my world view as well as to my scientific education. I would like to thank Dr. Karl Jansen  for discussions and many interesting ideas which stimulated me as a scientist.  Also I acknowledge discussions on specific topics with S. Capitani, T. Chiarappa, N. Christian, M. Hasenbusch, K.-I. Nagai, S. Necco, M. Papinutto, L. Scorzato, A. Shindler, C. Urbach, U. Wenger, I. Wetzorke. Further I would like to acknowledge Prof. J. J. M. Verbaarschot, Prof. P. Damgaard and  Prof. G. Akemann for useful discussions concerning the issues addressed in the thesis. I would also like to acknowledge all my colleagues: Dr. F. Hofheinz, Dipl. Phys. A. Sternbeck and Dipl. Phys. D. Peschka for creating a nice working atmosphere and helping me to figure out a number of small things which occur in everyday life. And last but not least I would like to thank my mother and grandmother for a kind support.

%------------------------
\tableofcontents
%\pagebreak
%\listoffigures
%\pagebreak
%\listoftables
\cleardoublepage
%-------------------------------
\pagenumbering{arabic}

\pagestyle{fancy}
\renewcommand{\chaptermark}[1]{\markboth{\chaptername\ \thechapter\ #1}{}}
\renewcommand{\sectionmark}[1]{}

%\lhead[\fancyplain{}{\bfseries\thepage}]{\fancyplain{}{\bfseries\rightmark}}
%\rhead[\fancyplain{}{\bfseries\leftmark}]{\fancyplain{}{\bfseries\thepage}} 
%\chead{} \lfoot{} \cfoot{} \rfoot{}
%--------------------------------

%Hier kommt jetzt Ihr Text ;)
\chapter{Motivation}
This dissertation is dedicated to the Quantum Chromodynamics (QCD), a theory that explained already many effects and processes in the hadron world.

QCD is a part of the Standard Model of elementary particles with the gauge group $SU(3)\otimes SU(2)\otimes U(1)$. It is the theory that describes the strong interactions of quarks and gluons.
Quarks are the fermion fields which belong to the  fundamental representation of the gauge group $SU(3)$ and gluons being in its Lie algebra. In QCD the strong interactions are mediated by the exchange of these gauge bosons. Due to the non-Abelian nature of the gauge group $SU(3)$ a self interaction of the gauge bosons is also present. There is experimental evidence that the quarks come in six flavors: up (u), down (d), strange (s), charm (c), bottom (b) and top (t).

 Many QCD calculations were successful in the domain of high energies where --- due to the asymptotic freedom --- the coupling constant becomes a convenient small expansion parameter. This is the perturbative regime where many quantities of interest can be expressed in terms of the asymptotic expansion in powers of the strong coupling constant $\alpha_s$.

 In the domain of energies of order $1$~GeV the coupling constant rises to order $1$, ${\cal O}(1)$, and the perturbative treatment fails. Therefore a non-perturbative approach must be applied to address the problems of interest in this domain.
This scale of energies corresponds to color singlet bound states of quarks when they form hadrons. One distinguishes between mesons, consisting of a quark and antiquark, and baryons, consisting of three quarks. An elegant classification of the mesons and baryons is given by the quark model. The particles are identified with the fields belonging to the irreducible representation of their flavor group. To this end the classification of the mesons for the flavor group $SU(2)$ is $\bar{2} \otimes 2=3\oplus 1$, i.e. there is a triplet of meson fields and a singlet. For the $SU(3)$ group the pattern is $\bar{3}\otimes 3=8\oplus 1$, i.e. an octet and a singlet.
    In experiments on the deep inelastic scattering the building blocks of hadrons, the partons, were identified. The quarks are, however, observed in nature only in such bound states and no free quark has ever been found. This phenomenon, known as confinement, cannot be derived by the perturbative analysis. 

From experimental data of particles one can see three lightest hadronic states $\pi^0$, $\pi^+$, $\pi^-$ belonging to the triplet of the $SU(2)$ multiplet of u, d flavors and $\pi^0$, $\pi^+$, $\pi^-$, $\eta$, $K^+$, $K^-$, $K^0$, $\bar{K}^0$  belonging to the octet of the $SU(3)$ multiplet of u, d, s flavors with the masses below the mass of the higher lying $\rho$ meson. At present it is generally assumed to be a result of the spontaneous chiral symmetry breaking which occurs in the massless QCD Lagrangian. The non-vanishing quark masses break the chiral symmetry of the Lagrangian explicitly which gives rise to non-zero masses of the quasi-Goldstone bosons.  The particles in the multiplets are identified with these quasi-Goldstone bosons. Their masses are proportional to the square root of the quark masses. This phenomenon, known as the chiral symmetry breaking, is a truly non-perturbative effect.  The chiral symmetry that is broken at low energies is, however, expected to be restored if the energies are increased. This scenario corresponds to a phase transition. An order parameter indicating this phase transition is the chiral condensate $\langle\bar\psi\psi\rangle$, which is zero in the chirally symmetric phase and picks up a non-zero value once we pass to the chirally broken phase.  

 Lattice QCD is a theory that provides a non-perturbative approach from the first principles. One introduces a space-time lattice and formulates the theory by a lattice regularized path integral. Quarks and gluons become regularized fields. Quarks are represented by Grassmann spinor fields attached to the lattice sites and the gauge fields are represented by parallel transporters defined on the lattice links. The regularization introduces a momentum cut-off $\pi/a$, where $a$ is the lattice spacing. This represents a statistical model that is subject to the Monte Carlo simulations on a computer. The continuum limit is recovered if one sends the lattice spacing to $0$ and the number of lattice sites to infinity, keeping the physical volume fixed. 

When addressing the phenomenon of the chiral symmetry breaking and the physics of the light quarks, it is vital to have a discretization of fermions which preserves the chiral symmetry at finite lattice spacing. In this thesis we will describe results obtained with various formulations of such chiral fermions. At present all these formulations are tedious to simulate in QCD due to their computational overhead compared to the non-chiral discretizations such as the Wilson fermions. Therefore all simulations are carried out at quite small physical volumes and the values of the meson masses higher than one would expect in nature. The latter is reflected for instance in the pion mass.  Hence the simulated data have to be necessari\-ly extrapolated to the small physical quark masses. In addition  the quark loops are also neglected which leads to quite difficult accountable systematic uncertainties. 
 A way to extract physical information from the QCD simulations is to use theoretical predictions provided by a low energy effective theory. In the context of QCD this is the Chiral Perturbation Theory ($\chi$PT).
   
Owing to the mass gap between the quasi-Goldstone bosons and the first heavier mesons (the $\rho$--meson) one can integrate out all higher excitations of the mass spectrum of QCD leaving the theory defined solely in terms of the quasi-Goldstone bosons. The resulting theory is a low energy effective theory, which is called the $\chi$PT. 

An attractive feature for the numerical simulations is that $\chi$PT can be formulated in a finite volume and some analytical predictions can be computed which are the same as in the infinite volume. Moreover since the present lattice simulations are done without quark loops this can also be accounted for in the quenched $\chi$PT which provides an analytical guide-line for quenched lattice simulations.
In particular the analytical predictions in the framework of $\chi$PT have been worked out for the regime of the soft pion where the Compton wavelength cannot be accommodated in the finite volume. This regime, though representing an unphysical situation, provides a powerful tool for the lattice simulations with chiral fermions. The main advantage is that the finite size effects can be calculated in this setting using the $\chi$PT and the low energy constants appearing in the corresponding analytical expressions are those of the infinite volume. The control over the finite size effects, which one otherwise wishes to suppress, provides us with physical information. Since in this regime one does not need to use large lattices, it is viewed to be promising for obtaining the low energy constants parameterizing the $\chi$PT Lagrangian. This can be achieved by using the formalism of the chiral fermions. The use of the chiral fermions allows us to perform calculations at small meson masses and thereby approach their physical values, i.e. the so-called chiral limit of observables of interest.

In Chapter~2 we give an introduction and fix our notations in QCD, $\chi$PT and chiral Random Matrix Theory. Lattice QCD will also be briefly discussed. For completeness the traditional formulations of lattice fermions will be reviewed.  We discuss various formulations of the chiral lattice fermions realized through the Ginsparg-Wilson relation. In particular we consider the conventional formulation of the overlap fermions and an improved formulation based on the perfect action --- the hypercube fermions. 
The Chapter~3 is concerned with simulation aspects of the Ginsparg-Wilson fermions and also quenched gauge simulations are briefly discussed. The Chapter~4 deals with the numerical aspects of simulation of the hypercube fermions. In Chapter~5 we present our results for the dependence of the pion mass versus the bare quark mass. In  Chapter~6 the results for the distributions of the eigenvalues of the overlap Dirac operator are considered in conjunction with the chiral Random Matrix Theory. In Chapter 7 the results for the topological susceptibility, computed with the chiral fermions, are addressed. In Chapter~8 the two-point mesonic functions are addressed. We discuss the feasibility to extract the physical information from the axial and pseudo-scalar correlators. In Chapter~9 we present a study of a gauge action which may be useful for the future lattice simulations in the $\epsilon$--regime. In Chapter~10 we draw the conclusions and give an outlook.

\chapter{Theoretical background}
\section{QCD at first glance. Gluons and quarks}
In this Section we introduce the basic notation of QCD. Its degrees of freedom are carried by quarks and gluons. Gluons are described by the non-Abelian $SU(3)$
gauge group while quarks belong to its fundamental representation.
We will consider QCD in Euclidean space  that will be the relevant theory for the results discussed in the thesis. 
 The Euclidean formulation is especially useful for the lattice simulations since the theory becomes a statistical ensemble with the Boltzmann weight $\exp{(-S_{QCD})}$, where $S_{QCD}$ is the QCD action.
The use of the Euclidean formulation is justified since there exists an analytical continuation of the Euclidean Green functions to the Minkowski space~\cite{Osterwalder:1973dx,Osterwalder:1975tc}.

The  quarks are represented by spinor fields in the fundamental representation of the $SU(3)$ color group, and the gluon vector fields $A_{\mu}$ ($\mu=1, \dots ,4$) take their values in the Lie algebra of the $SU(3)$ color group. The quarks come in six flavors $\psi_f$, where $f=1,\dots ,6$. We decompose the matrix valued vector potential $A_{\mu}$ in fields $A^a_{\mu}$, where $a=1,\dots , N^2_c-1$ with $N_c$ being the number of colors, 
\begin{equation}
  A_\mu=-ig\sum_{a=1}^{N_c^2-1}A_\mu^a\frac{T_a}{2}\:.
\end{equation}
$g$ is the bare coupling constant and  the matrices $T_a$ are the generators of the gauge group $SU(3)$, which are traceless and Hermitian. They are  normalized by
$$
{\rm Tr}(T^{a}T^{b})=\frac{1}{2}\delta_{a b}\ .
$$ 
The generators satisfy the commutation relation
\begin{equation}
\left[ \frac{T^{a}}{2}, \frac{T^{b}}{2} \right]=if_{abc}\frac{T^{c}}{2}\ ,
\end{equation}
where $f_{abc}$ are structure constants of the $SU(3)$ Lie group which can be chosen to be real. The values of the structure constants for the special choice of generators as the Gell-Mann matrices are given in Ref.~\cite{okun}. 

We split the QCD action into two parts, $S_{\rm QCD}=S_{\rm YM}+ S_{\rm F}$, with the action of quarks given by 
\begin{equation}
\label{FermAction}
S_{\rm F}=\sum_{f=1}^{N_f} \int_V d^{4}x\,\bar{\psi_f} ({\cal D}+m_{f}) \psi_f \ .
\end{equation}
$N_f$ is the number of quark flavors, $S_{\rm YM}$ is the
Euclidean Yang-Mills action,
${\cal D}$ is the massless Dirac operator and $m_f$ is the quark mass corresponding to the flavor $f$.
 The integration is done over the Euclidean four-volume $V=L^3 \times T$. The boundary conditions are taken to be periodic up to a gauge transformation  for the gluons, and antiperiodic for the quarks. This defines the theory on a four dimensional torus.
 The QCD Dirac operator is given
by
\begin{equation}
  \label{Dirac}
  {\cal D}=\gamma_\mu(\partial_\mu+A_\mu)\:.
\end{equation}
 The $\gamma_\mu$ are the Euclidean Dirac matrices with
$\{\gamma_\mu,\gamma_\nu\} = 2\delta_{\mu\nu}$. We chose them such that the Dirac operator is anti-Hermitian, ${\cal D}^\dagger=-{\cal D}$. In particular we use the
chiral representation in which $\gamma_5\equiv\gamma_1\gamma_2\gamma_3
\gamma_4 ={{\rm diag\:}}(1,1,-1,-1)$ (see for instance Ref.~\cite{bookMM}).

The Euclidean Yang-Mills gauge action is given by
\begin{equation}
\label{YMaction}
S_{\rm YM}=\frac{1}{4}\int_V d^4x\,F^{a}_{\mu \nu} F^{a}_{\mu \nu}\ .
\end{equation}
In this formula, and throughout this thesis, the summation over the repeated indices is assumed if not stated otherwise.
The field strength tensor $F^{a}_{\mu \nu}$ carries one color $a$ index running from $1 \dots N^2_c-1$ and two Lorentz indices $\mu, \nu =1 \dots 4$,
\begin{equation}
\label{StrengthTensor}
F^{a}_{\mu \nu}=\partial_{\mu} A^{a}_{\nu} -\partial_{\nu} A^{a}_{\mu} +gf_{abc}A^{b}_{\mu}A^{c}_{\nu}\ .
\end{equation}

The theory is finally defined by its partition function in the form of the functional integral over the fermion fields $\psi_f$ and $\bar{\psi}_f$  and the gauge vector potential $A_{\mu}$ in the Euclidean space, 
\begin{equation}
  \label{Z1QCD}
  Z^{\rm QCD}(m_f,\theta)=\int \prod_{\mu }[DA_\mu] \prod_{f=1}^{N_f}[D\psi_f D\bar{\psi_f}]\,  e^{-S_{\rm QCD}[\bar{\psi}_f, \psi_f, A] +i\theta\nu[A]}\; .
\end{equation}
The parameter $\theta$ is a real vacuum angle and the topological charge $\nu$ is defined by

\begin{equation}
\nu[A]=-\frac{1}{32 \pi^2}\int_V \, d^4x\, \epsilon_{\alpha \beta \rho \sigma} {\rm Tr}(F_{\alpha \beta}F_{\rho \sigma})
\label{gauge_index}
\end{equation}
where $\epsilon_{\alpha \beta \rho \sigma}$ is the antisymmetric unit tensor of rank four. $\nu[A]$ is the winding number of a field configuration and takes integer values for finite action field configurations.
The field $\psi_f$ is a four component spinor field and $\bar{\psi_f}$ is a conjugate spinor field.
Integrating out the fermion degrees of freedom we arrive at
\begin{equation}
  \label{Z2QCD}
  Z^{\rm QCD}(m_f, \theta)=\int \prod_{\mu}DA_\mu\prod_{f=1}^{N_f}\det({\cal D}+m_f)\:
  e^{-S_{\rm YM}[A] +i\theta\nu[A]} \, .
\end{equation}
The integral over all field configurations $\int \prod_{\mu}[DA_{\mu}]$ includes a sum over all topological sectors $\nu$.

The partition function in the sector of topological charge $\nu$ is obtained by the Fourier transform
\begin{equation}
Z^{\rm QCD}_{\nu}(m_f)=\frac{1}{2\pi}\int_0^{2\pi}d\theta \, e^{-i\nu\theta} Z^{\rm QCD}(m_f,\theta)\ .
\label{ZQCDnu}
\end{equation}
\section{Chiral symmetry breaking and eigenvalues of the Dirac operator}
\subsection{Implications of the chiral symmetry for the structure of the Dirac operator}
In this Section we discuss the chiral symmetry in QCD and  its implication for the spectrum of the Dirac operator.

We start from the definitions of right-handed and left-handed spinors $\psi^{R/L}$ which are defined as projections of $\psi$ and $\bar{\psi}$
\begin{eqnarray}
\psi^{R/L}&=&P^{(+/-)}\psi \ , \quad \bar{\psi}^{R/L}=\bar{\psi}P^{(-/+)}\ , \nonumber\\
P^{(+/-)}&=&\frac{1}{2}(1 \pm \gamma_5)\ ,\nonumber\\
1&=&P^{(+)} +P^{(-)}, \quad P^{(+)}P^{(-)}=0 \ ,
\end{eqnarray}
where the spinor $\psi$ describes $N_f$ Dirac spinor species.

The massless Dirac operator satisfies the anticommutation relation
\begin{equation}
  \label{anticomm}
  \{\gamma_5,{\cal D}\}=0\ .
\end{equation}
Using the decomposition of the fermion fields $\psi$, $\bar{\psi}$
\begin{equation}
\psi=\psi^R +\psi^L, \quad \bar{\psi}=\bar{\psi}^{R} +\bar{\psi}^{L}
\end{equation}
we obtain for the fermion action
\begin{equation}
S_F=\int d^4x \Bigr(\bar\psi^R{\cal D}\psi^R +
  \bar\psi^L{\cal D}\psi^L + \bar\psi^RM\psi^L +
  \bar\psi^LM\psi^R \Bigr)\ ,
\label{chiral_decom}
\end{equation}
where $M={\rm diag}(m_1, \dots , m_f)$.

Relation~(\ref{anticomm}) can be seen also as an expression of the chiral symmetry of the fermion action for every fermion flavor $\psi_f$, 
\begin{equation}
\psi_f \rightarrow e^{i\alpha \gamma_5}\psi_f \ , \quad \bar{\psi}_f \rightarrow \bar{\psi}_f e^{{i\alpha \gamma_5}}\ ,
\end{equation}
where $\alpha$ is a real angle of the chiral rotation. The quark action~(\ref{chiral_decom}) remains invariant under a chiral rotation if $M=0$. 

  One can write down an eigenvalue equation for ${\cal D}$ as
\begin{equation}
  \label{eveq}
  {\cal D}[A]\psi_n=i\lambda_n\psi_n\:, \quad \lambda_n\in \mathbb{R}\ ,
\end{equation}
where the eigenvalues and eigenfunctions depend on the gauge field.  Using Eq.~(\ref{anticomm}) one can show that the
nonzero eigenvalues of ${\cal D}$ occur in pairs $\pm i\lambda_n$ with
eigenfunctions $\psi_n$ and $\gamma_5\psi_n$.
Indeed, if $i\lambda_n$ is an eigenvalue of $\cal D$ and $\psi_n$ is the corresponding eigenfunction, then Eq.~(\ref{anticomm}) implies
\begin{equation}
{\cal D} \gamma_5\psi_n=-\gamma_5 {\cal D}\psi_n=-i\lambda_n \gamma_5 \psi_n \ ,
\end{equation}
hence $\gamma_5 \psi_n$ is also an eigenfunction of $\cal D$ with the eigenvalue $-i\lambda_n$.
  There can also be
eigenvalues equal to zero, $\lambda_n=0$.  Because of Eq.~(\ref{anticomm}) the corresponding
eigenfunctions can be arranged to be simultaneous eigenfunctions of
$\cal D$ and $\gamma_5$ with eigenvalues $\pm 1$ of $\gamma_5$, i.e.\ these states have a definite
chirality.  Denoting the number of zero eigenvalues per flavor $f$ with positive and
negative chirality by $N_+$ and $N_-$, respectively, the Atiyah-Singer
index theorem~\cite{Atiyah:1968mp} states that 
\begin{equation}
\nu= N_f(N_+ -N_-)
\end{equation}
 is a topological invariant that does not change under continuous deformations of the gauge
field. Note that the Atiyah-Singer index theorem identifies the definition of the index given by gauge fields~(\ref{gauge_index}) and the definition given by the index of the Dirac operator, i.e. the fermionic index. 

In a chiral basis with $\gamma_5\psi^{R/L}_n=\pm\psi^{R/L}_n$, one can use
Eq.~(\ref{anticomm}) to show that $\langle\bar\psi_m^R|{\cal
  D}|\psi_n^L\rangle=0 =\langle\bar\psi_m^L|{\cal D}|\psi_n^R\rangle$
for all $m$ and $n$.  From this property and the fact that ${\cal D}$ is anti-Hermitian, it follows that the projection of the Dirac operator onto the subspace spanned by the zero modes has the matrix structure
\begin{equation}
  \label{block}
  {\cal D}^{(0)}=\left( 
\begin{array}{cc} 
0 & iW \cr 
iW^\dagger &0 
\end{array}
 \right)\ .
\end{equation}

This off-diagonal block structure is characteristic for systems with
chiral symmetry. The matrix $W$ has dimension $N_fN_+\times N_fN_-$, and
the matrix ${\cal D}$ in Eq.~(\ref{block}) has $|\nu|$ eigenvalues
equal to zero.

Using formula~(\ref{block}) we can write down the contribution of the zero modes to the fermion action as
\begin{equation}
  S^{(0)}_F=\int d^4x\Bigr(\bar\psi^R{iW^\dagger}\psi^R +
  \bar\psi^L{iW}\psi^L + \bar\psi^RM\psi^L +
  \bar\psi^LM\psi^R \Bigr)\:.
\label{chiral_decomp_action}
\end{equation}

In the chiral limit where all $m_f=0$, the fermion actions~(\ref{chiral_decom}) as well as~(\ref{chiral_decomp_action}) are
invariant under the global transformations
\begin{eqnarray}
\begin{array}{l@{\hspace*{10mm}}l}
    \psi^L\to L\psi^L\ , & \bar\psi^{L}\to\bar\psi^{L}L^{-1}\ ,\\[2mm]
    \bar\psi^R\to\bar\psi^{R} R^{-1}\ , & \psi^R\to R\psi^R\ .
  \end{array}  
\end{eqnarray}
$ L \in U_L(N_f)$ and $R \in U_R(N_f)$, where $U_L(N_f)$ and $U_R(N_f)$ are independent unitary groups.
If the number of right-handed states $N_R=N_fN_+$ is equal to the number of
left-handed states $N_L=N_fN_-$, the symmetry is thus
${\rm U}_R(N_f)\otimes{\rm U}_L(N_f)$.  However, if $N_R\ne N_L$, i.e.\ if
$\nu\ne0$, the axial symmetry group $\rm U_A(1)$, which is a ${\rm U}(1)$ subgroup of ${\rm U}_R(N_f)\otimes {\rm U}_L(N_f) $, is broken 
by instantons.  Invariance under a second subgroup ${\rm U}(1)$ of ${\rm U}_R(N_f)\otimes {\rm U}_L(N_f) $, the group $\rm U_V(1)$,
corresponds to the conservation of the baryon number.  The full flavor
symmetry group in the chiral limit is thus given by $G = {\rm SU}_R(N_f)\otimes {\rm SU}_L(N_f)$.  

Since $G$ is a symmetry group for the massless fermion action one can write down the
Noether currents
\begin{eqnarray}
J^L_{a\mu}&=&\bar{\psi}\gamma_{\mu}\frac{(1-\gamma_5)}{2}T^a\psi \ ,\nonumber\\
J^R_{a\mu}&=&\bar{\psi}\gamma_{\mu}\frac{(1+\gamma_5)}{2}T^a\psi\ .
\end{eqnarray}
$T^a$ is a generator of the $SU(N_f)$ group, where $a=1,\dots , N^2_f-1$.
One obtains conserved Noether charges by the usual procedure,
\begin{equation}
Q_a^{R/L}=\int d^3x\, J^{R/L}_{a 4}(x)\ .
\end{equation}
We construct generators of the axial ${\rm SU}_A(N_f)$ and vector subgroup ${\rm SU}_V(N_f)$ by linear combination of the Noether currents $J^L$ and $J^R$,
\begin{eqnarray}
J^A_{a\mu}&=&J^R_{a\mu} -J^L_{a\mu}\ ,\nonumber \\
 J^V_{a\mu}&=&J^R_{a\mu} +J^L_{a\mu}\ .
\end{eqnarray}
The other two currents which are related to the ${\rm U}_V(1)$ and ${\rm U}_A(1)$ symmetries are given by
\begin{eqnarray}
J^A_{0\mu}&=&\bar{\psi} \gamma_5 \gamma_{\mu} \psi \ , \nonumber \\
 J^V_{0\mu}&=&\bar{\psi} \gamma_{\mu} \psi\ .
\end{eqnarray}
As we mentioned before, for $\nu \ne 0$ the axial current $J^A_{0\mu}$ is not conserved. Its divergence is given by
\begin{equation}
\partial_{\mu} J^A_{0\mu}(x) =-\frac{N_f}{32\pi^2}\epsilon_{\mu \nu \rho \sigma} {\rm Tr}[F_{\mu \nu}(x) F_{\rho \sigma}(x)]\ .
\end{equation}

One can also see that the axial ${\rm SU}(N_f)$ subgroup with
$L=R^{-1}$ is broken expli\-citly by the mass term. On the other hand the ${\rm SU}(N_f)$
vector subgroup (with $L=R$) is not broken for degenerate quark masses
($m_f=m$ for all $f$), but breaks explicitly for different quark
masses ($m_f\ne m_{f'}$ for some $f\ne f'$).

If the quark masses were zero, the action~(\ref{chiral_decomp_action}) would have an exact axial flavor ${\rm SU}(N_f)\otimes {\rm SU}(N_f)$ symmetry.
However, there is strong evidence that the axial ${\rm SU}(N_f)$ subgroup with $L=R^{-1}$
 is broken  spontaneously with the following spontaneous symmetry breaking pattern of the chiral group,  
\begin{equation}
{\rm SU}_R(N_f)\otimes {\rm SU}_L(N_f)\rightarrow {\rm SU}_V(N_f) \ .
\end{equation}
This implies that the ground state of the quantum system being invariant with respect to the ${\rm SU}_V(N_f)$ is, however, not invariant with respect to the ${\rm SU}_A(N_f)$ symmetry.
 According to the Goldstone theorem spontaneous breaking of a Lie group $G$ down to a Lie group $H$ gives rise to Goldstone boson fields~\cite{Nambu:1960xd,Goldstone:1961eq}.  The Goldstone manifold is a coset space $G/H={\rm SU}(N_f)$, and so there are $N_f^2-1$ Goldstone bosons. 

The fact that the pattern of the spontaneous symmetry breaking has the form ${\rm SU}_R(N_f)\otimes {\rm SU}_L(N_f)$ to ${\rm SU}_V(N_f)$  is supported firstly by the absence of the parity doublets in the hadron spectrum.
Indeed in the mass spectrum one can observe  three light pseudo-scalar particles, the pions (which are the triplet of an approximate ${\rm SU}(2)$ meson multiplet) as well as somewhat heavier pseudo-scalars ---the four kaons $K^{+}$, $K^{0}$, $\bar{K^{0}}$, $K^{-}$ and the $\eta$ -meson (belonging to the approximate ${\rm SU}(3)$ meson multiplet). Along with the eight approximate Goldstone bosons one observes a heavier $\eta^{\prime}$ meson.  Its mass is too large to be considered as a result of the spontaneous symmetry breaking of the axial ${\rm U}_A(1)$ subgroup. According to 't Hooft's idea its mass receives contributions from instantons due to a relation of the derivative of the anomaly axial current to the topological charge~\cite{'tHooft:1976up,'tHooft:1976fv}.

Secondly the value of the chiral condensate $\langle\bar\psi\psi\rangle=\langle\bar\psi^R\psi^L\rangle+\langle\bar\psi^L\psi^R\rangle$ would be zero
 if the ground state was
axial-flavor symmetric.  However, lattice QCD simulations indicate that $\langle\bar\psi\psi\rangle$ is non-zero.  The quantity
$\langle\bar\psi\psi\rangle$ is only invariant if $L=R$, i.e.\ the
vacuum state is symmetric under the flavor group $H={\rm SU}_V(N_f)$.
Thus, the vector symmetries are unbroken, whereas the axial symmetries
are maximally broken.
If we now add to this scenario of the spontaneous chiral symmetry breaking the explicit symmetry breaking by the mass terms of light quarks, then we arrive at a consistent picture for the observable approximate $SU(2)$ triplet and $SU(3)$ octet. 
Comparing masses of the pions in the triplet and pions, kaons and $\eta$ meson in the octet, we see that the explicit symmetry breaking is larger for the octet. This implies that the mass of the strange quark in the QCD Lagrangian is large compared to masses of the u and d quarks, which one often assumes to be degenerate.
%--------------------------------------------------------------------------------------
\subsection{Spectral density of the Dirac operator}
In terms of the eigenvalues of the Dirac operator, the QCD partition function~(\ref{Z2QCD}) can be rewritten as
\begin{equation}
Z^{\rm QCD}(m_f,\theta)=\sum_{\nu=-\infty}^{+\infty} e^{i\nu\theta}\prod_{f=1}^{N_f}m_f^{|\nu|}\int_{\nu}DA_{\mu}\prod_{k\ge 1}(\lambda_k^2+m_f^2)e^{-S_{\rm YM}[A]}\ ,
\label{Z4QCD}
\end{equation}
where $\int_{\nu}DA_{\mu}$ denotes the path integral over field configurations with topological charge $\nu$. The eigenvalues of the Dirac operator, $\lambda_k$, are introduced in Eq.~(\ref{eveq}).

The spectral density of the Dirac operator is given by
\begin{equation}
  \label{rho}
  \rho(\lambda)=\Bigl\langle\sum_n\delta(\lambda-\lambda_n) \Bigr\rangle\ ,
\end{equation}
where the expectation value is over gauge fields weighted by the full QCD
action.  The spectral density is important because of its relation to
the order parameter for spontaneous chiral symmetry breaking, the
chiral condensate $\Sigma=-\langle\bar\psi\psi\rangle$. 

 It was shown by Banks
and Casher~\cite{Banks:1980yr} that
\begin{equation}
  \label{BC}
  \Sigma =
\lim_{\varepsilon\to0} \lim_{m\to0} \lim_{V\to\infty} \pi
\rho(\varepsilon)/V\ .
\end{equation}
 It is important that the limits are taken in the order
indicated.  (In the normalization of Eq.~(\ref{rho}) the spectral
density is proportional to the volume, so the explicit factor of
$1/V$ in~(\ref{BC}) is compensated to yield a finite result.)
This relation can be readily derived. The chiral condensate is given by
\begin{equation}
\langle \bar{\psi}\psi\rangle=-\lim_{m\to 0}\lim_{V\to \infty}\frac{1}{VN_f}\partial_m {\rm log}Z^{\rm QCD}(m,\theta)\ .
\end{equation}
If we now apply this definition to the partition function~(\ref{Z4QCD}) then we arrive at
\begin{equation}
\langle \bar{\psi} \psi\rangle = -\lim_{m\to 0}\lim_{V\to \infty} \Bigl \langle \frac{1}{V} \sum_{n\ge 1} \frac{2m}{\lambda_n^2 +m^2}\Bigr\rangle\ ,
\label{cond_spectral}
\end{equation}
where the contribution of the zero modes, $|\nu|/(mV)$, was dropped assuming that it becomes asymptotically small in the $V\to \infty$ limit. Now taking the chiral limit we obtain the Banks-Casher relation~(\ref{BC}).

%-------------------------------------------------------------------------------------
\section{\texorpdfstring{$\chi$}{chiral }PT as a low energy effective theory of QCD}

Due to spontaneous chiral symmetry breaking, at low energies the QCD dynamics
is dominated by Goldstone bosons~\cite{Nambu:1960xd,Goldstone:1961eq} --- the pions for $N_f = 2$. 
It is possible to use a low-energy effective description that only involves 
the quasi-Goldstone boson fields~\cite{Coleman:1969sm,Callan:1969sn,Weinberg:1979kz}. Chiral perturbation 
theory~\cite{Gasser:1984yg,Gasser:1985gg} provides a systematic low-energy expansion 
that predicts the pion dynamics based on symmetry principles and a number of 
low-energy parameters (like the pion decay constant $F_\pi$ and the chiral condensate $\Sigma$) whose values can be determined 
either from experiments or from lattice QCD calculations.  
%-----------------------------------------------------------------------------------------------------
\subsection{The chiral Lagrangian and its low energy constants}
The Lagrangian of the low-energy effective theory is determined by the global symmetries of the underlying theory. Considering the low energy effective theories one has to think of an energy scale hierarchy and the corresponding theory which is relevant for a given energy interval. As we reduce the energy scale, some degrees of freedom, which were important for higher energies, may become irrelevant for the lower energies.
 
The low energy effective Lagrangian is constructed as a linear combination of all terms which are local, Lorentz invariant and compatible with the flavor symmetry. The constants in such an expansion parameterize our lack of knowledge about the underlying theory. 

If we write down a low energy effective theory for QCD then the corresponding flavor symmetry includes the chiral symmetry. It is broken down to ${\rm SU}(N_f)$ and therefore the Goldstone bosons belong to the coset space ${\rm SU}_R(N_f)\otimes {\rm SU}_L(N_f)/{\rm SU}_V(N_f)$, which is again ${\rm SU}(N_f)$. If we are interested only in the lightest excitations then the Lagrangian will contain only Goldstone fields $U(x)$. 
Under global chiral rotations they transform as
\begin{equation}
U(x)' = L U(x) R^{\dagger}\ ,
\label{for:chiral_trans_chpt}
\end{equation}
where $L$ and $R$ are matrices of the chiral transformation belonging to $SU_L(N_f)$ and $SU_R(N_f)$, respectively.
To fully determine the theory we need to write down its partition function and specify the measure.
The partition function of the chiral perturbation theory takes the following general form

\begin{equation}
Z_{\chi PT}= \int [DU(x)]\, \exp \left ( -\int d^4 x \, {\cal L}[U] \right )\ ,
\end{equation}
where $[DU(x)]$ represents the Haar measure and the integration is done over the flavor group $SU(N_f)$ in each space point $x$. 
The Haar measure is a left- and right-invariant measure, i.e.
\begin{equation}
\int_{SU(N_f)} dU \ f(\Omega U) = \int_{SU(N_f)} dU \ f(U \Omega) = 
\int_{SU(N_f)} dU \ f(U)\ ,
\end{equation}
for any function $f(U)$ and for any matrix $\Omega \in SU(N_f)$. It is convenient
to normalize the measure such that
\begin{equation}
\int_{SU(N_f)} dU = 1\ .
\end{equation}

The interaction among the Goldstone bosons is weak at low energies and decreases as the energy is reduced.
For the effective Lagrangian it implies that it can be represented  as an expansion in derivatives. On the other hand the quark masses 
can be viewed as parameters which characterize how strongly the chiral symmetry is broken.

To a good absolute accuracy the up and down quarks can be regarded to have nearly the same mass. So for $N_f=2$ we can choose the mass matrix ${\cal M}= m 1\! \!1$.
One can write a systematic low-energy expansion in powers of the pion momentum $p$ and mass $m_{\pi^0}^2 = 2 m \Sigma/F_\pi^2$ over the cut-off of the effective theory, $\Lambda_{QCD} \simeq 4\pi F_{\pi}$. In the standard chiral expansion both quantities, $\frac{p}{\Lambda_{QCD}}$ and $\frac{m_{\pi^0}}{\Lambda_{QCD}}$, are taken to be of the same order.
Because of these counting rules the chiral expansion cannot be considered as a Taylor expansion. 
 Here we give the result for the leading term of the pion effective Lagrangian in the $\theta$ vacuum,
\begin{equation}
{\cal L}^{(2)} =  \frac{F_{\pi}^2}{4} \mbox{Tr}(\partial_{\mu} U^{\dagger} \partial_{\mu} U)
- \frac{\Sigma}{2} 
\mbox{Tr}( e^{i\frac{\theta}{N_f}}{\cal M}U^\dagger + U e^{-i\frac{\theta}{N_f}}{\cal M}^{\dagger})\ .
\label{Chpt_lagr}
\end{equation}
The first term on the right-hand side is chirally invariant. Its prefactor is given by the constant $F_{\pi}$ related to the pion decay constant which determines the strength of the interaction between the Goldstone bosons. The second term is the chiral symmetry breaking mass term which contains the quark mass matrix. Under chiral 
transformations~(\ref{for:chiral_trans_chpt}) this term transforms as
\begin{equation}
\mbox{Tr}(e^{i\frac{\theta}{N_f}}{\cal M} {U'}^\dagger + U' e^{-i\frac{\theta}{N_f}}{\cal M}^\dagger) =
\mbox{Tr}(e^{i\frac{\theta}{N_f}}{\cal M} R U^\dagger L^{\dagger} + L U R{^\dagger} e^{-i\frac{\theta}{N_f}} {\cal M}^{\dagger})\ .
\end{equation}
Since we assume the two quarks to have identical masses, the Lagrangian is 
invariant under $SU(2)$ flavor rotations for which $R=L$. For a general 
diagonal mass matrix with $N_f$ flavors the flavor symmetry is reduced to 
$\prod_{f=1}^{N_f} U(1)_f$. 

The constants $F_\pi$ and $\Sigma$ determine the 
low-energy dynamics at leading order and enter the effective theory as free parameters.

It can be shown that in the chiral limit
\begin{equation}
\langle0|J^A_{0\mu}|\pi^+(p)\rangle=i\sqrt{2}F_{\pi}p_{\mu}\ .
\end{equation}
The same formula at physical values of quark masses is in fact the definition of the pion decay constant $F^{\rm decay}_{\pi}$.  So the physical value of  $F^{\rm decay}_{\pi}$ being $F_{\pi}$ in the leading order is modified by ${\cal O}(m_f^2) \sim {\cal O}(p^4)$ corrections in the next-to-leading order,
\begin{equation}
F^{\rm decay}_{\pi}=F_{\pi} + {\cal O}(m_f^2, p^4)\ .
\end{equation}

The pion decay constant is experimentally known from the process $\pi^{\pm} \rightarrow \mu^{\pm}+\nu$ and it is approximately $93$ MeV. If we formulate the chiral Lagrangian for the $SU(3)$ flavor group --- i.e. including the s-quark --- then in addition to the pion decay constant we will have also the kaon decay constant. In principal one could relate it to the prefactor of the kinetic term of the chiral Lagrangian. The kaon decay constant is about $114$ MeV. The difference between the pion decay constant and the kaon decay constant is a typical ${\cal O}(p^4)$ effect, which is beyond the lowest order. However, the effects of order ${\cal O}(p^4)$ are expected to be larger in the case of the kaon decay constant. Therefore the most natural determination of the prefactor at ${\cal O}(p^2)$ is still provided by the pion decay constant. 

Contrary to the decay constants, the quark condensate is not directly related to any physical observable. It is the product $m_q B$, where $B=\Sigma/F^2_{\pi}$, that can be constrained by means of experimental data.
To the leading order we have
\begin{eqnarray}
m^2_{\pi^0}&=&(m_u+m_d)B ,\nonumber\\
m^2_{K^0}&=&(m_d+m_s)B\ .
\label{pionmasses}
\end{eqnarray}
The analogous equation for $m^2_{\eta_8}$, where $\eta_8$ denotes the octet component of the $\eta$ meson, contains no new parameters and gives rise to a consistency relation,
\begin{equation}
3m^2_{\eta_8}=4m^2_{K^0} -m^2_{\pi^0}\ .
\end{equation}
This is the famous Gell-Mann-Okubo mass formula, well satisfied by experimental data under the assumption $m_{\eta}=m_{\eta_8}$. The validity of the  Gell-Mann-Okubo formula provides a significant a posteriori check that the ${\cal O}(m^2_q)$ corrections to Eq.~(\ref{pionmasses}) are small.

Up to these two low-energy constants the Goldstone boson dynamics at the leading order of the chiral perturbation theory is completely determined by chiral symmetry. At higher energies additional terms arise in the effective theory. 
Again, they are restricted by chiral symmetry and they contain new low energy
constants --- the Gasser-Leutwyler coefficients. 
%---------------------------------------------------------------------------------------------------
\subsection{The \texorpdfstring{$p$}{p}--  and \texorpdfstring{$\epsilon$}{epsilon}--expansion of \texorpdfstring{$\chi$}{chiral }PT}
Let the system now to be placed in a four dimensional box of length $L$.
We assume from now on that the volume is large with respect to the QCD scale set by the $1/\Lambda_{QCD}$. More precisely, the allowed momenta should not take values higher than the scale of the chiral symmetry breaking in the infinite physical volume given by $\Lambda_{QCD}=4\pi F_{\pi}$,
\begin{equation}
p\sim \frac{2\pi}{L}\ll 4\pi F_{\pi}\ .
\label{upper_relation}
\end{equation}
As a general property, no chiral symmetry breaking occurs in a finite volume. However, this condition separates the hard momenta from the soft ones and thereby provides the corresponding framework for calculations in the finite volume.

The inverse product $1/(F_{\pi}L)$ is not the only relevant parameter. Also the relative size of $m_{\pi}$ and $1/L$ is important, as it was discussed in the previous Section. Indeed the $\chi$PT considers an expansion in two parameters: the mass of the quasi-Goldstone boson $m_{\pi}$ and its momentum $p$ over the scale $4\pi F_{\pi}$. In the finite volume the corresponding momenta are quantized in units of $2\pi/L$. 
 The Euclidean partition function for the $\chi$PT,
\begin{equation}
Z(V,{\cal M})=\int [dU] \exp\left( -\int_{V} d^4x\, {\cal L}[U] \right)\ ,
\end{equation}
involves three energy scales: $\Lambda_{QCD}$, $\cal M$ and $1/L$ or $p=2\pi n/L$.

In the chiral expansion, $U=\exp(i\sqrt{2}\eta (x)/F_{\pi})$ is expanded around the classical solution 
\begin{equation}
U(x)=1+ i\sqrt{2}\eta (x)/F_{\pi}+ ...\ .
\label{vacuumexpansion}
\end{equation}
The field $\eta(x)$ belongs to the Lie algebra of $SU(3)$ and it describes fields of the pseudo-scalar mesons.

Inserting Eq.~(\ref{vacuumexpansion}) in the expression for the Lagrangian~(\ref{Chpt_lagr}) and integrating over the volume we arrive at the action
\begin{equation}
\int_V d^4x\, {\cal L}^{(2)}= -\frac{1}{2} N_{f} {F_{\pi}}^{2} m_{\pi}^{2}V + \frac{1}{2}\int_V d^4x\, \mbox{Tr} \, ((\partial_{\mu} \eta)^{2} +m_{\pi}^{2} \eta^{2})\ .
\label{Chpt_lag2}
\end{equation}

Since $\eta(x)$ is defined on the torus $V=L^4$ we may expand it in terms of periodic plane waves
\begin{eqnarray}
\eta(x)&=&\sum_{n_1= -\infty}^{\infty}\dots \sum_{n_4 =-\infty}^{\infty} q_{n} \, u_{n}(x) + \,h.c.\ , \quad n=(n_1,\dots , n_4)\ ,  \nonumber \\
 u_{n}(x)&=&\exp(2\pi i \, nx/L)\ ,
\end{eqnarray}
where h.c. represents the Hermitian conjugate of the preceding term.
Inserting this expression in Eq.~(\ref{Chpt_lag2}) we see that the zero mode degrees of freedom $q_{0}$ of course do not appear in the kinetic term. This fact manifests itself also in  the expression for the pion propagator which develops a pole,

\begin{equation}
G(x-y)=\frac{u_{0}(x) \, u_{0}(y)^{*} }{m_{\pi}^{2} \,V} + \dots
\label{Chpt_prop_exp}
\end{equation}
 
So now one can distinguish several cases. If the linear size $L$ is much larger than the Compton wavelength of the pions, $1/m_{\pi}$, then the system hardly feels the finite volume and we have the usual chiral expansion. As the size of the box or the pion mass becomes smaller, finite size effects become significant, but provided that $m_{\pi} L \geq 1$, ordinary perturbation theory is still applicable.  This is the so-called
{\it $p$-expansion}~\cite{Gas87} with the counting rules
$$
\frac{|q_n|}{F_{\pi}}~\sim~\frac{m_\pi}{\Lambda_{QCD}} ~\sim~ \frac{p}{\Lambda_{QCD}}
~\sim~ \frac{1}{L F_{\pi}}\ .
$$

If the chiral limit is approached further in such a way that the Compton
wavelength of the pion is larger than the box
size ($L< 1/m_\pi$), 
the conventional $p$-expansion eventually breaks down due to
the propagation of pions with zero momenta~\cite{Gas87}. Indeed, according to
Eq.~(\ref{Chpt_prop_exp}) the pion propagator contains a pole. Therefore in order to arrive at a representation which would be valid in this regime of small quark masses (or lengths $L$) one needs to reorder the chiral perturbation series by summing graphs which involve an arbitrary number of zero mode propagators.

An appropriate expansion for this
regime is the so-called {\it $\epsilon$-expansion}, in which~\cite{Gas87,Neuberger:1987zz,Neuberger:1987fd,Hasenfratz:1990pk,Hansen1990,Hansen:1990yg,Bietenholz:1992ix}
$$
\frac{m_\pi}{\Lambda_{QCD}} ~\sim~ \frac{p^2}{\Lambda_{QCD}^2} ~\sim~
\frac{1}{L^2 F_{\pi}^2} ~\sim~ \epsilon^2 ~.
$$
Note that $\Sigma$ and $F_{\pi}$, as well as other low energy constants, appearing in this counting scheme are those of the infinite volume.
Using Eq.~(\ref{upper_relation}) the above relation implies for the $\epsilon$--regime
$$
m_{\pi}< \frac{1}{L}\ll 4\pi F_{\pi} \ .\label{upper_and_lower_relation}
$$
The difficulty in this regime comes from the fact that
the zero modes have to be accounted for exactly, while the chiral perturbation theory applies to the non-zero mode integration, since $q_p^2 \sim O(\epsilon^2)$ for $|p|\ne 0$.
One treats the zero modes of the pion as a collective coordinate by factorizing $U(x)$ into a  constant collective field $U_0$ and  the pion fluctuations $\xi (x)$,
\begin{equation}
U(x) ~=~ U_0 \exp i \left (\frac{\sqrt{2}\xi(x)}{F_{\pi}} \right )\ .
\label{eq:fac}
\end{equation}
The field $\xi(x)$ can be represented in terms of the plane waves with momenta $|p|$ larger than zero. $U_0$ is a $SU(N_f)$ matrix independent of $x$.

To leading order, the partition function reads 

\begin{eqnarray}
{\cal Z}(\theta,m_q) = \int_{SU(N_f)} d\xi dU_0 \exp   {\frac{1}{2}}
\int_{V} d^4 x \, {\rm Tr} \left(\partial_\mu \xi(x)
\partial_\mu \xi(x) \right) \times  \\
 \exp \frac{\Sigma V}{2} {\rm Tr } \left( {\cal M}e^{i\theta/N_f} U_0 + h.c.\right) \ ,\nonumber\\
{\cal M}={\rm diag}(m_1,\dots ,m_{N_f})\ .\nonumber
\end{eqnarray}

It is also interesting to consider expectation values in sectors of fixed
topology~\cite{Leut92}.
Fourier-transforming in $\theta$ and dropping the non-zero mode contributions, we obtain
\begin{eqnarray}
\label{zchpt_eps}
{\cal Z}_\nu(m_1,\dots, m_{N_f}) \!&=&\! \int_{SU(N_f)} 
\hspace{-.5cm} dU_0 \left({\rm det}
U_0\right)^\nu \exp\! 
\left({\frac{\Sigma V}{2}} {\rm Tr } \left( {\cal M}U_0 +
U_0^{\dagger}{\cal M}^{\dagger}\right )\! \right)\nonumber
\end{eqnarray}
or for the degenerate quark masses with ${\cal M}={\rm diag}(m_q,\dots ,m_q)$
\begin{equation}
{\cal Z}_\nu(m_q) \!=\! \int_{SU(N_f)} 
\hspace{-.5cm} dU_0 \left({\rm det}
U_0\right)^\nu \exp\! 
\left({\frac{m_q \Sigma V}{2}} {\rm Tr } \left( U_0 +
U_0^{\dagger}\right )\! \right)\ .
\end{equation}
If $F_{\pi} L \gg 1$, the $p$- and $\epsilon$-expansions should match in the range
of quark masses such that $m_\pi L \sim 1$. In this regime $m_q \Sigma V \sim
(F_{\pi} L)^2 (m_\pi L)^2 \gg 1$, so the results in the $\epsilon$-expansion
should reproduce those of the $p$-expansion in the limit of large $m_q \Sigma V$.
%-----------------------------------------------------------------------------
\subsection{Quenched \texorpdfstring{$\chi$}{chiral }PT: first order expressions for the axial\\-vector correlation function}

We start by writing down the expression for the axial-vector current,

\begin{equation} 
\label{axialcurrent}
{\cal A}^{a}_{\mu}(t,\vec{x}) = \bar \psi (t,\vec x ) \gamma_{5}
\gamma_{\mu} T^{a} \psi (t, \vec x) \ .
\end{equation}

The two-point correlation function is as usually given by the expectation value

\begin{equation}
\langle{\cal A}^{a}_{\mu}(x) {\cal A}^{b}_{\nu}(y)\rangle=\frac{\int[ DA_{\mu}] [D\psi] [D\bar{\psi}]{\cal A}^{a}_{\mu}(x) {\cal A}^{b}_{\nu}(y) e^{-S_{\rm QCD} +i\theta \nu[A]} }{Z^{\rm QCD}(m_q,\theta)}
\label{axialcurrentcorr}
\end{equation}
or in terms of the propagator $({\cal D}+m_q)^{-1}(x,y)$,
\begin{equation}
\begin{array}{cc}
\langle{\cal A}^{a}_{\mu}(x) {\cal A}^{b}_{\nu}(y)\rangle=& \\
-{\rm Tr}(T^aT^b){\langle {\rm Tr} \left [\gamma_5 \gamma_{\mu} ({\cal D}+m_q)^{-1}(x,y) \gamma_5 \gamma_{\nu} ({\cal D}+m_q)^{-1}(y,x) \right ]}\rangle \ .&\label{axialcurrentcorr_spectral}
\end{array}
\end{equation}
This expression can be obtained using the partition function for QCD in the presence of auxiliary external fields $a_{\mu}(x)$, the so-called generating functional. For the sake of the following discussion we will also add auxiliary fields for the pseudo-scalar density $P(x)=\bar{\psi}i\gamma_5\psi$,
\begin{equation}
Z(a_{\mu},P)=\int DA_{\mu} D\psi D \bar{\psi} \exp\left [ -S_{\rm QCD} + \int_{V} dx\, \bar{\psi}(\gamma_{\mu} \gamma_{5} a_{\mu}+ i\gamma_5P) \psi \right ]\ .
\label{partfunccurrents}
\end{equation} 

If we compute the second functional derivative of the partition function $Z(a_{\mu},P)$ with respect to the axial current $a_{\mu}$ and put $a_{\mu}$ and $P$ to zero afterwards, we recover the expression~(\ref{axialcurrentcorr}) for the axial current correlation function.
This new partition function~(\ref{partfunccurrents}) is now subject to its low energy approximation, i.e. one integrates out the quark degrees of freedom leaving only the mesons.
The leading order expression of the chiral Lagrangian of the low energy effective theory in the presence of $a_{\mu}$ and $P$ takes the form

\begin{eqnarray}
{\cal L}^{(2)}[U, a_{\mu}, P] &=&  \frac{F_{\pi}^2}{4} \mbox{Tr}\left (\nabla_{\mu} U^{\dagger} \nabla_{\mu} U\right ) -\\
&-& \frac{\Sigma}{2} 
\mbox{Tr}\left [ e^{i\frac{\theta}{N_f}}({\cal M}+iP(x))U^\dagger + U e^{-i\frac{\theta}{N_f}}({\cal M}^{\dagger}-iP^{\dagger}(x))\right ]\ , \nonumber
\label{Chpt_lagr_curr}
\end{eqnarray}
 
where 
\begin{equation}
\nabla_{\mu} U=\partial_{\mu}U -ia_{\mu}U - iUa_{\mu}\nonumber \ , \quad {\cal M}={\rm diag}(m_q,\dots ,m_q)\ .
\end{equation}

We see that the partial derivatives in the expression~(\ref{Chpt_lagr}) were promoted to some kind of covariant derivatives. Note that $P(x)$ is assumed to be of the same order as $m_q$ in the chiral expansion, whereas  $a_{\mu}(x)$ is counted as the pion momenta $p$ .

Now from this expression one can construct the partition function. Taking then two functional derivatives with respect to the axial current $a_{\mu}(x)$ or two functional derivatives with respect to the pseudo-scalar density $P(x)$ we obtain the expression for the axial correlation function and pseudo-scalar correlation function in $\chi$PT.
As far as full QCD is concerned, all two-point functions had been calculated before for the $\epsilon$-regime in Ref.~\cite{Hansen1990}. 

The QCD partition function~(\ref{Z2QCD}) includes a fermion determinant. It accounts for the creation and annihilation of the quark-antiquark pairs. It turned out that in computer simulations the calculation of the determinant slows down the updating of the gauge fields tremendously since one has to recompute it for every gauge configuration. Therefore a crude approximation to QCD that goes under the name of {\em quenched approximation} is used in part of the contemporary simulations. It neglects the fermion determinant in the QCD partition function~(\ref{Z2QCD}) and in the expectation values of the observables calculated with it. Despite the crudeness of the approximation it is useful to employ it as an estimate for the observables of interest, keeping in mind the systematic uncertainties of about $10-15\%$~\cite{quench_success,Aoki:1999yr}. The latter depends of course on the observable of interest. We aim, however, at the quenched formulation of the chiral perturbation theory, i.e. a low energy effective theory where the quark loops were eliminated on the level of QCD. 
In order to get rid of the quark loops one can add quarks with bosonic statistics to the fermion part of the QCD Lagrangian~(\ref{FermAction}) keeping the external currents coupled to the quarks with the proper statistics. The latter are called the ``valence'' quarks. This results in the cancellation of the diagrams with internal loops. The corresponding partition function with the sources $J_i$ coupled to quark bilinears is,
\begin{equation}
Z_{\rm SUSY}(J_1, \dots , J_{N_v})=\int [DA_{\mu}]\, \frac{\prod_{i=1}^{N_v} {\rm det}({\cal D}+m_q +J_i)}{{\rm det}({\cal D} +m_q)^{N_v}}e^{-S_{\rm YM}[A] +i\theta \nu[A]}\ .
\end{equation}

The $2N_v$ quarks ($N_v$ with usual fermionic statistics and $N_v$ additional with bosonic statistics) lie in the fundamental representation of the graded group $SU(N_v|N_v)$. 
For massless quarks, the corresponding full QCD Lagrangian exhibits a graded symmetry $SU(N_v|N_v)_L \otimes SU(N_v|N_v)_R \otimes U(1)_V$ that is assumed to be spontaneously broken down to $SU(N_v|N_v)_V \otimes U(1)_V$.  This is the so--called  ``supersymmetric'' formulation of quenched QCD~\cite{Bernard:1992mk}. Using it as a starting point one obtains a low energy theory without quark loops.

The ``supersymmetric'' chiral Lagrangian in the leading order is given by
\begin{eqnarray}
{\cal L}_{\rm SUSY}^{(2)} ~=~\frac{F_{\pi}^2}{4} {\rm Str } \left(\partial_\mu U^{-1} \partial_\mu U \right) - {\frac{m_q \Sigma}{2}} {\rm Str }\,
 \!\!\left( U_{\theta} U + U^{-1} U^{-1}_{\theta}\right) +\\
+{\frac{m_0^2}{2 N_c}} \Phi^2_0 + \frac{\alpha}{2N_{c}}\partial_{\mu}
\Phi_0(x)\partial^{\mu}\Phi_0(x)\nonumber \ ,
\label{Lrep}
\end{eqnarray}
where $\rm Str$ denotes the ``supertrace'', $\Phi_0 \equiv \frac{F_{\pi}}{2} {\rm Str}[-i \log(U)]$ and
\begin{equation}
U_\theta\equiv\exp(i \theta/N_v) I_{N_v} + \tilde{I}_{N_v}\ .
\end{equation}
Here, $I_{N_v}$
is the identity matrix in the fermion--fermion block of ``physical''
Goldstone bosons and zero otherwise, while
$\tilde{I}_{N_v}$ is the identity in the boson--boson block
and zero elsewhere. Apart from the familiar low energy constants $F_{\pi}$ and $\Sigma$ --- which now, however, take their quenched values --- we have also two new parameters $m_0$ and $\alpha$, which are artifacts of the quenched approximation. 

We note that there is also another method to formulate the quenched $\chi$PT --- the replica method~\cite{Damgaard:2000gh}. Here one considers $N_v$ valence quarks to be part of $N$ quarks in total. Considering now $N$ as a parameter and sending it to $0$ we obtain the replica limit. The equivalence of the two methods was verified at the perturbative level of the chiral Lagrangian~\cite{Damgaard:2000gh}. 

An important difference of quenched $\chi$PT from the full theory is
the unavoidable presence of a singlet scalar field $\Phi_0$, which cannot be
decoupled in this case.
The singularities of the propagator of the $\Phi_0$ field are responsible for the quenching artifacts.
Also the formulation of the counting rules for the quenched $\chi$PT becomes more involved.  
These issues, as well as the two-point functions for the $\epsilon$-- regime in the quenched approximation, were addressed in Refs.~\cite{Damgaard2002,Damgaard2001}. 

For the purpose of this thesis we consider the expression for the axial correlation function with one flavor.
We define the bare axial current with one flavor at momentum
$\vec p = \vec 0$ and Euclidean time $t$,
\begin{equation} 
\label{axialcurrent2}
{\cal A}_{\mu}(t) = \sum_{\vec x} \bar \psi (t,\vec x ) \gamma_{5}
\gamma_{\mu} \psi (t, \vec x) \ .
\end{equation}
The formula for the axial correlation function in a volume $L^{3}
\times T$, to the first order in quenched $\chi$PT, reads~\cite{Damgaard2002}\footnote{We
  have an extra factor of 2 compared to Ref.~\cite{Damgaard2002} because this
  definition of the current is considered only for one flavor.}
\begin{eqnarray}  
\langle {\cal A}_{0}(t) \, {\cal A}_{0}(0) \rangle_{\nu} &=& 
2\cdot \left(\frac{F_{\pi}^{2}}{T} 
 + 2 m_{q} \Sigma_{\vert \nu \vert}(z_q) 
T  \cdot h_{1}(\tau )\right)\ , 
\nonumber\\
h_{1}(\tau ) &=& \frac{1}{2} \left( \tau^{2} - \tau + \frac{1}{6} \right),
\qquad \tau = \frac{t}{T}\ , 
\nonumber \\
\Sigma_{\nu}(z_q) &=& \Sigma \left( z_q \left[ I_{\nu}(z_q) K_{\nu}(z_q) + 
I_{\nu +1}(z_q) K_{\nu -1}(z_q) \right] + \frac{\nu}{z_q} \right)\ .
\label{AA}
\end{eqnarray}
$I_{\nu}$ and $K_{\nu}$ are modified Bessel functions, $z_q=m_q \Sigma V$, $\nu$ is again the topological charge, and
$h_{1}$ is a purely kinematic function. The latter shows that this
correlation function is given by a parabola in $t$ with its minimum at
$T/2$. This behavior is qualitatively different from the infinite
volume, where the correlations decay exponentially (resp.\ as a
$\cosh$ function in a periodic volume).  However, the small and the
infinite volume have in common that in both cases the quenched axial
correlation function to the first order only depends on the low energy
constants $F_{\pi}$ and $\Sigma$.
%------------------------------------------------------------------------
\subsection{Zero mode contributions to the pseudo-scalar correlation function}
We start off by writing the expressions of the pseudo-scalar densities ${\cal P}^0(x),{\cal P}^a(x) $
\begin{eqnarray}
{\cal P}^0&=&\bar{\psi}i\gamma_5\psi\ ,\nonumber \\
{\cal P}^a&=&\bar{\psi}T^ai\gamma_5\psi \ , \quad a=1,\dots , N^2_f-1\ ,
\label{eq:pseudo-scalardensity}
\end{eqnarray}
where $T^a$ are generators of the group $SU(N_f)$ and $a$ runs from $1$ to $N^2_f-1$.
Let $I, J$ run from $0$ to $N^2_f-1$ and $T^0=1$, then the correlation function $\langle{\cal P}^I(x){\cal P}^J(y)\rangle_{\nu}$ takes the general form
\begin{eqnarray}
\langle{\cal P}^I(x){\cal P}^J(y)\rangle_{\nu} =-{\rm Tr}(T^I T^J) \,{\cal P}_1(x,y)+ 
 {\rm Tr}(T^I) \,{\rm Tr}(T^J) \,{\cal P}_2(x,y)\ .
\label{eq:pseudoscalarcorrfunction}
\end{eqnarray}
${\cal P}_1$ and ${\cal P}_2$ are the connected and disconnected parts of the correlation function,
\begin{eqnarray}
{\cal P}_1(x,y)&=&{\rm Tr}[i\gamma_5({\cal D}+m_q)^{-1}(x,y) \,i\gamma_5({\cal D}+m_q)^{-1}(y,x)] \ ,\nonumber \\
{\cal P}_2(x,y)&=&{\rm Tr}[i\gamma_5 ({\cal D}+m_q)^{-1}(x,x)] \,{\rm Tr}[i\gamma_5 ({\cal D}+m_q)^{-1}(y,y)] \ ,
\end{eqnarray}
where ${\cal D}^{-1}(x,y)$ is the propagator of the Dirac operator.
The correlation function can be computed from the generating functional which we discussed in
the previous Subsection. Now one has to calculate two derivatives of the generating functional with respect to the auxiliary pseudo-scalar field $P(x)$ introduced in Eq.~(\ref{partfunccurrents}) and put all the auxiliary fields to zero at the end. Applying the same technique to the generating functional of the low energy effective theory we obtain the corresponding pseudo-scalar correlation function in the quenched $\chi$PT (q$\chi$PT).

We consider the pseudo-scalar correlation function for the definite topological sector $\nu$.
Application of the spectral representation of the quark propagator to the pseudo-scalar correlation function was first considered in Ref.~\cite{Giusti:2003iq}.
By employing  the spectral representation of the quark propagator, it is clear that the correlator $\langle{\cal P}^I(x){\cal P}^J(y)\rangle_{\nu}$ contains a pole in $m^2_q$, due to the exact zero modes. Its residue is
\begin{equation}
 \lim_{m_q\rightarrow 0} (m_qV)^2\langle{\cal P}^I(x){\cal P}^J(0)\rangle_{\nu}
 =  {\rm Tr} [T^I T^J]~{\cal{C}}_{|\nu|} (x) + {\rm Tr} [T^I] {\rm Tr} [T^J] {\tilde {\cal{C}}}_{|\nu|} (x) \ ,
\label{eq_zeromode}
\end{equation}
where 
\begin{eqnarray}
 {\cal{C}}_{|\nu|}(x-y) & = & \Bigl\langle 
 \sum_{i,j =1}^{  K} v^\dagger_j(x) v_i(x) v^\dagger_i(y) v_j(y) 
 \Bigr\rangle_\nu 
 \;,\label{eq_connected}
 \\ 
 {\tilde{\cal{C}}}_{|\nu|}(x-y) &= & - ~ \Bigl\langle 
 \sum_{i=1}^{K} v^\dagger_i(x) v_i(x) 
 \sum_{j=1}^{K} v^\dagger_j(y) v_j(y) 
 \Bigr\rangle_\nu 
 \; , \label{eq_disconnected}
\end{eqnarray}
and the sums are over the set 
of $K=N_{+} + N_{-}$ zero modes $v_i$ of the Dirac operator,
${\cal D} v_i =0 \; , \forall \; i =1,\dots , K$.
We recall that $N_{+}$ zero modes have positive chirality while $N_{-}$ have negative chirality. The modes are normalized
so that $\int \! {\rm d}^4 x \, v^\dagger_i(x) v_i(x) = V$.

It is important to note that in writing 
Eqs.~(\ref{eq_zeromode})---(\ref{eq_disconnected})
it is assumed that poles arise only from exact zero modes, 
i.e. that taking the limit $m_q\to0$ and performing the 
average over the full space of configurations commute. 
At fixed volume the only potential danger 
arises from the average distribution of eigenvalues near zero;
this assumption holds if the density of eigenvalues
vanishes for gauge configurations at fixed non-zero topological charge.  In 
$\chi$PT, as well as in random matrix theory (RMT)~\cite{Damgaard:1998ye,Damgaard:2000ah,Wilke:1998gf,Nishigaki:1998is}, which will be discussed in the next Section, 
the densities behave as $\rho_\nu(\lambda)\sim\lambda^{(2|\nu|+1+2 N_f)}$,
and no contribution from the non-zero modes is thus expected
in the considered observables.

The zero mode contribution to the pseudo-scalar correlation functions $\langle P^I(x) P^I(y) \rangle_{\nu}$ at next-to-next-to-leading order (NNLO) in the $\epsilon$--regime for a given topological sector $\nu$
was computed in Ref.~\cite{Giusti:2003iq}, both for full and quenched QCD. In this computation the chiral expansion in momentum and mass was combined with the $1/N_c$ expansion
\begin{equation}
p^2 \sim \frac{1}{N_c} \sim \epsilon^2 \ .
\end{equation}
The corresponding Lagrangian for the quenched $\chi$PT to the NNLO reads
\begin{eqnarray}
{\cal L}^{(2)}_{q\chi PT}&=& \frac{F_{\pi}^2}{4} \mbox{Str}\left [ \partial_{\mu} U \partial_{\mu} U^{-1}\right] -\frac{m_q\Sigma}{2}\mbox{Str}\left [ U_{\theta}U + U^{-1} U^{-1}_{\theta} \right ] - \nonumber\\
&-&im_qK\Phi_0\mbox{Str}\left [ U_{\theta}U -U^{-1} U^{-1}_{\theta} \right ] + \frac{m_0^2}{2N_c}\Phi^2_0 +\frac{\alpha}{2N_c}(\partial_{\mu} \Phi_0)^2 \ .
\end{eqnarray}
It has an additional low energy constant $K$ coupling the singlet scalar field $\Phi_0$ to the pseudo-scalar meson field $U$. 

In this particular expansion the mass $m_0$ was chosen to be a small parameter of order $\epsilon$, in contrast to the standard counting where it is treated to be of order ${\cal O}(1)$. It is also assumed that $F_{\pi}$, $\Sigma$ and $K$ scale as
\begin{equation}
F_{\pi}^2=\bar{F}_{\pi}^2N_c, \quad \Sigma=\bar{\Sigma}N_c, \quad K=\frac{\bar{K}}{\sqrt{N_c}}\ . 
\label{eq:counting_Nc}
\end{equation}
This implies that ${\cal O}(\bar{F}_{\pi})\sim {\cal O}(\bar{\Sigma})\sim 1$. Given the fact that ${\cal O}(m_q \Sigma V)\sim 1$ we are led to the relation
\begin{equation}
m_q\sim \epsilon^6\ .
\end{equation}
It is further assumed that ${\cal O}(\alpha) \sim {\cal O}(\bar{K}) \sim 1 $ and ${\cal O}(m_0^2)\sim \epsilon^2$.
In this framework the pseudo-scalar density $P^I$ corresponding to ${\cal P}^I$ in QCD takes the form
\begin{equation}
 {P}^I  =  
 \;\; i \frac{\Sigma}{2} {\rm Str} \Bigl[ T^I 
  \Bigl( U_{\theta} U - U^{-1}U_{\theta}^{-1}  \Bigr)\Bigr] -K\Phi_0 \mbox{Str}\Bigl [ T^{I} \Bigl ( U_{\theta} U + U^{-1} U^{-1}_{\theta} \Bigr )\Bigr ]
 \ .
\end{equation}
The authors of the Ref.~\cite{Giusti:2003iq} worked out the expressions for the first derivative of the pseudo-scalar as well as the scalar correlators at non-zero time separations in quenched as well as in full $\chi$PT. It was pointed out that taking the first derivative helps to get rid of the NNLO Gasser-Leutwyler coefficients for the expressions in full $\chi$PT,
 \begin{eqnarray}
 {\frac{1}{2}} C_{|\nu|}'(t)&=&\lim_{m_q\to 0} (m_qV)^2 \frac{{\rm d}}{{\rm d}t}
 \int\! {\rm d}^3 x \, \langle P^a(x) P^a(0) \rangle_{\nu} 
 \;,  \nonumber\\
   {N_f}\, C_{|\nu|}'(t) + 
        {N_f^2}\, \tilde C_{|\nu|}'(t)&=& \lim_{m_q\to 0} (m_q V)^2 \frac{{\rm d}}{{\rm d}t}
 \int\! {\rm d}^3 x \,  \langle P^0(x) P^0(0) \rangle_{\nu}  \ . 
\end{eqnarray}
 $C_{|\nu|}(x)$ and $\tilde C_{|\nu|}(x)$ are quantities which correspond to ${\cal C}_{|\nu|}(x)$ and $\tilde{{\cal C}}_{|\nu|}(x)$, but calculated in the framework of $\chi$PT. 
If we define the volume average ${\cal C}_{|\nu|}(t)=\int d^3 x\, {\cal C}_{|\nu|}(x)$ and $\tilde{{\cal C}}_{|\nu|}(t)=\int d^3 x\, \tilde{{\cal C}}_{|\nu|}(x)$, then at large time $t$
\begin{equation}
{\cal C}_{|\nu|}(t)=C_{|\nu|}(t)\ ,\quad \tilde{{\cal C}}_{|\nu|}(t)=\tilde{C}_{|\nu|}(t) \ ,
\end{equation}
which represent the matching of the results in quenched QCD and in quenched $\chi$PT.

In the following we redefine $\alpha$ as
\begin{equation}
\frac{\alpha}{2N_c} \rightarrow \frac{\alpha}{2N_c} -\frac{2KF_{\pi}}{\Sigma} \ .
\end{equation} 
This allows us to simplify the resulting expressions for the zero mode contributions to the pseudo-scalar correlator in a way that only two low energy constants will appear instead of the combination of $K$, $F_{\pi}$, $\Sigma$ and $\alpha$.
The expressions for q$\chi$PT in the volume $V=L^3T$ read

\begin{eqnarray}
 F_{\pi}^2 C_{|\nu|}'(t)
 & = & 
 2 |\nu|
 \biggl[ 
 |\nu|  h_1' (\tau) 
 + \biggl( {\frac{\alpha}{2 N_c}} 
% - \frac{2 K F_{\pi}}{ \Sigma} 
 - {\frac{\beta_1}{F_{\pi}^2 \sqrt{V}}} 
  \biggr)  h_1' (\tau) 
 \nonumber \\& &   
  + \frac{T^2}{F_{\pi}^2 V} \biggl( 2 \nu^2 + \fr73  
  - 2 \langle\nu^2\rangle \biggr) h_2'(\tau)   
  + \frac{T^2}{2 F_{\pi}^2 V} h_3'(\tau) 
 \biggr] 
 \;, \\
 F_{\pi}^2 {\tilde C}_{|\nu|}'(t) 
 & = & - 2 |\nu|
 \biggl[ 
  h_1' (\tau) 
  +|\nu| \biggl( {\frac{\alpha}{2 N_c}} 
% - \frac{2 K F_{\pi}}{ \Sigma} 
 - {\frac{\beta_1}{F_{\pi}^2 \sqrt{V}}}
  \biggr)  h_1' (\tau) 
 \nonumber \\& & 
 +  
 \frac{T^2}{F_{\pi}^2 V} \biggl( \fr{13}3 |\nu| - 2 |\nu| \langle\nu^2\rangle
  \biggr)  h_2'(\tau)   
   +|\nu| 
 \frac{T^2}{2 F_{\pi}^2 V} h_3'(\tau) 
  \biggr] 
 \ , \hspace{0.5cm}\label{zero_mode_contr}
 \end{eqnarray}
where $h_1(\tau)$ and $\tau$ were defined in Eq.~(\ref{AA}) and $h_2(\tau)=\frac{1}{24}\left [ \tau^2(\tau-1)^2-\frac{1}{30}\right ]$. $\beta_1$ is a geometrical shape coefficient depending on the form of the volume. $\beta_1$ as well as $h_1$, $h_2$ and $h_3$ were introduced in Ref.~\cite{Hasenfratz:1990pk} in context of computation of the two-point functions in the $\sigma$-model at finite volume. $h_3(\tau)$ is defined by
\begin{eqnarray}
h_3(\tau)&=&[h_1(\tau)]^2 + \sum_{\vec{n}}{'}{\left [  \frac{\cosh(|\vec{p}|T(\tau-1/2))}{2|\vec{p}|T\sinh(|\vec{p}|/2)} \right ]}^2 \nonumber \ , \\
|\vec{p}|&=&\frac{2\pi}{L}\left [  \sum_{i=1}^{3}n_i^2\right ]^{\frac{1}{2}}\ ,\label{g1}
\end{eqnarray}
where the prime in the sum indicates that the zero mode term, with $\vec{n}=\vec{0}$, is left out. The expectation value $\langle \nu^2 \rangle$ is related to the mass $m_0$ in analogy to the Witten-Veneziano formula which will be discussed in Chapter 7.

%-----------------------------------------------------------------------
\section{Chiral random matrix theory}

The appearance of the theory of random matrices 
in the mathematical literature goes back to 1928 and was first
applied to physics in the context of nuclear resonances by Wigner
almost 50 years ago.  At that time, theoretical
approaches such as the shell model had proven to be very successful in
describing the low-lying excitations of complex nuclei.  However,
highly excited resonances, which can be observed experimentally by
neutron scattering, could not be described by the microscopic theory.
The problem is generic: for any complex
quantum system containing many degrees of freedom with complicated
dynamics, it is very hard, if not impossible, to obtain exact results
for the energy levels far above the ground state of the system.

Even though the highly excited states cannot yet be predicted
individually, one can try to find in the experimental data some
generic statistical features that can be described theoretically.
This is where RMT comes in. It aims at a statistical description of the energy levels of the system using only basic symmetry considerations as its theoretical input.

 In the following discussion we would like to restrict our attention to the case of QCD. One can write a Hamilton operator for QCD in matrix form. 
Since QCD is not solved analytically one may follow an 
approach whereby we assume that all interactions
which are consistent with the symmetries of the system are equally
likely. This means that we replace the Hamilton matrix by a matrix which is compatible with the global symmetries of QCD and with elements which become now uncorrelated and distributed according to some distribution of random numbers.  To obtain definite results,
observables such as the level density must then be averaged over the
random matrix elements.  This defines a statistical theory of energy
levels, which is known as RMT.

In this Section we will discuss the {\em chiral Random Matrix Theory}
($\chi$RMT).
 The structure of the $\chi$RMT partition function is inspired by the instanton gas model. Although it is not necessary,  it is instructive
to think of the field configurations as a superposition of $N_+$ instantons
and $N_-$ anti-instantons. Each isolated instanton or anti-instanton
has exactly one fermionic zero mode
with a definite chirality.
In total we have $\nu = N_+ - N_-$ for the index of the gauge field configuration. At finite separation of instantons and anti-instantons
the remaining modes are no longer exact zero modes of the
Dirac equation and give rise to nonzero overlap matrix elements in the submatrix $W$
of the Dirac operator~\cite{Diakonov:1985nj}.

A model describing the zero mode part of the QCD partition function in the Euclidean volume $V$ is defined by~\cite{Shur93_a}
\begin{equation}
Z(m_f,\theta) = \sum_{N\ge 0} \mu(N)\sum_{N_+= 0}^{N}
\left( \begin{array}{c} N\\ N_+ \end{array} \right) e^{i\theta(2N_+-N)}
\int  DW\, P(W)\prod_{f=1}^{N_f}\det \left (
\begin{array}{cc} m_f & iW\\
                 iW^\dagger & m_f
\end{array} \right )\ .
\end{equation}
$N=N_++N_-$ is the number of zero modes, and the matrix $W$ is a $N_+\times N_-$
matrix. The integration measure ${D}W$ is given by the Haar measure.
The binomial factor arises since zero modes of each chirality are treated as independently distributed identical
particles. $\mu(N)$ is the distribution function of the total number of zero modes.
$P(W)$ denotes the probability distribution for field configurations. The structure of the overlap matrix, with off-diagonal blocks $W$ and $W^\dagger$ and diagonal blocks equal to the quark masses $m_f$ times the identity,
is dictated by the chirality of the zero modes, see Eq.~(\ref{block}). 
 The density $\overline{N}/V$ of the total number of zero modes
is kept fixed as $V$ changes. As in the instanton liquid approximation
to the QCD partition function~\cite{Shuryak:1982ff,Shuryak:1982dp,Shuryak:1982hk}, it
is considered to be an external parameter.

We now replace the average over all gauge field configurations with an average over Gaussian distributed overlap matrix elements with the distribution function
\begin{equation}
P(W) = \exp\left (-\frac {N \Sigma^2 \beta}{4} {\rm Tr}(W W^{\dagger}) \right )\ .
\end{equation}
$\beta$ is the so-called Dyson index which is defined as the number of independent variables per matrix element. To this end the matrix elements of $W$ can be either real [$\beta =
1$, chiral Gaussian Orthogonal Ensemble ($\chi$GOE)], complex [$\beta =
2$, chiral Gaussian Unitary Ensemble ($\chi$GUE)], or quaternion real
[$\beta = 4$, chiral Gaussian Symplectic Ensemble ($\chi$GSE)]. In the
latter case, the eigenvalues of ${\cal D}$ are doubly degenerate, and
the use of Majorana fermions is implemented by replacing the
determinant by its square root. 

We will restrict ourselves to the $\chi$RMT partition function in a specific topological sector,
\begin{equation}
  Z_{N_f,\nu}^{(\beta)}(m_1,\ldots, m_{N_f}) = 
  \int DW \prod_{f= 1}^{N_f} \det({\cal D} +m_f)
  e^{-\frac{N \Sigma^2 \beta}4 {{\rm Tr\:}} (W^\dagger W)}\ ,
\label{zrandom1}
\end{equation}
\begin{equation}
  {\cal D} = \left (\begin{array}{cc} 0 & iW\\
      iW^\dagger & 0 \end{array} \right )\ .
\end{equation}
The parameter $\Sigma$ will be later identified with the chiral condensate and $N$ with the spacetime volume. 
 For large $N$ we assume that $|\nu|$ does not exceed $\sqrt N$ so that, to a good
approximation, $N_+ = N/2$ for large $N$. These assumptions motivated by the instanton liquid model are, however, not fulfilled in the lattice simulations with the Ginsparg-Wilson fermions which are the subject of the thesis. In fact one observes zero modes either with positive or negative chirality. Therefore an aim of this work is also to prove the viability of the predictions by the RMT in our quenched QCD simulations.

This model inherits the following symmetries of the QCD partition
function:
\begin{itemize}
\item The $\rm U_A(1)$ symmetry. All eigenvalues of the random matrix
  Dirac operator occur in pairs $\pm i\lambda_n$, or they are zero.
\item The topological structure of the QCD partition function. The
  Dirac matrix ${\cal D}$ has index $\nu=N_+-N_-$. This
  identifies $\nu$ as the topological charge of the model.
\item The flavor symmetry, which is the same as in QCD. For $\beta = 2$ it is
  ${\rm SU}(N_f) \otimes {\rm SU}(N_f)$, for $\beta = 1$ it is ${\rm
    SU}(2N_f)$, and for $\beta = 4$ it is ${\rm SU}(N_f)$ (two
  Majorana flavors count as one Dirac flavor).
\item The chiral symmetry, which is broken spontaneously with a chiral
  condensate given by
  \begin{equation}                                                 
    \Sigma = \lim_{m_f\rightarrow 0} \lim_{N\rightarrow \infty} {\pi \rho(0)}/N\:.
  \end{equation}
  ($N$ is interpreted as the dimensionless volume of spacetime.) The
  symmetry-breaking pattern is~\cite{SmV} ${\rm SU}(N_f) \otimes {\rm
    SU}(N_f)\to{\rm SU}(N_f)$, ${\rm SU}(2N_f)\to{\rm Sp}(2N_f)$, and
  ${\rm SU}(N_f)\to{\rm O}(N_f)$ for $\beta = 2$, 1, and 4,
  respectively --- the same as in QCD~\cite{Peskin:1980gc,Dimopoulos:1980sp,Kogan:1985nb}.

\end{itemize}
The chiral ensembles are, however, only a part of a larger classification scheme that also
includes ensembles for the description of disordered superconductors~\cite{Altland1}. In total, 10 different families of random matrix
ensembles have been identified. They correspond one-to-one 
to the Cartan classification of symmetric spaces.\\

We consider the spectral representation of the $\chi$RMT partition function.
The partition function~(\ref{zrandom1}) is invariant under the transformation
\begin{equation}
  W \rightarrow U^\dagger W V\ ,
\label{inv}
\end{equation}
where the $n\times n$ matrix $U$ and the $m\times m$ matrix $V$ are
orthogonal matrices for $\beta=1$, unitary matrices for $\beta = 2$,
and symplectic matrices for $\beta = 4$.
Using this property the matrix $W$ can be rewritten as 
\begin{equation}
  W = U^\dagger \Lambda V\ .
\end{equation}
Here, $\Lambda$ is a diagonal matrix with real positive matrix
elements $\lambda_k$.  In terms of the eigenvalues, the partition
function~(\ref{zrandom1}) is given by
\begin{equation}
  Z_{N_f, \nu}^{(\beta)}(m_1, \ldots, m_{N_f}) = 
  \int \prod_k d\lambda_k \,  |\Delta(\lambda^2)|^\beta
  \prod_k \lambda_k^{2N_f+\beta|\nu|+\beta-1} 
  e^{-\frac{N \Sigma^2 \beta}4 \lambda_k^2}
  \prod_f m_f^{|\nu|}(\lambda_k^2 + m^2_f)\ ,
  \label{zeig}
\end{equation}
where the Vandermonde determinant $\Delta$ is defined by
\begin{equation}
  \Delta(\lambda^2) = \prod_{k<l} (\lambda_k^2-\lambda_l^2)\ .
  \label{vandermonde}
\end{equation} 

In the following we will need the $\sigma$ model representation of the $\chi$RMT partition function~\cite{Shur93_a,Halasz:1995qb}. One arrives at it by writing the fermion determinant of Eq.~(\ref{zrandom1})  as a Grassmann integral. Then one averages over the Gaussian distribution function. The resulting expression is subject to the saddle point approximation in the large $N$ limit. We give  the expression for the partition function at $\beta=2$ 

\begin{eqnarray}
  \label{Znum}
  Z^{(2)}_{N_f,\nu}( m_1,\dots ,m_{N_f}) &=& \int_{U \in {\rm U}(N_f)} DU\: ({\det} U)^{\nu}\:
  e^{ \frac{N\Sigma}{2}  {{\rm Tr\:}} ({\cal M}U + {\cal M}^\dagger U^{-1})}\ ,\nonumber  \\
 {\cal M}&=&{\rm diag}(m_1,\dots ,m_{N_f})\ .
\end{eqnarray}

One can see that the partition function for the $\chi$RMT in the large $N$ limit coincides with the leading order expression of the $\chi$PT partition function in the $\epsilon$-- regime characterized by Eq.~(\ref{zchpt_eps})~\cite{Shur93_a}. This fact leads to profound consequences for the observables calculated in the $\epsilon$--regime. We are going to discuss the applications for the so-called microscopic eigenvalue density in the next Subsection.
%--------------------------------------------------------------------
\subsection{Microscopic spectral properties}
We start off by discussing the Leutwyler-Smilga sum rules~\cite{Leut92}. 
The  microscopic spectral density of the Dirac eigenvalues arises naturally from them.
The  Leutwyler-Smilga sum rules are obtained by expanding the partition function ${\cal Z}_{\nu}(m_f)$ in powers of $m_f$  for the expression in the $\epsilon$--regime~(\ref{zchpt_eps}), and $Z^{QCD}_{\nu}(m_f)$ given in Eq.~(\ref{ZQCDnu}), and then equating the coefficients.
We take the topological sector to be $\nu=0$. Comparing coefficients at ${\cal O}(m_f^2)$ we obtain
\begin{equation}
\langle \sum_{\lambda_k>0} \frac{1}{\lambda_k^2}\rangle=\frac{\Sigma^2V^2}{4N_f}\ ,
\end{equation}
where the sum is restricted to nonzero positive eigenvalues.
The Leutwyler-Smilga sum rules can be expressed as an integral over the eigenvalue spectral density $\rho(\lambda)$ and spectral correlation functions,
\begin{equation}
\frac{1}{V^2\Sigma^2}\int \frac{\rho(\lambda)d\lambda}{\lambda^2}=\frac{1}{4N_f}\ .
\end{equation}
If we introduce the microscopic variable $z$,
\begin{equation}
  \label{scale}
  z=\lambda V\Sigma\ ,
\end{equation}
the integral can be rewritten as
\begin{equation}
\frac{1}{V\Sigma}\int \frac{\rho(\frac{z}{V\Sigma})dz}{z^2}=\frac{1}{4N_f}\ .
\end{equation}

Also according to the Banks-Casher relation  the
small eigenvalues are spaced as
\begin{equation}
  \label{spacing}
  \Delta\lambda=\frac{1}{\rho(0)}=\frac{\pi}{V\Sigma}\ ,
\end{equation}
provided that $\rho(0)/V>0$.  These arguments naturally define a scale
which is suitable for the study of the distribution of individual eigenvalues.  For this
purpose, it is convenient to define the so-called microscopic spectral
density
\begin{equation}
\rho_{s}(z) = \lim_{\begin{array}{c}  ^{ V\rightarrow \infty} \\
{ ^{\Sigma Vm_f} \,\,^{\rm fixed}}    \end{array} } \, \frac{1}{V\Sigma} \rho
\left(\frac{z}{V\Sigma}\right)=\lim_{\begin{array}{c}  ^{ V\rightarrow \infty} \\
{ ^{\Sigma Vm_f} \,\,^{\rm fixed}}    \end{array} } \sum_n \langle \delta(z-\lambda_n \Sigma V) \rangle_{\rm QCD}\ ,
\end{equation}
where the average is taken over the QCD partition function.
This limit exists if chiral symmetry is broken. 
Ab initio calculation of this quantity in QCD is a challenging task which requires a non-perturbative treatment and could not be solved yet analytically. The microscopic spectral density, however, arises as a natural quantity for the $\chi$PT in a finite volume since it is defined at fixed $m_f\Sigma V$ in the dynamical case. By using the relation between the $\chi$PT at the leading order and the $\chi$RMT in the large $N$ limit, we can reduce the problem to the calculation of the microscopic spectral density for the $\chi$RMT~\cite{Damgaard:1998ye,Damgaard:2000ah,Wilke:1998gf,Nishigaki:1998is}.
The corresponding definition of the microscopic spectral density in $\chi$RMT is given by
\begin{equation}
\rho^{\chi {\rm GUE}}_{s}(x) =  \lim_{\begin{array}{c}  ^{ N\rightarrow \infty} \\
{ ^{\Sigma Nm_f} \,\,^{\rm fixed}}    \end{array} } \frac{1}{N \Sigma} \, \rho^{\chi {\rm GUE}}
\left(\frac{x}{N\Sigma}\right)=\lim_{\begin{array}{c}  ^{ N\rightarrow \infty} \\
{ ^{\Sigma Nm_f} \,\,^{\rm fixed}}    \end{array} }\sum_{n}\langle \delta(x-\lambda_n N \Sigma) \rangle_{\chi {\rm GUE}}\ ,
\end{equation}
where $N$ is the size of the random matrix, $\langle \dots \rangle_{\chi {\rm GUE}}$ denotes averaging with respect to the $\chi$GUE partition function and $\lambda_n$ are the eigenvalues of the random matrices. If $N$ corresponds to the lattice volume then $\Sigma$ is assumed to be taken in dimensionless units.  
   
The conjecture is that $\rho_{s}(z)$ is a universal function that only depends on the global symmetries of the QCD partition function and can be therefore computed by means of the $\chi$GUE,
\begin{equation}
\rho_{s}(z) =\rho^{\chi {\rm GUE}}_{s}(z)\ .
\end{equation}

Below we illustrate how this statement fits in the framework of $\chi$PT and $\chi$GUE:
$$
\begin{array}{ccc}
Z_{\nu}^{\rm QCD}(m_f)=
\int_{\nu}  DA_\mu
\,e^{-S[A]}\times
&
\stackrel{\mbox{(b)}}{\longrightarrow}&
\hspace{-0.5cm} Z_{\chi{\rm UE}}=\int D W\,e^{-Nv(W)}\,\prod_f \det
({\cal D}+m_f)\\
\ & \ & \ \\
 \prod_f \det ({\cal D}+m_f)  & &\\
\ & \ & \ \\
\downarrow \mbox{(a)}&
\ &
\mbox{(c)}
\downarrow 
\\
\ & \ & \ \\
\hspace{-1cm}Z_{\nu}({\cal M})=\int_{{\rm SU}(N_f)} D U\,
(\det U)^{\nu} \times&
\stackrel{\mbox{(d)}}{\longleftarrow} &
Z^{(2)}_{N_f,\nu}(m_1,\dots ,m_{N_f})=\\ 
\ & \ & \ \\
 e^{\frac{V\Sigma}{2}\,{\rm Tr}\,({\cal M} U  + {\cal M} \,  U^{\dagger})}&&=\int D W\,e^{-\frac{N\Sigma^2}{2}\,{\rm Tr}\,WW^{\dagger}} \prod_f\det({\cal D}+m_f)\nonumber
\end{array}
$$

\vspace{12pt}\noindent
We show the relation between the partition functions of QCD~(\ref{Z2QCD}), $\chi$PT~(\ref{zchpt_eps}) and $\chi$RMT~(\ref{zrandom1}) with GUE. 

If we formally substitute the integration over gauge fields in the QCD partition function by integration over the Dirac matrices then we come to the formulation in terms of the random matrices $W$ and the partition function of the chiral random unitary ensemble (link (b)). We introduce the potential 
\begin{equation}
v(W)=\frac{1}{N}(S_{YM}[A] -\log(J)),
\end{equation}
 where $J$ is the corresponding Jacobian and $N$ is the size of the random matrices, that can be related to the physical volume $V$. This reduction is, however, not well theoretically founded.

The link (a) corresponds to a reduction of the QCD partition function $Z^{\rm QCD}$ in the finite volume to the low energy effective theory for the soft pions, which amounts to neglecting the derivative terms in the chiral Lagrangian. 

Once we have a formulation in terms of the unitary random matrices $Z_{\chi {\rm UE}}$ we can apply the result of Ref.~\cite{Akemann:1997vr}, where it has been proven that the microscopic spectral densities, and all microscopic spectral correlators are universal within the given classes of matrix model ensembles. Hence, the link (c) leads to the $\chi$GUE partition function.

The link (d) corresponds to the reduction of the $\chi$GUE in the large $N$ limit to the $\chi$PT partition function for the soft pions~\cite{Shur93_a}.

All this pattern can be reformulated for quenched QCD. Since RMT is well defined for any number of flavors, one needs only to substitute the $\chi$PT by its quenched formulation. Also the formalism of the q$\chi$PT is known.

There is now mounting evidence that the described scenario is correct.
The microscopic spectral densities, derived from large $N$ random matrix ensembles with Gaussian weights~\cite{Verbaarschot:1993pm,Verbaarschot:1994qf,Verbaarschot:1994gr,Verbaarschot:1994ia,Verbaarschot:1994ip}, have been shown to consistently reproduce the exact spectral sum rules of Leutwyler and Smilga. However there is no similar analytical check for the probability distributions of the individual eigenvalues.

In particular we want to highlight the analytical results for the microscopic spectral density and the probability distributions of the individual eigenvalues. The latter were calculated in the framework of the $\chi$RMT with the Gaussian unitary chiral ensemble.

The microscopic spectral density was calculated for the $\chi$GUE, as well as for the $\epsilon$--regime of quenched $\chi$PT. (Also some other special cases were computed in $\chi$PT which agree well with $\chi$RMT results~\cite{Damgaard:1998xy,Verbaarschot:1994qf,Verbaarschot:1994gr,Verbaarschot:1994ia}.)
The microscopic spectral density in the topological sector $|\nu|$ for the $\chi$GUE reads
\begin{equation}
 \rho_{s}^{(\nu)}(z)=\frac{z}{2}\left ( J_{N_f+|\nu|}^2(z)-J_{N_f+|\nu|+1}(z)J_{N_f+|\nu|-1}(z)\right) +|\nu|\delta(z)\ ,
\label{micr_spectral_den}
\end{equation}
where the $J$'s are Bessel functions.
These expressions can be reassembled in the microscopic spectral density $\rho_{s} (z)$,
\begin{equation}
\rho_{s} (z) = \sum_{\nu = -\infty}^{\infty} \rho_{s}^{(\nu )}(z)\,,
\qquad \nu = \hbox{topological charge}\ .
\end{equation}

These functions $\rho_{s}^{(\nu )}(z)$   can be expressed in series of distributions of the $k^{\rm th}$  individual lowest eigenvalue for each topological sector.
Refs.~\cite{Damgaard:1998ye,Damgaard:2000ah,Wilke:1998gf,Nishigaki:1998is} present the general formula for the probability distributions of the individual eigenvalues  $\rho^{(\nu )}_{k}(z)$ calculated for the $\chi$GUE,
\begin{equation}
\rho_{s}^{(\nu )}(z) = \sum_{k \ge 1} {\rho}^{(\nu )}_{k}(z) +|\nu|\delta(z) \ ,
\label{micr_den_expan}
\end{equation}
where the sum runs over the non-zero modes.
The distribution of the lowest eigenvalue of the Dirac operator for an arbitrary topological sector and $N_f=0$ reads 
\begin{equation}
{\rho}^{(\nu )}_{1}(z) =  \frac{z}{2}{\rm e}^{-z^2/4}
\det[I_{2+i-j}(z)]_{i,j=1,\dots , |\nu|} \ .
\label{eq:RMTpredictions012}
\end{equation}

\begin{figure}[htbp]
\centering
%\begin{turn}{-90}
\includegraphics[width=0.6\textwidth, angle=-90]{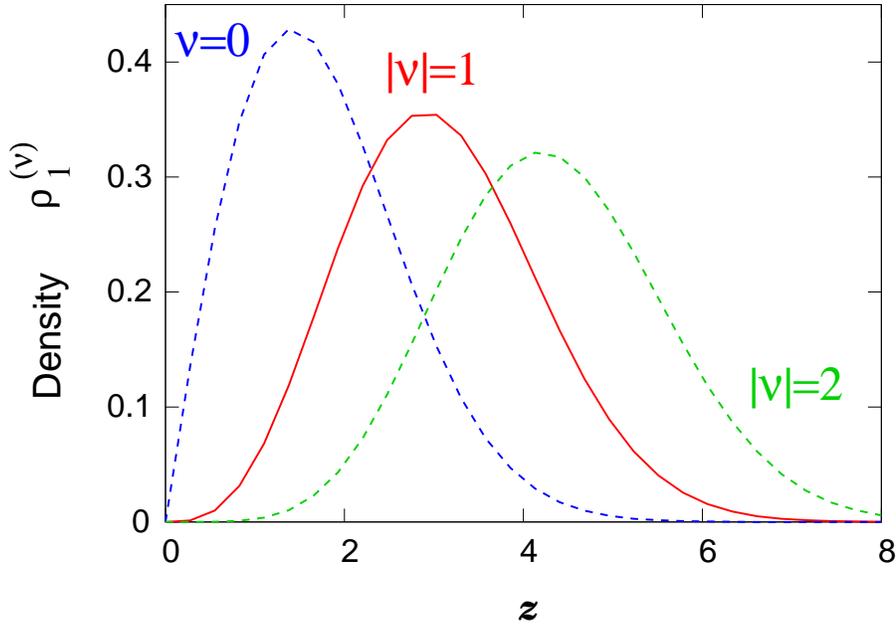}
%\end{turn}
%\epsfig{file=inc.figures/rmtcurves.eps,angle=270,width=.6\textwidth,clip=}
\caption{The distributions for the lowest eigenvalue at $\vert \nu
\vert = 0,$ $1$ and $2$, as predicted by $\chi$RMT in Eq.~(\ref{eq:RMTpredictions012}).}
\label{RMT}
\end{figure}

The predictions for the
lowest eigenvalue at $\vert \nu \vert = 0,$ 1 and 2 are depicted in
Figure~\ref{RMT}.  We see in particular that the density peak moves to
larger values of $z$ as $\vert \nu \vert$ increases.

%-----------------------------------------------------------------------
\subsection{Bulk eigenvalues. Unfolding of the spectra}
The spectra of many complex quantum systems exhibit a universal behavior. To see this, however, one needs to unfold the spectrum.  
  The unfolding can be done if the scale of the variations of the average spectral density is much larger than the scale of the spectral fluctuations. 
Then we can separate the spectral density in an average density $\bar\rho(\lambda)$ and a fluctuating part,
\begin{equation}
  \rho(\lambda) = \bar \rho(\lambda) + \rho_{\rm fl}(\lambda)\ .
\end{equation}
The details of the separation are specific for the observable of interest. The only common feature is that the resulting unfolded spectrum $\{x_n\}$ has average spectral density equal to $1$ and is given by
\begin{equation}
  x_n=\int_{-\infty}^{E_n}\bar\rho(\lambda) d\lambda \ ,
\end{equation}
where $\{E_n\}$ represents the original spectrum.
There are several ways to obtain the average spectral density. In some cases it can be conjectured analytically by some polynomial approximations. In most cases, however, we have to average over many level spacings. There are two different procedures to achieve this. The first is the spectral averaging where one takes averages over many level spacing around the energy of interest. The second is the ensemble average where one performs averages over many different level spacings drawn from all configurations of the ensemble. If we calculate the spectral correlation functions from the unfolded spectrum then these two procedures do not necessarily yield the same behavior at long distances. The equivalence of the two procedures with respect to the correlation functions is known as ``spectral ergodicity''. 

The distribution of the unfolded eigenvalues takes the usual form
\begin{equation}
\rho_{\rm un}(x ) =  \sum_{n} \delta (x - x_{n}) \ .
\end{equation}

In this context one usually considers the unfolded level spacing  $\Delta_{n}$ 
as a tool to probe the short range correlations in the spectrum
\begin{equation}
  \Delta_{n}=x_{n} -x_{n-1}\ .
\end{equation}
Its distribution is given by
\begin{equation}
\rho(s) = \sum_{n} \delta (s - \Delta_{n}) \ .
\end{equation}
This function is equal to the probability density for two neighboring levels  $n$ and  $n+1$ having the spacing $s$.

For the uncorrelated case it is expected that the level spacings are distributed according to the Poisson distribution $\rho(s)=\exp(-s)$. A good ansatz for the correlated case is provided by the Wigner distributions, which are the exact ``spacing distribution'' for a $2\times 2$ matrix model~\cite{Wigner}
 \begin{equation}
  \rho (s) =  \left\{ \begin{array}{ll}
{\frac{\pi}{2}} s \,{\rm e}^{-{\frac{\pi}{4}}s^2} & \mbox{orthogonal ensemble} \\
&\\
 {\frac{32}{\pi^2}} s^2 \,{\rm e}^{-{\frac{4}{\pi}}s^2} & \mbox{unitary ensemble} \\
&\\
 {\frac{262144}{729\pi^3}} s^4 \,{\rm e}^{-{\frac{64}{9\pi}} s^2} & \mbox{symplectic ensemble .}
\end{array} \right.
 \end{equation}
%------------------------------------------------------------------------

\section{Lattice QCD}
\subsection{The Wilson gauge action and the Wilson fermions}
QCD is a renormalizable quantum field theory that must be regularized.
A way to achieve this is to put the system on a cubic space-time lattice with some lattice spacing $a$~\cite{Wils1974}. 
This introduces a momentum cut-off of $\pi/a$. Now one has to introduce gauge and quark degrees of freedom on the lattice. 
Maintaining manifest gauge invariance is essential when gauge theories are regularized on the lattice. In order to achieve this one needs a convenient way to represent the gauge fields on the lattice. Wegner and Wilson, as well as Smit, independently introduced the concept of a {\em parallel transporter} $U_{x, \mu} \in SU(N_c)$ connecting neighboring lattice points $x$ and $x+ a\hat{\mu}$. Here $\hat{\mu}$ is a unit vector pointing in the $\mu$ direction.
The parallel transporter is related to an underlying continuum
gauge field $A_\mu(x) = i g A^a_\mu(x) T^a$ by
\begin{equation}
U_{x,\mu} = {\cal P} \exp \int_0^a ds \ A_\mu(x+\hat\mu s)\ ,
\end{equation}
where ${\cal P}$ denotes path-ordering for non-Abelian gauge theory. Under a gauge 
transformation $\Omega_x$ the parallel transporter transforms as
\begin{equation}
U_{x,\mu}' = \Omega_x U_{x,\mu} \Omega_{x+\hat\mu}^\dagger \ .
\end{equation}
One tries to preserve as many symmetries as possible at the finite lattice spacing. The hope is that the rest will be recovered in the continuum limit. A gauge action that respects the gauge symmetry was first put forward by K. G. Wilson. It is a sum over the so-called plaquette variables $U_{x,\mu\nu}^{(P)}$, which are the products of link variables around a closed square path,
\begin{eqnarray}
S_{\rm YMW}[U]&=&\beta\sum_{x, \mu < \nu}S_P(U^{(P)}_{x,\mu \nu}) \ , \nonumber \\
S_P(U^{(P)}_{x,\mu \nu})&=&1-\frac{1}{N_c}{\rm Re}\, {\rm Tr}\,U^{(P)}_{x,\mu \nu} \ ,\nonumber \\
U^{(P)}_{x,\mu \nu}&=&U_{x,\mu}U_{x+\mu,\nu}U_{x+\nu,\mu}^{\dagger}U_{x,\nu}^{\dagger}\ .
\label{eq:wilson_action}
\end{eqnarray}
$U_{x+\mu,\nu}$ represents a link attached to a site $x+a\hat{\mu}$ and pointing in the direction $\nu$ and $U^{\dagger}_{x,\nu}$ represents a link pointing in the opposite direction to $\nu$ and attached to the site $x+a\hat{\nu}$.
$\beta$ is related to the bare coupling constant as $\beta=\frac{2N_c}{g^{2}}$.
The Wilson gauge action reduces to the continuum Yang-Mills action~(\ref{YMaction}) in the naive continuum limit $a \rightarrow 0$. This is demonstrated e.g. in Refs.~\cite{bookMM,Rothe}.

To fully define the path integral we must also consider the measure. The 
lattice functional integral is obtained as an integral over all configurations 
of parallel transporters $U_{x,\mu}$, i.e.
\begin{equation}
Z = \prod_{x,\mu} \int_{SU(N_c)} dU_{x,\mu} \ \exp(- S_{\rm YMW}[U])\ .
\end{equation}
One integrates independently over all link variables using locally the Haar 
measure $dU_{\mu,x}$ for each parallel transporter.

For compact groups like $SU(N_c)$ the integration is limited to a finite domain. This
makes it unnecessary to fix the gauge in lattice QCD because the functional 
integral is finite even without gauge fixing. This is an important
advantage of the formulation using parallel transporters.

The Yang-Mills functional integral contains a single parameter ---
the bare gauge coupling $g$. The continuum limit is taken by
searching for values of $g$ for which the correlation length of the
lattice theory diverges in lattice units. In the language of statistical 
mechanics one looks for a second order phase transition. Due to asymptotic
freedom, in lattice QCD one expects a second order phase transition at 
$g \rightarrow 0$.

The physical observables must be formulated in a gauge invariant way. In fact, the path integral average over gauge non-invariant observables vanish. This statement is a consequence of Elitzur's theorem~\cite{Elitzur} which in particular implies
\begin{equation}
\langle U_{x,\mu}\rangle=0\ 
\end{equation}
for each link variable $U_{x,\mu}$. \\

Now we would like to discuss how the fermion fields are put on the lattice. The quark lattice fields $\Psi_{x}, \, \Psibar_x$ are Grassmann variables attached to the sites $x$. The Dirac operator contains a continuum derivative which has to be regularized on the lattice. It turned out that it is a highly non-trivial problem to regularize it in a way that preserves the chiral symmetry. The latter becomes of great importance when addressing the physics of light quarks.  

A simple way to regularize the Dirac operator is the so-called naive discretization of fermions on a lattice. One substitutes the partial derivatives with finite differences,
\begin{equation}
S[\Psibar,\Psi] = a^4\sum_{x,y} \bar{\Psi}_x(D)_{xy}\Psi_y = a^4 \sum_{x,\mu} \frac{1}{2a}
(\Psibar_x \gamma_\mu \Psi_{x+\hat\mu} - \Psibar_{x+\hat\mu} \gamma_\mu
\Psi_x) + a^4 \sum_x m \Psibar_x \Psi_x \ .
\label{for:naive_discretization}
\end{equation}
This action has the so-called doubling problem which is also present in the interacting case, where the action reads,
\begin{eqnarray}
S[\overline \Psi,\Psi,U]&=&a^4 \sum_{x,\mu} \frac{1}{2a} 
(\bar{\Psi}_x \gamma_\mu U_{x,\mu} \Psi_{x+\hat\mu} -
\bar{\Psi}_{x+\hat\mu} \gamma_\mu U_{x,\mu}^\dagger \Psi_x) + 
a^4 \sum_x m \bar{\Psi}_x \Psi_x \ . \nonumber 
\end{eqnarray}
The free quark propagator has poles at momenta $p_{\mu}=\{ 0, \pi/a \}$, which gives rise to a
total of $2^4$ species in $4$ dimensions.   

It was Wilson who proposed to  eliminate the unwanted doubler 
fermions by breaking chiral symmetry explicitly~\cite{Wil75}.
The unwanted $2^4-1$
species can be sent to the cut-off scale by adding to the Dirac operator an
additional term, the Wilson term, which removes the doublers in the
continuum limit, 
\begin{eqnarray}
\label{intwilson}
S_{W}[\overline \Psi,\Psi,U]&=& a^4\sum_{x,y} \bar{\Psi}_x(D_W)_{xy}\Psi_y~ \nonumber \\
&=&a^4 \sum_{x,\mu} \frac{1}{2a} 
(\bar{\Psi}_x \gamma_\mu U_{x,\mu} \Psi_{x+\hat\mu} -
\bar{\Psi}_{x+\hat\mu} \gamma_\mu U_{x,\mu}^\dagger \Psi_x) + 
a^4 \sum_x m \bar{\Psi}_x \Psi_x \nonumber \\
&+&a^4 \sum_{x,\mu} \frac{1}{2a} (2 \bar{\Psi}_x \Psi_x -
\bar{\Psi}_x U_{x,\mu} \Psi_{x+\hat\mu} -
\bar{\Psi}_{x+\hat\mu} U_{x,\mu}^\dagger \Psi_x)\ .
\end{eqnarray}
 The doubling problem is expressed by the Nielsen and Ninomiya No-Go theorem~\cite{Nielsen:1981rz,Nielsen:1981xu,Karsten:1981gd}
which states that it is not possible to simultaneously solve
the doubling problem and have exact chiral symmetry on the lattice
with a local, discretely translation invariant and real Euclidean action for lattice fermions. The locality of a lattice action is defined by requiring the couplings to decay exponentially with the distance. This provides a safe continuum limit for a lattice Dirac operator.
 
Many lattice Dirac operators, such as the Wilson Dirac operator, satisfy the following condition, which is called $\gamma_5$--Hermiticity,
\begin{equation}
\label{g5herm}
\mathcal{D^{\dagger}}=\gamma_5 \mathcal{D} \gamma_5\ .
\end{equation}
If the lattice is finite, one can simulate
the theory on a computer.  This can be done efficiently only in
Euclidean space, where the gluonic weight function is $\exp(-S_{\rm
  YM})$ with $S_{\rm YM}$ real and positive.  The
full weight function of QCD also contains the fermion determinant,
which can be expressed in terms of the gauge fields.  Observables are
computed by generating gauge field configurations at some value of $\beta$ in a Monte Carlo
update procedure and averaging an observable over many configurations.

Since the inclusion of the fermion determinant is very
time consuming, it is a well-known simplification to use only the gluonic part of the
weight function in the Monte Carlo updates, i.e. to perform  the
{\em quenched} simulations, which correspond in some sense to the limit $N_f=0$. To make
contact with continuum physics, the results of lattice simulations
must be extrapolated to the continuum limit, i.e to the infinite lattice volume and zero lattice spacing keeping the physical volume fixed. 

%---------------------------------------------------------------------------------
\subsection{Staggered fermions}
For the sake of the following discussion we would like to introduce in this Subsection the staggered fermions.
Staggered fermions are obtained from naive doubled lattice fermions~(\ref{for:naive_discretization}) by the so-called spin diagonalization~\cite{Kogut:1974ag,Susskind:1977jm,Kogut:1979wt,Sharatchandra:1981si,Kluberg-Stern:1983dg}. It consists in performing a local change of fermionic variables
\begin{eqnarray}
\Psi_x\rightarrow A_x\Psi_x \ ,\quad \Psibar_x \rightarrow \Psibar_x A^{\dagger}_x\ ,\\
A^{\dagger}_x\gamma_{\mu} A_{x+a\hat{\mu}}= \Delta_{\mu}(x)\in U(1)^{\otimes d} \ .
\end{eqnarray}
$A_x$ is a $d\times d$ unitary matrix which performs the spin diagonalization and $\Delta_{\mu}(x)$ is a diagonal unitary matrix belonging to the direct product of $U(1)$ groups denoted by $U(1)^{\otimes d}$.
In four space-time dimensions the size of the Dirac matrices 
is $4 \times 4$. By spin diagonalization one can reduce the fermion 
multiplication factor $2^d$ to $2^d/4$, where $d$ is the spacetime dimension. Hence, for $d = 4$, staggered 
fermions represent $2^4/4=4$ pseudo-flavors of mass-degenerate fermions.  
 The new fermionic variables $\chi_x$ become a $3$(colors)$\times 1$(spin) component object.
The corresponding lattice action for free staggered fermions takes the form
\begin{equation}
S[\chibar,\chi] = 
a^4 \sum_{x,\mu} \frac{1}{2a} 
(\chibar_x \eta_{x,\mu} \chi_{x+\hat\mu} - 
\chibar_{x+\hat\mu} \eta_{x,\mu} \chi_x) + a^4 \sum_x m \chibar_x \chi_x \ ,
\end{equation}
where
\begin{equation}
\eta_{x,1} = 1, \ \eta_{x,2} = (-1)^{x_1/a} \ , \ 
\eta_{x,3} = (-1)^{(x_1+x_2)/a}, \ \eta_{x,4} = (-1)^{(x_1+x_2+x_3)/a}\ .
\end{equation}
The $4$ spin degrees of freedom for the 4 pseudo-flavors are made from the 16 components of the $\chi$ field on a $2^4$ hypercube~\cite{Golterman:1984cy}.

For $m = 0$ the interacting staggered fermion action has an exact remnant $U(1)_e \otimes U(1)_o$ chiral symmetry
\begin{eqnarray}
\chi'_x &=& \exp(i \varphi_e) \chi_x\ , \ 
\chibar'_x = \chibar_x \exp(- i \varphi_o)\ , \ 
\mbox{for} \ (x_1 + x_2 + x_3 + x_4)/a \ \mbox{even}\ , \nonumber \\
\chi'_x &=& \exp(i \varphi_o) \chi_x \ , \ 
\chibar'_x = \chibar \exp(- i \varphi_e) \ , \ 
\mbox{for} \ (x_1 + x_2 + x_3 + x_4)/a \ \mbox{odd}\ ,\nonumber \\
\end{eqnarray}
which is a subgroup of the $SU(4)_L \otimes SU(4)_R \otimes U(1)_B$ chiral 
symmetry of the corresponding continuum theory. These symmetries are expected to be recovered in the continuum limit. 

Due to the remnant chiral symmetry the staggered fermions are protected from some problems which are inherent for the Wilson fermions. These are the additive mass renormalization and the problem of the exceptional configurations in the quenched simulations, i.e. the occurrence of the configurations with accidentally small eigenvalues of the lattice Dirac operator. The staggered fermions are also order $a^2$ improved i.e. the naive discretization errors are expected to behave like ${\cal O}(a^2)$ which is opposed to the Wilson fermions where the errors are of order ${\cal O}(a)$. 

The first down-side of staggered quarks is the fact that we have to deal with exactly four pseudo-flavors. Secondly, the discretization errors induce flavor-changing interactions and so are rather dangerous. To overcome the first problem it is tempting to consider a fourth square root of the staggered Dirac operator to obtain only one pseudo-flavor theory. This operation, however, is dangerous since the locality of the resulting action is questionable~\cite{Bunk:2004kf}.  
%------------------------------------------------------------------------------------
\subsection{The Ginsparg--Wilson relation and the Neuberger overlap operator}
It were Ginsparg and Wilson who first put forward a relation on the lattice Dirac operator which helps to circumvent the Nielsen and Ninomiya No-Go theorem. One gives up the chiral symmetry in its conventional formulation,
\begin{equation}
\{D,\gamma_5 \}=0 \ ,
\label{eq:chiralsymmetrylattice}
\end{equation}
and  requires for the lattice Dirac operator to satisfy the remnant chiral symmetry condition of Ginsparg and Wilson~\cite{Ginsparg:1982bj},
\begin{equation}  \label{GWR}
D \gamma_{5} + \gamma_{5} D = \frac{a}{\mu} D \gamma_{5}R D \ .
\end{equation}
where $R_{x,y}$ is a local operator for the $x,y$ indices and its structure in the spinor space is dictated by $\{\gamma_5, R\}=0$. 
The Ginsparg-Wilson relation states that the continuum condition for the right-hand-side to vanish is
relaxed to a term of ${\cal O}(a)$, where $a$ is the lattice spacing.  $\mu$
is a mass parameter of order ${\cal O}(1)$, see below.  The Ginsparg-Wilson relation (GWR) implies that 
\begin{equation}
D^{-1}\gamma_{5} + \gamma_{5} D^{-1} = (a/\mu) \gamma_{5} R
\end{equation}
is local. It is a formulation of the chiral symmetry after performing the Wilson renormalization group transformation from the continuum theory to the lattice. Hence the Ginsparg-Wilson fermions necessarily have exact zero modes, which occur with positive or negative chirality. From now on we will consider $R_{x y}=\delta_{x y}$.
We follow the idea of M. L\"uscher and write down a modified infinitesimal chiral transformation on the lattice
\begin{eqnarray}
\label{Lue}
&&\Psi' = \Psi + \delta \Psi = 
\left(1 + i \varepsilon^a T^a\gamma_5(1 - \frac{a}{2\mu} D)\right) \Psi\ , 
\nonumber \\
&&\Psibar' = \Psibar + \delta \Psibar = 
\Psibar \left(1 + i \varepsilon^a T^a (1 - \frac{a}{2\mu} D)\gamma_5\right)\ .
\end{eqnarray}
where $\varepsilon^a$ are parameters of the chiral rotation~\cite{Luscher:1998pq}.

 Through $D$ L\"uscher's lattice version of a chiral transformation 
depends on the gluon field. Still, in the continuum limit $a \rightarrow 0$ it
reduces to the standard chiral symmetry of the continuum theory. Any lattice fermion action of the form $S_F=a^4 \Psibar  D\Psi$ is invariant under transformation~(\ref{Lue}) to order ${\cal O}(\varepsilon)$, provided that $D$ obeys 
the Ginsparg-Wilson relation Eq.~(\ref{GWR}). Indeed,
\begin{eqnarray}
\delta S_F&=&a^4(\delta \Psibar D\Psi + \Psibar D\delta\Psi)\nonumber \\
&\simeq&a^4\Psibar Di\varepsilon^a T^a \gamma_5(1-\frac{a}{2\mu}D) + a^4 \Psibar i\varepsilon^a T^a (1-\frac{a}{2\mu}D)\gamma_5 D\Psi\nonumber\\
&=&a^4\Psibar i\varepsilon^a T^a \{\gamma_5, D \}\Psi -a^4\Psibar i\varepsilon^a T^a \frac{a}{\mu}D\gamma_5 D \Psi =0 \ .
\end{eqnarray} 

Under this transformation the fermion measure $D\overline{\Psi} D\Psi$ is not $U(1)_A$ invariant. This is analogous to continuum QCD where this property of the fermion measure in the partition function leads to the anomaly for the axial current. In this way the axial anomaly is also correctly reproduced by the Ginsparg-Wilson fermions~\cite{Hasenfratz:1998ri,Hasenfratz:1998jp,Luscher:1998pq}.

 We now define the topological charge
$\nu$ simply by the index
\begin{equation}
\nu = N_{+} - N_{-} \ ,
\end{equation}
where $N_{+}$ ($N_{-}$) is the number of positive (negative) chiral
zero modes. Here one adopts the continuum index theorem and uses it to
define the topological charge of the lattice gauge
configurations~\cite{Hasenfratz:1998ri,Hasenfratz:1998jp,Luscher:1998pq}.
Since the Wilson operator $D_{W}$
obeys $\gamma_{5}$-Hermiticity~(\ref{g5herm}) this allows for a simple solution of the GWR by inserting $D_{W}$  into the overlap formula~\cite{Kikukawa:1997qh,Neuberger:1998fp,Neuberger:1998wv},
\begin{equation}
D_{\rm ov} = \frac{\mu}{a} \left[ 1 + {A}/{\sqrt{A^{\dagger}A}}\right]\ ,
\qquad A = a\hat{D} - \mu \ ,
\label{overlap_operator}
\end{equation}
\begin{equation}
\hat{D}=D_{W} \ .
\end{equation}
$D_{\rm ov}$ is a solution to condition~(\ref{GWR})
and it is $\gamma_5$-Hermitian as well. 
$\mu $ represents a negative mass of the Wilson fermion, which 
can be chosen in some interval as long as the gauge fields are smooth.
This particular solution with the kernel $\hat{D}=D_{W}$ is often referred to as the Neuberger overlap operator. In this thesis we will denote it also as the overlap Wilson Dirac operator.

If we assume that the operator $\hat{D}$ already satisfies the GWR and also is $\gamma_5$-Hermitian, then
\begin{eqnarray}
A^\dagger A&=&(a\hat{D}^\dagger -\mu)(a\hat{D}-\mu)=a^2\hat{D}^\dagger \hat{D}-\mu a (\hat{D}^\dagger +\hat{D}) +\mu^2=\nonumber \\
&=&a\mu \gamma_5 \left (\frac{a}{\mu}\hat{D}\gamma_5 \hat{D}- \{ \hat{D},\gamma_5 \}\right ) +\mu^2=\mu^2 \ .
\end{eqnarray}
This implies for the overlap operator $D_{\rm ov}$
\begin{equation}
D_{\rm ov}=\frac{\mu}{a}\left [1+ \frac{a\hat{D} -\mu}{\mu} \right ]=\hat{D} \ ,
\label{eq:help1}
\end{equation} 
i.e. an operator satisfying the GWR is reproduced by the overlap formula~\cite{Bie1998ut}.

Note that the operator $A/\sqrt{A^{\dagger}A}$ is unitary, hence the
spectrum of $D_{\rm ov}$ is located on a circle in the complex plane
through zero, with center and radius $\mu /a$, which we denote as the GW circle.~\cite{Hasenfratz:1998ri}.

%---------------------------------------------------------------------------------------
\subsection{The hypercube Dirac operator}
 There is a natural way to proceed from the continuum formulation of the action to a lattice form. In this approach one defines the lattice fermion fields as block averages of continuum fields integrated over hypercubes. The lattice action of these lattice fermion fields is related to the continuum one by the Wilson renormalization group transformation.  
The resulting lattice theory, which is called the perfect lattice fermions, is equivalent to the underlying continuum theory with respect to the physical observables, i.e. it is completely free of lattice artifacts.
Similarly one can also start from a given lattice formulation on a fine lattice and --- by using the renormalization group transformation --- arrive at another lattice formulation at coarser lattice which is physically equivalent. Iterating this procedure infinitely many times we arrive at the fixed point action (here we assume that this limit exists). The fixed point action of the quantum renormalization group trajectory represents perfect lattice fermions since it is insensitive to a change of the lattice spacing.

To simplify the notation we will set the lattice spacing to $a=1$. The hypercube Dirac operator is a free fixed point action with its couplings being truncated to the unit hypercube. In this Subsection we consider the construction of the  hypercube Dirac operator.  The discussion will closely follow Ref.~\cite{Bietenholz1998wx}.
We will start from the case of free fermions and then explain how gauge fields can be introduced.

Let us divide the (infinite) lattice into
disjoint hypercubic blocks of $n^{d}$ sites each and introduce new
variables living on the centers of these blocks (block factor $n$
RGT).  Then the RGT relates
\begin{equation}
\Psi '_{x'} \sim \sum_{x\in x'} \Psi_{x}\ ,
\end{equation}
where $\Psi$  represents the fermions on the fine lattice and the $\Psi '$ represents fermions on the coarser lattice. The points $x \in \mathbb{Z}^{d}$ are the sites of
the original fine lattice and $x'$ are those of the new lattice with
spacing $n$. $x\in x'$ means that the site $x$ belongs to the block with center $x'$.

Now the original action $S[\bar{\Psi},\Psi ]$ transforms into a new action
$S'[\bar{\Psi} ',\Psi ']$ on the coarse lattice. The latter is determined by
the functional integral
\begin{equation}
e^{-S'[\bar{\Psi}' , \Psi ']} = \int D \bar{\Psi} D \Psi \ K[\bar{\Psi}', \Psi ' ,\bar{\Psi}, \Psi ] e^{-S[\bar{\Psi}, \Psi ]}\ .
\end{equation}
The kernel $K[\bar{\Psi}', \Psi ',\bar{\Psi}, \Psi ]$ has to be chosen such that the partition
function and all expectation values remain invariant under the RGT. At
the end, one usually rescales the lattice spacing back to 1.  In any
case, the correlation length in lattice units gets divided by $n$.

For the kernel functional there are many possible choices~\cite{WilsKogut1974,Kadanoff77}. We will use the Gaussian type kernel

\begin{eqnarray}
\lefteqn{e^{-S'[\bar \Psi ' , \Psi ' ]} 
= \int D \bar \Psi D \Psi\,
e^{-S[\bar \Psi , \Psi ]}} \label{trafo}\nonumber \\
&\times& \exp \Big\{ - \frac{1}{\alpha} \sum_{x'} \Big[
\bar \Psi^{'}_{x'} - \frac{1}{n^{(d+1)/2}} \sum_{x\in x'} \bar \Psi_{x} \Big]
\Big[ \Psi^{'}_{x'} - \frac{1}{n^{(d+1)/2}} \sum_{x\in x'} \Psi_{x} \Big]
\Big\} \ .
\end{eqnarray}
This type of transformation with non-vanishing $\alpha > 0$ is not chirally invariant, which will lead to the formulation with broken chiral symmetry in its standard form given by Eq.~(\ref{eq:chiralsymmetrylattice}).

Assume that we are on a ``critical surface'', where the correlation
length is infinite. After an infinite number of RGT iterations we obtain a
finite fixed point action (FPA) $S^{*}[\bar \Psi, \Psi ]$. 
The critical surface requires a fermion mass $m=0$, but one can generalize
the consideration to a finite mass.

Assume that we want to perform a number $N$ of RGT
iterations. If we start from a small mass $m/(nN)$,
then the final mass will be $m$. In the limit $nN \to \infty$ (i.e. we start from an infinitesimal mass) we obtain a perfect action at finite mass.
In this context, ``perfect'' means that dimensionless quantities
do not depend on the lattice spacing, hence they are identical
to the continuum values.

For the above transformation~(\ref{trafo}), this perfect action
can be computed analytically in momentum space~\cite{BW96}.
The computation simplifies if we let $n\to \infty$, so
that $N=1$ is sufficient. Hence we start from the continuum
action now, and the perfect action takes the form
\begin{eqnarray}
S^{*}[\bar \Psi , \Psi ] &=& \frac{1}{(2\pi )^{d}}
\int_{-\pi}^{\pi} d^{d}p \ \bar \Psi (-p) \Delta^{*} (p)^{-1} \Psi (p) \ , 
\nonumber \\
\Delta^{*}(p) &=& \sum_{l \in \Z^{d}} \frac{\Pi (p+ 2\pi l)^{2}}
{i (p_{\mu}+2\pi l_{\mu}) \gamma_{\mu}+m} + \alpha \ , \quad \nonumber \\
\Pi (p) &=& \prod_{\mu =1}^{d} \frac{2 \sin (p_{\mu} /2)}{p_{\mu} }\ ,
\end{eqnarray}
where $\Delta^{*}$ is the free perfect propagator. The same perfect action
is obtained starting from a variety of lattice actions~\cite{Biet94FPA}, in particular
from the Wilson fermion action.

In coordinate space we write this action as
\begin{equation}
S^{*}[\bar \Psi , \Psi ] = \sum_{x,r} \bar \Psi_{x}
[ \rho_{\mu}(r)\gamma_{\mu} + \lambda (r) ] \Psi_{x+r} \ .
\label{lam}
\end{equation}

At the cost of broken chiral symmetry  we obtain a theory with its couplings in $\rho_{\mu}$ and $\lambda$ decaying exponentially as $\vert r \vert  $ increases. 
An exception is the case $d=1$, where they are confined to one
lattice spacing for the special choice
\begin{equation}
\alpha = \frac{e^{m}-m -1}{m^{2} }\ .
\label{alpha_choice}
\end{equation}
It is also in agreement with the Nielsen-Ninomiya theorem that, in order to obtain a local perfect action, one must break chiral symmetry explicitly.
Although the chiral symmetry is not manifest in the perfect action, all chiral properties are still correctly reproduced by it. This is due to famous Ginsparg-Wilson relation, introduced before in~(\ref{GWR}), to which the perfect action obeys,

\begin{equation}
 \{ \Delta^{*}(p), \gamma_5  \}=2\alpha \gamma_5\ .
\end{equation}
Here we can identify $\alpha$ with the mass parameter $\mu$, namely,
\begin{equation}
\alpha=\frac{1}{2\mu} \ .
\end{equation}
 
It turns out that for the choice of $\alpha$ in Eq.~(\ref{alpha_choice}) the locality is also
excellent in higher dimensions, \ie, the exponential decay of the
couplings is very fast.  This is important, because for practical
purposes the couplings have to be truncated to a short range, and the
truncation should not distort the perfect properties too much. Ref.~\cite{BIE96} proposed a nice truncation scheme that uses periodic boundary conditions over 3
lattice spacings and thus confines the couplings to a unit hypercube.
It was pointed out that the spectral and thermodynamic properties of the HF are still drastically improved compared to Wilson fermions~\cite{BIE96,BIE98}.

\begin{figure}[htb]
\centerline{\includegraphics[width=.5\textwidth]{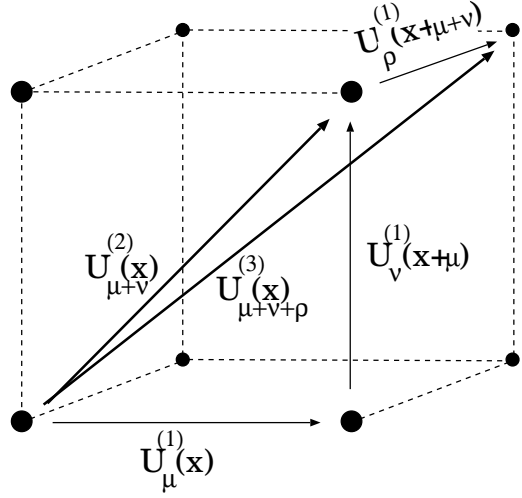}}
\caption{1-space, 2-space and 3-space hyper-links.
\label{links}}
\end{figure}

It is far more difficult to construct an approximately
perfect action for a complicated interacting theory like QCD.
However, as it was proposed in Ref.~\cite{BIE96,Orginos:1997fh,Bietenholz1998wx}  one can just use a simple ansatz for HF together with the standard gauge link variables.  Apart from nearest neighbors, one also
has couplings over 2, 3 and 4-space diagonals in the unit hypercube.
One connects all these coupled sites by all possible {\em shortest}
lattice paths, by multiplying the compact gauge fields on the path
links. This procedure was called ``minimal gauging''.  Note that
one can connect two sites $x$ and $y$ lying on $d$-space
diagonals via $d!$ such shortest lattice paths.  One averages over all
of them to construct the hyper-link, see Figure~\ref{links} which we adopted from Ref.~\cite{Bietenholz1998wx}.
The hyper-link $U^{(1)}_{\mu}(x)$ between site $x$ and
$x+\hat{\mu}$ is identified with $U_{x,\mu}$, and we denote the hyper-link in
plane, cube and hyper-cube as $U^{(2)}_{\mu+\nu}(x)$,
$U^{(3)}_{\mu+\nu+\rho}(x)$, and $U^{(4)}_{\mu+\nu+\rho+\sigma}(x)$,
respectively.  Then we can write the corresponding fermion matrix in
terms of the hyper-links which are constructed recursively starting
from the gauge links $U^{(1)}_{\mu}$,
\begin{eqnarray}
U^{(d)}_{\mu_1+\mu_2+\dots+\mu_d}(x)
=&\frac{1}{d}\Big[&
U^{(1)}_{\mu_1}(x)\,
U^{(d-1)}_{\mu_2+\mu_3+\dots+\mu_d}(x+\hat{\mu}_1) \nonumber\\
&+&U^{(1)}_{\mu_2}(x)\,
U^{(d-1)}_{\mu_1+\mu_3+\dots+\mu_d}(x+\hat{\mu}_2) \nonumber\\
&+&\dots\nonumber\\
&+&U^{(1)}_{\mu_d}(x)\,
U^{(d-1)}_{\mu_1+\mu_2+\dots+\mu_{d-1}}(x+\hat{\mu}_d)\;\;\Big]\ .
\end{eqnarray}

It is convenient to introduce prefactors which are functions of the
HF hopping parameters $\kappa_i$ and $\lambda_i$, $i=1,\dots,4$, and
sums of $\gamma$-matrices%
\begin{eqnarray}
\Gamma_{\pm\mu} &=& \lambda_1 +\kappa_1(\pm\gamma_{\mu})\nn\\
\Gamma_{\pm\mu\pm\nu} &=& \lambda_2 +\kappa_2(\pm\gamma_{\mu}\pm\gamma_{\nu})\nn\\
\Gamma_{\pm\mu\pm\nu\pm\rho} &=& \lambda_3 +\kappa_3(\pm\gamma_{\mu}\pm\gamma_{\nu}\pm\gamma_{\rho})\nn\\
\Gamma_{\pm\mu\pm\nu\pm\rho\pm\sigma} &=& \lambda_4
+\kappa_4(\pm\gamma_{\mu}\pm\gamma_{\nu}\pm\gamma_{\rho}\pm\gamma_{\sigma})\, .
\label{lamm}
\end{eqnarray}
Note that the $\lambda_i$ in Eq.~(\ref{lamm}) differ from $\lambda(r)$ in
Eq.~(\ref{lam}) by a normalization factor $\frac{1}{\lambda_0}$. The
$\kappa_i$ arise from $\rho_{\mu}(r)$ by the same normalization. The couplings are given in Ref.~\cite{BIE96} and in Chapter 4.

The hypercube operator is organized in sums which run over four different directions for two 1-space links, six directions for four 2-space links, four directions
for eight 3-space links, and one direction for the sixteen 4-space
links. Altogether 80 hyper-links plus the term $\delta_{x y}$ contribute.
With each path of the free HF, a $\gamma$-matrix is associated.  The $\gamma$-matrices are chosen so that they add up to produce a prefactor $\Gamma$, see Eq.~(\ref{lamm}), which is associated with a given hyper-link.

The 1-space links $U^{(1)}_{\mu}(x)$ are identical to the
Hermitian-conjugate links in negative direction,
$U^{\dagger}_{x+\hat\mu, -\mu}$, and this feature also holds for
the hyper-links, \eg
\begin{equation}
U^{\dagger(d)}_{\mu_1+\mu_2+\dots+\mu_d}(x)=
U^{(d)}_{-\mu_1-\mu_2-\dots-\mu_d}(x+\hat\mu_1+\hat\mu_2+\dots+\hat\mu_d)\ .
\end{equation}
Therefore, only one half of the $3^d-1$ hyper-links has to be computed and stored in
the implementation of the HF. 

As in the case of Wilson fermions, the HF matrix exhibits the ``$\gamma_5$-Hermiticity''
\begin{equation}
\gamma_5\,D_{HF}\gamma_5=D_{HF}^{\dagger}\ ,
\end{equation}
i.e.  $D_{HF}$ is non-Hermitian but its eigenvalues come in
complex-conjugate pairs.
\renewcommand{\arraystretch}{1}
\newcommand{\pls}{{\!\!\!+\!}}
\newcommand{\summm}{\!\sum}
The corresponding HF matrix is defined by~\cite{Bietenholz1998wx}

\begin{eqnarray}
D_{HF}(x,y)& =& \lambda_{0}\Bigg\{ 
\delta_{x,y} \nonumber\\
+\summm\limits_{\mu}\Big[
\Gamma_{+\mu} U^{(1)}_\mu(x)         \delta_{x,y-\hat \mu}&\pls&
\Gamma_{-\mu}
U^{(1)}_{-\mu}(x)       \delta_{x,y+\hat\mu}\;\;\Big]\nonumber\\
\nonumber\\
+\summm\limits_{{\mu}}\summm\limits_{{\nu>\mu}}\Big[\nonumber\\
\Gamma_{+\mu+\nu}
U^{(2)}_{+\mu+\nu}(x)   \delta_{x,y-\hat\mu-\hat\nu}&\pls&
\Gamma_{+\mu-\nu}
U^{(2)}_{+\mu-\nu}(x)   \delta_{x,y-\hat\mu+\hat\nu}\;\;\nonumber\\
                              +
\Gamma_{-\mu+\nu}
U^{(2)}_{-\mu+\nu}(x)   \delta_{x,y+\hat\mu-\hat\nu}&\pls&
\Gamma_{-\mu-\nu}
U^{(2)}_{-\mu-\nu}(x)   \delta_{x,y+\hat\mu+\hat\nu}\;\;\Big]\nonumber\\
\nonumber\\
+\summm\limits_{{\mu}}
\summm\limits_{{\nu>\mu}}
\summm\limits_{{\rho>\nu}}\Big[\nonumber\\
\Gamma_{+\mu+\nu+\rho}
U^{(3)}_{\mu+\nu+\rho}(x)
\delta_{x,y-\hat\mu-\hat\nu-\hat\rho}&\pls&
\Gamma_{+\mu+\nu-\rho}
U^{(3)}_{\mu+\nu-\rho}(x)   \delta_{x,y-\hat\mu-\hat\nu+\hat\rho}\nonumber\\
+
\Gamma_{+\mu-\nu+\rho}
U^{(3)}_{\mu-\nu+\rho}(x)   \delta_{x,y-\hat\mu+\hat\nu-\hat\rho}&\pls&
\Gamma_{+\mu-\nu-\rho}
U^{(3)}_{\mu-\nu-\rho}(x)   \delta_{x,y-\hat\mu+\hat\nu+\hat\rho}\;\;\nonumber\\
+
\Gamma_{-\mu+\nu+\rho}
U^{(3)}_{-\mu+\nu+\rho}(x)
\delta_{x,y+\hat\mu-\hat\nu-\hat\rho}&\pls&
\Gamma_{-\mu+\nu-\rho}
U^{(3)}_{-\mu+\nu-\rho}(x)   \delta_{x,y+\hat\mu-\hat\nu+\hat\rho}\nonumber\\
+
\Gamma_{-\mu-\nu-\rho}
U^{(3)}_{-\mu-\nu+\rho}(x)
\delta_{x,y+\hat\mu+\hat\nu-\hat\rho}&\pls&
\Gamma_{-\mu-\nu-\rho}
U^{(3)}_{-\mu-\nu-\rho}(x)   \delta_{x,y+\hat\mu+\hat\nu+\hat\rho}\;\;\Big]\nonumber\\
\nonumber\\
+\summm\limits_{{\mu}}
\summm\limits_{{\nu>\mu}}
\summm\limits_{{\rho>\nu}}
\summm\limits_{{\sigma>\rho}}\Big[\nonumber\\
\Gamma_{+\mu+\nu+\rho+\sigma}
U^{(4)}_{\mu+\nu+\rho+\sigma}(x)\delta_{x,y-\hat\mu-\hat\nu-\hat\rho-\hat\sigma}&\pls&
\Gamma_{+\mu+\nu+\rho-\sigma}
U^{(4)}_{\mu+\nu+\rho-\sigma}(x)\delta_{x,y-\hat\mu-\hat\nu-\hat\rho+\hat\sigma}\nonumber\\
+\Gamma_{+\mu+\nu-\rho+\sigma}
U^{(4)}_{\mu+\nu-\rho+\sigma}(x)\delta_{x,y-\hat\mu-\hat\nu+\hat\rho-\hat\sigma}&\pls&
\Gamma_{+\mu+\nu-\rho-\sigma}
U^{(4)}_{\mu+\nu-\rho-\sigma}(x)\delta_{x,y-\hat\mu-\hat\nu+\hat\rho+\hat\sigma}\nonumber\\
+\Gamma_{+\mu-\nu+\rho+\sigma}
U^{(4)}_{\mu-\nu+\rho+\sigma}(x)\delta_{x,y-\hat\mu+\hat\nu-\hat\rho-\hat\sigma}&\pls&
\Gamma_{+\mu-\nu+\rho-\sigma}
U^{(4)}_{\mu-\nu+\rho-\sigma}(x)\delta_{x,y-\hat\mu+\hat\nu-\hat\rho+\hat\sigma}\nonumber\\
+\Gamma_{+\mu-\nu-\rho+\sigma}
U^{(4)}_{\mu-\nu-\rho+\sigma}(x)\delta_{x,y-\hat\mu+\hat\nu+\hat\rho-\hat\sigma}&\pls&
\Gamma_{+\mu-\nu-\rho-\sigma}
U^{(4)}_{\mu-\nu-\rho-\sigma}(x)\delta_{x,y-\hat\mu+\hat\nu+\hat\rho+\hat\sigma}\;\;\nonumber\\
+
\Gamma_{-\mu+\nu+\rho+\sigma}
U^{(4)}_{-\mu+\nu+\rho+\sigma}(x)\delta_{x,y+\hat\mu-\hat\nu-\hat\rho-\hat\sigma}&\pls&
\Gamma_{-\mu+\nu+\rho-\sigma}
U^{(4)}_{-\mu+\nu+\rho-\sigma}(x)\delta_{x,y+\hat\mu-\hat\nu-\hat\rho+\hat\sigma}\nonumber\\
+\Gamma_{-\mu+\nu-\rho+\sigma}
U^{(4)}_{-\mu+\nu-\rho+\sigma}(x)\delta_{x,y+\hat\mu-\hat\nu+\hat\rho-\hat\sigma}&\pls&
\Gamma_{-\mu+\nu-\rho-\sigma}
U^{(4)}_{-\mu+\nu-\rho-\sigma}(x)\delta_{x,y+\hat\mu-\hat\nu+\hat\rho+\hat\sigma}\nonumber\\
+\Gamma_{-\mu-\nu+\rho+\sigma}
U^{(4)}_{-\mu-\nu+\rho+\sigma}(x)\delta_{x,y+\hat\mu+\hat\nu-\hat\rho-\hat\sigma}&\pls&
\Gamma_{-\mu-\nu+\rho-\sigma}
U^{(4)}_{-\mu-\nu+\rho-\sigma}(x)\delta_{x,y+\hat\mu+\hat\nu-\hat\rho+\hat\sigma}\nonumber\\[0pt]
+\Gamma_{-\mu-\nu-\rho+\sigma}
U^{(4)}_{-\mu-\nu-\rho+\sigma}(x)\delta_{x,y+\hat\mu+\hat\nu+\hat\rho-\hat\sigma}&\pls&
\Gamma_{-\mu-\nu-\rho-\sigma}
U^{(4)}_{-\mu-\nu-\rho-\sigma}(x)\delta_{x,y+\hat\mu+\hat\nu+\hat\rho+\hat\sigma}
\;\;\nonumber\\
&\Big]\Bigg\}\ .
&\qquad
\qquad
\qquad
\qquad
\qquad
\qquad
\qquad
\label{HFMATRIX}
\end{eqnarray}

%---------------------------------------------------------------------
\subsection{The overlap hypercube Dirac operator}

In this Section we would like to present motivations for a non-standard choice of the kernel in the overlap formula.

The generalization of the overlap fermions to a whole class of solutions of the GWR and the motivation for considering alternative kernels $\hat{D}$ were given in Ref.~\cite{Bie1998ut}. In particular it was pointed out that the Ginsparg-Wilson fermions reproduce themselves if the kernel already obeys GWR as we showed in Section 2.5.3. To put it the other way: if the kernel does not obey the GWR then it gets corrected by the overlap formula~\cite{Bie1998ut}. 
This additional freedom of having different kernels can be used in order to improve such properties of the overlap operator as locality and get rid of the lattice artifacts which reside,  for example, in approximate rotational symmetry. Also the dependence of the observables on different lattice spacings, that is referred to as the scaling behavior, is expected to be improvable.  To this end we start from the kernel that already has good properties in rotational invariance, locality, scaling behavior and approximately satisfies the GWR. If we insert it into the overlap formula then we can expect that due to its already good chiral properties the chiral corrections to the operator will be small and therefore the properties of locality, approximate rotational symmetry and scaling behavior will be essentially inherited by the resulting overlap operator~\cite{Bie1998ut}. A good candidate for that would be the approximate perfect fermions introduced in the previous Subsection. 

When constructing the overlap operator the parameter $\mu$ must be chosen in some interval to ensure that the doublers are projected to the right arc of the GW spectrum. This interval is $\mu \in [0,2]$ for the free Wilson fermions and shrinks from both sides in the interacting case. This procedure can be realized once the gauge configurations are smooth enough (or $\beta$ is sufficiently large) to allow for the spectrum of the kernel develop such a window. It is conceivable that using the hypercube operator as a kernel in the overlap formula one can admit lower values of $\beta$ than it is the case for the Wilson kernel. This is based on the fact that the hypercube operator approximately obeys GW relation and this should lead to a better behavior in its spectrum.  Having the doublers projected to the cut-off scale one can optimize the overlap operator further by requiring the best locality properties or the least condition number for the $A^\dagger A$ operator. This will specify $\mu$ and also impose additional constraints on the couplings in the hypercube operator.

%----------------------------------------------------------------------------
\chapter{Lattice simulations}
\section{Quenched simulations of the gauge fields}
The QCD partition function in the formalism of the path integral is defined by an integral with an infinite number of dimensions which cannot be analytically solved so far. However its regularization on a finite lattice is reduced to an integral over a finite number of dimensions and can be used to address non-perturbative physics. To this end one can use the Monte Carlo (MC) simulations to evaluate the lattice regularized partition function.

In lattice MC simulations of quenched QCD one usually generates equilibrium gauge configurations of $SU(3)$ fields $U$ with a probability distribution proportional to $W[U]= e^{-S_{YM}[U]}$ at a certain value of the strong gauge coupling. This procedure is referred to as importance sampling. It can be simulated by generating a Markov chain of configurations. One parameterizes the Markov chain with $\tau$ and introduces the transition probability matrix $P([U_{\tau+1}]\leftarrow [U_\tau])$ connecting the field configuration at simulation time $\tau$ and $\tau+1$. A condition to arrive at the equilibrium distribution in the limit of large $\tau$ is detailed balance
\begin{equation}
P(\,[U_{\tau+1}]\leftarrow [U_\tau]\,) \,W[U_\tau]=P(\, [U_{\tau}]\leftarrow [U_{\tau+1}]\, )\, W[U_{\tau+1}]\ .
\end{equation}
  For quenched simulations it is numerically favorable to perform a link by link update of the given configuration instead of the update of it as a whole. The process when all links of the field configuration have been once updated is referred to as a configuration update or one sweep.
In quenched $SU(3)$ gauge simulations two transition probability distributions are mostly used. These are the heat bath~\cite{Creutz:1980zw} and Metropolis~\cite{metropolis} transition probability distributions. The heat bath was conventionally designed for the generation of $SU(2)$ pure gauge configurations. To adopt it for quenched simulations with the $SU(3)$ gauge group one needs to use the Cabbibo-Marinari update~\cite{Cabibbo:1982zn} whereby one splits an $SU(3)$ link variable in three $SU(2)$ link variables and performs the heat bath update over each of them. Then these three $SU(2)$ link variables are reassembled again in the updated $SU(3)$ link variable.
In Markov processes one has to consider the autocorrelation of the configurations. Indeed since every next configuration is produced from the previous one, it contains information about it. This spoils the sampling of observables which in the spirit of MC simulations are supposed to be computed on an uncorrelated set of configurations picked out with the probability distribution $W[U]= e^{-S_{YM}[U]}$.
 Since the quenched updates are computationally quite fast we can arrange for many of them in-between to make sure that the resulting configurations become uncorrelated.
 To cope with the autocorrelations and thereby to speed up the updating process it was proposed~\cite{Adler:1981sn,Adler:1988ce,Brown:1987rr,Creutz:1987xi} also to choose the trial elements for the update in a particular way. This is the so-called overrelaxation update. It is achieved by choosing the new link $U'_{x,\mu}$ as far as possible from $U_{x,\mu}$. The overrelaxation update for $SU(N)$ gauge fields relies on the particular form of the lattice gauge action.
The part of the gauge action containing one link variable $U_{x,\mu}$ must have the form
\begin{equation} 
\Delta S[U_{x,\mu}]=\mbox{const}\cdot \, {\rm Re\, Tr}(U_{x,\mu} V)\ ,
\label{general_struc_action}
\end{equation}
where the matrix $V$ does not necessarily belong to $SU(N)$.
This is the case in particular for the Wilson gauge action~(\ref{eq:wilson_action}) that we are using.
%------------------------------------------------------------------------------------
 \section{A numerical treatment of the overlap Dirac operator}
In this Subsection we discuss numerical details of simulations of the overlap Dirac operator.

 The formula of the overlap Dirac operator contains the square root of a Hermitian operator.
It is instructive to rewrite it as
\begin{eqnarray}
D_{ov}&=&\mu \left [ 1+ \gamma_5\frac{Q}{\sqrt{Q^2}}  \right]\ , \nonumber \\
 Q&=&c\gamma_5(\hat{D}-\mu) \ , \quad \hat{D}=D_{W} \ .\label{forQ2oper}
\end{eqnarray}
$Q$ is a Hermitian operator and constant $c$ was chosen to ensure that the spectrum of $Q^2$ is bounded by $1$. 
  For our treatment of the square root we used the Chebyshev polynomials $T_n(x)$ to approximate the function $1/\sqrt{x}$ in the interval ($\epsilon$,1), where $\epsilon$ is the inverse condition number of the operator $Q^2$. This method was proposed in Ref.~\cite{Hernandez:1998et}%,Hernandez:1999cu}.

The Chebyshev polynomials $T_i$ are the orthonormal polynomials with respect to the inner product
\begin{equation}
\langle T_i,\, T_j\rangle =\int_{-1}^{1}\frac{1}{\sqrt{(1-t^2)}}T_i(t)T_j(t)\, dt\ .
\end{equation}
We thus have
\begin{equation}
\langle T_i,\,T_j \rangle=\delta_{ij}\ .
\end{equation}
Every function $f$ for which $\langle f,\,f \rangle$ exists can be expanded into its Chebyshev series
\begin{equation}
f(x)=\sum_{i=0}^{\infty} c_i T_i(x) \quad \mbox{where} \quad c_i=\langle f,\,T_i \rangle \ .
\end{equation}

Truncating the series at the $N$-th summand gives the polynomial approximation of degree $N-1$
\begin{equation}
P_{N-1}(x)=\sum_{i=0}^{N-1}c_iT_i(x) \ .
\label{chebyshev_serie}
\end{equation}

The Chebyshev polynomials $T_n(x)$ have exactly $n$ zeros on the interval $[-1,1]$, and they are located at the points

\begin{equation}
x_k=\cos \left (\frac{\pi (k-0.5)}{n} \right ) \quad \mbox{where} \quad k=1,2,\dots ,n \ .
\end{equation}

In the same interval there are $n + 1$ extrema (maxima and minima), located at
\begin{equation}
x_k=\cos \left (\frac{\pi k}{n} \right ) \quad \mbox{where} \quad k=0,1,\dots ,n\ .
\end{equation}
At all of the maxima we have $T_n(x) = 1$, while at all of the minima $T_n(x) =  -1$. This property  makes the Chebyshev polynomials so useful in polynomial approximation of functions. It provides the basis for deducing the bounds of the maximal error in the approximation. 

The coefficients $c_j$ of the series~(\ref{chebyshev_serie}) are given by
\begin{equation}
c_j=\frac{2}{N}\sum_{k=1}^{N}f(x_k)T_j(x_k)\ , \quad \mbox{where} \quad j=0, \dots, N-1\ ,
\label{cheb_coeff}
\end{equation}
where $x_k$ are the $N$ zeros of $T_N(x)$.
Then for an arbitrary function $f$ on the interval $[-1,1]$ the approximation formula 
\begin{equation}
f(x)\simeq P_{N-1}(x) = \left [  \sum_{k=0}^{N-1} c_k T_k(x) \right ] -\frac{1}{2} c_0
\label{cheb_appr}
\end{equation}
is exact at $x_1, \dots ,x_N$.

For any point $x\in [-1,1]$ this gives the Chebyshev approximation of degree $N-1$ for function $f(x)$.
The Chebyshev approximation is not the best approximation with respect to all polynomials of the same degree, i.e. it does not have the smallest maximum deviation from the true general function. The latter is true for the so-called minimax polynomials. However the Chebyshev approximation is almost as good as the minimax polynomials. 
The advantage of the Chebyshev polynomials is that the formulae~(\ref{cheb_coeff}) and~(\ref{cheb_appr}) can be arranged in one recurrence process, the Clenshaw recurrence formula~\cite{bookNR}. This property makes the Chebyshev polynomials attractive for numerical simulations since one can avoid the calculation and use of the $c_j$ coefficients which can take large magnitudes and bring in numerical instabilities.

In our simulations we made the following approximation

\begin{equation}
\frac{1}{\sqrt{t}}=P_{N, \epsilon}(t)\ , \quad \mbox{where}\quad t\in [\epsilon,1] \ ,
\end{equation}
and $\epsilon=\lambda_{\rm min}/\lambda_{\rm max}$, where $\lambda_{\rm min}$ and $\lambda_{\rm max}$ are the minimal and maximal eigenvalues of $Q^2$, respectively.

Inserting the approximation to the square root we obtain for the overlap Dirac operator
\begin{equation}
D_{\rm ov}\simeq \mu \left [ 1+ \gamma_5{Q}P_{N,\epsilon}(Q^2)  \right]\ .
\end{equation}
 
The degree $N$ of the polynomial approximation was chosen in a way to obtain a specific precision $\epsilon_{\rm overlap}$ for the square root approximation

\begin{equation}
\frac{\Vert X- Q^2P_{N,\epsilon}(Q^2)^2X\Vert}{4\Vert X \Vert}\le \epsilon_{\rm overlap}\ ,
\end{equation} 
where $X$ is a random spinor vector and $\epsilon_{\rm overlap}$ was usually chosen between $10^{-12}$ and $10^{-15}$. We checked that this condition is insensitive to different choices of the random vector $X$.
It also implies a similar precision for the GWR.

The computational cost of the approximation is roughly proportional to the polynomial degree. The latter is proportional to the square root of the condition number.
In order to decrease the condition number of the $Q^2$ operator, $\lambda_{\rm max}/\lambda_{\rm min}$, and thus to decrease the required polynomial degree, we projected out the lowest eigenmodes,
\begin{equation}
Q^2_{\rm proj}=Q^2 -\sum_{k=1}^{m} \lambda_k^2|k\rangle \langle k| \ , \quad \mbox{where} \quad Q^2|k\rangle =\lambda_k^2 |k\rangle \ . 
\end{equation}    
Then the overlap formula takes the form
\begin{equation}
D_{ov}=\mu \left [ 1+ \gamma_5{Q_{\rm proj}}P_{N,\epsilon}(Q_{\rm proj}^2) + \gamma_5\sum_{k=1}^{m}|k \rangle\langle k|\,  {\rm sign}(\lambda_k)  \right]\ ,
\end{equation} 
and the condition number has to be evaluated with respect to the operator $Q^2_{\rm proj}$.  
To render the chiral symmetry reliable and assure the convergence of the polynomial approximation, the eigenvectors of the operator $Q^2$ should be computed to a very good accuracy. We take it again to be between $10^{-12}$ and $10^{-15}$.  
 
For $\beta=5.85$ and $\beta=6$ on lattices of $V=8^4$, $10^4$ we projected out $10$ eigenvectors. It turned out that projecting out more of them does not help much any more. This is due to the fact that above a certain energy the eigenvalues are densely packed and further increasing the number of projected out eigenvectors does not influence much the condition number.  
\begin{figure}%[htbp]
\centering
\includegraphics[width=0.6\textwidth, angle=-90]{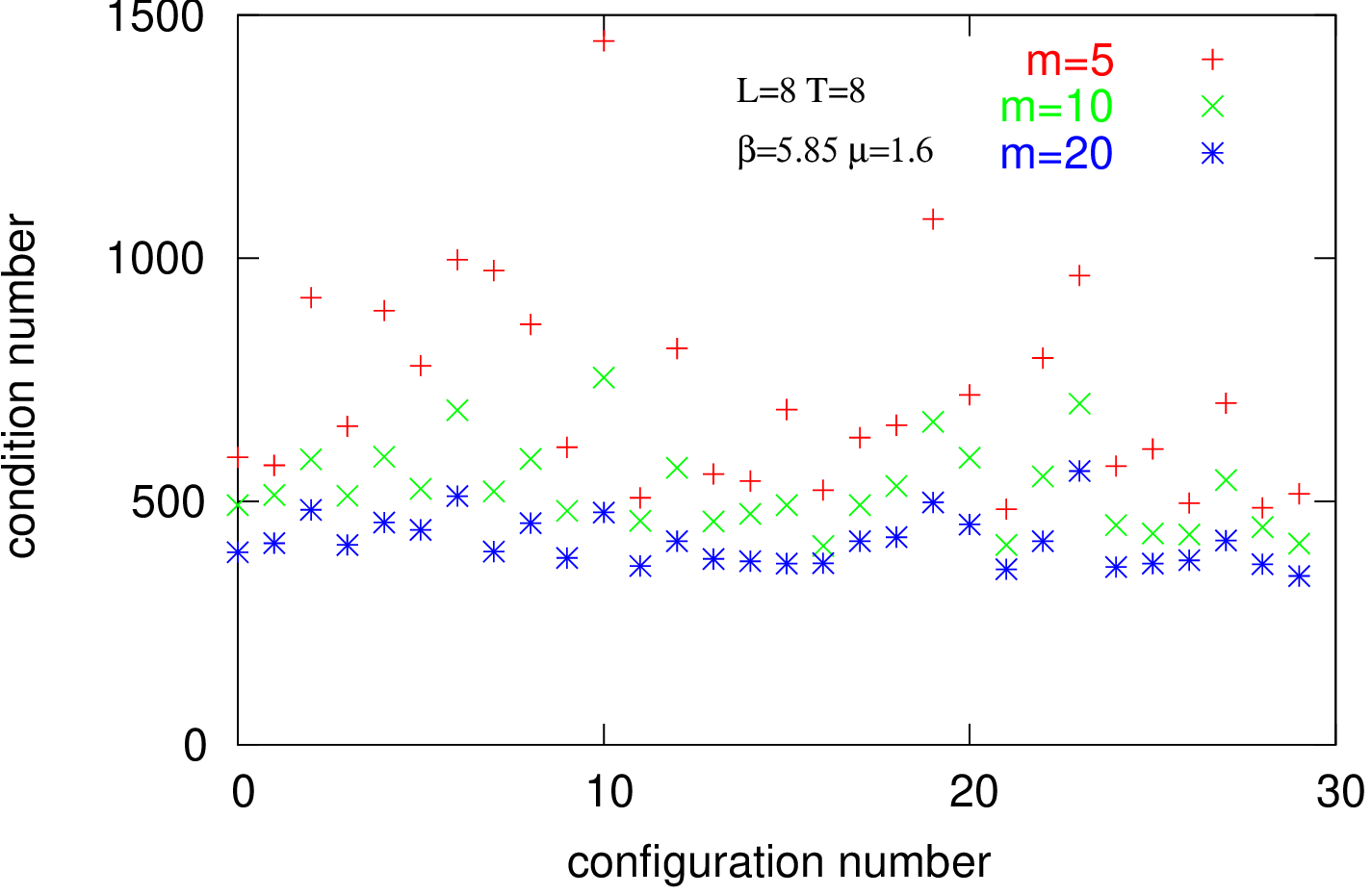}
\includegraphics[width=0.6\textwidth,angle=-90]{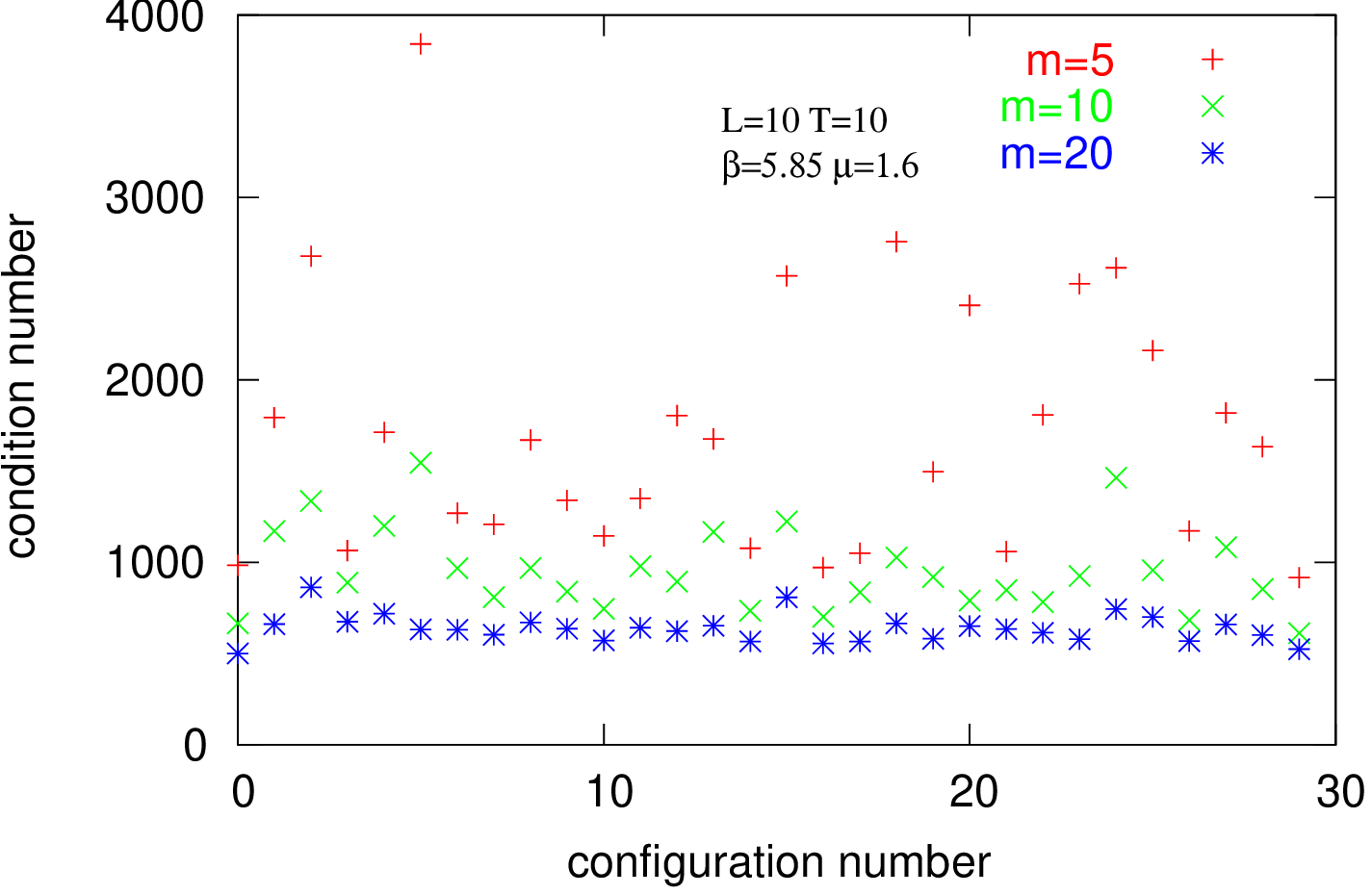}
%\end{rotate}
\caption{The condition number of the $Q^2$ operator in $V=8^4$ and $10^4$ volumes at $\beta=5.85$ and $\mu=1.6$ after $m=5$, $10$ and $20$ eigenvectors being projected out.}
\label{condnumber_L8b5.85mu1.6}
\end{figure}
In Figure~\ref{condnumber_L8b5.85mu1.6} we show the condition number for the lattice volumes $V=8^4$ and $10^4$ at $\beta=5.85$ and $\mu=1.6$. Here we projected out $5$, $10$ and $20$ eigenvectors, respectively.
The corresponding degree of the polynomial approximation varies for different lattice volumes and $\beta$. For instance on the $10^4$ lattice at $\beta=5.85$, for $44$ projected out eigenvectors it was about $180$ at $\epsilon_{\rm overlap}=10^{-15}$.

In addition one has to assure that one deals with  a local formulation of the overlap fermions.
The locality of the overlap Wilson fermions is related to the spectrum of its $Q^2$ operator.
Let the spectrum be bounded by $u$ from below and by $1$ from above 
\begin{equation}
u \le \lambda(Q^2) \le 1 \ .
\end{equation}
If $u$ comes too close to the origin and the spectrum develops also very dense eigenvalues then this will lead to strong fluctuations in the inverse square root and locality will be lost. Note, however, that the presence of a few near-zero eigenvalues does not spoil the locality.
The issue of locality was addressed in Ref.~\cite{Hernandez:1998et} where also a numerical study at $\beta=6, \, 6.2, \, 6.4$ was presented.  Analytically the locality properties of the overlap operator can be established for smooth configurations. Once each plaquette variable $U^{(P)}$ defined in Eq.~(\ref{eq:wilson_action}) obeys an inequality of the form
\begin{equation}
|1-\frac{1}{3} {\rm Re \,Tr}\, U^{(P)}| \le \epsilon \quad \mbox{for all plaquettes} 
\label{plaquet_smothness}
\end{equation}
then the operator $Q^2$ is uniformly bounded from below and the locality is guaranteed. In particular, for $\mu=1$ Ref.~\cite{Hernandez:1998et} found the condition
\begin{equation}
\frac{Q^2}{c^2}\ge 1-30\epsilon\ .
\label{AdaggerAcond}
\end{equation}
Even if this condition is not obeyed, numerical results can still reveal locality.
In particular in the simulations at $\beta=6, \, 6.2, \, 6.4$  the corresponding overlap operator can be made local~\cite{Hernandez:1998et}. The optimal locality is achieved by varying the parameter $\mu$. The optimized values of $\mu$ at $\beta=6, \, 6.2, \, 6.4$ were presented in  Ref.~\cite{Hernandez:1998et} and in Ref.~\cite{Hernandez:1999cu} the value of $\mu$ that guarantees a good locality at $\beta=5.85$ was pointed out.

\begin{figure}%[htbp]
\hspace{-1cm}\includegraphics[width=0.55\textwidth]{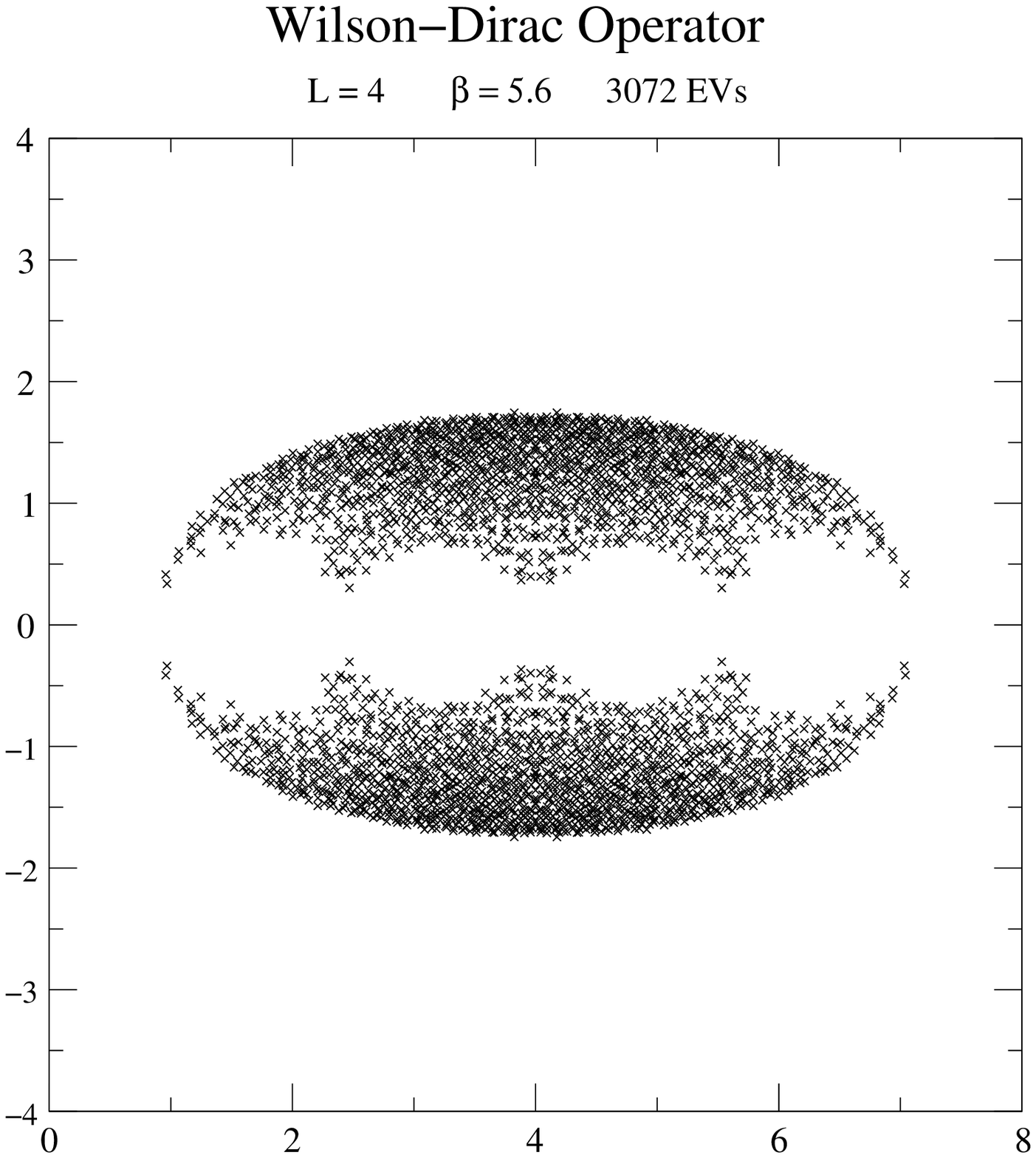}\hspace{-1cm}%
\includegraphics[width=0.55\textwidth]{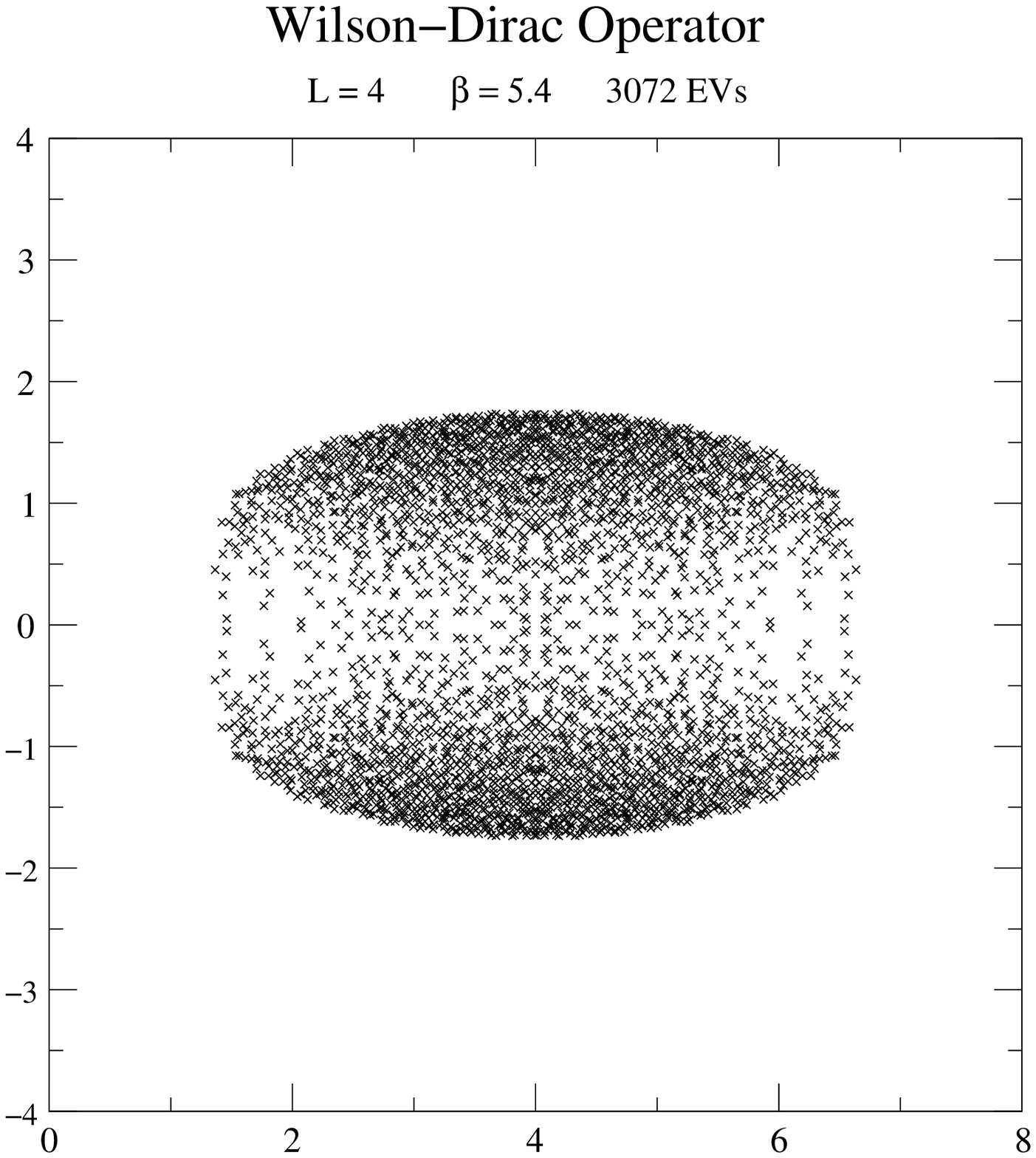}
%\end{rotate}
\caption{Eigenvalues of the Wilson operator at $\beta=5.6$ and $\beta=5.4$ on the lattice $4^4$. We see that at $\beta=5.4$ the spectrum does not develop a window.}
\label{eigenvalues_wilson}
\end{figure}

The mass $\mu$ separates doublers from the ``physical'' eigenvalues of the Wilson operator, which are projected onto the left arc of the GW fermion spectrum. This imposes a restriction on $\mu$ to be between $0\le \mu \le 2$ in the free case. In the interacting case the interval shrinks from both sides. The separation can be done unambiguously only if the spectrum of the Wilson operator has a window with a negligible eigenvalue density. The latter cannot be achieved for the Wilson operator at strong $\beta$. Therefore one rather has to work in a range of moderate values of $\beta$. For illustration of this behavior we show in Figure~\ref{eigenvalues_wilson}~\cite{Bietenholz:2000iy}  two plots of the distribution of eigenvalues of the Wilson operator for quenched QCD configurations at $\beta=5.6$ and $\beta=5.4$ on the lattice of volume $4^4$. The left plot shows a spectrum with such a window while the right plot does not have it. Note that these distributions are done for quite small lattices and due to the boundary conditions the configurations are smother than they would be in a larger volume at the same $\beta$. Hence the window in the spectrum opens up at somewhat lower values of $\beta$ than it would be expected for large lattice volumes. 

Once the window was identified one can use the freedom of varying the value of $\mu$ within the window to optimize the locality properties further. This can be done by comparing the exponential decays of the couplings of the overlap operator at different $\mu$.  It turned out that the locality breaks down once $\beta$ is below the value about $5.7$~\cite{Jansenprivate}.

\begin{figure}%[htbp]
\centering
\includegraphics[width=0.4\textwidth,angle=-90]{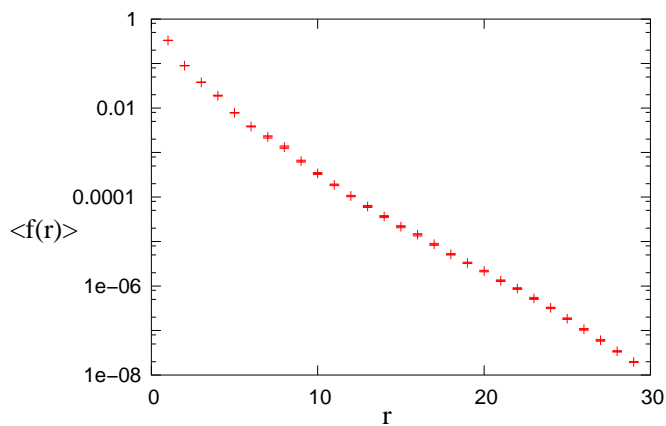}
%\end{rotate}
\caption{Locality of the overlap Wilson operator at $\beta=5.85$, $\mu=1.6$ on the lattice of volume $12^3\times 24$. The exponential decay is clearly pronounced.}
\label{locality_ovwilson}
\end{figure}

According to Ref.~\cite{Hernandez:1998et} it is optimal for locality to put $\mu=1.4$ at $\beta=6$ and from Ref.~\cite{Hernandez:1999cu} $\mu=1.6$ at $\beta=5.85$. 
In Figure~\ref{locality_ovwilson} we show the locality property of the overlap Wilson operator at $\beta=5.85$, $\mu=1.6$. We take the "taxi driver metrics" $\Vert x-y\Vert_{\rm taxi}$ to measure the distance on the lattice
\begin{equation}
\Vert x-y\Vert_{\rm taxi}=\sum_{\mu=1}^{4}|x_{\mu}-y_{\mu}|\ .
\end{equation}
The locality is measured in terms of $f(r)$

\begin{equation}
f(r)=\max_{x}||\sum_y   {(D_{\rm ov})}_{xy} \psi_{y}|| \ ,\quad \mbox{where} \quad r=||x-x_0||_{\rm taxi} \quad \mbox{and} \quad \psi_y=\delta_{yx_0}\ .
\end{equation}
%\section{Error estimates}
We see in Figure~\ref{locality_ovwilson} that the decay is safely exponential and therefore the operator is local.

\chapter{Numerical treatment of the hypercube fermions}
In this Chapter we discuss the simulation of the hypercube Dirac operator.
First we consider the hypercube operator itself. We tune it to decrease the remaining additive mass renormalization and minimize the violation of the GWR, i.e. to improve its chiral behavior.
Secondly we use the hypercube operator as a kernel for the overlap operator.  
Since the hypercube operator is a truncated fixed point operator we expect that the favorable properties such as locality become much better than for the overlap Wilson Dirac operator. 

Note that the hypercube operator would be perfect if we considered it for the free case on the lattice of size $3^4$.
This provides us with an ansatz for the parameters ($\lambda_0, \dots, \lambda_4, \kappa_1, \dots, \kappa_4$) in Eq.~(\ref{HFMATRIX}). We give the corresponding values in Table~\ref{hf_param}.
\begin{table}[H]
\centering
\begin{tabular}{|c|c|}
\hline
$\lambda_0$&1.8527205471652000 \\
\hline
$\lambda_1$&-0.0327938644185406 \\
\hline
$\lambda_2$&-0.0162118524304803 \\
\hline
$\lambda_3$ &-0.0086184721387295 \\
\hline
$\lambda_4$ &-0.0045483448675513 \\
\hline
$\kappa_1$& 0.0738626203107888 \\
\hline
$\kappa_2$& 0.0173136117866899 \\
\hline
$\kappa_3$& 0.0059685910389885 \\
\hline
$\kappa_4$& 0.0025632525141047 \\
\hline
\end{tabular}
\caption{The parameters of the HF.\label{hf_param}}
\end{table}
 On a larger lattice and for the interactive case we adopted the ``minimal gauging'' described in Subsection 2.5.4.  As an ansatz for the couplings we take the values from the free case to start with. However this creates a large additive mass renormalization that was pointed out in Ref.~\cite{Bietenholz:2002ks}. In order to account for it we rescaled the parameters $\lambda_i$ and $\kappa_j$ as it was prescribed in Ref.~\cite{Bietenholz:2002ks}. 
\begin{eqnarray}
\lambda_0&\rightarrow& \lambda_0\ ,\nonumber\\
\lambda_1&\rightarrow &u\lambda_1\ , \quad \kappa_1 \rightarrow uv\, \kappa_1\ ,\nonumber \\
\lambda_2&\rightarrow &u^2\lambda_2\ ,  \quad \kappa_2 \rightarrow (uv)^2\kappa_2\ ,\nonumber \\
\lambda_3&\rightarrow &u^3\lambda_3\ , \quad \kappa_3 \rightarrow (uv)^3\kappa_3\ ,\nonumber \\
\lambda_4&\rightarrow & u^4\lambda_4\ , \quad \kappa_4 \rightarrow (uv)^4\kappa_4\ ,
\end{eqnarray}
where $1/u \le v\le 1$.

If we represent the hypercube operator as $D_{HF}[x,y, U]=\gamma_{\mu} \rho_{\mu} (x,y,U) +\lambda (x,y,U)$ then the rescaling of the couplings would imply a rescaling of the link variable $U_{\mu}(x)$ as follows
\begin{eqnarray}
U_{x,\mu}&\rightarrow &uv\, U_{x,\mu} \quad \mbox{in} \quad \rho_{\mu}(x,y,U) \ , \nonumber \\
U_{x,\mu}&\rightarrow &u\, U_{x,\mu}\ \ \quad \mbox{in} \quad \lambda (x,y,U) \ .
\end{eqnarray}
The factor $u$ is used to compensate the (mean) link suppression due to the gauge field and therefore reduce the mass renormalization. The factor $v$ controls mostly the imaginary part of the eigenvalues and serves to move them closer to the GW circle which we described below Eq.~(\ref{eq:help1}). 

\begin{figure}[H]
\centering
%\begin{displaymath}
%\xymatrix{{x+\nu} \ar[r]_{U_{\mu}(x+\nu)} & {} \ar[dd]|{U^{\dagger}_{\nu}(x+\mu)}\\
 % & \\
%x\ar [uu]|{U_{\nu}(x)}  \ar @{.>}[r]_{U_{\mu}(x)} &
%  {x+\mu}}
%\end{displaymath}
\includegraphics[width=0.3\textwidth]{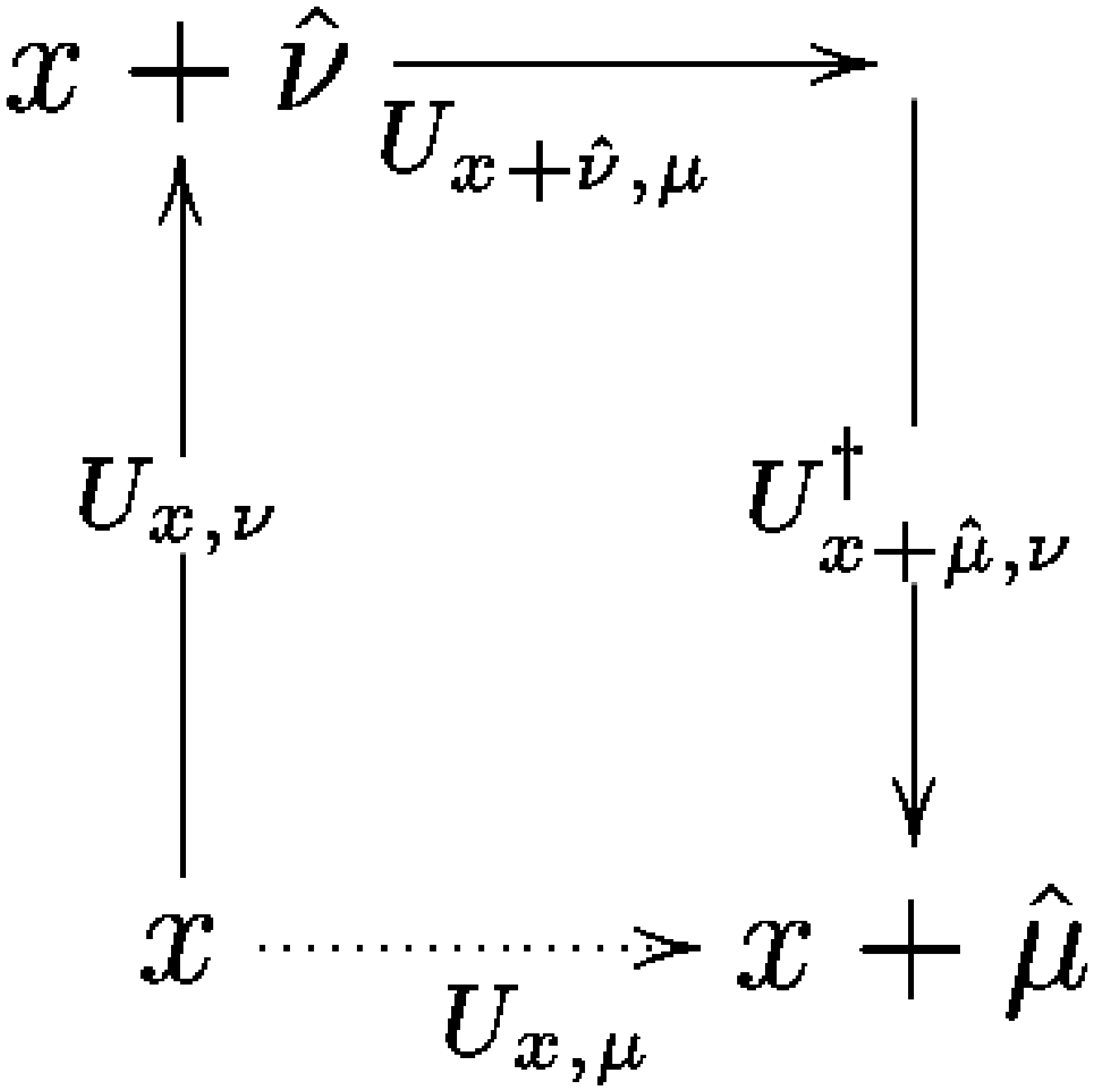}
\caption{A staple in the positive direction for the link variable $U_{x,\mu}$.}
\label{fig:staples}
\end{figure}

A further chiral improvement which was outlined in Ref.~\cite{Bietenholz:2002ks} is the use of the fat link. For a given configuration we substitute each link variable $U_{x,\mu}$ by the following expression
\begin{equation}
U_{x,\mu} \rightarrow (1-\alpha)U_{x,\mu} + \frac{\alpha}{6}\left [ \sum \mbox{staples}\right ]\ ,
\end{equation}
where the staples for the link $U_{x,\mu}$ are depicted on the Figure~\ref{fig:staples}. The staples are calculated as a product of the three links and are given by: $U_{x,\nu}\, U_{x+\hat{\nu},\mu} \, U^{\dagger}_{x+\hat{\mu},\nu}$ in positive direction and $U^{\dagger}_{x-\hat{\nu},\nu}\, U_{x-\hat{\nu},\mu} \, U_{x-\hat{\nu} +\hat{\mu},\nu}$ in negative direction. In the four dimensional space there are six staples for each link variable, hence the factor $6$ in the definition.

\begin{figure}%[htbp]
\centering
\includegraphics[width=0.5\textwidth,angle=-90]{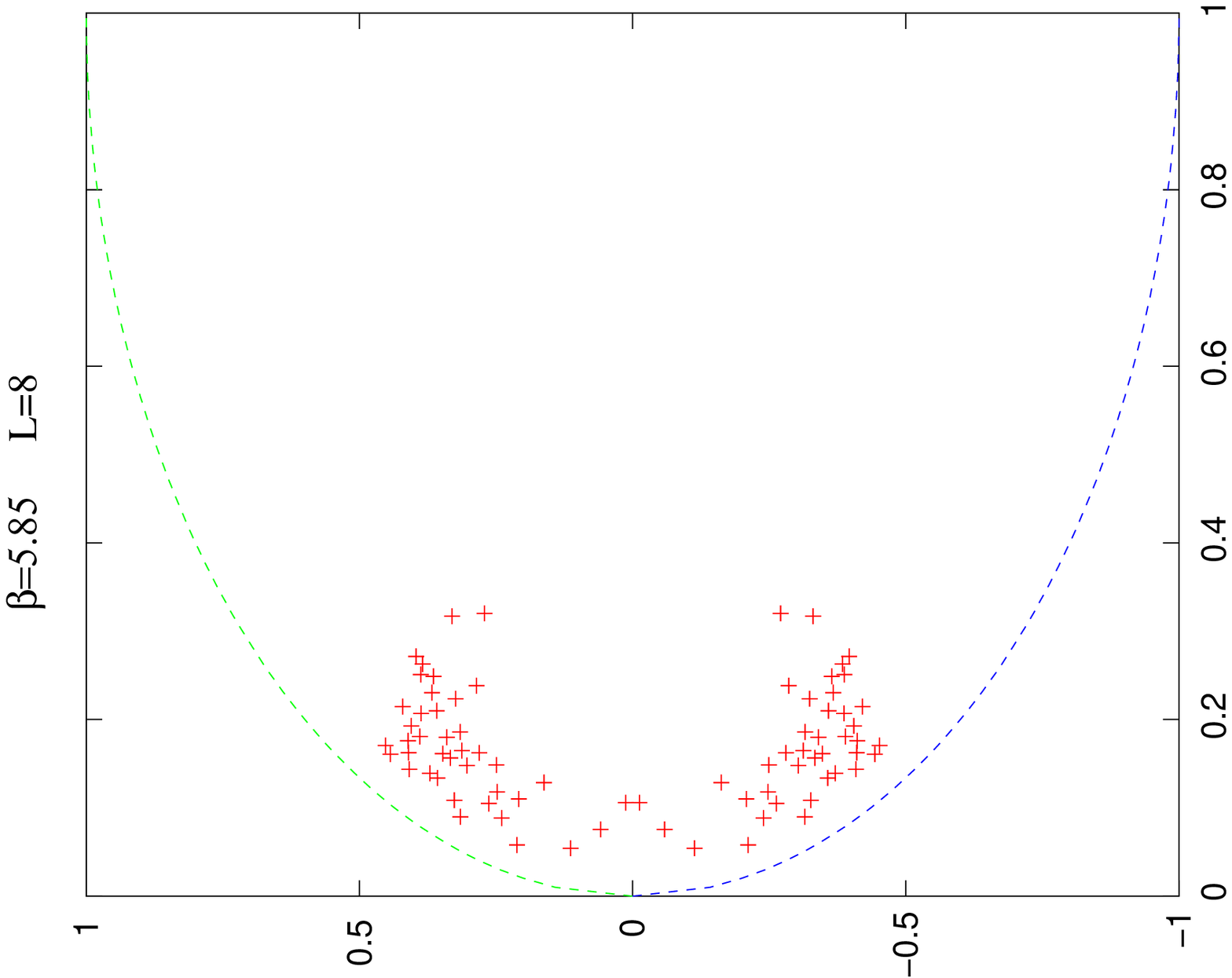}
\caption{The lowest $80$ eigenvalues of the hypercube operator $D_{\rm HF}$ with $\alpha=0.3$, $u=1.32$, $v=0.76$ on a $8^4$ lattice, at $\beta=5.85$. For comparison we also show the corresponding GW circle.}
\label{fig:evhfb5.85_L8}
\end{figure}

We applied a step by step procedure to tune couplings of the hypercube operator for quenched configurations generated at $\beta=5.85$. We started from a $8^4$ lattice and worked out a first approximation of $\alpha$, $u$ and $v$ by plotting the eigenvalue spectrum of the hypercube operator $D_{HF}$ and comparing it with the GW circle. In this way we obtained the following approximations
\begin{equation}
\alpha = 0.3\ ,\quad u = 1.32 \ , \quad v = 0.76\ .  
\end{equation}
In Figure~\ref{fig:evhfb5.85_L8} we plot the lowest $80$ eigenvalues of the hypercube operator for a typical configuration on $8^4$ lattice at $\beta=5.85$ for the mentioned parameters. 
To improve the chiral properties of the hypercube operator further we decided to look at the condition number of the $Q^2$ operator resp. the polynomial degree of the overlap hypercube Dirac operator after projecting out $44$ eigenvectors as a function of $\alpha$, $u$ and $v$ for a given lattice volume $L^3\times T$. The accuracy of the $Q^2$ eigenvectors was taken $10^{-12}$ and the accuracy for the polynomial approximation was of $10^{-15}$. These accuracies were sufficient to obtain a good convergence of the Chebyshev polynomial approximation. The parameter $\mu$ of an auxiliary mass in the definition of the operator $Q=c\gamma_5 (D_{\rm HF}-\mu)$ c.f. Eq.~(\ref{forQ2oper}) becomes redundant in this context since we can account for it by the proper combination of $u$ and $v$. Therefore we set $\mu=1$.
We calculated the average value of the $Q^2$ condition number over $30$ configurations --- which we found to be sufficient for this purpose --- and built the gradient of it in the three dimensional parameter space of $\alpha$, $u$ and $v$. By going in the direction of the gradient, i.e. the direction of the steepest descent, we found the parameters optimized with respect to the condition number of $Q^2$. This procedure was iterated until it converged. The optimal parameters are given in Table~\ref{tab:HFparam}.

\begin{figure}%[htbp]
\centering
\includegraphics[width=0.5\textwidth,angle=-90]{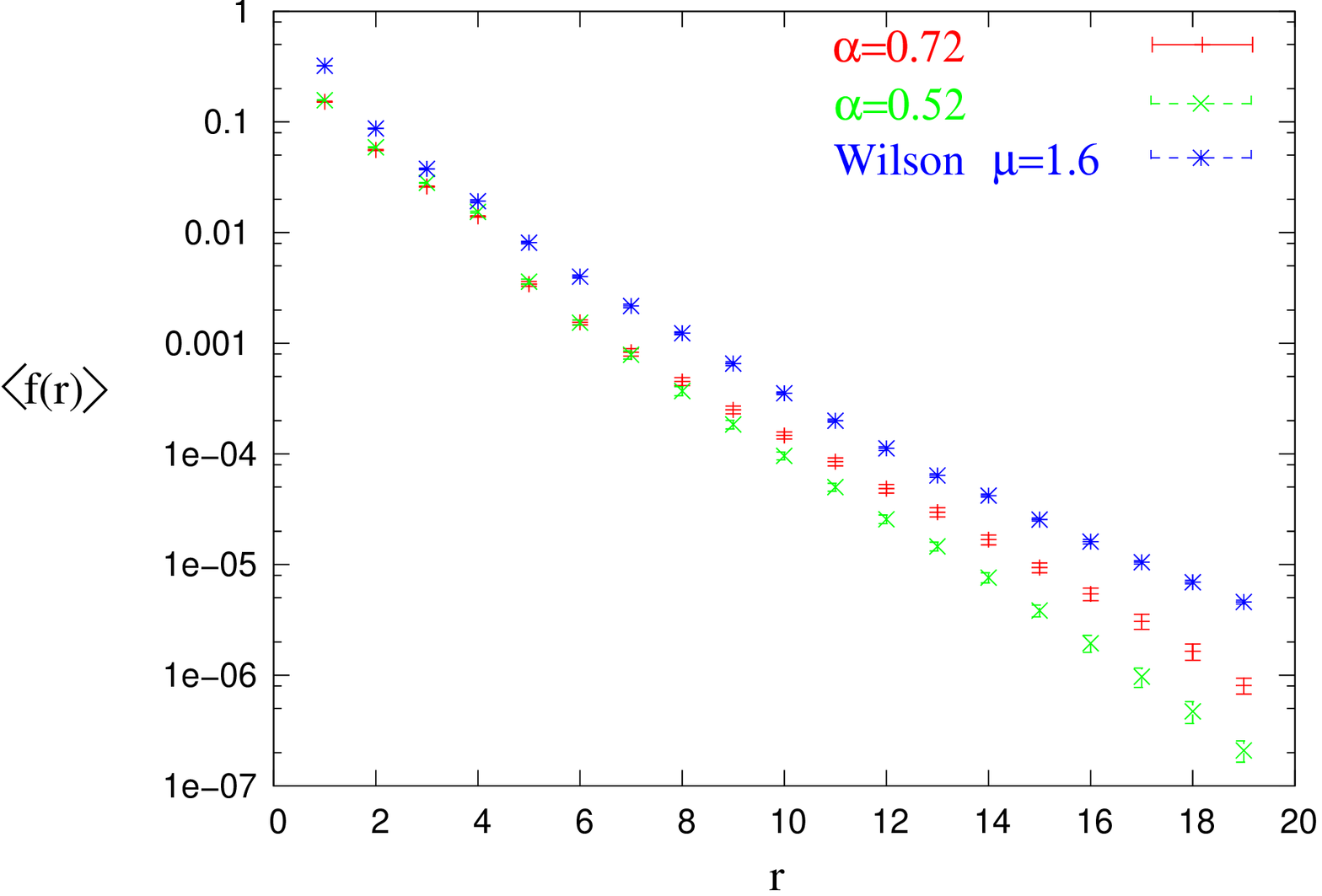}
\includegraphics[width=0.5\textwidth,angle=-90]{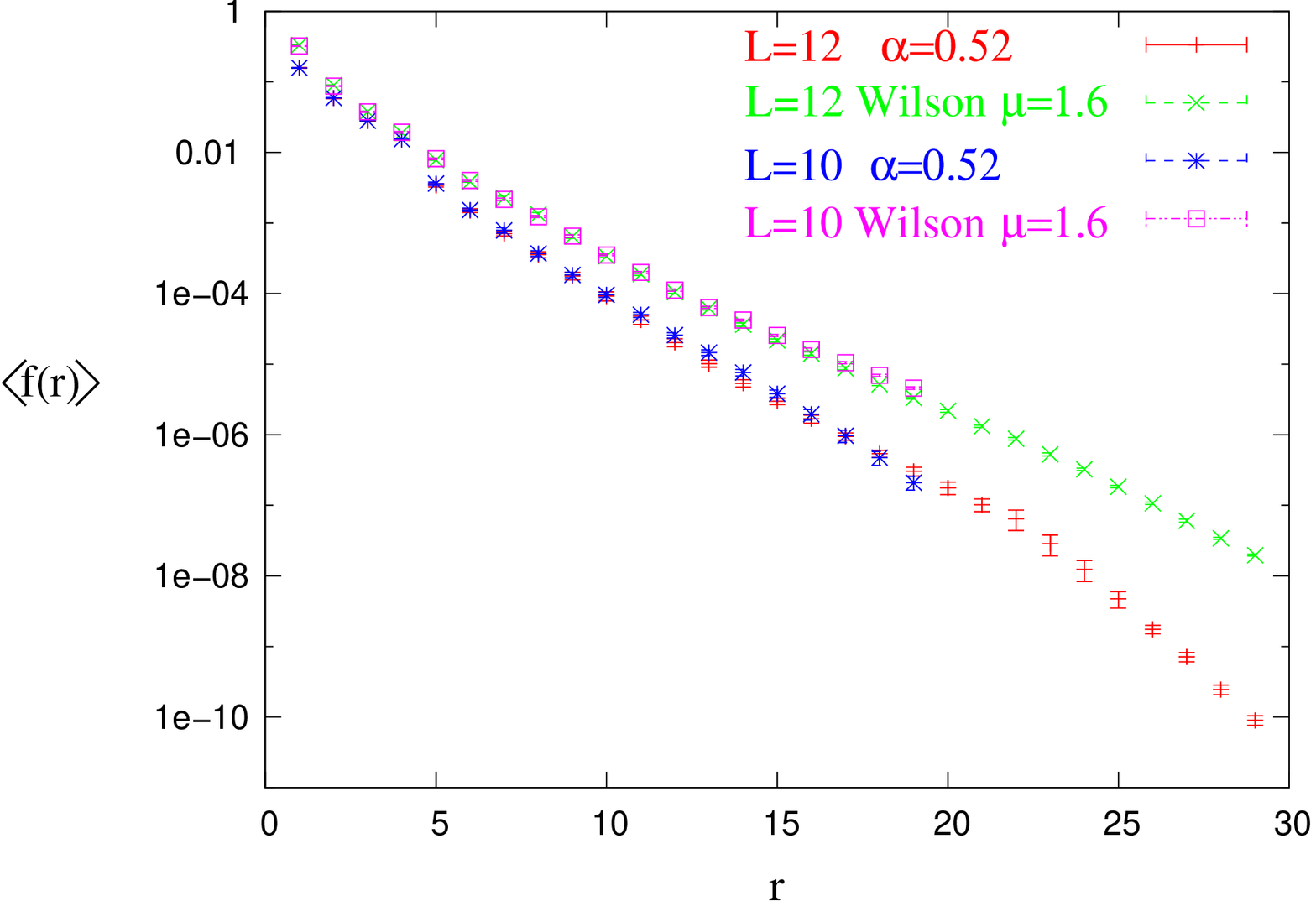}
\caption{The locality of the overlap HF and the overlap Wilson fermions vs. the taxi driver distance $r$. The upper Figure shows the results on a $10^4$ lattice, and the lower Figure shows two results on $12^3\times 24$ lattice along with $10^4$ lattice. We see the same exponential decay as we increase the lattice volume.}
\label{fig:localityHFvsWilson}
\end{figure}

 Alternatively, one can also think of an optimization with respect to the locality properties. On the lattice $10^4$ we found that if we increase the parameter $\alpha$ then the locality properties of the corresponding overlap hypercube Dirac operator become worse. Therefore we decided to relax the optimal parameter $\alpha$ for the condition number in favor of the locality of the resulting operator. We want to point out that the parameter $\alpha$ was decreased only mildly to provide the better locality property and still keeping the condition number at a good value. The suitable parameter $\alpha$ in this respect is $\alpha=0.52$ on $10^4$ lattice at $\beta=5.85$. When we increased the lattice volume to $12^3\times 24$ we found out that the trend established  on a smaller lattice is reproduced to a good extent. Therefore we can safely adopt the parameters from the study on the $10^4$ lattice. This we first confirmed by the recalculating the gradient for the optimal parameters on the $12^3\times 24$ lattice. Secondly it is supported by the behavior of the locality properties on $10^4$ and $12^3\times 24$ lattices.
To illustrate this we show in Figure~\ref{fig:localityHFvsWilson} the locality of the overlap hypercube Dirac operator at two different values of $\alpha$ together with the locality of the overlap Wilson Dirac operator. We see that the larger value of $\alpha$, which corresponds to a lower value of the condition number, shows a weaker slope and thus results in a worse locality. We therefore decide to stick in our simulations of the overlap hypercube Dirac operator to $\alpha=0.52$. We also observe that the locality behavior on the $10^4$ lattice is reproduced almost identically on a larger $12^3\times 24$ lattice.  
\begin{table}[H]
\centering
\begin{tabular}{|c|c|c|c|}
\hline
$V$ & $u$&$v$&$\alpha$ \\
\hline
 $8^4$& $1.22$&$0.76$&$0.4$\\
\hline
 $10^4$& $1.28$ & $0.96$ & $0.72$\\
\hline
\end{tabular}
\caption{The parameters of the HF optimized with respect to the condition number at $\beta=5.85$.\label{tab:HFparam}}
\end{table}

In Table~\ref{tab:HFcondnumber} we give the condition numbers for the hypercube $Q^2$ operator for two optimized sets of parameters: with respect to the best condition number and the locality properties. We include the polynomial degree of the overlap hypercube operator after $44$ eigenvectors being projected out at accuracy $\epsilon_{\rm overlap}=10^{-15}$.
In Table~\ref{tab:Wcondnumber} we add for comparison the condition number for the Wilson $Q^2$ operator.
Note that the degree of the Chebyshev polynomial approximation is proportional to the square root of the condition number.
\begin{table}[H]
\centering
\begin{tabular}{|c|c|c|c|c|c|c|c|}
\hline
$V$ & $u$&$v$&$\alpha$ & polynomial&\multicolumn{3}{|c|}{condition number}\\
    &    &   &         &degree&$m=20$ & $m=30$ &$m=40$\\
\hline
 $8^4$& $1.22$&$0.76$&$0.4$&&$25.2\pm 0.6$ &$22.2\pm 0.4$ &$20.2 \pm 0.3$\\
\hline
 $10^4$& $1.28$ & $0.96$ & $0.52$&$40\pm 1$&$26.8\pm 0.9$&$22\pm 0.5$&$19.4\pm 0.4$\\
\hline
 $10^4$& $1.28$ & $0.96$ & $0.72$ &$39\pm 1$&$26\pm 1$ & $20.9 \pm 0.6$&$18.2\pm 0.3$\\
\hline
 $12^3\times 24$& $1.28$ & $0.96$ & $0.52$& $53\pm 1$&$81 \pm 4$ &$52 \pm 3$ & $41 \pm 1$ \\
\hline
\end{tabular}
\caption{The condition number for the hypercube $Q^2$ operator at $\mu=1$ after $m=20$, $30$, $40$ eigenvectors being projected out and the polynomial degree for the overlap hypercube Dirac operator after $44$ eigenvectors being projected out.\label{tab:HFcondnumber}}
\end{table}

\begin{table}[H]
\centering
\begin{tabular}{|c|c|c|c|c|}
\hline
$V$ & polynomial&\multicolumn{3}{|c|}{condition number}\\
            &degree&$m=20$ & $m=30$ &$m=40$\\
\hline
 $8^4$&&$417\pm 10$ &$363\pm 8$ &$330 \pm 4$\\
\hline
 $10^4$&$172\pm 1$&$636\pm 14$ &$529\pm 10$ &$467 \pm 7$\\
\hline
 $12^3\times 24$&$247\pm 4$&$2100\pm 200$ &$1260\pm 60$ &$996 \pm 36$\\
\hline
\end{tabular}
\caption{The condition number for the Wilson $Q^2$ operator at $\mu=1.6$ after $m=20$, $30$, $40$ eigenvectors being projected out and the polynomial degree of the overlap Wilson Dirac operator after $44$ eigenvectors being projected out.\label{tab:Wcondnumber}}
\end{table}

\chapter{Results on the pion dispersion relation}
The $p$--expansion can be used once the pion Compton wavelength is well accommodated inside the physical volume. This implies that in order to reach the chiral limit, i.e. very small quark masses, one has to simulate in extremely large lattice volumes. On the other hand the $\epsilon$--regime is defined when the pion Compton wavelength extends over the sizes of the physical volume. One uses the fact that the finite size effects appearing in this regime are controlled by the low energy constants of the infinite volume. In this Chapter we explore the product $Lm_{\pi}$ in the definition of the $p$-- and $\epsilon$--regime.

The pion mass in the infinite volume is extracted from a time exponential decay of the pseudo-scalar correlator $a^3\sum_{\vec{x}}\langle {\cal P}^b(\vec{x},t) {\cal P}^b(\vec{0},0)\rangle$. Here summation over $b$ is not taken and the pseudo-scalar density ${\cal P}^b(x)$ was defined in Eq.~(\ref{eq:pseudo-scalardensity}). In the $p$--regime on a lattice of size $L^3\times T$ this is modified to
\begin{equation}
a^3\sum_{\vec{x}}\langle {\cal P}^b(\vec{x},t){\cal P}^b(\vec{0},0) \rangle=\frac{|\langle 0| {\cal P}^b|\mbox{pion} \rangle |^2}{m_{\pi}}e^{-m_{\pi}\frac{T}{2}}\cosh\left [ m_{\pi}\left ( t-\frac{T}{2}\right )\right ]\ .
\end{equation}
As the bare quark mass $m_q$ is decreased, the contribution of the zero modes to the $\langle {\cal P} \cal{P}\rangle$ rises as $|\nu|/(m_qV)^2$, which is seen from the spectral representation of the pseudo-scalar correlator in Eqs.~(\ref{eq_zeromode}-\ref{eq_disconnected}). On the other hand $|\nu| \sim \sqrt{V}$ (we anticipate the formula~(\ref{for:top_sus_sqrtV}) for the topological susceptibility). In total the contribution of the zero modes is reduced as the physical volume $V$ is increased. In full QCD simulations the configurations with topological charge $\nu$ would be suppressed compared to the quenched case  due to the fermionic determinant by a factor $\prod^{N_f}_{q=1}m_q^{|\nu|}$. Therefore in our quenched simulations the sampling of the pseudo-scalar correlator has a systematic error hidden in the contribution of the zero modes. This error is increased for smaller $m_q$ and decreased for larger $V$. As it was pointed out in Ref.~\cite{Blum:2000kn}, the problem can be avoided by subtracting the scalar correlator $\langle {\cal S}^b(x){\cal S}^b(y)\rangle$,
\begin{equation}
{\cal S}^b(x)=\bar{\psi}(x)iT^b\psi(x)\ ,
\end{equation}
which leads to the cancellation of the zero mode contributions. This happens due to their exact chirality. The proposed method introduces also an additional scalar mass to the spectrum of the $\langle {\cal P}{\cal P}-{\cal S}{\cal S}\rangle$ correlator. It is, however, expected to be much larger than the pion mass and therefore it should be visible only among the excited states which are usually suppressed at least by a factor of $\exp{(m_{\pi}-M_{\mbox{excited}})t}$.

\begin{figure}%[htbp]
\centering
\includegraphics[width=0.6\textwidth, angle=-90]{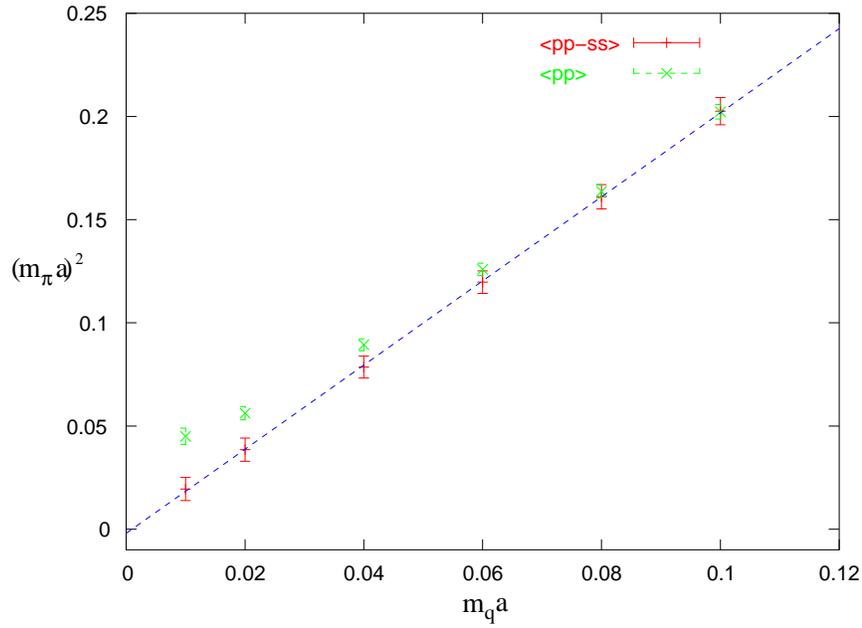}
%\end{rotate}
\caption{The square of the pion mass $m^2_{\pi}$ vs. the bare quark mass computed both for $\langle {\cal P}{\cal P}-{\cal S}{\cal S}\rangle$ and $\langle {\cal P}{\cal P}\rangle$ correlators. The data for $\langle{\cal PP- SS}\rangle$ can be linearly extrapolated to $m_qa=0$ with an intercept compatible with $m^2_{\pi}a^2=0$.}
\label{fig:ov_pion_mass}
\end{figure}

We used the Neuberger operator as a regularization for fermions to extract the pion mass in quenched QCD simulations on $12^3\times 24$ lattice at $\beta=5.85$. To calculate the pseudo-scalar correlation function the Dirac operator has to be inverted as in Eq.~(\ref{eq:pseudoscalarcorrfunction}) at a bare quark mass $m_q$ which is introduced as
\begin{equation}
D_{\rm ov}(m_q)=\left ( 1-\frac{m_q}{2\mu} \right )D_{\rm ov} +m_q\ .
\end{equation}
We approximated the square root appearing in  $D_{\rm ov}$ with the Chebyshev polynomials to accuracy $10^{-15}$. Descriptions of highly optimized algorithms to invert the overlap operator is given in Refs.~\cite{vandenEshof:2002ms,Giusti:2002sm,Arnold:2003sx,Cundy:2004pz}. In particular we use the multiple mass solver~\cite{Chiarappa:2004ry}.
The quark masses were chosen to lie in the range where the $p$--regime is applicable so that the exponential decay of the pseudo-scalar correlator can be used. The lowest considered quark mass corresponds to the cross-over of the $p$--regime to the $\epsilon$--regime. In Table~\ref{tab:mpivsmq} and in Figure~\ref{fig:ov_pion_mass} the quark masses and the pion masses extracted from the ${\cal P}{\cal P}$ as well as ${\cal P}{\cal P}-{\cal S}{\cal S}$ correlators are reported~\cite{BietenholzPP}. We see that the deviation between the two methods becomes larger as the quark mass is decreased. We also give the product $m_{\pi}L$ to quantify the the $p$--regime. In Figure~\ref{fig:ov_pion_mass} we also show the chiral extrapolation by a linear fit of the square of the pion mass vs. the quark mass down to $m_q=0$. We see that the linear behavior is reproduced well for the pion mass calculated with the $\langle {\cal PP-SS}\rangle$ correlator. The intercept amounts to $-0.002(6)$. For the lowest quark mass $m_q=0.01$, which is in the domain of the cross over to the $\epsilon$--regime, we found $m_{\pi}L=1.68$. This is quite small and therefore the finite size effects on the pion mass cannot be neglected.% although they should be on the level of a few percent. 

To fix the physical scale we used the Sommer scale $r_0$, see Refs.~\cite{Sommer:1993ce,Necco:2001xg}. It was computed for the considered range of $\beta$ in Ref.~\cite{Guagnelli:1998ud}

\begin{table}[H]
\centering
\begin{tabular}{|c|c|c|c|}
\hline
$m_qa$ & $m^{\cal P}_{\pi}a$& $m^{{\cal P}-{\cal S}}_{\pi}a$ &$m^{\cal P-S}_{\pi}L$ \\
\hline
\hline
$0.01$ & $0.212(9)$&$0.140(20)$& $1.68$\\
$0.02$& $0.237(7)$&$0.196(14)$& $2.35$\\
$0.04$& $0.299(5)$&$0.280(10)$& $3.36$\\
$0.06$& $0.355(4)$& $0.346(8)$& $4.15$\\
$0.08$& $0.405(4)$& $0.401(7)$&$4.81$\\
$0.10$& $0.450(4)$& $0.451(6)$&$5.41$\\
\hline 
\end{tabular}
\caption{ The pion mass $m_{\pi}$ calculated from $140$ configurations for the $\langle {\cal PP}\rangle $ and for the $\langle {\cal PP-SS}\rangle$ correlators.  The products $m_{\pi}L$ are also shown. At $m_qa=0.01$  we obtain the pion mass $m_{\pi}^{\cal P-S}\approx 225$ MeV.\label{tab:mpivsmq}}
\end{table}

\chapter{Eigenvalue distributions of the overlap Dirac operator}
Since the individual distributions of the eigenvalues of the Dirac operator were only conjectured in the framework of the $\chi$RMT, it is of interest to compare these predictions to the corresponding eigenvalue distributions obtained from QCD lattice simulations with chiral fermions.

The Wilson Dirac operator seems hardly suitable in this respect; because of the mass renormalization it would be highly problematic to identify the zero modes and the relative values of the remaining modes. In addition, quenched simulations at small quark masses are plagued by the occurrence of accidentally extremely small unphysical eigenvalues which is known as exceptional configurations.

Staggered fermions do not suffer from additive mass renormalization because they have an exact remnant chiral symmetry $U(1)\otimes U(1)$. Indeed such comparisons exist. The staggered fermion spectrum agrees well with the $\chi$RMT prediction in the sector $\nu = 0$. However, it turns out that all other sectors, $\nu \ne 0$, yield the same distributions, in particular the same histogram for $ \rho^{(\nu)}_1 (z)$ in QCD~\cite{Damgaard:1999gj,Damgaard:1999bq}. It seems that staggered fermions are generally insensitive to the topology, and therefore not adequate in their conventional formulation for this purpose, at least for moderate and strong gauge coupling. However, latest studies with improved staggered fermions, where the mixing of the pseudo-flavors is suppressed, showed that the sensitivity can be recovered~\cite{Durr:2003xs,Follana:2004sz,Wong:2004ai}. There is also an indication that the sensitivity of the conventional staggered fermions becomes better as we approach the continuum limit. This was suggested in Ref.~\cite{Farchioni:1999yy} where a study with the Schwinger model was presented.

The lattice studies of the distributions of individual eigenvalues with the GW fermions were pioneered in Ref.~\cite{Farchioni:1998jc} where the eigenvalues of the overlap operator and the fixed point operator were considered for the Schwinger model. In both cases the results agreed with the $\chi$RMT formulae.

Such studies were also performed with the overlap operator in 4d QED~\cite{Berg:2001nn} and again the predictions were successful within the statistical errors.

Finally QCD was considered, but initially just on tiny lattices of size $4^4$. Ref.~\cite{Edwards:1999ra} used the overlap operator at strong coupling of $\beta = 5.1$ which in our opinion corresponds to too strong coupling regime where the overlap formula cannot be safely applied, as we pointed out in Chapter 3. Ref.~\cite{Hasenfratz:2002rp} applied a truncated fixed point action and obtained a decent agreement in a volume of V = $(1.2 \fm)^4$. However, when the physical lattice spacing is decreased so that the volume shrinks to $(0.88 \fm)^4$, the leading non-zero eigenvalue distributions of the different topological sectors are on top of each other, in contrast to the $\chi$RMT prediction.

%---------------------------------------------------------------------------------------
\section{Microscopic regime}
In this Section we discuss our results for the individual eigenvalue distributions of the overlap Wilson Dirac operator.
Most of them were published in Ref.~\cite{Bietenholz:2003mi}.

Due to the computational time required by the overlap operator we had to use the quenched approximation, as it was also the case in all previous studies mentioned above.
We generated quenched $SU(3)$ gauge fields with the Wilson gauge action~(\ref{eq:wilson_action}) with periodic boundary conditions. 
We used moderate gauge couplings, $\beta\ge \, 5.85$ and chose the mass $\mu$ in the overlap formula in a way to render the operator optimally local. In particular we took $\mu=1.4$ for $\beta=6$ and $\mu=1.6$ for $\beta=5.85$.
 We approximated the inverse square root by Chebyshev polynomials to an accuracy of $\epsilon_{\rm overlap}=10^{-12}$ for lattices of volume $8^4, \, 10^4, \, 12^4$. Then we used the PARPACK routines~\cite{PARPACK1,PARPACK2} to evaluate up to $100$ eigenvalues of the overlap Wilson Dirac operator with the least absolute value. The index for each configuration was identified by computing eigenvalues of the operator $\gamma_5$ for the leading modes. They came out to be $\pm 1$ to a very high accuracy for the zero modes. The corresponding zero eigenvalues on the $10^4$ lattice at $\beta=5.85$ were of order $10^{-10}$ to $10^{-6}$, and the first non-zero eigenvalues were in the range about $10^{-3}$ to $10^{-2}$. So there was a large gap between non-zero and zero eigenvalues, which allowed us to safely identify the zero modes. Then the index was read off from the chirality of the corresponding eigenvectors. To illustrate this we show in Table~\ref{tab:EVsample} an example of the eigenvalues and the chirality of the corresponding eigenvectors of the Neuberger operator in topological sectors $\nu=0$ and $-2$.
\begin{table}[h]
\centering
\begin{tabular}{|c|c|}
\hline
EV of $D^{\dagger}_{\rm ov} D_{\rm ov}$ & EV of $\gamma_{5}$ \\
\hline
\multicolumn{2}{|c|}{example for charge $\nu =0$} \\
\hline
$6.23013 e-3$  & $~~0.3608201$ \\
$6.23014 e-3$  & $-0.3608198$ \\
$9.90174 e-3$  & $-0.7622260$ \\
$9.90177 e-3$  & $~~0.7622260$ \\
$2.69086 e-2$  & $-0.7528350$ \\
\hline
\multicolumn{2}{|c|}{example for charge $\nu = -2$} \\
\hline
$6.03761 e-16$ & $-1.0000000$ \\
$9.43764 e-11$ & $-0.9999999$ \\
$3.68203 e-3$  & $-0.9240370$ \\
$3.73623 e-3$  & $~~0.9205659$ \\
$7.04176 e-3$  & $-0.7016518$ \\
\hline
\end{tabular}
\caption{\it{Typical examples for the lowest eigenvalues
of $D^{\dagger}_{\rm ov} D_{\rm ov}$ on a $10^3 \times 24$ lattice at $\beta=6$.}}
\label{tab:EVsample}
\end{table}
Of course the non-zero eigenvalues occur in complex conjugate pairs, hence we will consider only one sign for the imaginary part below.
Our statistics in various topological sectors is collected in Table~\ref{tab1}. To fix the physical scale we again used the Sommer scale $r_0$ introduced in Chapter 5.
\begin{table}[t]
\centering
\begin{tabular}{|c|c|c|c|c|c|c|}
\hline
lattice & & &physical &
\multicolumn{3}{|c|}{number of configurations} \\
size & $\beta$&$\mu$ & volume & $\nu = 0$ & $\vert \nu \vert = 1$ 
& $\vert \nu  \vert = 2$  \\
\hline
\hline
$8^{4}$ & 5.85 & 1.6&$(0.98 \fm)^{4}$ & 80 & 63 & 28 \\
\hline
$12^{4}$ & 6 & 1.4&$(1.12 \fm)^{4}$ & 44 & 70 & 24  \\
\hline 
$10^{4}$ & 5.85 & 1.6&$(1.23 \fm)^{4}$ & 74 & 112 & 84 \\
\hline
$16^{3}\times 32$ & 6 & 1.4&$(1.49 \fm)^{3}\times 2.98 \fm$ & 49 & 87 & 63  \\
\hline
\end{tabular}
\caption{The statistics of our simulations on four
lattice sizes.\label{tab1}}
\end{table}
%----------------------------------------------------------------------------------
\subsection{Distributions of individual eigenvalues}
In this Subsection we discuss the probability distributions of the individual lowest eigenvalues of the overlap Wilson Dirac operator.

 In order to relate the eigenvalues found on the Ginsparg-Wilson circle to the continuum eigenvalues $\lambda_n$, we map the circle stereographically onto the imaginary axis. Requiring $f(z) = z + {\cal O}(z^2)$ and $f(2µ) =\infty$   singles out the M{\"o}bius transform 
\begin{equation}
f(z) = \frac{z}{ 1  -z/(2µ)} \ . 
\end{equation}
This mapping has been suggested before in Refs.~\cite{Farchioni:1999se,Capitani:1999uz}, for instance, in connection with the Leutwyler-Smilga sum rules~\cite{Farchioni:1999se}.

The results for the distribution of the lowest eigenvalue of the overlap Wilson Dirac operator can be represented by a histogram, which then could be compared to Figure~\ref{RMT}. However, such a picture depends on the arbitrary choice of the bin size in the histogram. This can be avoided by calculating the cumulative density, which sums up all the entries up to the considered value of $z$   --- it is normalized so that the full number of entries corresponds to the cumulative density $1$. The resulting cumulative densities can be compared to the analytical curves of the cumulative density of the $\chi$RMT prediction for the $k$th eigenvalue in the topological sector $\nu$,
\begin{equation}
\rho^{(\nu)}_{k,c}(z)=\frac{\int_{0}^{z} \rho^{(\nu)}_{k}(z')\, dz'}{\int_{0}^{\infty} \rho^{(\nu)}_{k}(z')\, dz'}\ .
\end{equation}
The chiral condensate enters these analytical formulae as the only free parameter. Therefore, applying a one parameter fit to the data in various topological sectors it is possible to obtain an estimate for its bare quenched value.

We note in passing that the absolute values of the eigenvalues of the operator $\gamma_5 D_{\rm ov}$ are equal to absolute values of the  $\sqrt{D_{\rm ov}^{\dagger}D_{\rm ov} }$. Moreover the stereographically projected eigenvalues of the operator $D_{\rm ov}$ can be related to the Hermitian eigenvalues by the simple formula~\cite{Arnold:2003sx,Bietenholz_private},
\begin{eqnarray}
\left \vert \frac{\lambda_j}{1-\lambda_j/(2\mu)}\right \vert&=&\frac{|\lambda_j^H|}{\sqrt{1-(\frac{1}{2\mu}\lambda_j^H)^2}}\ ,\,  \mbox{where}\\
D_{\rm ov}\psi_j&=&i\lambda_j\psi_j\ , \quad \gamma_5D_{\rm ov}\phi_j=i\lambda^H_j\phi_j\ .
\end{eqnarray}

\begin{figure}%[htbp]
\centering
\includegraphics[width=0.4\textwidth,angle=-90]{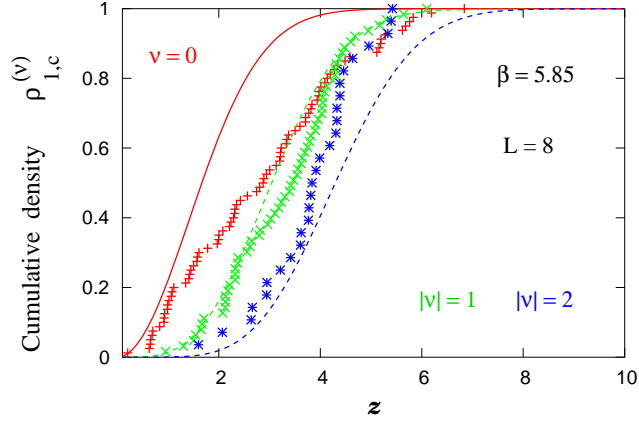}
\caption{The cumulative distribution of the lowest non-zero eigenvalue of the overlap Wilson Dirac operator on $8^4$ lattice, at $\beta=5.85$ in the topological sectors $|\nu|=0$, $1$, $2$ vs. $\chi$RMT predictions (lines). No agreement is found.}
\label{RMT_8}
\end{figure}

We start the discussion of the cumulative distributions for our smallest physical volume of $(0.98 \fm)^4$ in Figure~\ref{RMT_8}. 
Here the lines represent the predictions by $\chi$RMT for the lowest eigenvalue of the Dirac operator in the topological sectors $|\nu|=0$, $1$ and $2$, and the symbols the cumulative distribution of the overlap Wilson Dirac operator.
One can see a clear disagreement with the $\chi$RMT curves. In particular the sensitivity of the data to the topological charge is lost to a large extent.  Notice that for very small values of $z$ the data are closer to the analytical predictions. However, this does not allow for a reasonable fitting procedure to extract the chiral condensate $\Sigma$.

\begin{figure}%[htbp]
\centering
\includegraphics[width=0.4\textwidth,angle=-90]{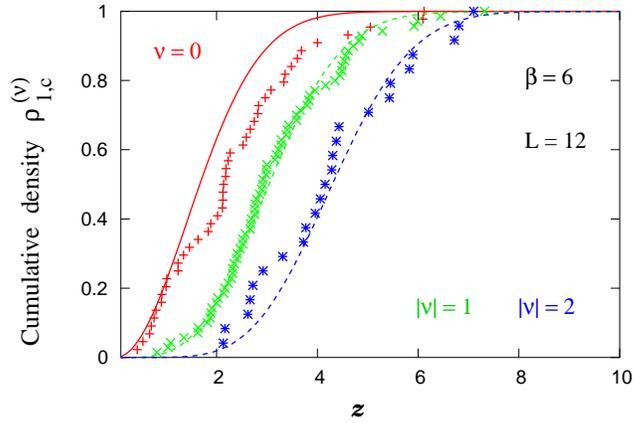}
\caption{The cumulative distribution of the lowest non-zero eigenvalue of the overlap Wilson Dirac operator on $12^4$ lattice, at $\beta=6$ in the topological sectors $|\nu|=0$, $1$, $2$ vs. $\chi$RMT predictions at $\Sigma=(256 \, \mbox{MeV})^3$ (lines). A reasonable agreement is observed.}
\label{RMT_12}
\end{figure}

Now we consider a larger physical volume of $(1.12 \fm)^{4}$. It corresponds in our simulations to a $12^4$ lattice volume at $\beta=6$.
The plot is depicted in Figure~\ref{RMT_12}.
One can see that the agreement starts to set in. We interpret this picture as a satisfactory agreement with the predictions by the $\chi$RMT. 
In particular, we do see the effect that the peak of the density, resp. the interval of steepest ascent of the cumulative density, moves to larger values of $z$ for increasing topological charge $|\nu|$.  The plot is shown for the optimal value of the chiral condensate $\Sigma=(256\ \mbox{MeV})^3$.

\begin{figure}%[htbp]
\centering
\includegraphics[width=0.4\textwidth,angle=-90]{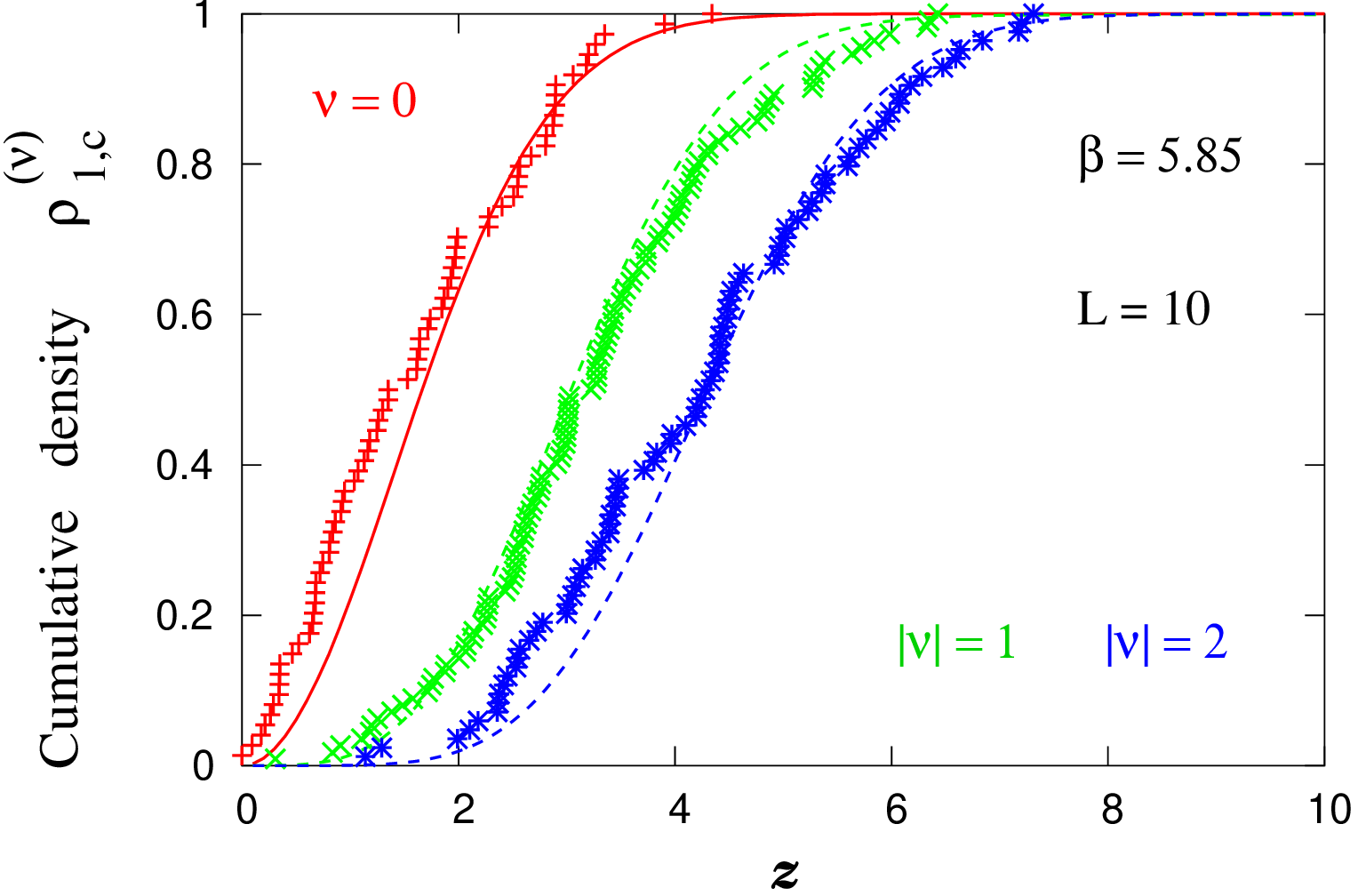}
\caption{The cumulative distribution of the lowest non-zero eigenvalue of the overlap Wilson Dirac operator on $10^4$ lattice, at $\beta=5.85$ in the topological sectors $|\nu|=0$, $1$, $2$ vs. $\chi$RMT predictions at $\Sigma=(253 \, \mbox{MeV})^3$ (lines). A good agreement is observed.}
\label{RMT_10}
\end{figure}

Next we studied a $10^4$ lattice at $\beta = 5.85$, which corresponds to a somewhat larger physical volume of $V = (1.23 \fm)^4$. Again the results for the leading non-zero eigenvalue are shown for the topological sectors $|\nu| = 0$, $1$ and $2$ in Figure~\ref{RMT_10}.
We interpret the results to be in a good agreement with the $\chi$RMT predictions. 
The optimal value of the chiral condensate is modified to $\Sigma=(253 \mbox{MeV})^3$.

\begin{figure}%[htbp]
\centering
\includegraphics[width=0.4\textwidth,angle=-90]{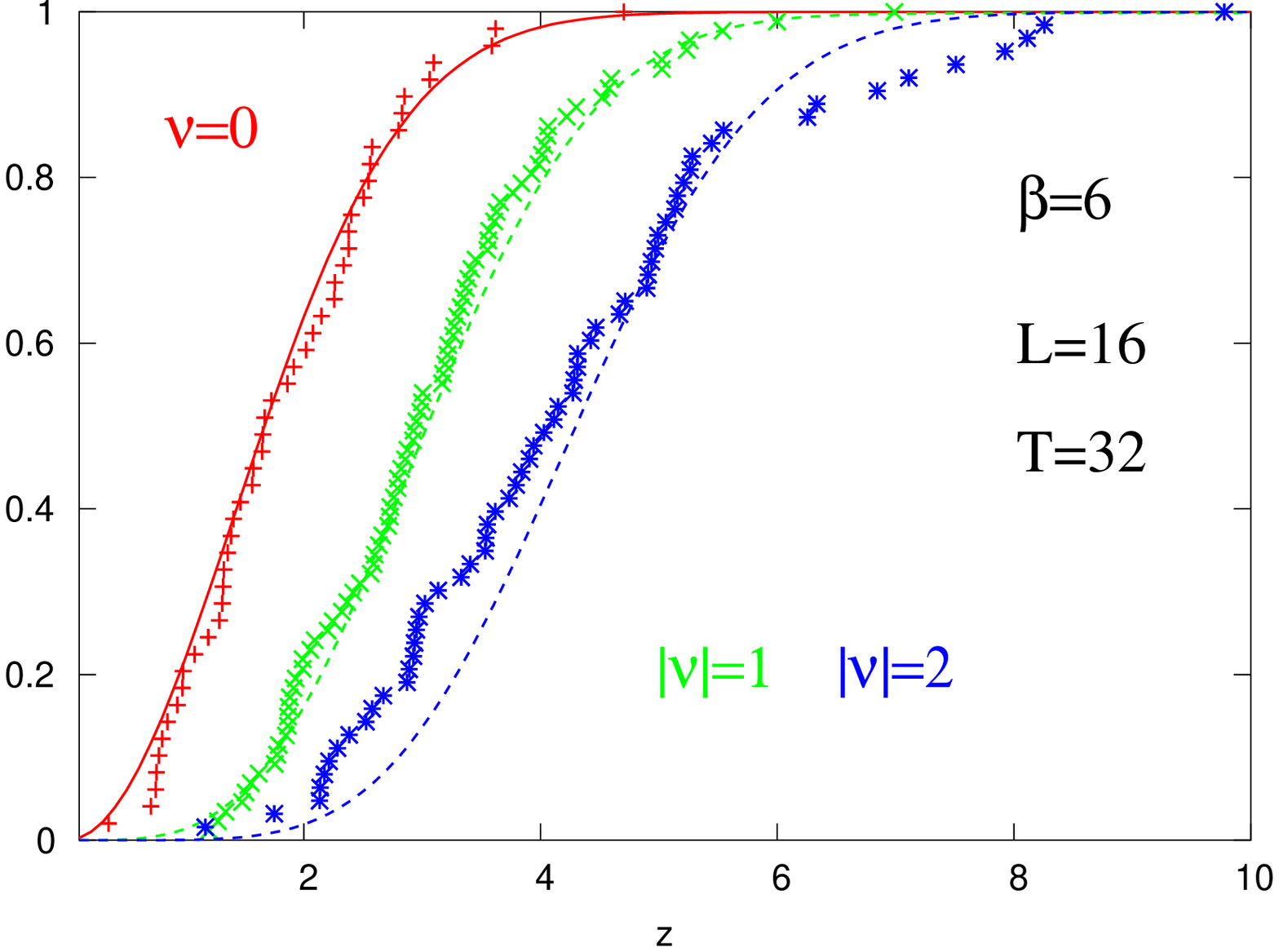}
\caption{The cumulative distribution of the lowest non-zero eigenvalue of the overlap Wilson Dirac operator on $16^3\times 32$ lattice, at $\beta=6$ in the topological sectors $|\nu|=0$, $1$, $2$ vs. $\chi$RMT predictions at $\Sigma=(286 \mbox{MeV})^3$ (lines). A good agreement is observed.}
\label{RMT_16}
\end{figure}
\begin{figure}%[htbp]
\centering
\includegraphics[width=0.4\textwidth,angle=-90]{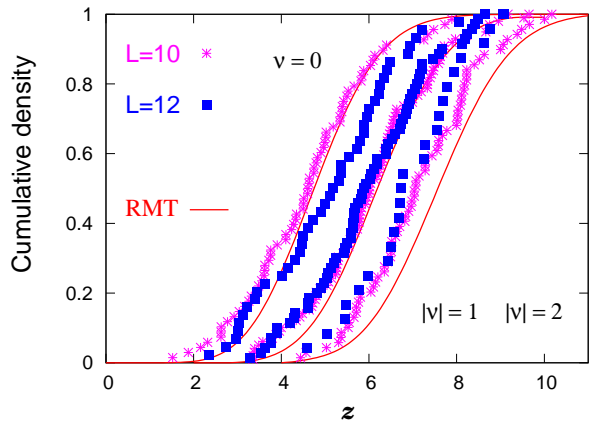}
\caption{The results for the cumulative distribution of the next to lowest eigenvalue in topological sectors $|\nu|=0$, $1$, $2$ for two lattice volumes $10^4$ and $12^4$ vs. the $\chi$RMT predictions (lines). No convincing agreement is observed.}
\label{RMT_NLE}
\end{figure}

Finally we looked at the lattice size $16^3\times 32$, at $\beta=6$ which corresponds to a yet larger volume of $(1.49\fm)^3\times 2.98 \fm$. In Figure~\ref{RMT_16} we show the cumulative densities of the lowest eigenvalue in topological sectors $|\nu|=0$, $1$ and $2$ for $\Sigma=(286 \, {\rm MeV})^3$. We interpret it again as a good agreement.

$\chi$RMT predicts a (logarithmically) increasing  $\Sigma(V)$ due to quenching~\cite{Damgaard:2001xr}. A conclusive verification of this behavior (beyond possible lattice artifacts) would require further simulations.

In Figure~\ref{RMT_NLE} we show the cumulative distributions of the next-to-lowest eigenvalue in the topological sectors $|\nu|=0$, $1$ and $2$.
The data we take from simulations on $10^4$ and $12^4$ lattices. Here we do not find a convincing  agreement with the $\chi$RMT predictions; note that the relevant values of $z$ are larger compared to Figure~\ref{RMT_12}. Also the second non-zero eigenvalue moves to larger values of $z$ if $|\nu|$ increases.

%At this point we want to show results with eigenvalues of $\gamma_5 D_{ov}$ operator which are real numbers and therefore one does not need to use the stereographic projection. In the Figure~\ref{herm} one can see that the cumulative distributions of the eigenvalues of $\gamma_5 D_{ov}$ operator are closed to the corresponding data from $D_{ov}$ operator in Figure~\ref{RMT_10}.

%\begin{figure}%[htbp]
%\centering
%\includegraphics[width=0.4\textwidth,angle=-90]{inc.figures/k-s_test_hn_10-2}
%\caption{The results for the cumulative distribution of the lowest eigenvalue of the $\gamma_5 D_{ov}$ operator in topological sectors $|\nu|=0$, $1$, $2$ for the lattice volume $10^4$ vs. the RMT predictions (lines).}
%\label{herm}
%\end{figure}
To quantify the quality of the fit we used the Kolmogorov-Smirnov test~\cite{bookNR}. This test provides a confidence level that the given data set is drawn from the probability distribution of interest. 
In Table~\ref{tab2} we collect the confidence levels and the optimal values for $\Sigma$ on the lattice volumes where we found a reasonable agreement with the $\chi$RMT. We see that the box length of $1.2 \fm$ constitutes the lower bound of applicability of the $\chi$RMT predictions. We do not observe a full agreement. However, one can argue that the separation of the topologies for the lowest eigenvalue has set in to a large extent, and the box length is sufficient to obtain a good estimate for the chiral condensate.

\begin{table}[t]
\centering
\begin{tabular}{|c|c|c|c|c|c|}
\hline
lattice & & optimal &
\multicolumn{3}{|c|}{confidence level} \\
size & $\beta$ & $\Sigma$ & $\nu = 0$ & $\vert \nu \vert = 1$ 
& $\vert \nu  \vert = 2$  \\
\hline
\hline
$12^{4}$ & 6 & $(256 \mbox{MeV})^{3}$ & 0.003 & 0.73 & 0.79  \\
\hline 
$10^{4}$ & 5.85 & $(253 \mbox{MeV})^{3}$ & 0.03 & 0.48 & 0.10 \\
\hline
$16^{3}\times 32$ & 6 & $(286 \mbox{MeV})^{3}$ & 0.7 & 0.8 & 0.12 \\
\hline
\end{tabular}%
\caption{ The results for the chiral condensate and the statistical confidence level
according to the Kolmogorov-Smirnov test.\label{tab2}}
\end{table}

The observation that there is a minimal physical box length where the $\chi$RMT predictions set in can be interpreted in terms of the so-called Thouless energy, a term that originates from mesoscopic physics (see the Review~\cite{Verbaarschot:2000dy} on more details).  The $\chi$RMT predictions become valid once we ensure that the simulations are performed in the $\epsilon$--regime of the $\chi$PT. This can be estimated if we require the potential term in the Lagrangian of the $\chi$PT~(\ref{Chpt_lagr}) to dominate over the kinetic term. Applying this argument to the characteristic scale of Dirac eigenvalues we arrive at
\begin{equation}
\lambda \ll \lambda_{\rm Thouless}=\frac{F^2_{\pi}}{\sqrt{V} \Sigma}\ ,
\label{Thouless}
\end{equation}
or converting this to the dimensionless variable: $z_{\rm Thouless}=F^2_{\pi}\sqrt{V}$.
The above relation can also be obtained from the lower bound in the relation~(\ref{upper_and_lower_relation}) by taking its square and recalling the formula for the pion mass through the quark mass, see Subsection 2.3.1.  
The determination of the value of $z$ where the $\chi$RMT predictions fail can be considered as an empirical value for the Thouless energy. Thus the Thouless energy sets a limit on the value of eigenvalues where the $\chi$RMT applies. On the other hand one has to ensure that the physical volume is large enough to be in the confinement phase where all assumptions for the $\chi$PT are made. In addition one has to keep in mind the relation~(\ref{upper_relation}) ensuring that the $\chi$PT expansion is defined. The latter imposes on the physical extent the requirement $L\gg 1/(2F_{\pi})\sim 1\fm$.

To check the condition~(\ref{Thouless}) one can perform the same eigenvalue study at rather large fixed physical volume and different lattice spacings $a$. Due to the renormalization of $F_{\pi}$ and $\Sigma$ this will lead to different Thouless energies and therefore it should be seen in the behavior of the corresponding cumulative distributions. In fact such a study was done in Ref.~\cite{Giusti:2003gf}, but no ultimate confirmation of this behavior was found.

%----------------------------------------------------------------------------
\subsection{Spectral density}
\begin{figure}%[htbp]
\centering
\includegraphics[width=0.5\textwidth,angle=-90]{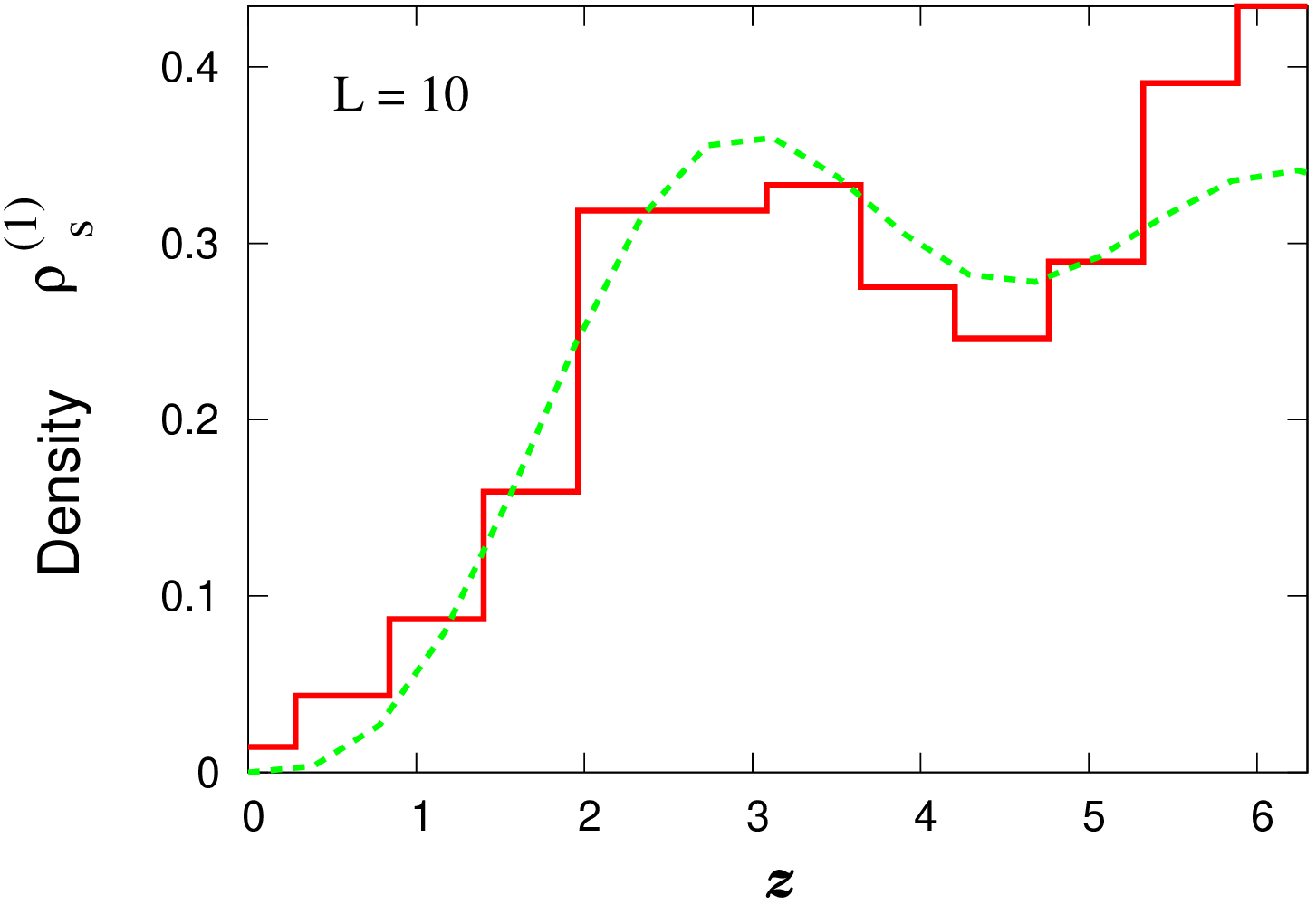}

\includegraphics[width=0.5\textwidth,angle=-90]{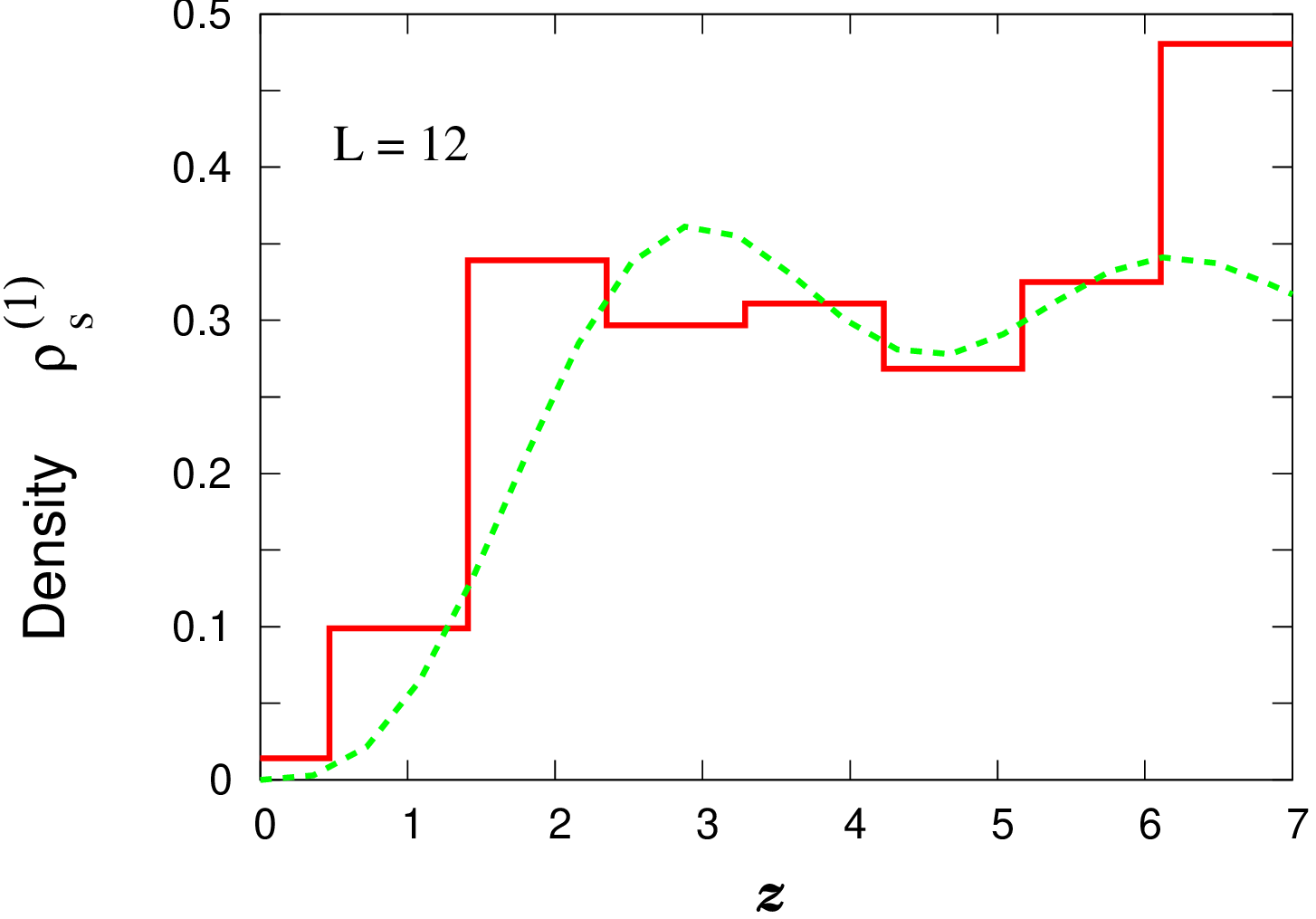}
\caption{The results on the microscopic spectral density $\rho_s(z)$ in topological sectors $|\nu|=1$ for two lattice volumes $10^4$ and $12^4$ vs. the $\chi$RMT predictions (lines).}
\label{RMT_MSDENS}
\end{figure}

In this Subsection we discuss the microscopic spectral density of the overlap Wilson Dirac operator. We still separate topological sectors, but now we consider the distributions of all computed eigenvalues. We show the eigenvalue distributions for topological sector $|\nu|=1$ and two lattice volumes, $10^4$ and $12^4$, in Figure~\ref{RMT_MSDENS}. $\chi$RMT predicts an oscillating behavior~(\ref{micr_spectral_den})~\cite{Verbaarschot:1993pm}, which is also plotted for comparison.  However, here we omit the zero eigenvalues, hence the term with the delta function drops out. We find a good agreement roughly up to the second peak,as expected from Figures~\ref{RMT_12},~\ref{RMT_10} and~\ref{RMT_NLE}. Then we are leaving the microscopic regime and turn to the bulk. The Thouless energy appears in this context as a value of $z$ where we leave the microscopic regime. 

If the oscillation is averaged to a plateau, its height agrees well with the eigenvalue density at zero according to the Banks-Casher relation, $\rho(z = 0) = \rho(\lambda = 0)/ \Sigma V = 1/\pi$.
%---------------------------------------------------------------------------------------
\section{Bulk eigenvalues}

\begin{figure}%[htbp]
\centering
\includegraphics[width=0.5\textwidth,angle=-90]{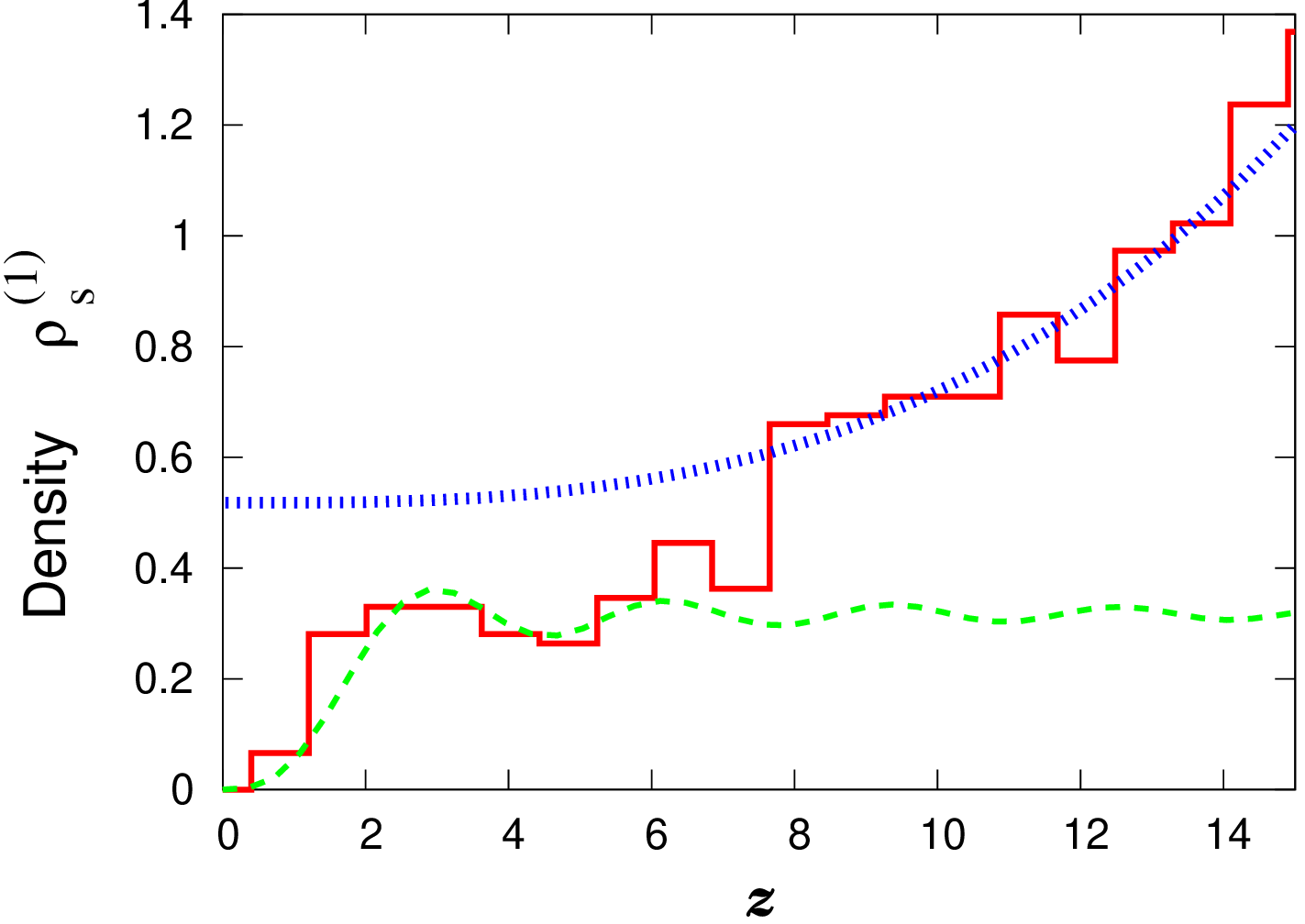}

\includegraphics[width=0.5\textwidth,angle=-90]{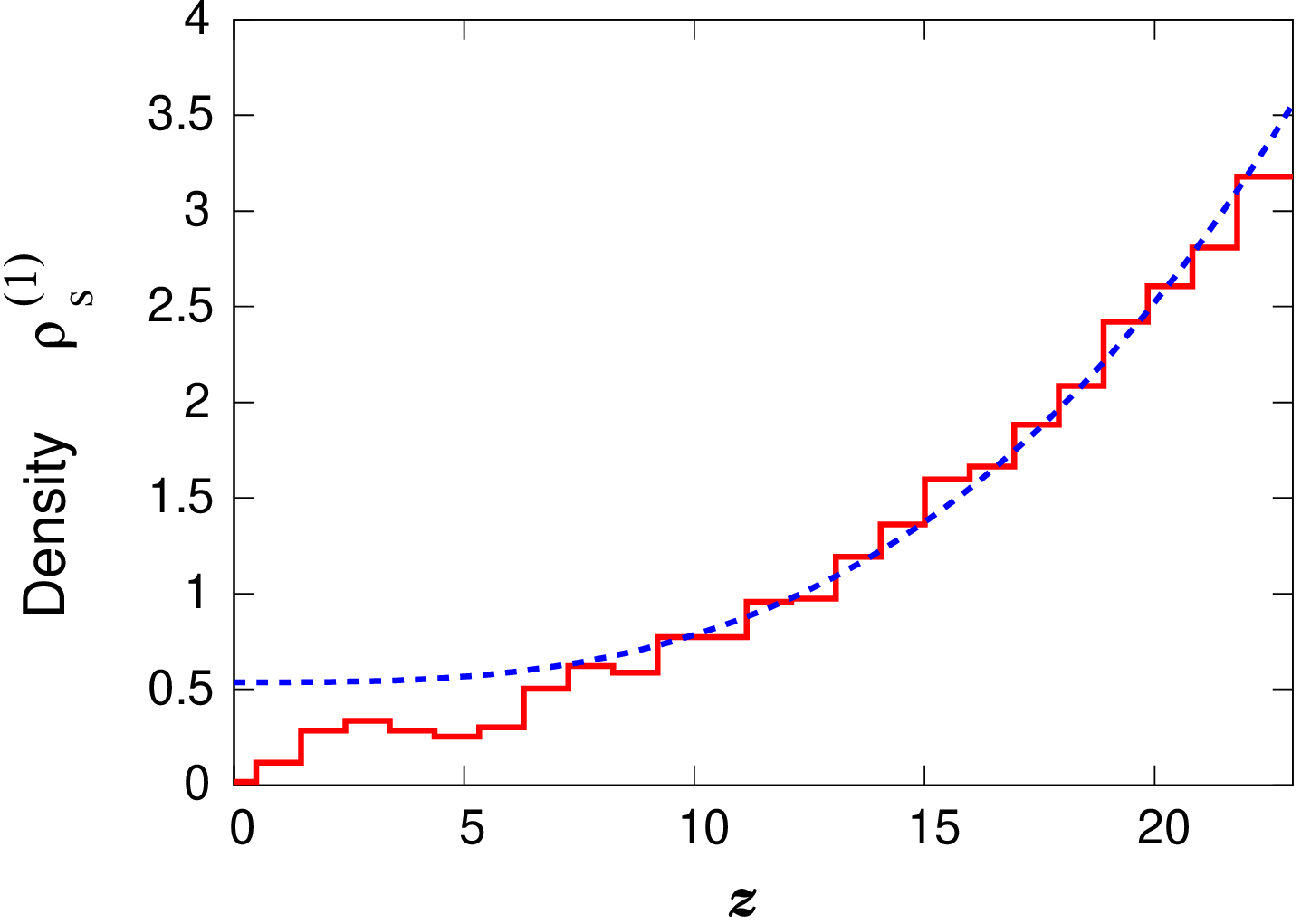}
\caption{The upper plot shows the spectral density $\rho_s(z)$ in the bulk vs. the microscopic regime. One line corresponds to $a+bz^3$ and another to the microscopic spectral density. The plot at the bottom  shows the spectral density in the bulk. The cross-over from the microscopic regime to the bulk is nicely seen.}
\label{RMT_BULK}
\end{figure}

In this Section we discuss the eigenvalues in the bulk of the Dirac spectrum. It is expected that the spectral density in the deep bulk domain  has a $\lambda^{3}$ behavior. Since we were able to compute as many as $50$ conjugate pairs of eigenvalues of the overlap Wilson Dirac operator we can analyze the spectral density in the bulk. The largest eigenvalue corresponds to an energy of about $800 \ \mbox{MeV}$. However, it is difficult to argue whether the considered domain is sufficiently large to accommodate the $\lambda^3$ behavior. In Figure~\ref{RMT_BULK} at the bottom we plot the eigenvalue spectral density for the topological sector $\nu=1$ on the lattice $12^4$ against the $\lambda^3$ curve, which we take to be $a+bz^3$. The data follow amazingly well the predicted behavior. 

Now we want to make a remark on the Banks-Casher relation.
 If we were not able to obtain information from the microscopic regime and we had only the eigenvalues in the bulk, then the chiral condensate which would follow from the approximation of the eigenvalue density would be overestimated. This can be seen from the plateau extrapolated to low values of $z$ on the Figure~\ref{RMT_BULK} on the top. For illustration we plot two curves for the bulk and the microscopic regime.

%-----------------------------------------------------------------------------------------
\subsection{The unfolded spectrum}

We now turn to another way of comparing our lattice data to a conjecture from $\chi$RMT. This additional evaluation allows us to take all our non-zero eigenvalues into account (for one sign of the imaginary part, i.e.~up to 50 for each configuration).
We build from all the eigenvalues of all configurations the {\em hierarchically unfolded distribution} as described for instance in Ref.~\cite{Edwards:1999ra}. For a comprehensive review and other ways of unfolding we refer for instance to Ref.~\cite{Verbaarschot:2000dy}.

To construct the hierarchically unfolded spectrum  we first numerate all available non-zero eigenvalues with positive imaginary part in each configuration in ascending order, given by the angle in the Ginsparg-Wilson circle. Then we put all these eigenvalues from all configurations together and numerate them again in ascending order. Now we consider pairs of eigenvalues from the same configuration, which follow immediately one after the other in the original numeration. If they differ in the global numeration by $k$, then $k/N_{\rm conf}$ is the hierarchically unfolded level spacing, where $N_{\rm conf}$ is the number of configurations involved.
\begin{figure}%[htbp]
\centering
\includegraphics[width=0.5\textwidth,angle=-90]{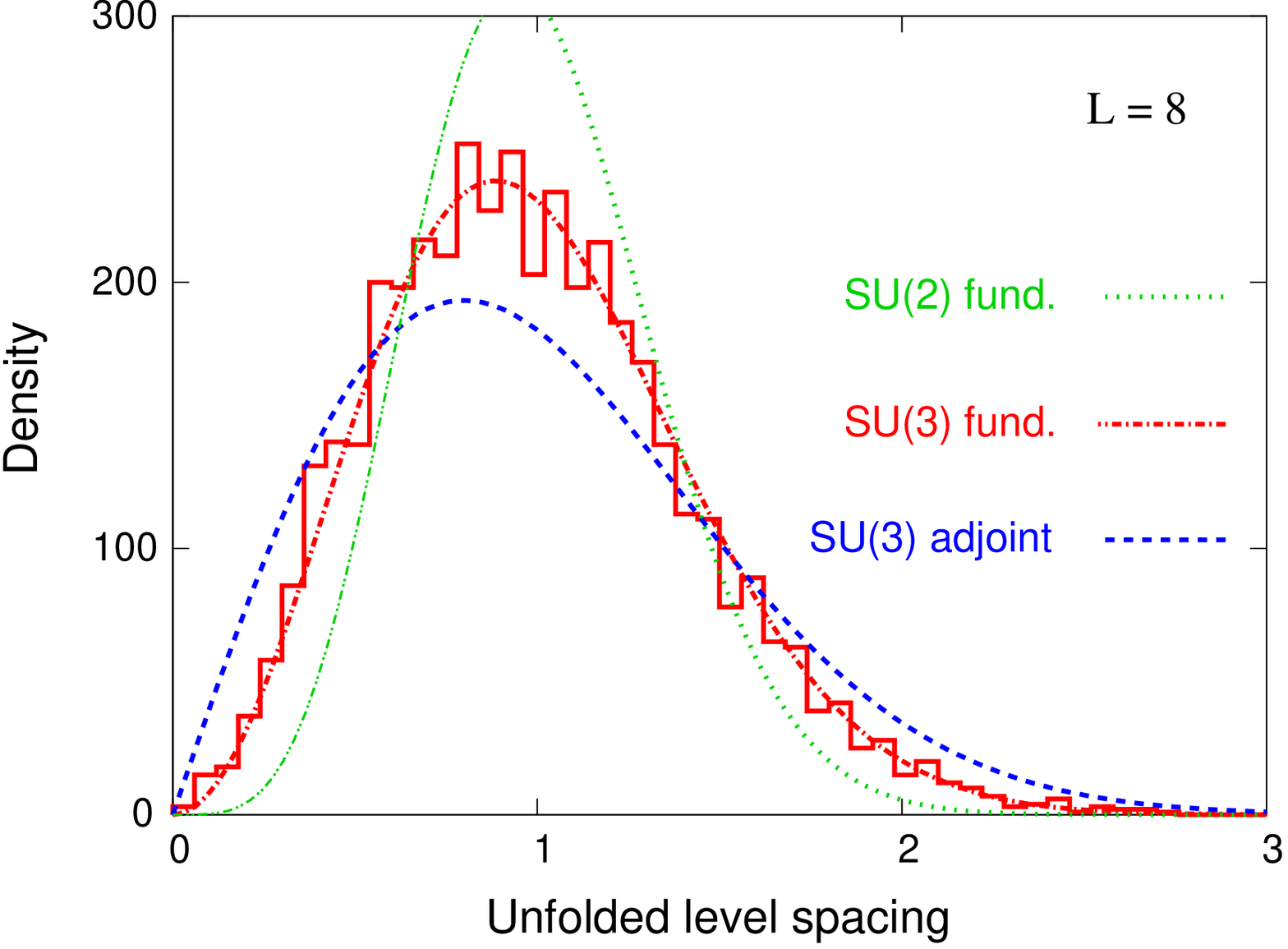}

\includegraphics[width=0.5\textwidth,angle=-90]{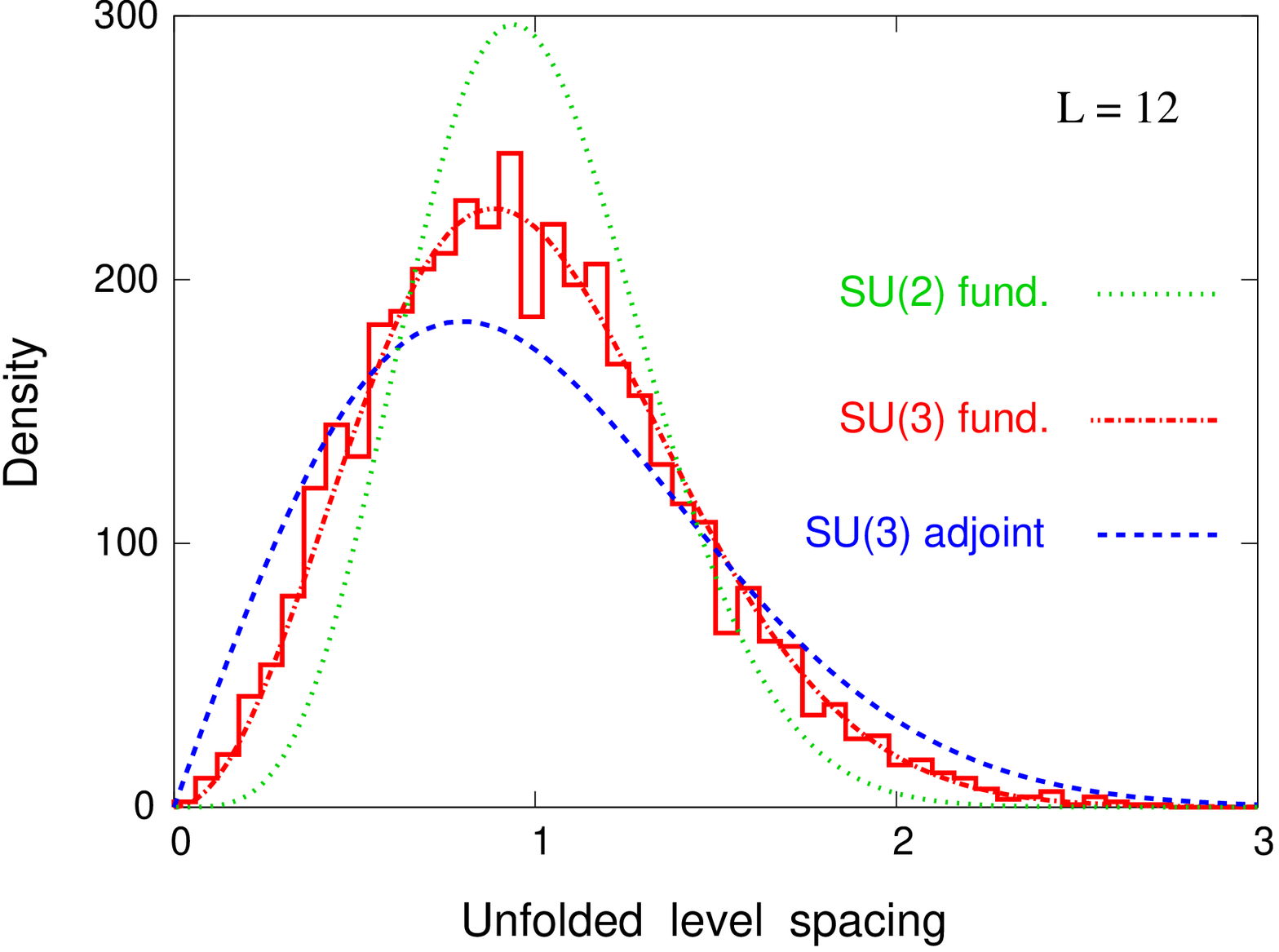}
\caption{The hierarchically unfolded level spacing distributions. For comparison we show the curves for $SU(2)$ and $SU(3)$ fundamental representation and for $SU(3)$ adjoint representation.}
\label{RMT_UNF}
\end{figure}
In Figure~\ref{RMT_UNF} we show our results for the corresponding level spacing distribution in $V = 12^4$, $\beta = 6$, and in $V = 8^4$, $\beta = 5.85$. The histograms are compared to RMT conjectures for different groups and representations, and we can confirm a good agreement with the Wigner distribution predicted for $SU(3)$ in the fundamental representation.  We conclude that this property holds over a large range of volumes. Since we include all our eigenvalues here, our statistics is much larger than in the plots for the microscopic regime. Note, however, that this analysis is not sensitive to the topology any more. There is no physics information such as values for the condensate or the pion decay constant, because one does not keep track of a physical scale. The hierarchically unfolded distribution only tests the correct symmetry group that is to be taken in RMT.  Nevertheless, the hierarchically unfolded distribution provides a non-trivial and successful test for $\chi$RMT.

\chapter{Topological susceptibility}
\section{Motivation}
The topological susceptibility for a Yang-Mills gauge theory is defined by
\begin{equation}
\chi_t=\frac{1}{V}\int d^4x d^4y\, \langle q(x)q(y)\rangle_{\rm YM} \ ,
\label{chiralsusc}
\end{equation}
where  $q(x)$ is a density of the topological charge $Q=\int d^4x\, q(x)$ and the subscript YM means that the average is normalized by the partition function. We refer to quenched QCD, although the topological susceptibility also exists in the dynamical case.
The main reason to look at the quenched topological susceptibility comes from the Witten-Veneziano formula~\cite{Witten:1979vv,Veneziano:1979ec}. It is motivated by the inability of the constituent quark model to describe a large mass of the $\eta'$ meson. The Witten-Veneziano formula predicts that at the leading order in the $N_f/N_c$ expansion, with $N_f$ being the number of flavors and $N_c$ the number of colors, the contribution to the $\eta '$ mass due to the $U_A(1)$ anomaly is given by  
\begin{equation}
m^2_{\eta '}=\frac{2N_f}{F^2_{\pi}}\chi_t \ .
\label{wittenven}
\end{equation}

 In the absence of the anomaly one could argue that $U_A(1)$ symmetry would be broken spontaneously leading to a massless Goldstone boson for massless quarks. However, a particle with the corresponding quantum numbers is not found, which is attributed to the existence of the $U_A(1)$ anomaly.

To interpret Eq.~(\ref{chiralsusc}) one has to find a properly normalized topological density $q(x)$ and to subtract from the quantity $q(x)q(0)$ an appropriate contact term to define it properly in the continuum limit. 

In the case of Ginsparg-Wilson fermions the topological density can be determined by
\begin{equation}
q_x=\frac{1}{2a^3}{\rm Tr}[\gamma_5 D[U]_{x,x}] \ ,
\label{charge_GW}
\end{equation}
where the trace is taken over color and spinor indices of a Ginsparg-Wilson operator $D$. It can be shown that this definition leads to a properly defined topological charge and to the Atiyah-Singer index theorem~\cite{Hasenfratz:1998ri,Luscher:1998pq,Giusti:2004qd,Luscher:2004fu}.
In Ref.~\cite{Giusti:2001xh} it was also demonstrated that this discretization naturally enters the Witten-Veneziano formula if we consider Ginsparg-Wilson regularization for the fermions
\begin{equation}
m^2_{\eta '}=\frac{2N_f}{F^2_{\pi}}\lim_{\substack{ V\to \infty \\ a \to 0}}a^4 \frac{1}{V}\sum_{x,y} \langle q_y q_x \rangle_{\rm YM}\ .
\end{equation}
Note that due to the chiral symmetry of the fermions no additional subtraction is necessary~\cite{Chandrasekharan:2004cn}. 
Using Eq.~(\ref{charge_GW}) we obtain for the topological susceptibility
\begin{equation}
\chi_t=\lim_{\substack{ V\to \infty \\ a \to 0}}\frac{\langle (N_+ -N_-)^2 \rangle_{\rm YM}}{V}\ ,
\label{for:top_sus_sqrtV}
\end{equation}
where $N_+$ and $N_-$ is the number of zero modes of the Ginsparg-Wilson fermions with positive and negative chirality, respectively.
This formula will be a starting point for the discussion of the results which we obtained using the overlap operator.

The study of the continuum limit of the topological susceptibility with the GW fermions was pioneered in Refs.~\cite{Hasenfratz:2002rp,DelDebbio:2003rn,Giusti:2003gf,DelDebbio:2004ns}. It is argued that the continuum extrapolated value of the topological susceptibility is about $\chi_t r_0^4=0.059\pm 0.003$~\cite{DelDebbio:2004ns}, where $r_0=0.5 \fm$.
%------------------------------------------------------------------------------------------------------
\section{Results with overlap fermions}  

We simulated quenched QCD on $L^3\times T$ lattices  with periodic boundary conditions and used the overlap Wilson Dirac operator, as well as the overlap hypercube Dirac operator, to determine the index of the gauge configurations. The configurations were simulated using the usual Monte Carlo procedure where we employed the heat bath and the overrerlaxation method as described in Subsection 3.1. We assume the configurations to be statistically independent, which we checked by measuring the autocorrelation time of the topological charge. It turned out to be below 0.5, which characterizes a completely decorrelated sample.

 We considered several lattice volumes and in Table~\ref{tab:susc1} we collect the topological susceptibility for the Neuberger operator as well as for the overlap hypercube operator.  

\begin{figure}[t]
\centering
\includegraphics[width=0.6\textwidth,angle=-90]{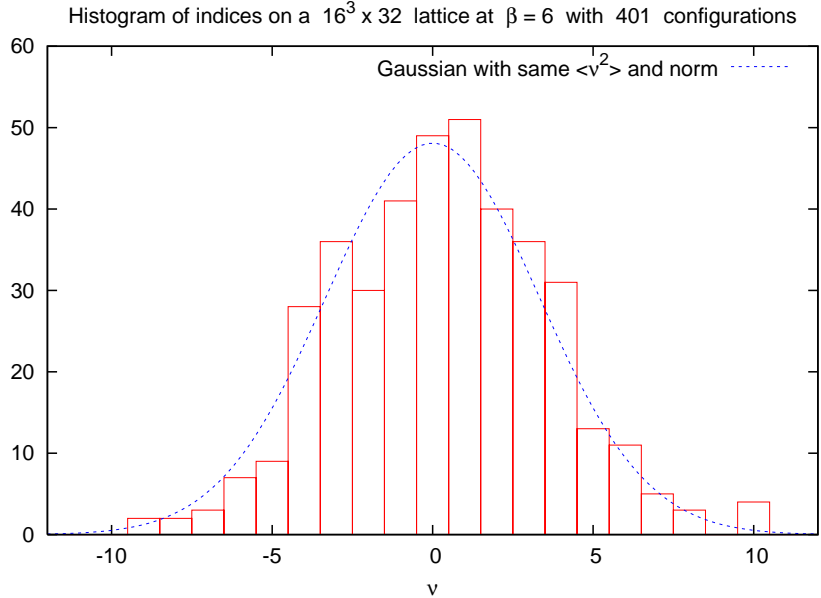}
\caption{The charge histogram and the corresponding Gaussian distribution with the same $\langle \nu^2\rangle$ and norm.}
\label{fig:gaussiancharge}
\end{figure}

One expects a Gaussian charge distribution which is confirmed by our data, see Figure~\ref{fig:gaussiancharge}.
In order to render the values of the topological susceptibility stable against the fluctuations due to a few entries of high charges ($|\nu|\ge 8$) on the lattice volumes $12^3\times 24$ and $16^3\times 32$ we applied the following method: we counted the number of entries $q_{\nu}$ in the interval $[-\nu-0.5,\nu+0.5]$ and normalized them to the total number of entries. Then we employed the formula
\begin{eqnarray}
q_{\nu}&=&\frac{1}{\sqrt{2\pi}}\int_{-u}^{u}\exp{(-y^2/2)}\,dy\ ,\nonumber \\
u&=&\frac{|\nu|+0.5}{\langle \nu^2 \rangle} \ ,
\label{eq:errfun}
\end{eqnarray}
to extract the value of $\langle \nu^2 \rangle$ and thereby the topological susceptibility. We calculated it for $|\nu|=1,2,\dots ,5 $. Out of these values  we computed a mean value and a standard deviation.  In Table~\ref{tab:susc2} the topological susceptibility calculated with this method is shown.
\begin{table}[H]
\centering
\begin{tabular}{|c|c|c|c|c|c|c|}
\hline
operator&lattice & & &physical &&\\
&size & $\beta$ &$\mu$& volume & $\chi_t r_0^4$& $N_{\rm conf}$ \\
\hline
\hline
Neuberger&$12^{4}$ & $6$&$1.4$ & $(1.12 \fm)^{4}$ & $0.063 \pm 0.008$&$143$  \\
\hline 
Neuberger&$10^{4}$ & $5.85$ &$1.6$& $( 1.23\fm)^{4}$ &  $0.078\pm 0.006$&$308$\\
\hline 
Neuberger&$12^{3}\times 24$ & $5.85$&$1.6$ & $( 1.48\fm)^{3}\times 2.96\fm$ & $0.068\pm 0.0048$&$400$ \\
\hline 
Neuberger&$16^{3}\times 32$ & $6$ &$1.4$& $( 1.49\fm)^{3}\times 2.98\fm$ &  $0.073  \pm 0.0053$ &$401$\\
\hline
overlap HF&$12^{3}\times 24$ & $5.85$ &$1$& $( 1.48\fm)^{3}\times 2.96\fm$ &  $0.066\pm 0.0043$ &$391$\\
\hline
\end{tabular}
\caption{ The topological susceptibility calculated with the overlap hypercube Dirac operator and with the Neuberger operator using the standard deviation.\label{tab:susc1}}
\end{table} 
    
\begin{table}[h]
\centering
\begin{tabular}{|c|c|c|c|c|c|c|}
\hline
operator &lattice &&& physical &&\\
 &volume&$\beta$ &$\mu$& volume & $\chi_t r_0^4$& $N_{\rm conf}$ \\
\hline
\hline
Neuberger&$12^{3}\times 24$ & $5.85$&$1.6$ & $( 1.48\fm)^{3}\times 2.96\fm$ & $0.069\pm 0.0013$&$400$ \\
\hline 
Neuberger&$16^{3}\times 32$ & $6$ &$1.4$& $( 1.49\fm)^{3}\times 2.98\fm$ &  $0.072  \pm 0.0054$ &$401$\\
\hline 
overlap HF&$12^{3}\times 24$ & $5.85$ &$1$& $( 1.48\fm)^{3}\times 2.96\fm$ &$0.065\pm 0.0027$ &$391$ \\
\hline
\end{tabular}
\caption{The topological susceptibility calculated with the overlap hypercube Dirac operator and the Neuberger operator using Eq.~(\ref{eq:errfun})\label{tab:susc2}}
\end{table} 

There are two issues when considering the topological susceptibility calculated on the lattice. First the finite size effects dependence of the topological susceptibility, which is governed by $\exp{(-LM_{\rm glueball})}$, where $M_{\rm glueball}\approx 1.5$ GeV is the mass of the lightest glueball. Thus for lattices of $L \ge 1\fm$ the finite size effects on the topological susceptibility are expected to be far below our statistical errors. We see from Table~\ref{tab:susc1} that at $\beta=5.85$ and lattice volumes $V=10^4$ and $12^3\times 24$ the corresponding values of the topological susceptibility coincide within the error bars. Second, at finite lattice spacing $a$ the topological susceptibility suffers from discretization effects starting from ${\cal O}(a^2)$. 

\begin{figure}[t]
\centering
\includegraphics[width=0.6\textwidth,angle=-90]{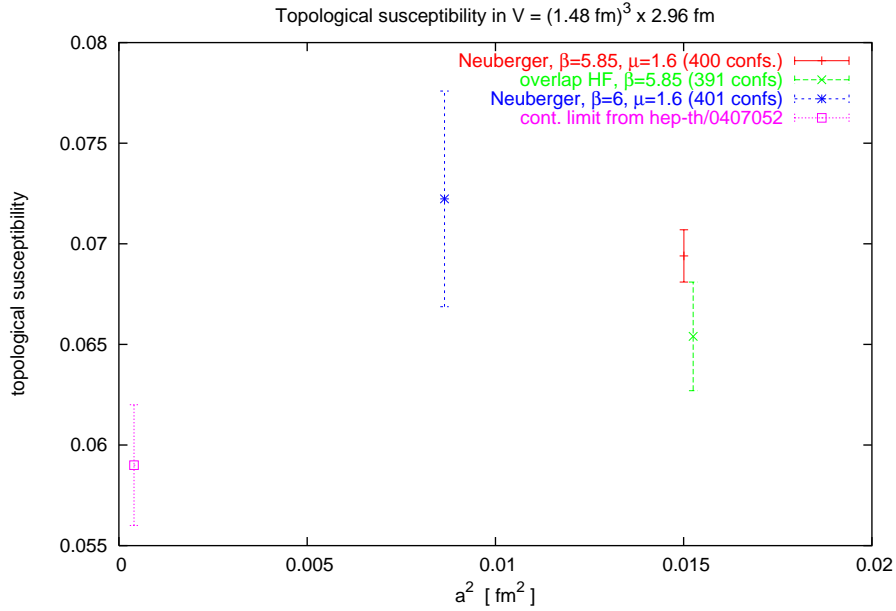}
\caption{The topological susceptibility $\chi_t r_0^4$, calculated with the Neuberger operator and the overlap hypercube Dirac operator, vs. the square of the lattice spacing.}
\label{fig:sus_scaling}
\end{figure}

In Figure~\ref{fig:sus_scaling} we plot the topological susceptibility calculated with the Neuberger operator and the overlap hypercube operator using Eq.~(\ref{eq:errfun}) vs. the square of the lattice spacing. We fixed the physical volume to $(1.48\fm)^3\times 2.96\fm$. We show the results for the Neuberger operator for $12^3\times 24$ volume at $\beta=5.85$ and $16^3\times 32$ at $\beta=6$. The result for the overlap hypercube fermions is given for the volume $12^3\times24$ at $\beta=5.85$. For comparison we also show the continuum extrapolated value from Ref.~\cite{DelDebbio:2004ns}. We see that the value of the topological susceptibility calculated with the overlap hypercube Dirac operator is closer to the continuum value than the corresponding value of the Neuberger operator. This result hints at an improved scaling behavior of the overlap HF, though further statistics and observables are needed to certify this trend.

\chapter{Mesonic two-point functions}

%-------------------------------------------------------------------------------------------------
\section{The axial-vector correlator}
In this Section we discuss results on the axial-vector correlator. Most of these results were published in Ref.~\cite{Bietenholz:2003bj}. We simulated quenched QCD with the Wilson gauge action on the lattices of size $10^3\times24$ and $12^4$ at $\beta=6$ with periodic boundary conditions. As a regularization scheme for the fermions we chose the overlap Wilson Dirac operator. We calculated the axial-vector correlator according to formula~(\ref{axialcurrentcorr_spectral}). To this end we needed to invert the overlap Wilson Dirac operator at a small quark mass $m_q$.
We approximated the square root appearing in  $D_{\rm ov}$ with Chebyshev polynomials to accuracy $10^{-16}$. 
The statistics of the configurations for various topological sectors is given in Table~\ref{tabcorr}.
 To compute the index we used PARPACK~\cite{PARPACK1,PARPACK2} and Ritz functional methods~\cite{Jansen_notes,Kalkreuter:1996mm}. We kept the quark masses light enough in order to stay in the $\epsilon$--regime. To quantify this condition, $z_q=m_q\Sigma V$ must be well below $1$. In particular the considered quark masses were chosen in a way that $z_q\le 0.34$.  
\begin{table}[H]
\centering
\begin{tabular}{|c|c|c|c|c|c|c|}
\hline
lattice & & &physical &
\multicolumn{3}{|c|}{number of configurations} \\
size & $\beta$&$\mu$ & volume & $\nu = 0$ & $\vert \nu \vert = 1$ 
& $\vert \nu  \vert = 2$  \\
\hline
\hline
$12^{4}$ & $6$ & $1.4$&$(1.12 \fm)^{4}$ & 47 & 78 & 24  \\
\hline 
$10^{3}\times 24$ & $6$ & $1.4$&$(0.93 \fm)^{3}\times 1.86 \fm$ & 20 & 24 & 17 \\
\hline
\end{tabular}
\caption{The statistics of our simulations on two
lattice sizes.\label{tabcorr}}
\end{table}
 We determined the parameter $F_{\pi}$  by making a fit with an additive constant in Eq.~(\ref{AA}) at $\tau=0.5$. Thus it can easily be computed. The chiral condensate $\Sigma$, on the other hand, is proportional to the curvature of the parabola. However, as the value of $\Sigma$ varies in the range of interest, the corresponding parabola changes only very little which makes the extraction of $\Sigma$ extremely difficult. 

\begin{figure}%[htbp]
\centering
\includegraphics[width=0.4\textwidth,angle=-90]{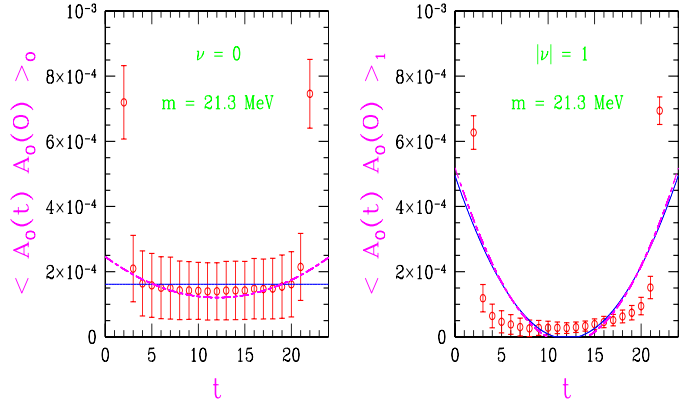}
\caption{The axial-vector correlator computed at quark mass $21.3\, \mbox{MeV}$ for a lattice volume $10^3\times 24$ at $\beta=6$. The left plot shows the axial-vector correlator for topological charge $\nu=0$ and the right plot for $|\nu|=1$. The result at $\nu=0$ suffers from large error bars. In the sector $|\nu|=1$ we see a clear failure to reproduce the q$\chi$PT predictions (curves).}
\label{axialcorr_2vol}
\end{figure}

We find that the lattice size $10^3\times 24$ at $\beta=6$ is too small to accommodate the predictions by the $\chi$PT. In Figure~\ref{axialcorr_2vol} we show the data for the axial-vector correlator at quark mass $m_q=21.3\,\mbox{MeV}$, lattice  volume $V=10^3\times24$, at $\beta=6$ for topological sectors $|\nu|=1,2$ along with the corresponding analytical predictions. We plot the analytical curves for two very distinct values of the chiral condensate, $\Sigma=0$ and $\Sigma=(250\,\mbox{MeV})^3$. We see that the resulting curves lie too close to each other to allow for $\Sigma$ to be evaluated. The fact that those two curves are almost indistinguishable is valid for topological sectors $|\nu| \ne 0$. It is due to the last term in the expression~(\ref{AA}). However in the topological sector $0$ the sensitivity is restored. At $|\nu|=1$ we see that the analytical prediction for this lattice volume does not model the data. The failure of the $\chi$PT to describe the result on volumes with box length less than $1.2 \fm$ has been observed before in Section 6.1 where the distributions of eigenvalues of the overlap Wilson Dirac operator were confronted with the $\chi$RMT predictions. Thus this conclusion is rather general and sets a consistent constraint in the lattice simulations where the results can be confronted with $\chi$PT.

\begin{figure}%[htbp]
\centering
\includegraphics[width=0.4\textwidth,angle=-90]{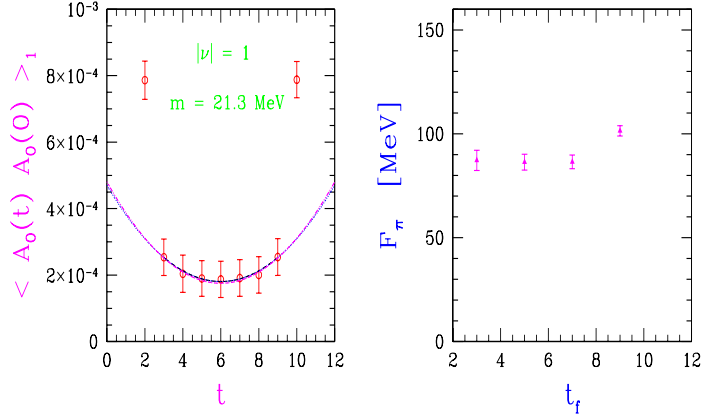}
\caption{The fit of the quenched QCD simulation data for the axial-vector correlator vs. the analytical predictions by $\chi$PT (left plot). The curves represent the q$\chi$PT predictions for $\Sigma=0$ and $\Sigma=(250\, {\rm MeV})^3$. The extracted value of $F_{\pi}$ vs.  number of points $t_f$ for the fit range around $t/T=0.5$ (is shown on the right plot). The lattice volume is $12^4$, $\beta=6$, $|\nu|=1$ and $m_q=21.3\,\mbox{MeV}$.}
\label{axialcorr_fit}
\end{figure}

\begin{figure}%[htbp]
\centering
\includegraphics[width=0.4\textwidth,angle=-90]{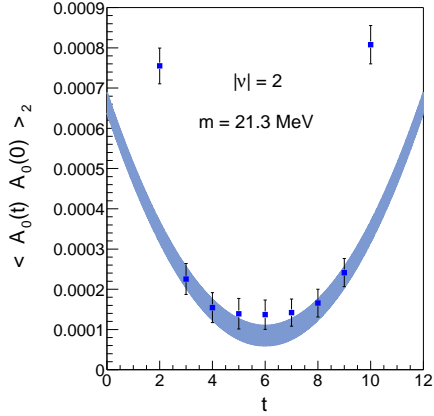}
\caption{The quenched QCD simulation data for the axial-vector correlator vs. the analytical predictions with the parameter $F_{\pi}$ in the range that was found by fitting in the topological sector $|\nu|=1$ (shaded area). The lattice volume is $V=12^3$, at $\beta=6$, for topological charge $|\nu|=2$. The quark mass is $m_q=21.3\, \mbox{MeV}$.}
\label{axialcorr_fit_2}
\end{figure}

Next we consider the fit by the analytical prediction of the $\chi$PT to the simulation data on the volume $12^4$ and topological sector $|\nu|=1$. Our $\chi$RMT study of Dirac eigenvalues in Chapter 6 suggests that this volume should be at least at the threshold where the predictions by $\chi$PT  become valid.
The result is shown in Figure~\ref{axialcorr_fit} (on the left) where we fit Eq.~(\ref{AA}) to the simulation data, omitting the points located far from the center which are strongly affected by the contributions of the excited states. We can compute the additive constant that corresponds to $F_{\pi}$ quite accurately. Figure~\ref{axialcorr_fit} (on the right) shows the fitted value of $F_{\pi}$ against the extension of the time interval where the fit was done. We find a decent plateau which suggests a value $F_{\pi}=87 \pm 4\, \mbox{MeV}$ for the quenched, bare pion decay constant. Again the chiral condensate cannot be extracted from these data. This we demonstrate in Figure~\ref{axialcorr_fit} (on the left) where we plot the analytical prediction for two cases $\Sigma=0$ and $\Sigma=(250\,\mbox{MeV})^3$. We see that the resulting curves reside within the error bars and hence make the determination of $\Sigma$ impossible.  Finally we look at the simulation data on $12^4$ lattice at $\beta=6$ for the topological sector $|\nu|=2$. To check whether the data agree with the analytical predictions we varied $F_{\pi}$ in the range that we found from the fit in the topological sector $|\nu|=1$. The resulting shaded area is shown in Figure~\ref{axialcorr_fit_2}. We conclude that the data for the topological sectors $|\nu|=1$ and $2$ agree within the errors. Note that the agreement at $|\nu|=2$ is, however, somewhat worse than for the topological sector $|\nu|=1$ (the data are more flat). This is, in fact, in agreement with the $\chi$RMT study of the eigenvalues of the Dirac operator, where we found that the agreement with the $\chi$RMT is expected to be worse as we increase the topological sector $|\nu|$.    

We obtained of course the bare values for $F_{\pi}$ which are still subject to the renormalization. The latter is expected to increase the bare values substantially.~\cite{Berruto:2003rt,BietenholzPP}. 

%--------------------------------------------------------------------------------   
\section{Subtleties of numerical simulations in the \texorpdfstring{$\epsilon$--}{epsilon--}regime}
In this Section we address the practical side of the simulations in the $\epsilon$--regime. 

\begin{figure}%[htbp]
\centering
\includegraphics[width=0.4\textwidth,angle=-90]{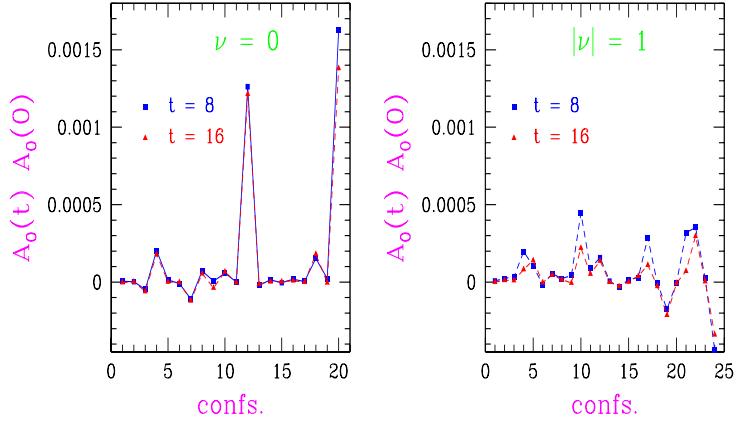}
\caption{Monte Carlo histories of the axial-vector correlator for topological sector $\nu=0$ (left plot) and topological sector $|\nu|=1$ (right plot), at $V=10^3\times 24$, $\beta=6$. }
\label{axialcorr_hist}
\end{figure}

\begin{figure}%[htbp]
\centering
\includegraphics[width=0.5\textwidth]{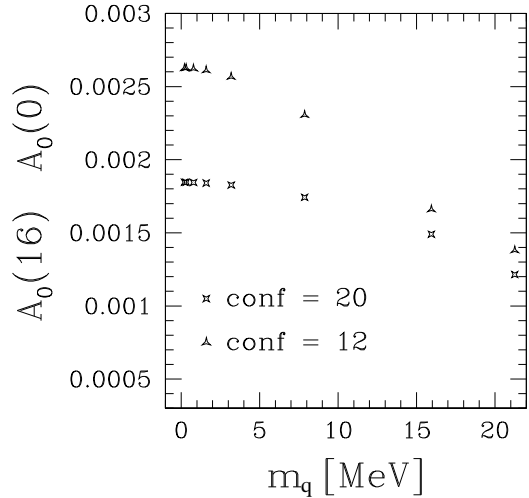}
\caption{The height of the spikes vs. the bare quark mass.}
\label{axialcorr_spikes_height}
\end{figure}

In the $\epsilon$--regime the sector of topological charge $\nu=0$ may appeal as a good candidate where the chiral condensate can be extracted since the expression~(\ref{AA}) does not show such an insensitivity to $\Sigma$ as it is the case for non-zero topological sectors.   Unfortunately, this sector is not favorable for the numerical simulations because of the frequently occurring small eigenvalues. As we saw in Section 2.4 and in Chapter 6, where we studied the probability distribution of the lowest eigenvalues, it is the topological sector $\nu=0$ where we observe a high probability of very small eigenvalues $\lambda_1$. As we increase the topological charge, the corresponding probability is suppressed, see for instance Figures~\ref{RMT} and \ref{RMT_10}. This behavior is reflected in the relatively large error bars for the axial-vector correlator in the topological sector $0$, illustrated in Figure~\ref{axialcorr_2vol}. We can demonstrate this explicitly by showing how the axial-vector correlator is sampled for the topological sectors $0$ and $1$. In Figure~\ref{axialcorr_hist} the Monte Carlo histories of $A_0(t) A_0(0)\vert_{\nu=0} $ and   $A_0(t) A_0(0) \vert_{|\nu|=1}$ are shown.
In particular we see strong spikes in the history of the axial-vector correlator for the neutral topological sector. We checked the configurations where they show up. They correspond to cases of small eigenvalues of the overlap Wilson Dirac operator. The heights of the spikes become larger for smaller quark masses. This is shown in Figure~\ref{axialcorr_spikes_height}.
The height of the spikes rises monotonously in $\frac{1}{|\lambda_1|}$, where $\lambda_1$ is the lowest eigenvalue of the Dirac operator. This also holds for the spikes in the non-trivial sectors, where, however, the occurrence of such very small modes is suppressed. In practice one has to trade off the quark mass in a way first to ensure that it is small enough to be in the $\epsilon$--regime and secondly to have it large enough to damp down the values of the spikes.

\begin{figure}%[htbp]
\centering
\includegraphics[width=0.5\textwidth, angle=-90]{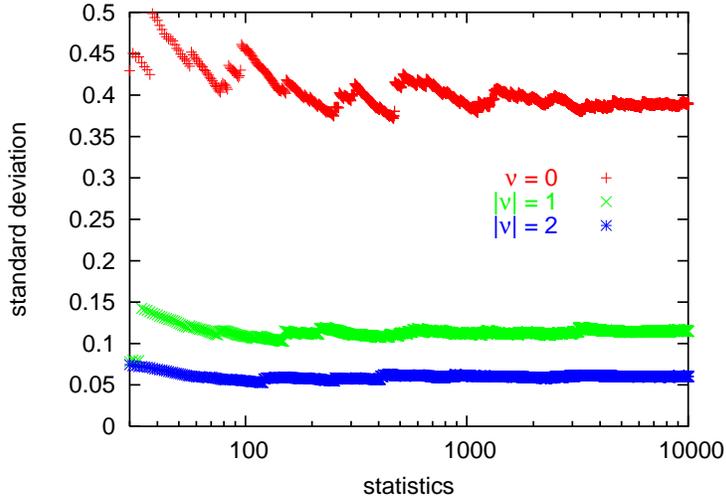}
\caption{The standard deviation of $\Sigma^{(\nu)}_{\rm min}$ for topological sectors $|\nu|=0$, $1$ and $2$ vs. pseudo statistics. The stabilization is observed much earlier for the topological sectors $|\nu|\ne 0$ than for $\nu=0$.}
\label{fake_sim}
\end{figure} 
Now we want to trace the effect of the spikes on the statistics necessary to sample the Monte Carlo averages of observables of interest. In particular we consider the chiral condensate, which is related to the density of the lowest eigenvalues of the Dirac operator by the Banks-Casher relation~(\ref{BC}). 
The starting point for us will be the spectral representation of the chiral condensate~(\ref{cond_spectral}).
If we consider the dependence of the chiral condensate on the mass of the valence quark, $m_q$, i.e. the quark which does not show up in the path integral average --- then we can rewrite
\begin{equation}
  \Sigma^{(\nu)}(m_q)=\int_{0}^{\infty} dz \,\frac{2m_q\rho^{(\nu)}_s(z)/V}{m_q^2+(z/(\Sigma V))^2}\ , \quad z=\lambda \Sigma V \ .
\end{equation}
Now we take the first term in the expansion~(\ref{micr_den_expan}) and consider only the contribution of the lowest eigenvalue to the chiral condensate
\begin{equation}
\Sigma_{\rm min}^{(\nu)}(m_q)=\int_0^{\infty} dz \,\frac{2m_q\rho^{(\nu)}_1(z)/V}{m_q^2+(z/(\Sigma V))^2}\ .
\label{sigmamin}
\end{equation}
We evaluate this integral by a Monte Carlo method, whereby we sample the integral at the values $z$ distributed according to the $\chi$RMT predictions for $\rho^{(\nu)}_1(z)$ given in Eq.~(\ref{eq:RMTpredictions012}). The resulting value can be trusted only if the standard deviation stabilizes. In Figure~\ref{fake_sim} we show the simulation of the integral at quark mass $m_q=21.3\,\mbox{MeV}$ for various topological charges. In particular we show the standard deviation vs. the statistics.
We find that one needs ${\cal O}(10^4)$ configurations to sample $\Sigma^{(0)}_{\rm min}$, whereas this number reduces to ${\cal O}(10^2)$ for topological charges $|\nu|=1$ and $2$.
Therefore it is advantageous to simulate in topological sectors $|\nu|>0$.

As a possible remedy for the spikes there was proposed the low mode averaging method in Ref.~\cite{Giusti:2004yp}.
%----------------------------------------------------------------------------------------------------
\section{Zero mode contributions to the pseudo-scalar correlator}
In this Section we discuss the results for the zero mode contribution to the pseudo-scalar correlator. The necessary theoretical introduction, as well as the notation, were fixed in  Subsection 2.3.4. The method employed here aims to consider only the zero mode contributions to the pseudo-scalar correlation function. The latter allows to compute the low energy constants as we saw in Subsection 2.3.4. Here we discuss the numerical evaluation. 

We simulated quenched QCD on a $12^3\times 24$  at $\beta=5.85$ and on a  $16^3\times 32$  at $\beta=6$ lattices with periodic boundary conditions with the Wilson gauge action and computed the lowest eigenmodes of the Neuberger operator. On the $12^3\times 24$ lattice at $\beta=5.85$  we also calculated the lowest eigenmodes of the overlap hypercube Dirac operator. Out of the eigenvectors we evaluated the zero mode contribution to the pseudo-scalar correlator according to formulae~(\ref{eq_zeromode}~--~\ref{eq_disconnected}). We use the next--to--next--to--leading order expression~(\ref{zero_mode_contr}) in the $\epsilon$--regime to confront the time derivative of the zero mode contribution to the pseudo-scalar correlator with the corresponding quantity built up from the simulation data. This concept follows Ref.~\cite{Giusti:2003iq}, which also presents the first numerical study.

\begin{figure}%[htbp]
\centering
\includegraphics[width=0.6\textwidth,angle=-90]{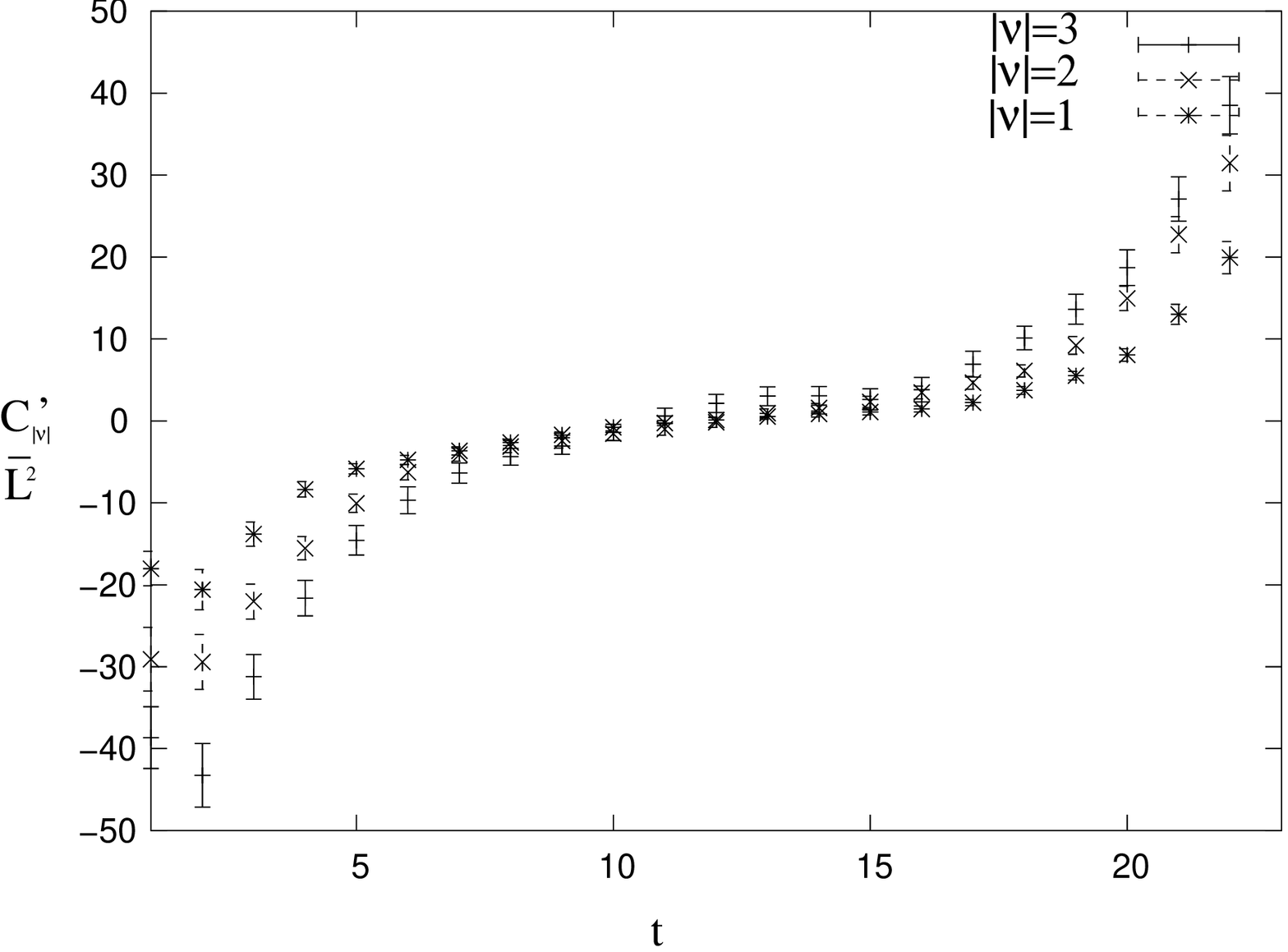}

\includegraphics[width=0.6\textwidth,angle=-90]{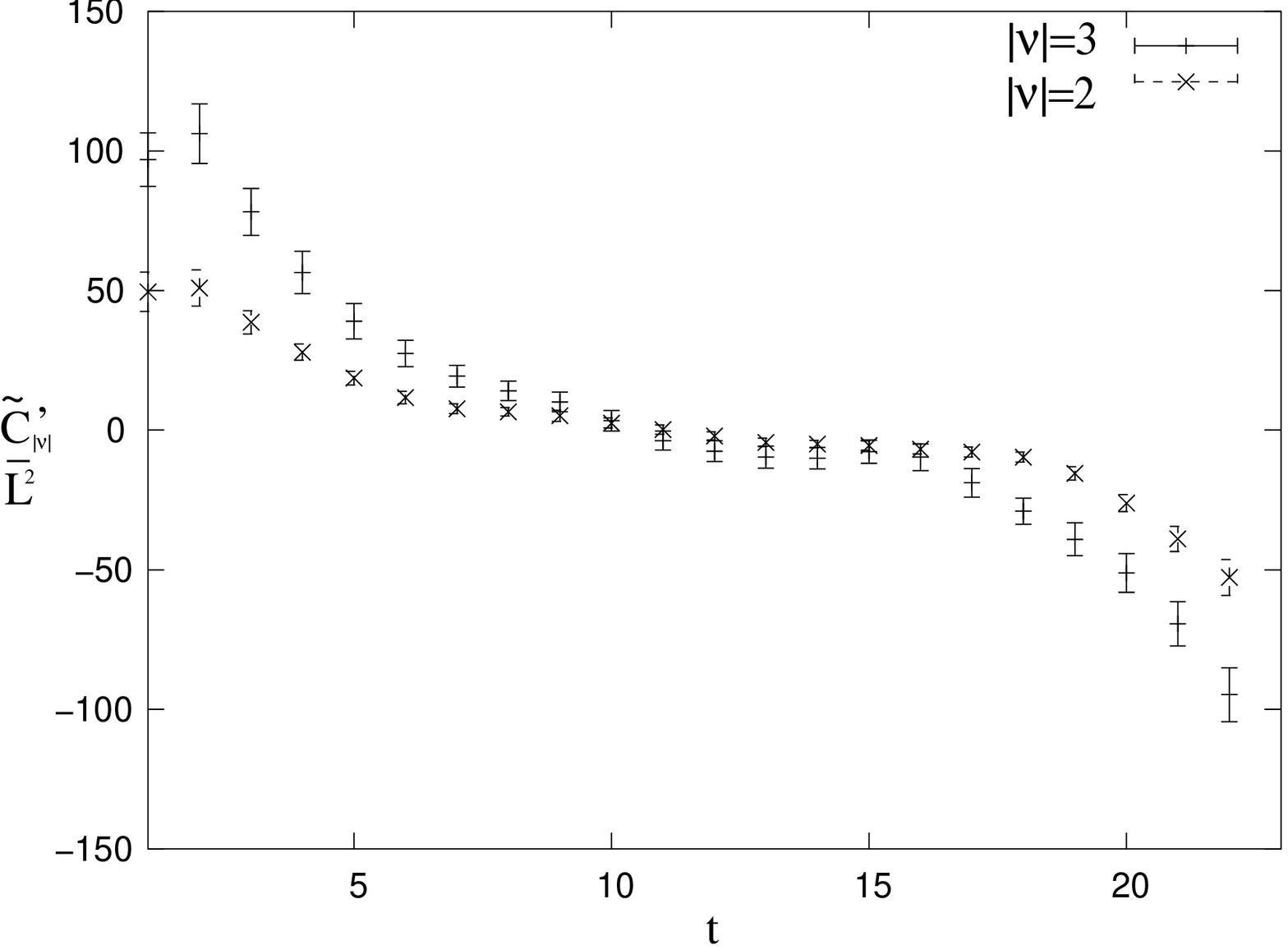}
\caption{The plot on top shows the connected contributions of zero modes to the pseudo-scalar correlator for topological sectors $|\nu|=1$, $2$ and $3$. The plot at the bottom represents the disconnected contributions to the pseudo-scalar correlator for topological sectors $|\nu|=2$ and $3$. We use the Neuberger operator on a lattice volume  $V=12^3\times24$, at $\beta=5.85$.}
\label{zeromode_contr_data}
\end{figure}
\begin{figure}%[htbp]
\centering
\includegraphics[width=0.6\textwidth,angle=-90]{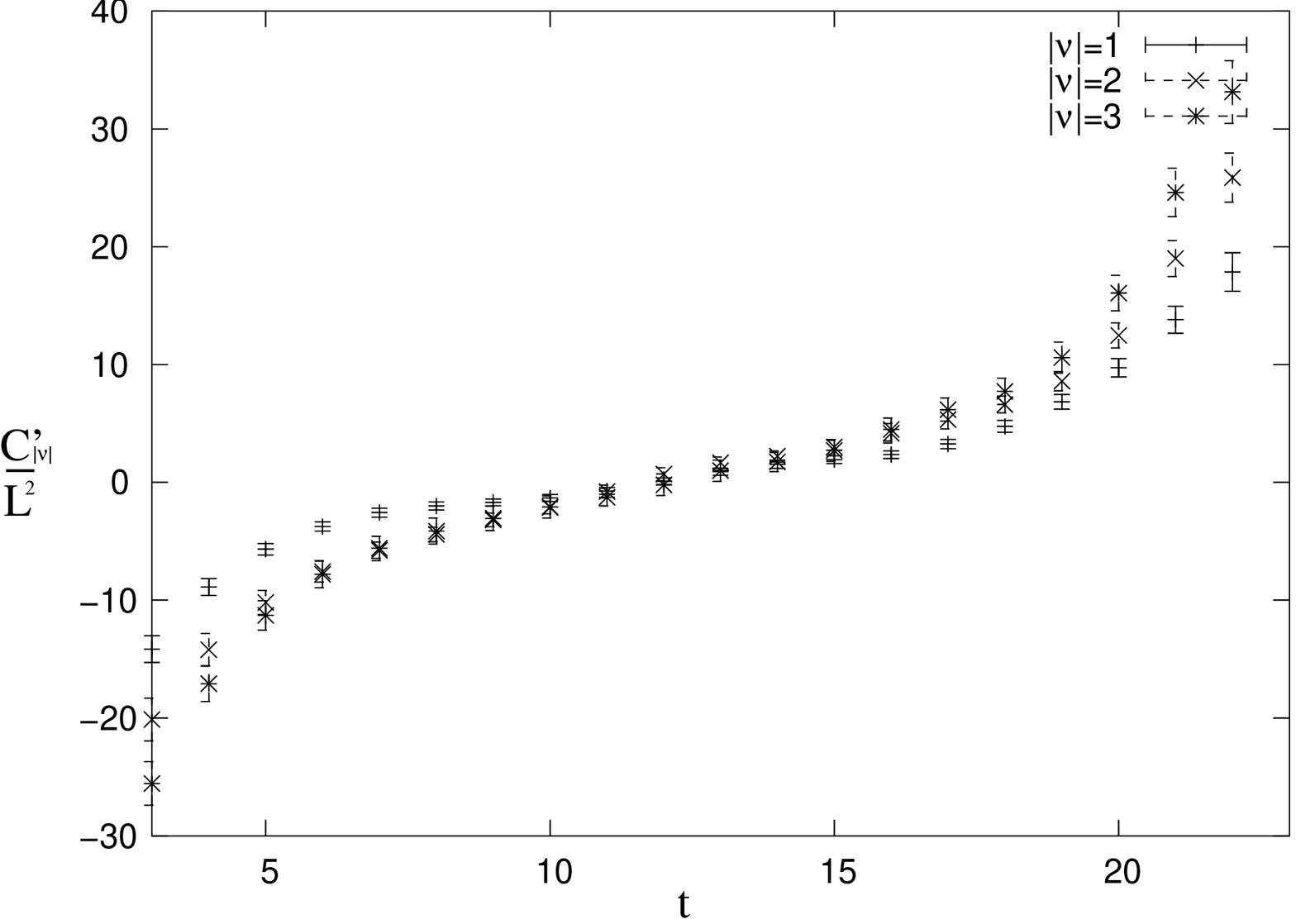}

\includegraphics[width=0.6\textwidth,angle=-90]{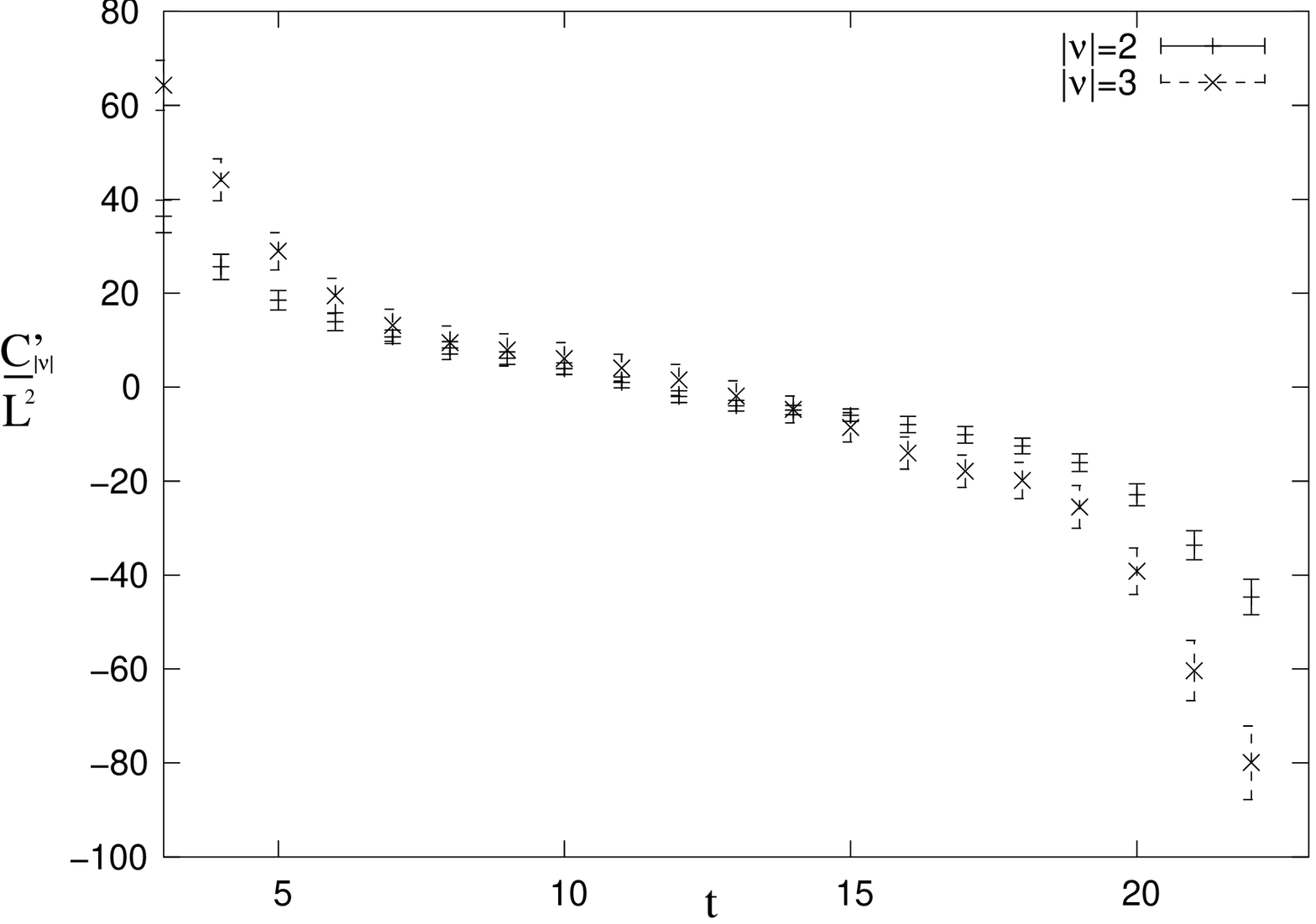}
\caption{The plot on top shows the connected contributions of zero modes to the pseudo-scalar correlator for topological sectors $|\nu|=1$, $2$ and $3$. The plot at the bottom represents the disconnected contributions to the pseudo-scalar correlator for topological sectors $|\nu|=2$ and $3$. We use the overlap hypercube operator on a lattice volume  $V=12^3\times24$, at $\beta=5.85$.}
\label{zeromode_contr_data_HF}
\end{figure}
\begin{figure}%[htbp]
\centering
\includegraphics[width=0.6\textwidth,angle=-90]{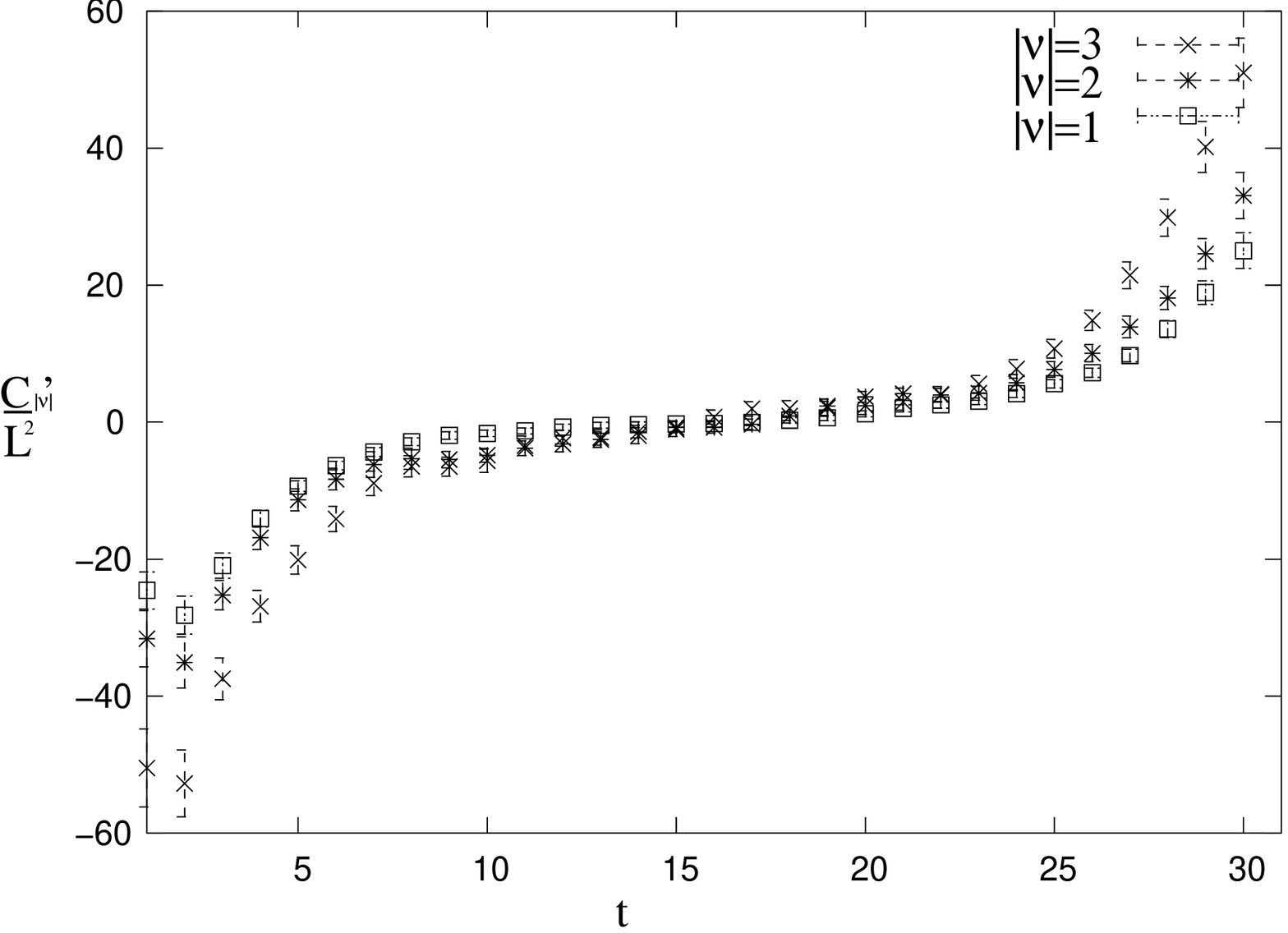}

\includegraphics[width=0.6\textwidth,angle=-90]{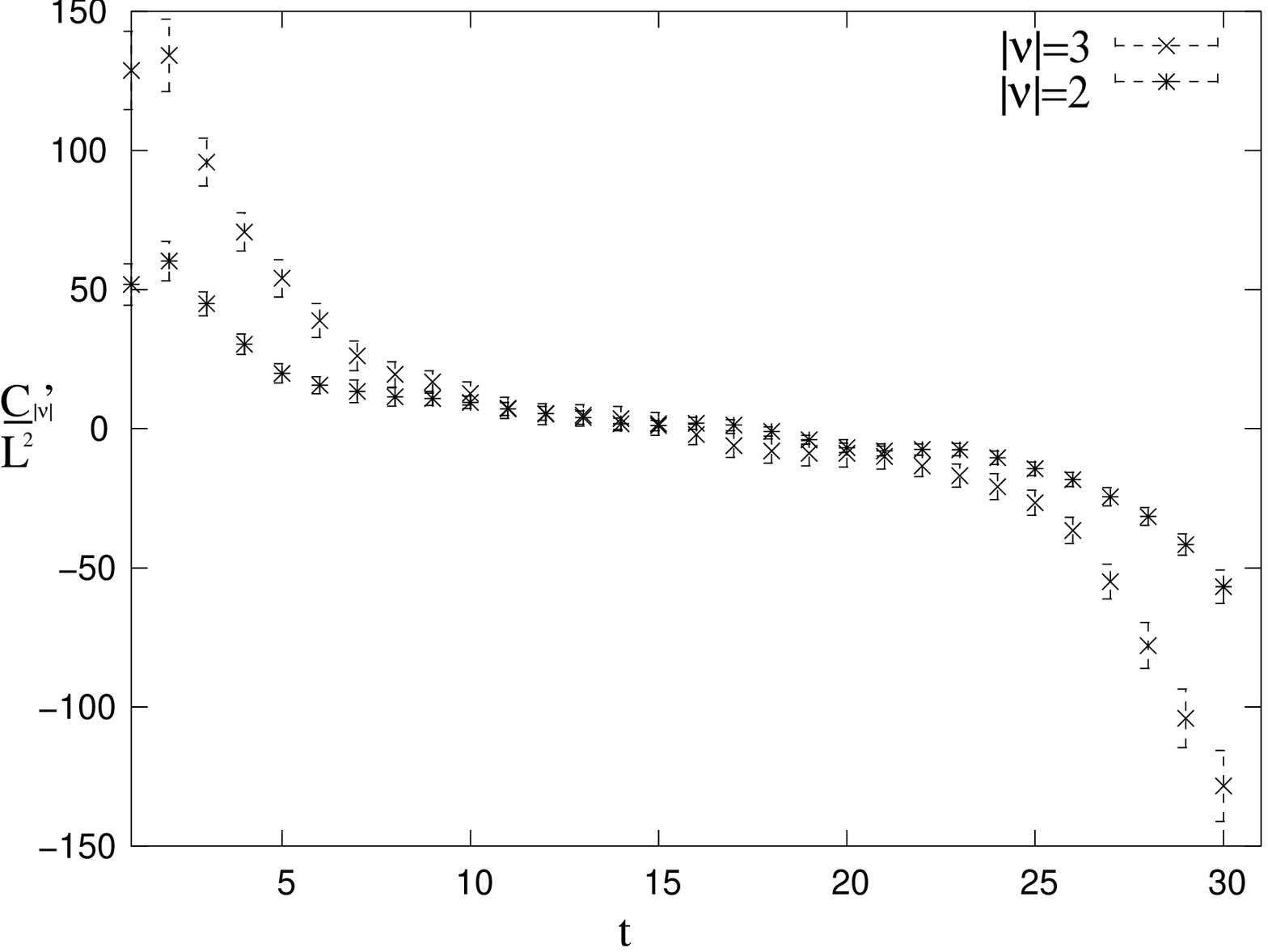}
\caption{The plot on top shows the connected contributions of zero modes to the pseudo-scalar correlator for topological sectors $|\nu|=1$, $2$ and $3$. The plot at the bottom represents the disconnected contributions to the pseudo-scalar correlator for topological sectors $|\nu|=2$ and $3$. We use the Neuberger operator on a lattice volume  $V=16^3\times32$, at $\beta=6$.}
\label{zeromode_contr_data_L16T32}
\end{figure}

We looked at topological sectors $|\nu|=1$, $2$ and $3$ and we found that only the data from the topological sectors $1$ and $2$ can be used to obtain the low energy constants by fitting to the $\chi$PT predictions. 
Figures~\ref{zeromode_contr_data} and~\ref{zeromode_contr_data_L16T32} show the simulation data for the connected contribution ${\cal C}_{|\nu|}  /L^2$ and for the disconnected contribution $\widetilde {\cal C}_{|\nu|}/L^2$ computed with the Neuberger operator. For the disconnected part we show data only for topological sectors $|\nu|=2$ and $3$ since the data for the topological sector $1$ are the same as for the connected contribution up to the negative sign.  In Figures~\ref{zeromode_contr_data_HF} we show the contribution of the zero modes calculated with the overlap hypercube fermions.  

The time derivative of the zero mode contribution to the pseudo-scalar correlator was approximated by the symmetric lattice derivative 
\begin{equation}
\frac{df}{dt}\simeq \frac{f(t+1)-f(t-1)}{2}\ .
\end{equation}

First we fitted the data using the general formula~(\ref{eq_connected}). Since that formula contains the parameters $F_{\pi}$, $\alpha$ and $\langle \nu^2 \rangle$ as unknown coefficients, we have to perform a $3$--parameter fit. The shape coefficient $\beta_1$  for our anisotropic lattice volume was calculated according to the prescription in Ref.~\cite{Hasenfratz:1990pk}, 
$$
\beta_1=0.131465 \ .
$$
The fit failed to produce reasonable estimates for the parameters. Therefore we followed the idea outlined in Ref.~\cite{Giusti:2003iq} and considered the expansion of the formula~(\ref{eq_connected}) in powers of $t/T-0.5$,
\begin{eqnarray}
\frac{1}{L^2} C_{|\nu|}'(t)=D_{|\nu|}\cdot(t/T-0.5) +{\cal O}((t/T-0.5)^3)\nonumber\ ,\\
\frac{1}{L^2}\widetilde{C'}_{|\nu|}(t)=\widetilde{D}_{|\nu|}\cdot(t/T-0.5) +{\cal O}((t/T-0.5)^3)\ ,
\label{formula1}
\end{eqnarray}
where the coefficients $D_{|\nu|}$ and $\widetilde D_{|\nu|}$ are given by
\begin{eqnarray}
 D_{|\nu|}  &=&
 \frac{2|\nu|}{(F_{\pi}L)^2}
 \biggl\{
 |\nu|  + \frac{\alpha}{2 N_c} - \frac{\beta_1}{F_{\pi}^2\sqrt{V}}+ \nonumber \\
&&+\biggl[ 
 \biggl( 
 \fr73 + 2 \nu^2 - 2 \langle \nu^2 \rangle
 \biggr) \zeta_2 + \fr12 \gamma_1
 \biggr] \frac{T^2}{F_{\pi}^2 V} 
 \biggr\} 
 \;, \label{qDnu} \nonumber \\ \bigskip
 \widetilde D_{|\nu|}  &=& - 
 \frac{2|\nu|}{(F_{\pi}L)^2}
 \biggl\{
 1 +|\nu| \biggl( \frac{\alpha}{2 N_c}  - \frac{\beta_1}{F_{\pi}^2 \sqrt{V}} 
 \biggr) + \nonumber \\
 &&+|\nu| \biggl[ 
 \biggl( \frac{13}{3} - 2 \langle \nu^2 \rangle \biggr)\zeta_2 + 
 \frac{ \gamma_1}{2} 
 \biggr] \frac{T^2}{F_{\pi}^2 V}
 \biggr\} \ .\label{qDnup}
\end{eqnarray}
Here 
\begin{eqnarray}
h_2'(\tau)&=&\zeta_2 (\tau-0.5) +{\cal O}((\tau-0.5)^3)\ ,\nonumber \\ 
h_3'(\tau)&=&\gamma_1 (\tau-0.5)+ {\cal O}((\tau-0.5)^3)\ ,\nonumber \\
\zeta_2&=&-\frac{1}{24} \quad \mbox{and} \quad \gamma_1=-\frac{1}{12}+ \frac{1}{2}\sum_{\vec{n} \ne \vec{0}} \frac{1}{\sinh^2(T|\vec{p}|/2)}=-0.083291\ .
\end{eqnarray}
with $|\vec{p}|=(2\pi/L)|\vec{n}|$, as in Eq.~(\ref{g1}). The coefficients $D_{|\nu |}$ and $\widetilde D_{|\nu |}$ are arranged as sums of terms of the leading order ${\cal O}(\epsilon^4)$, next-to-leading order ${\cal O}(\epsilon^6)$ and next-to-next-to-leading order ${\cal O}(\epsilon^8)$, where $\epsilon=1/(L\bar{F}_{\pi})$ and $\bar{F}_{\pi}=F_{\pi}/\sqrt{3}$ was introduced in Eq.~(\ref{eq:counting_Nc}).

\begin{figure}%[htbp]
\centering
\includegraphics[width=0.6\textwidth,angle=-90]{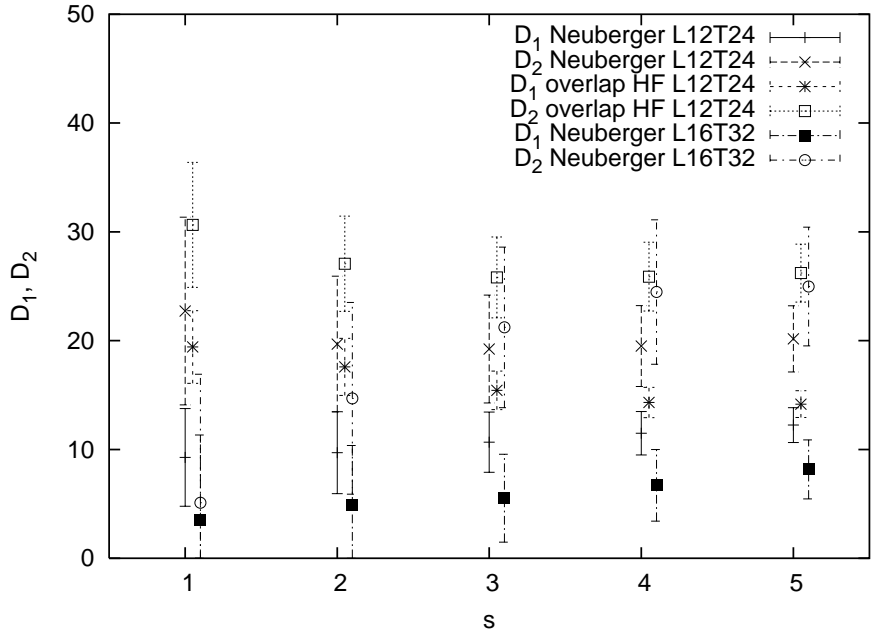}
\caption{The slopes $D_1$ and $D_2$ vs. the fitting range $[T/2-s,T/2+s]$. We used the Neuberger operator on a $12^3\times 24$ lattice, at $\beta=5.85$ and on a $16^3\times 32$ lattice, at $\beta=6$ as well as the overlap hypercube Dirac operator on a $12^3\times 24$ lattice, at $\beta=5.85$}
\label{fig:slopes_vs_range}
\end{figure}

We employ these analytical predictions up to next--to--next--to--leading order to analyze our data. Applying one parameter fits of the data in topological sectors $|\nu|=1$, $2$ and $3$ to the linear part of the formula~(\ref{formula1}), we extracted the slopes of $C_{|\nu |}'(t)/L^2$ from the data with jack-knife errors~\cite{Gottlieb:1985rc,Gupta:1987zc}. The results for different fitting ranges are given in Tables~\ref{tab:slopes}, \ref{tab:slopes_HF}, \ref{tab:slopes_L16T32} and in Figure~\ref{fig:slopes_vs_range}. 
 \begin{table}[H]
\centering
\begin{tabular}{|c|c|c|c|c|c|c|}
\hline
fitting & \multicolumn{6}{|c|}{$D_{|\nu |}$} \\
range & $\vert \nu \vert = 1$ & $\vert \nu \vert = 2$& $\vert \nu  \vert = 3$&$\chi^2/d.o.f.$& $\chi^2/d.o.f.$ & $\chi^2/d.o.f.$ \\
\hline
\hline
$[11,13]$ &$9.2\pm 4.5$& $22.7\pm 8.6$&$20.3\pm 12.7$&$0.2$ &$0.05$ &$4.9$\\
\hline
$[10,14]$ &$9.7\pm 3.8$& $19.7\pm 6.2$&$24.3\pm 9.9$&$0.1$ &$0.05$ &$2.8$\\
\hline
$[7,17]$ &$12.2 \pm 1.6 $& $20.2 \pm 3.0 $&$27.0 +\pm 5.0 $&$1.3$ &$0.1$ &$1.3$\\
\hline
\end{tabular}
\caption{The slopes $D_{|\nu |}$ vs. the fitting range computed with the overlap Wilson Dirac operator at $\beta=5.85$ on $12^3\times24$ lattice. \label{tab:slopes}}
\end{table}

\begin{table}[H]
\centering
\begin{tabular}{|c|c|c|c|c|c|c|}
\hline
fitting & \multicolumn{6}{|c|}{$D_{|\nu |}$} \\
range & $\vert \nu \vert = 1$ & $\vert \nu \vert = 2$& $\vert \nu \vert = 3$ &$\chi^2/d.o.f.$& $\chi^2/d.o.f.$& $\chi^2/d.o.f.$\\
\hline
\hline
$[11,13]$ &$19.4\pm 3.3$& $30.6\pm 5.7$&$27.7 \pm 11.4$ & $0.16$ & $1.6$ & $0.05$ \\
\hline
$[10,14]$ &$17.6\pm 2.6 $& $27.1\pm 4.4$& $23.7 \pm 8.9$ & $0.17$ &$0.86$ & $0.07$\\
\hline
$[7,17]$ &$14.2 \pm 1.2$& $26.2 \pm 2.7$&$26.2 \pm 4.5$ &$0.6$ &$0.4$& $0.15$ \\
\hline
\end{tabular}
\caption{The slopes $D_{|\nu |}$ vs. the fitting range computed with the overlap hypercube Dirac operator at $\beta=5.85$ on $12^3\times24$ lattice. \label{tab:slopes_HF}}
\end{table}

\begin{table}[H]
\centering
\begin{tabular}{|c|c|c|c|c|c|c|}
\hline
fitting & \multicolumn{6}{|c|}{$D_{|\nu |}$} \\
range & $\vert \nu \vert = 1$ & $\vert \nu \vert = 2$& $\vert \nu \vert = 3$ &$\chi^2/d.o.f.$& $\chi^2/d.o.f.$& $\chi^2/d.o.f.$\\
\hline
\hline
$[15,17]$ &$3.5 \pm 7.8$& $5.1 \pm 11.8$ &$48.1 \pm 14.5$ & $0.21$ & $0.87$ & $0.47$ \\
\hline
$[14,18]$ &$4.9 \pm 5.4$& $14.7 \pm 8.8$& $35.4 \pm 12.9$ & $0.12$ &$0.58$ & $0.32$\\
\hline
$[11,21]$ &$8.2 \pm 2.7$& $25.0 \pm 5.5$&$22.0 \pm 7.4$ &$0.44$ &$0.47$& $0.3$ \\
\hline
\end{tabular}
\caption{The slopes $D_{|\nu |}$ vs. the fitting range computed with the overlap Wilson Dirac operator at $\beta=6$ on $16^3\times 32$ lattice. \label{tab:slopes_L16T32}}
\end{table}

First we see that the data for the topological charge $|\nu|=3$ are statistically not very stable and thus we do not use it to extract physical information. Note that in the case of the overlap hypercube Dirac operator the data for the topological sector $|\nu|=3$ tends to lie on the top of the data for $|\nu|=2$. Therefore we concentrate on the slopes for topological charges $|\nu|=1$ and $2$.

Secondly, we observe in Figure~\ref{fig:slopes_vs_range} that the data stabilize on a fitting range $[T/2-s,T/2+s]$ starting from $s\ge 3$. Note that we used all data points within this interval. Since the term of ${\cal O}((\tau-0.5)^3)$ in the expansion~(\ref{formula1}) is a NNLO term in the counting, it will be much more complicated to extract it. Therefore we consider only the linear term. Note that we neglect the ${\cal O}((\tau-0.5)^3)$ term although we still allow the coefficients $D_{|\nu|}$ to be corrected by the NNLO terms in Eq.~(\ref{qDnup}). 

We now take the values for the topological susceptibility (resp. the expectation value of the topological charge squared, $\langle \nu^2 \rangle$) from our simulations on the $12^3 \times 24$ lattice at $\beta=5.85$ as an input for our method,
\begin{eqnarray}
\langle \nu^2 \rangle &=&10.4\pm 0.7\ ,\quad \mbox{for the overlap Wilson Dirac} \ ,\nonumber \\
\langle \nu^2 \rangle &=&9.8\pm 0.7\ , \quad \mbox{for the overlap hypercube Dirac}\ .
\end{eqnarray} 
Then we end up with a system of two equations
\begin{eqnarray}
D_1&=&D_1(F_{\pi}, \alpha, \langle \nu^2 \rangle)\nonumber \ ,\\
D_2&=&D_2(F_{\pi}, \alpha, \langle \nu^2 \rangle)\ .
\label{system_eq}
\end{eqnarray}
Here the left hand side is the numbers extracted from Figure~\ref{fig:slopes_vs_range} at $s=5$ given in Tables~\ref{tab:slopes}~-~\ref{tab:slopes_L16T32}. We are aware that this choice is quite arbitrary but we have to postpone a more systematic analysis until larger statistics is available.  
Solving them we determine $\alpha$ and $F_{\pi}$. Unlike Ref.~\cite{Giusti:2003iq}, where a three parameter fit to the linear part of the expressions~(\ref{formula1}) was used to extract the low energy constants, in this method we employ only the values of the slopes and their jack-knife errors to obtain the estimates on $F_{\pi}$ and $\alpha$. The corresponding error bars will follow as a result of the error propagation on the fitted values of $D_{|\nu|}$ and the error on the topological susceptibility, i.e.
\begin{eqnarray}
\delta F_{\pi}&=&\sqrt{\sum_{i=1}^{2}\left (\frac{\partial{F_{\pi}(D_1,D_2,\langle \nu^2 \rangle)}}{\partial D_i} \delta D_i \right )^2 +\left (\frac{\partial F_{\pi}(D_1,D_2,\langle \nu^2 \rangle)}{\partial{\langle \nu^2 \rangle}} \delta \langle \nu^2 \rangle\right )^2}\ , \nonumber \\
\delta \alpha &=&\sqrt{\sum_{i=1}^{2}\left (\frac{\partial{\alpha (D_1,D_2,\langle \nu^2 \rangle)}}{\partial D_i} \delta D_i \right )^2 +\left (\frac{\partial \alpha (D_1,D_2,\langle \nu^2 \rangle)}{\partial{\langle \nu^2 \rangle}} \delta \langle \nu^2 \rangle\right )^2}\ .
\label{errorFpi}
\end{eqnarray}
$F_{\pi}(D_1,D_2,\langle \nu^2 \rangle)$ and $\alpha (D_1,D_2,\langle \nu^2 \rangle)$ are implicitly determined by the system of Eqs.~(\ref{system_eq}). We would like to note that the jack-knife error estimation would be quite difficult to apply for this method since the solutions to Eqs.~(\ref{system_eq}) are not always real. Therefore one has to consider the resulting error bars with some caution.

The results of the solution of the Eqs.~(\ref{system_eq}) are given in Table~\ref{tab:Falpha}. The first line represents $F_{\pi}$ and $\alpha$ extracted from our data with the Neuberger operator, the values for the second line are obtained using the overlap hypercube Dirac operator. The third line represents a three parameter fit to the data from  Ref.~\cite{Giusti:2003iq} (Table~\ref{tab:Falpha}) for the Neuberger operator on lattices which are there denoted as $C_0$ and $C_1$.

\begin{table}[H]
\centering
\begin{tabular}{|c|c|c|c|c|c|c|}
\hline
$\beta$&$V$&$D_1$&$D_2$&$F_{\pi}$ [MeV] & $\alpha$ &remarks \\
\hline
\hline
$5.85$&$12^3\times 24$&$13.5\pm 1.4$ &$21.8\pm 2.6$&$80.7\pm 9.1$ &$-12.6 \pm 6.6 $& {\small Neuberger}\\
\hline
$5.85$&$12^3\times 24$&$14.2\pm 1.2$ &$26.2\pm 2.7$&$85.7\pm 11$ &$-6.2 \pm 8.2 $& {\small overlap HF}\\
\hline
 $6$&$20^4$ &$3.9 \pm 0.5$ &$9.4 \pm 0.7$&&&$C_1$\\
 $5.8784$ &$16^4$ & $3.4\pm 0.3$ &$8.9\pm 0.5$&$(80.8,107.4)$ &$(-1.8,7.8)$ &$C_0$ \\
\hline
\end{tabular}
\caption{The first two lines show $F_{\pi}$ and $\alpha$ as a solution of the Eq.~(\ref{qDnup}), extracted from our data. The third line gives for comparison the values quoted in Ref.~\cite{Giusti:2003iq}. \label{tab:Falpha}}
\end{table}
The results for the $F_{\pi}$ and $\alpha $ are compatible for different overlap fermions within the error bars. At this stage we only conclude that the computation of $\alpha$ poses a challenge and further investigations are needed. 

For completeness we show the statistics of our data together with the statistics for the $C_0$ lattice in Table~\ref{tab:stat2}.
\begin{table}[H]
\centering
\begin{tabular}{|c|c|c|c|c|c|c|}
\hline
operator&lattice & & physical &
\multicolumn{3}{|c|}{number of configurations} \\
kernel&size & $\beta$ & volume & $|\nu| = 1$ & $\vert \nu \vert = 2$ 
& $\vert \nu  \vert = 3$  \\
\hline
\hline
Wilson&$12^{3}\times24$ & $5.85$ & $(1.48 \fm)^{3}\times 2.96\fm$ &90  &72  &64   \\
\hline
Wilson&$16^{3}\times 32$ & $6$ & $(1.49 \fm)^{3}\times 2.98\fm$ &92  &70  &71   \\
\hline
 hypercube&$12^{3}\times24$ & $5.85$ & $(1.48 \fm)^{3}\times 2.96\fm$ &91  &83  & 69   \\
\hline
Wilson&$16^{4}$ & $5.8784$ & $(1.86 \fm)^{4}$ & 229 & 186 & -- \\
\hline
\end{tabular}
\caption{The statistics for our data ($1$st, $2$nd and $3$rd lines) and for the $C_0$ lattice in Ref.~\cite{Giusti:2003iq} (fourth line).\label{tab:stat2}}
\end{table}

\begin{figure}%[htbp]
\centering
\includegraphics[width=0.6\textwidth,angle=-90]{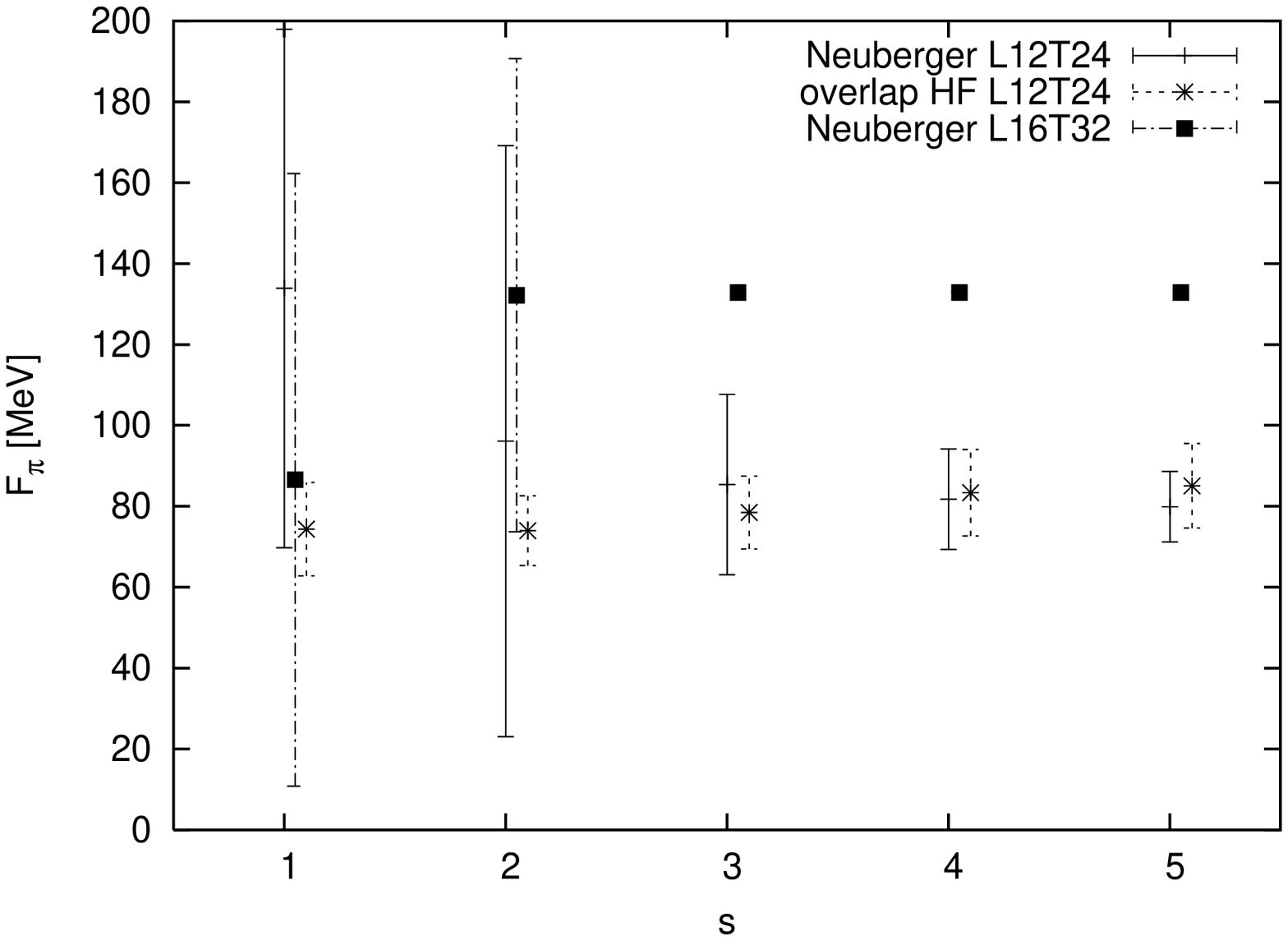}

\includegraphics[width=0.6\textwidth,angle=-90]{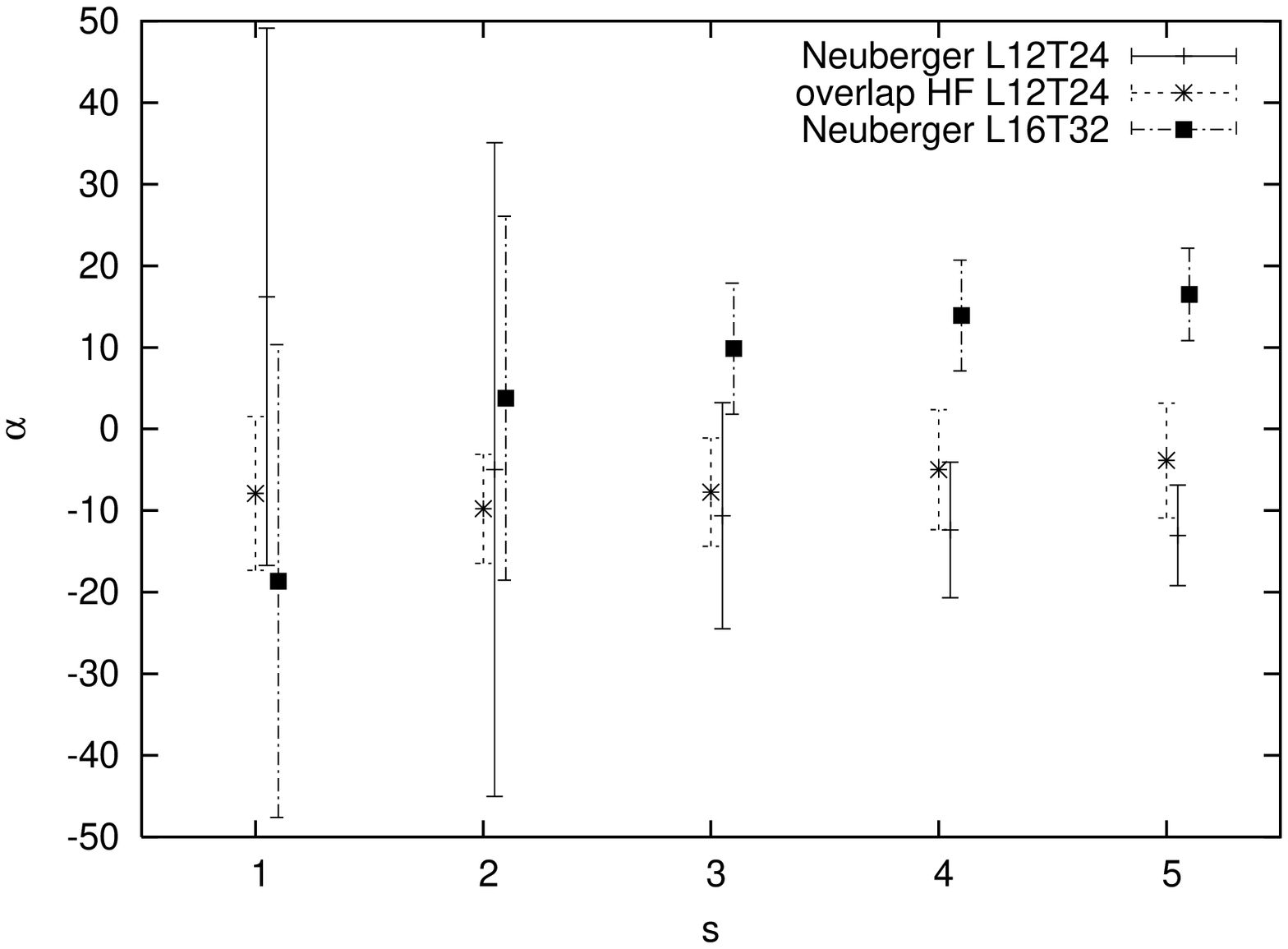}
\caption{$F_{\pi}$ and $\alpha$ extracted from a two parameter fit to the combined data from topological sectors $|\nu|=1$ and $2$ vs. the fitting range $[T/2-s,T/2+s]$. We used the Neuberger operator on a $12^3\times 24$ lattice at $\beta=5.85$ and on a $16^3\times 32$ lattice at $\beta=6$, as well as the overlap hypercube Dirac operator on a $12^3\times 24$ lattice, at $\beta=5.85$.}
\label{fig:fpi_vs_range}
\end{figure}

Now we fix the value of $\langle \nu^2 \rangle=10.4$ for the calculations with the overlap Wilson Dirac operator and $\langle \nu^2 \rangle=9.8$ for the overlap hypercube Dirac operator. Then we perform a two parameter fit of the expression~(\ref{formula1}) to the combined data for two topological sectors $|\nu|=1$ and $2$ on different intervals $t \in [T/2-s,T/2+s]$. The results for $F_{\pi}$ and $\alpha$ vs. the fitting range are collected in Figure~\ref{fig:fpi_vs_range}.
We see that the values of $F_{\pi}$ and $\alpha$ stabilize at $s=3$.  The values of $F_{\pi}$ and $\alpha$  for the Neuberger operator are consistently larger at finer lattice spacing.  $F_{\pi}$ is still subject to renormalization which can increase it substantially.  The value of $\alpha$ seems to scale within $2\sigma$. We also see that the results for the Neuberger operator are compatible with those of the overlap hypercube operator. Note, however, that the data for the overlap hypercube operator tend to be less noisy than for the Neuberger operator on the lattice $12^3\times 24$, at $\beta=5.85$. 

The estimates and the jack-knife error bars of  $F_{\pi}$ and $\alpha$, as well for the re-calculated $D_{|\nu|}$, for the fitting range $[T/2-5,T/2+5]$  are collected in Table~\ref{tab:Falphafit}.
\begin{table}[H]
\centering
\begin{tabular}{|c|c|c|c|c|c|}
\hline
$D_1$&$D_2$& $D_3$&$F_{\pi}$ & $\alpha$ &$\chi^2 /\mbox{d.o.f.}$ \\
\hline
\hline
$12.4 \pm 1.6$ &$20.2 \pm 3.1$&$7.4\pm 17$ &$79.9 \pm 8.7 \, \mbox{MeV}$ &$-13.0 \pm 6.1 $& $0.7$ \\
\hline
$14.3 \pm 1.3$ &$26.2 \pm 2.7$ &$23.3\pm 15$&$85.1 \pm 10.5 \, \mbox{MeV}$ &$-3.9 \pm 7.0 $& $0.5$ \\
\hline
$9.9 \pm 1.9$ &$21.8 \pm 3.8$ &$33.7\pm 5.7$&$132.9 \pm 0.0 \, \mbox{MeV}$ &$16.5 \pm 5.7 $&$0.6$ \\
\hline
\end{tabular}
\caption{ Results of the two parameter fit in the range $[T/2-5,T/2+5]$ for the combined data in topological sectors $|\nu| =1$ and $2$. The first line represents the results for the overlap Wilson Dirac operator and the second line represents the results for overlap hypercube Dirac operator on a lattice of size $12^3\times 24$, at $\beta=5.85$.The third line represents the results for the overlap Wilson Dirac operator  on a lattice of size $16^3\times 32$, at $\beta=6$.  \label{tab:Falphafit}}
\end{table}
First of all we can confirm again that the given statistics is not sufficient to obtain a stable value of the slope $D_3$ at least on $12^3\times 24$ lattice at $\beta=5.85$.
Next we see that the two methods that we discussed lead to compatible results. However the two parameter fit produces significantly more reliable error bars. Note that if we change the topological susceptibility in our two parameter fit by $10\%$ this will lead only to a ${\cal O}(10\%)$ change in $\alpha$ while $F_{\pi}$ remains almost the same.  We conclude that performing the two parameter fit is still the best method to obtain the estimates for $F_{\pi}$ and $\alpha$ with a control over the errors. We argue that the calculation of the mere zero mode contribution is potentially capable of the extraction of $F_{\pi}$, $\alpha$ and topological susceptibility $\chi_t$ once a good statistics is available. Its obvious virtue is that it is not affected by the problems of small quark masses and small eigenvalues of the Dirac operator which magnify the noise of the propagator as we discussed in Section 8.2.

\chapter{Topology conserving L\"{u}scher gauge action}
\section{Motivation}
 We employ the overlap operator as a regularization for the fermion fields. It obeys the Ginsparg-Wilson relation~(\ref{GWR}) and thus the chiral symmetry is preserved at a finite lattice spacing as discussed in Ref.~\cite{Luscher:1998pq} and we demonstrated in Section 2.5.3. This enables us to perform simulations at small quark masses and to define the topological charge as the index of the overlap operator. In particular it is applied to explore the $\epsilon$--regime. 
In the previous Chapter we saw that in order to perform the precision calculation of the low energy constants in the $\epsilon$--regime one has to collect as much statistics as possible for a given topological sector. A computer time consuming part of these calculations is that for each configuration one needs to compute the index of the overlap operator. In particular we saw that the simulations in the topological sector $\nu=0$, which occurs quite often in the simulations, are plagued by the large statistical fluctuations and therefore it is favorable to evaluate observables in the sectors of $|\nu|\ne 0$. As we learned from the $\chi$RMT study of the probability distributions of the individual eigenvalues, the topological sectors $|\nu|=1$ and $2$ are of primary interest for the simulations in moderate physical volumes. In simulations on a large lattices the charges fluctuate between $\nu=-10$ and $10$, see Figure~\ref{fig:gaussiancharge}. Therefore it would be highly profitable for the simulations if a method was found to sample the statistics within a certain topological sector without recomputing the index. In other words, one seeks a method which provides a large autocorrelation time for the topological charge and still small autocorrelation times for the observables of interest.

The change of the topological sector for the overlap Dirac operator implies that an eigenvalue flips from the origin of the Ginsparg-Wilson circle to $2\mu/a$, i.e. to the cut-off scale, or vice versa as the gauge field is deformed. If the corresponding deformation is done continuously then the inverse square root of $A^{\dagger}A$ will develop a singularity. As it was already mentioned it is guaranteed under the condition~(\ref{AdaggerAcond})  that  $A^{\dagger}A$ remains positive. It was shown later by H. Neuberger~\cite{Neuberger:1999pz} that this condition can be relaxed to   
\begin{equation}
 \epsilon \le \frac{1}{(1+\frac{1}{\sqrt{2}})d(d-1)}\simeq\frac{1}{20.5}\ .
\label{for:neuberger_condition}
\end{equation}
 If it is fulfilled then the topological charge is guaranteed not to change under a continuous deformation of the gauge field. A non-continuous deformation can, however, still change it but the hope is that the probability of such a change is small in simulations. Applying the weak coupling expansion for the Wilson gauge action we found that the above relation is satisfied for the mean plaquette value at $\beta\sim 40$. Obviously simulations are not possible with such a large value of $\beta$ due to the strong autocorrelations and tiny lattice spacings. However one hopes that the topological charge will be still conserved over a large number of configurations even if the gauge action only favors the plaquettes obeying the further relaxed condition~(\ref{for:neuberger_condition}).

   It is possible to introduce a lattice gauge action which will suppress large values of $S_P(U_P)=1-(1/3) {\rm Re \, Tr} \, U_P$ and thereby the condition~(\ref{plaquet_smothness}) will be fulfilled. This will in turn lead to conservation of the topological charge along the simulation trajectory.  To this end we define
\begin{equation}
S_P(U^{(P)}_{x,\mu \nu})\rightarrow S_{\alpha}(U^{(P)}_{x,\mu \nu})=\left\{  \begin{array}{c@{\quad \quad}l}
\frac{S_P(U^{(P)}_{x, \mu \nu})}{(1-S_P(U^{(P)}_{x, \mu \nu})/\epsilon)^\alpha} & S_P(U^{(P)}_{x, \mu \nu}) < \epsilon \tabularnewline
\infty &  S_P(U^{(P)}_{x, \mu \nu}) \ge \epsilon
  \end{array}    \right . \ ,
\label{admis_action}
\end{equation}
where $\alpha >0$.
In particular, $\alpha=1$ was used by M. L\"uscher~\cite{Luscher:1998du,Luscher:1999un} for conceptual purposes and was also applied in Ref.~\cite{Fukaya:2003ph} to the Schwinger model. In the latter the authors used $\epsilon=1$ as opposed to the strict theoretical bound of $1/5$ for two dimensions, in order to let the gauge fields fluctuate more while still suppressing topological transitions. This was sufficient to stabilize the topological history over hundreds of configurations.
A potential problem for the simulations with this action can be that due to the stringent constraint the resulting lattice spacing may become too small. This would require very large lattice volumes to obtain a reasonable physical volume for simulations. Yet another caveat was raised by M. Creutz that such an action may lead to a non-positive transfer matrix which will make the transition to Minkowski space at finite lattice spacing impossible~\cite{Creutz:2004ir}. However, one can think of first taking the continuum limit and then going to Minkowski space. 
%------------------------------------------------------------------------------------------------------------------
\section{Results on the topological histories}
We used the local Hybrid Monte Carlo (HMC) algorithm --- for a description we refer to Appendix A --- to simulate the action~(\ref{admis_action}) with $\alpha=1$ in quenched QCD. The standard tools like heat bath and overrelaxation are not applicable in this case since they very much rely on the properties of the gauge action~(\ref{general_struc_action}) which do not hold for the action~(\ref{admis_action}). For HMC one has to compute the ``force'' $F_{x,\alpha}$,
\begin{eqnarray}
F_{x,\mu,\alpha}&=&\frac{\delta S_{\alpha}(U^{(P)}_{x, \mu \nu})}{\delta U_{x, \mu }}=F^{\rm Wilson}_{x,\mu,\alpha}\frac{1+\frac{\alpha-1}{\epsilon}S_P}{(1-S_P/\epsilon)^{\alpha+1}}\ , \nonumber \\
F^{\rm Wilson}_{x,\mu,\alpha}&=&\frac{\delta S_P(U^{(P)}_{x, \mu \nu})}{\delta U_{x, \mu}}\ .
\end{eqnarray}

\begin{figure}%[htbp]
\centering
\includegraphics[width=0.5\textwidth, angle=-90]{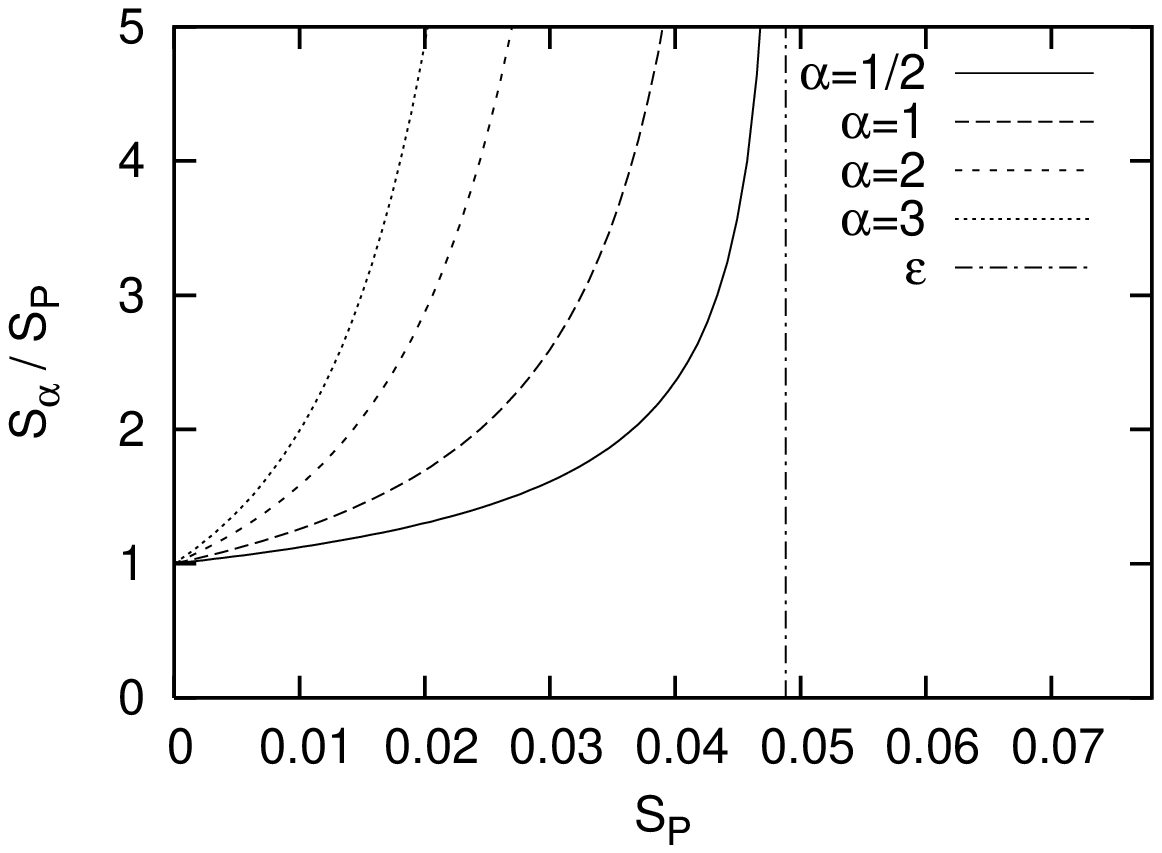}
\includegraphics[width=0.5\textwidth, angle=-90]{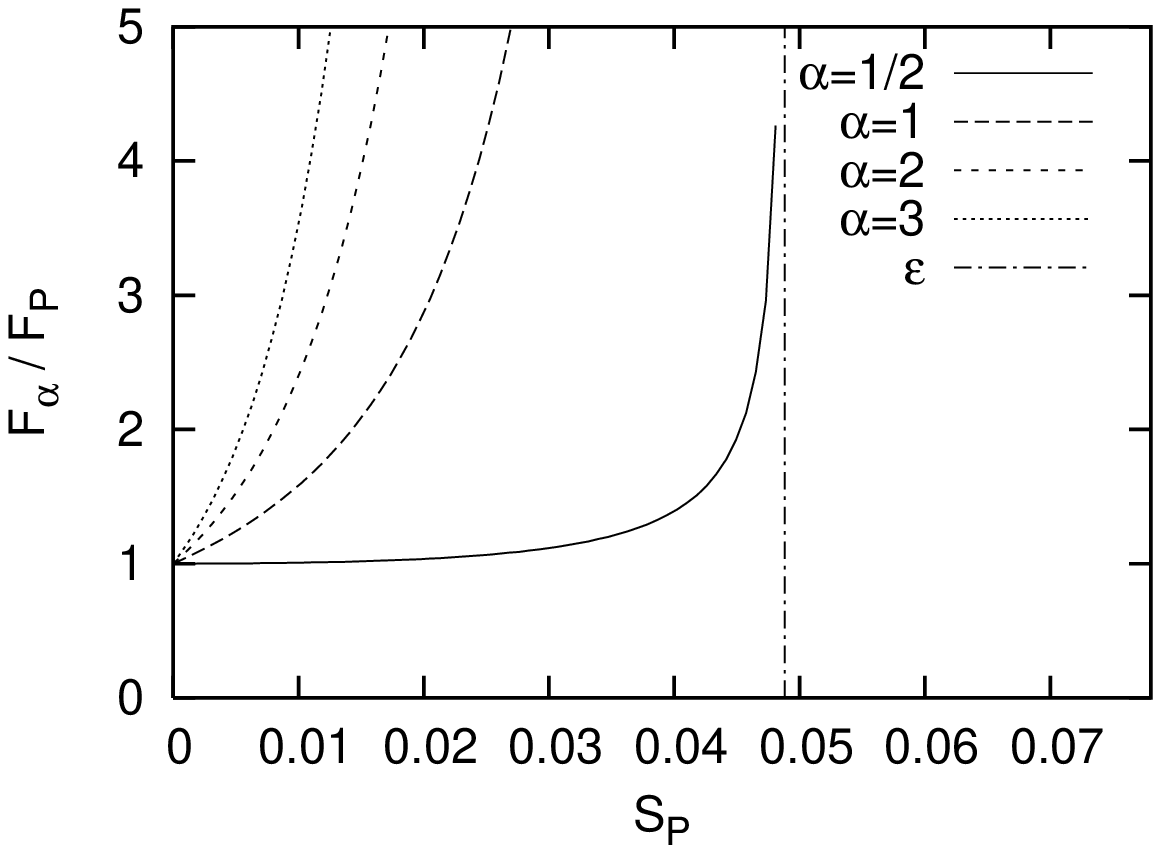}
%\end{rotate}
\caption{The plot on top shows the dependence of $S_{\alpha}/S_P$ vs. $S_P$. The plot at the bottom shows the dependence of the ``force'' $F_{\alpha}/F_P$ vs. $S_P$. The theoretical bound is shown as an asymptotic vertical line at $S_P=1/\epsilon$.}
\label{fig:action_force}
\end{figure}

Figure~\ref{fig:action_force}  shows the dependence of $S_{\alpha}/S_P$ vs. $S_P$ as well as $F_{\alpha}/F_P$ vs. $S_P$.  As $\alpha$ decreases the ``force'' becomes more flat in the domain of $S_P$ close to the free field.

\begin{figure}[h]
\centering
\includegraphics[width=0.6\textwidth]{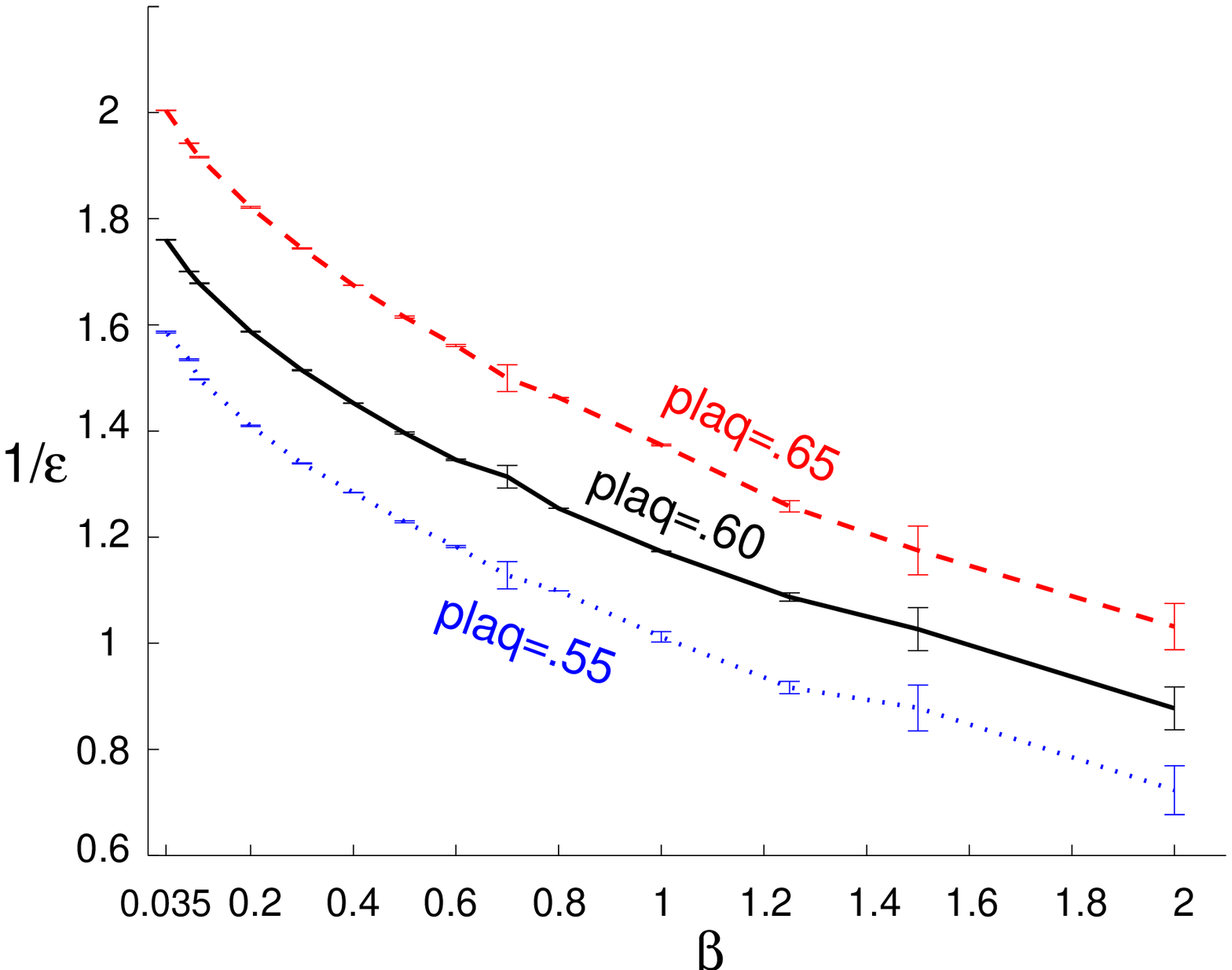}
%\end{rotate}
\caption{The lines of constant expectation value of the plaquette, $\langle S_P \rangle$, in the plane spanned by the action parameters $1/\epsilon$ and $\beta$.  }
\label{fig:plaquette_isoline}
\end{figure}
If the new action is defined with quite a stringent constraint this will reduce the freedom for the gauge field to fluctuate which could imply a very small lattice spacing as we mentioned before. To investigate this we start off by searching a set of values of $\{ \epsilon,\beta \}$ which will provide the same $\langle S_P \rangle$ in simulations on $4^4$ lattice. The obtained curve of the ``constant physics'' was used as a guideline for the further analysis where we computed $r_0$ for the specific values of $\epsilon$, $\beta$ on the curve. The results for the curve of the ``constant physics'' are reported in Figure~\ref{fig:plaquette_isoline}. As $\epsilon$ decreases the value of $\beta$ must be also decreased in order to obtain the same expectation value for the plaquette $\langle S_P \rangle$. Another problem is the rejection of a link variable due to the violation of the boundary in definition of the action~(\ref{admis_action}). It turned out that the rejection rate is increased as $\beta$ is decreased, see Table~\ref{tab:top_histories_luescher}~\cite{Shcheredin:2004xa}. 

\begin{figure}[h]
\centering
\includegraphics[width=0.4\textwidth, angle=-90]{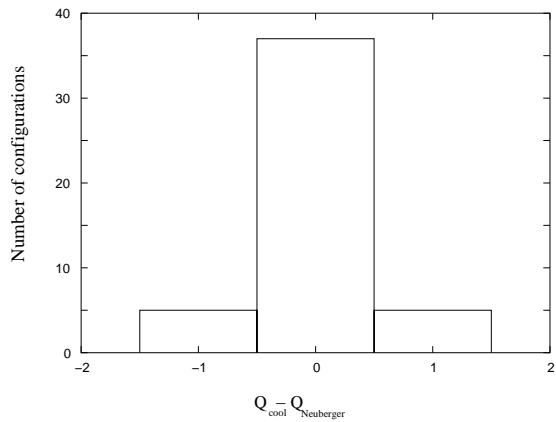}
%\end{rotate}
\caption{The histogram of the  difference between topological charge defined by the cooling, $Q_{\rm cool}$, and the topological charge defined by the index of the Neuberger operator, $Q_{\rm Neuberger}$. For our parameters $V=10^3\times 24$, $\beta=6$ we observe $80$ \% agreement.}
\label{fig:qcool-qneu}
\end{figure}

Next we picked out a few points from the curve and computed $r_0$ on a $16^4$ lattice. The results for $r_0$ are given in Table~\ref{tab:top_histories_luescher}. We see that $r_0$ slightly increases as $\beta$ is decreased. As a pilot test for the topological stability we employed the {\em cooling method}~\cite{Ilgenfritz:1985dz}. This is a relatively fast way compared to the overlap index to estimate the topological charge of a gauge configuration at moderate $\beta$. In this approach one deforms a gauge configuration by minimizing its gauge action locally to reach the approximate solutions of the lattice Yang-Mills equations of motion, which are the Euler equations for the Wilson Yang-Mills gauge action. The resulting gauge configurations, which are picked out from the first stable plateau, dissociate in objects with topological charges $\nu=\pm 1$. They can be interpreted as instantons and anti-instantons. Since the cooled configurations are smooth, these charges can be determined using a simple lattice discretization of the continuum formula~(\ref{gauge_index}) given for instance in Ref.~\cite{Ilgenfritz:2002qs}. Hence the topological charge of a cooled gauge configuration can be determined by the difference of number of instantons $N_+$ and anti-instantons $N_-$. To quantify the topological charge conservation we computed the autocorrelation time of  $Q_{\rm cool}=N_+-N_-$. The main conjecture is that the autocorrelation time of this quantity is a good estimate for the autocorrelation time of the topological charge given by the index of the overlap operator for the equilibrium configurations. The latter is supported by the fact that the corresponding equilibrium configurations are quite smooth ($\beta_W\simeq 6.19$). Hence the definition of the index given by cooling matches the definition of the index given by the overlap operator at least up to $80 \% $~\cite{Nagai:2003gq}. This is verified by our study at stronger coupling $\beta=6$, shown in Figure~\ref{fig:qcool-qneu}. It shows the distribution of the difference of the topological charge defined with  cooling and the topological charge defined by the Neuberger index.         

\begin{figure}
\centering
\includegraphics[width=0.5\textwidth]{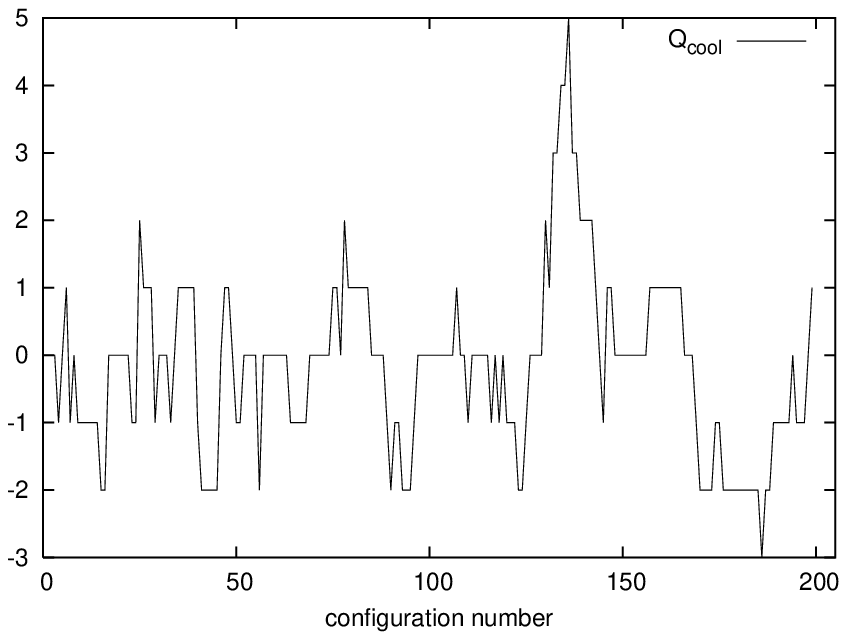}

\vspace{1cm}
\includegraphics[width=0.5\textwidth]{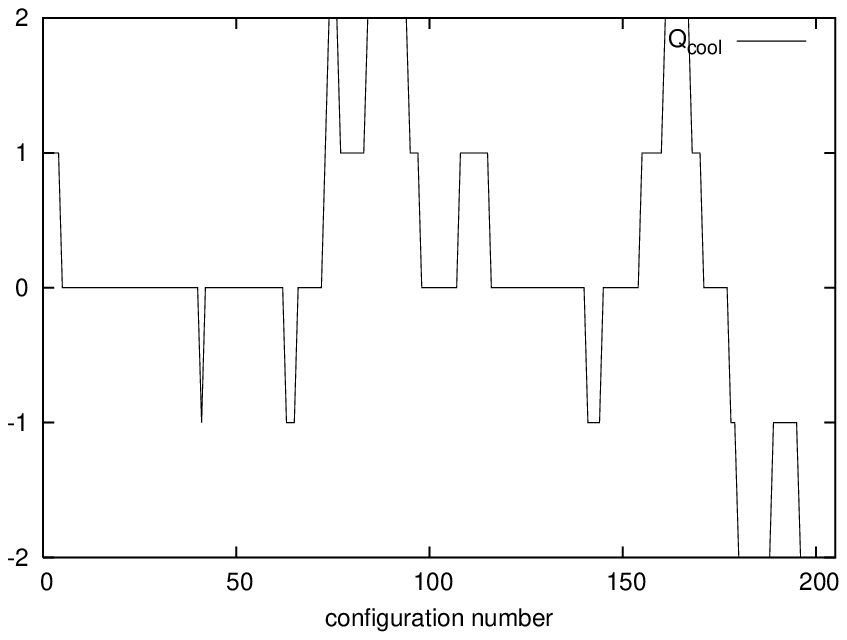}

\vspace{1cm}
\includegraphics[width=0.5\textwidth]{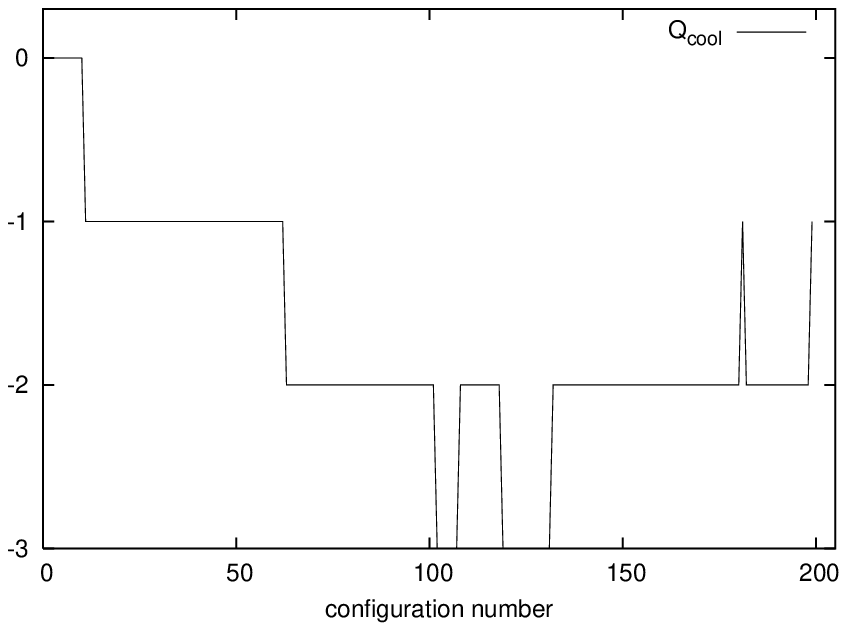}
\caption{Histories of $Q_{\rm cool}$ at  $\{ \beta , 1/\epsilon \}=\{6.19, 0\}$, $\{0.8, 1.25 \}$, $\{0.3, 1.52\}$  for the plot on top, in the middle and at the bottom, respectively.}
\label{fig:qcool_histories}
\end{figure}

\begin{table}[h]
\centering
\begin{tabular}{|c|c|c|c|c|c|c|}
\hline
$1/\epsilon$ & $\beta$ & rejection rate &$r_0/a$&$\beta_W$ & $\tau^{\rm cool}_{\rm aut}$& $\tau^{\rm plaq}_{\rm aut}$\\
\hline
\hline
     $0$        &  $6.19$  &$0 $& $7.25$    & $6.19$       &          $3.94$&   $7.27$   \\
\hline 
     $1.25$        &  $0.8$  &$<0.1 \%$& $7.0(1)$    & $6.17$       &          $5.07$&   $1.15$   \\
\hline 
     $1.52$        &  $0.3$  &$5 \%$& $7.3(4)$    & $6.19$       &          $21.03$&   $0.85$   \\
\hline 
\end{tabular}
\caption{Results for the parameter $r_0/a$ computed in simulations on $16^4$ lattice for various values of $\epsilon$ and $\beta$. We give the corresponding value of $\beta$ for the Wilson action, which we denote as $\beta_W$. The autocorrelation times for $Q_{\rm cool}$ and for the plaquette are reported. We also show the the rejection rate for the links.\label{tab:top_histories_luescher}}
\end{table}
In Figures~\ref{fig:qcool_histories} we show the histories of $Q_{\rm cool}$ for the parameters $\beta$ and $\epsilon$ given in Table~\ref{tab:top_histories_luescher}. 

The autocorrelation times for the quantity $Q_{\rm cool}$, $\tau_{\rm aut}^{\rm cool}$, as well as for the plaquette $S_P$, $\tau_{\rm aut}^{\rm plaq}$, are given in Table~\ref{tab:top_histories_luescher}. We see that $\tau_{\rm aut}^{\rm cool}$ becomes larger as $\epsilon$ is decreased, while the autocorrelation times for the plaquette become even smaller. This gives confidence that the pair $1/\epsilon=1.52$ and $\beta=0.3$ constitutes a good parameter set for the action to conduct simulations in the $\epsilon$-regime.

\chapter{Conclusions}
In this thesis we presented a study of the feasibility to extract physical information in the $\epsilon$--regime using various formulations of chiral lattice fermions. In this final Section we briefly summarize the main results. 

The investigations were done in the framework of quenched lattice QCD. To discretize the fermion fields we chose the Ginsparg-Wilson fermions. In particular we applied the Neuberger operator and the overlap hypercube operator. These formulations provide a regularization scheme which preserves the chiral symmetry at a finite lattice spacing. As a consequence the topological charge is well defined by the index of the overlap Dirac operator. Therefore by simulating quenched QCD in the $\epsilon$--regime one can reach the domain of very small quark masses. The computations are, however, quite tedious and therefore at present only the quenched approximation can be studied. Controlling the finite size effects in QCD simulations in the $\epsilon$--regime one can obtain the low energy constants of the chiral Lagrangian, which are also those of the infinite volume up to quenching effects. 

One of the main topics of these investigations was the analysis of the probability distributions of the individual eigenvalues of the Dirac operator in the QCD simulations in the $\epsilon$--regime. We confronted the simulation data with the analytical prediction conjectured in the framework of the $\chi$RMT. The conjecture is that the $\chi$RMT is a theory which can describe eigenvalue densities in the $\epsilon$--regime. It is mainly based on the fact that in the limit of large matrices the $\chi$RMT partition function coincides with the $\chi$PT partition function to the leading order in the $\epsilon$--regime. Also some quantities like the microscopic spectral density were calculated for the special cases both, in $\chi$RMT and in $\chi$PT, leading to the same expressions. The ability of $\chi$RMT to explain the simulations in the $\epsilon$--regime being yet a conjecture received a convincing support in our investigations. To this end we found that the behavior of the probability distributions of the individual eigenvalues can be reliably described by the $\chi$RMT if the box length of the physical volume exceeds about $L\ge 1.12\fm$.  The chiral condensate enters as the only parameter for the probability distributions of the individual eigenvalues. Therefore it can be obtained by fitting the $\chi$RMT curves to the simulation data.

A next result obtained from this study was the determination of the lower bound for the size of the physical volume where the data can be described by $\chi$PT.
It has proven to help in understanding the behavior of the two-point mesonic functions. In particular we could explain why the correlators calculated in quenched simulations on $(0.93 \fm)^3\times 1.86 \fm$ are not modeled by the $\chi$PT. It turned out that the size of the physical volume of $L\approx 0.93 \fm$ is too small for the simulations in the $\epsilon$--regime and a size of $L\approx 1.12 \fm$ lies on the threshold. When trying to extract the low energy constants $F_{\pi}$ and $\Sigma$ from the axial-vector two-point function we found that $F_{\pi}$, which enters the formula in an additive constant, can be obtained by fitting. The chiral condensate $\Sigma$, on the contrary, enters in a way that leaves the analytic formula almost insensitive to drastic changes in $\Sigma$ for the topological sectors $|\nu|> 0$. The sensitivity is restored only for $\nu = 0$.  This makes it prohibitively difficult to extract $\Sigma$ using the axial-vector two-point function at $|\nu|\ne 0$. On the other hand, the signal for the axial-vector correlator in the topological sector $\nu=0$ turned out to be very noisy. This we attributed to a high probability of occurrence of near-zero modes. The latter was also supported by our $\chi$RMT study of the eigenvalues. 

To estimate the statistics needed for the topological sectors $|\nu|=0$, $1$ and $2$ we considered the contribution to the chiral condensate given by the distribution of the lowest eigenvalue of the Dirac operator $\Sigma^{(\nu)}_{\rm min}$. Then we simulated fake eigenvalues with these distributions and obtained the necessary statistics to stabilize the standard deviation of the $\Sigma^{(\nu)}_{\rm min}$. It turned out that one needs a tremendous statistics of ${\cal O}(10^4)$ configurations in the topological sector $\nu=0$ (for this stabilization). For higher topological sectors ($|\nu|=1$, $2$) this number is decreased to ${\cal O}(10^2)$ configurations.
     
As an alternative method to calculate the low energy constants one can consider only the contribution of the zero modes to the mesonic two-point functions. We computed such contributions to the pseudo-scalar two-point function. The analytical expressions for them are known and we fitted them to our data to explore the possibility of extracting the parameters $F_{\pi}$ and $\alpha$ --- yet another low energy parameter appearing in the chiral quenched Lagrangian. We were able to obtain reasonable bounds on these values with our statistics. Therefore we conclude that this is a promising method to compute the low energy constants.

Next we considered the topological susceptibility. This quantity was computed using two formulations of the overlap operator: the Neuberger fermions (with the usual Wilson action as its kernel) and the hypercube fermions (with the truncated fixed point action as its kernel). The overlap hypercube operator is expected to be superior to the conventional formulation.  It turned out that the locality is improved and the topological susceptibility comes closer to the continuum value at the same lattice spacing. 

As a theoretical development of the simulations in the $\epsilon$--regime one can think of a method to simulate a set of configurations belonging to a certain topological charge without recomputing the index. The problem can be reformulated as to simulate configurations with the topological charge being strongly autocorrelated while other observables of interest remaining decorrelated. This we studied using a gauge action proposed by M. L\"uscher.  We are confident that we found a good set of parameters for the L\"uscher gauge action to meet these requirements.

%------------------------------------------------------
% Der (x)emacs oder auch andere LaTeX-Systeme unter Windows besitzen einen Bibtex-Modus, 
% der auf Knopfdruck die entsprechenden Felder liefert. 
% Empfehlenswert ist unter UNIX/LINUX die Benutzung des grafischen Frontends bibview.  
%------------------------------------------------------
% You should be using BibTeX as follows;
\pagestyle{empty}
%\renewcommand{\chaptermark}[1]{\markboth{\chaptername\ \thechapter\ #1}{}}
%\renewcommand{\sectionmark}[1]{\markright{\thesection\ #1}}

%\lhead[\fancyplain{}{\bfseries\thepage}]{\fancyplain{}{\bfseries\rightmark}}
%\rhead[\fancyplain{}{\bfseries\leftmark}]{\fancyplain{}{\bfseries\thepage}} 
%\chead{} \lfoot{} \cfoot{} \rfoot{}
\bibliography{bibliography}

\begin{thebibliography}{100}

\bibitem{Osterwalder:1973dx}
K.~Osterwalder and R.~Schrader.
\newblock Axioms for euclidean green's functions.
\newblock {\em Commun. Math. Phys.}, 31:83, 1973.

\bibitem{Osterwalder:1975tc}
K.~Osterwalder and R.~Schrader.
\newblock Axioms for euclidean green's functions. 2.
\newblock {\em Commun. Math. Phys.}, 42:281, 1975.

\bibitem{okun}
L.~B. Okun.
\newblock {\em Leptons and quarks}.
\newblock Nauka, Moscow, 1981.

\bibitem{bookMM}
I.~Montvay and G.~M{\"u}nster.
\newblock {\em Quantum Fields on a Lattice.}
\newblock Cambridge University Press, Cambridge, U.K., 1994.

\bibitem{Atiyah:1968mp}
M.~F. Atiyah and I.~M. Singer.
\newblock The index of elliptic operators. 1.
\newblock {\em Annals Math.}, 87:484, 1968.

\bibitem{Nambu:1960xd}
Y.~Nambu.
\newblock Axial vector current conservation in weak interactions.
\newblock {\em Phys. Rev. Lett.}, 4:380, 1960.

\bibitem{Goldstone:1961eq}
J.~Goldstone.
\newblock Field theories with 'superconductor' solutions.
\newblock {\em Nuovo Cim.}, 19:154, 1961.

\bibitem{'tHooft:1976up}
G.~'t~Hooft.
\newblock Symmetry breaking through bell-jackiw anomalies.
\newblock {\em Phys. Rev. Lett.}, 37:8, 1976.

\bibitem{'tHooft:1976fv}
G.~'t~Hooft.
\newblock Computation of the quantum effects due to a four- dimensional
  pseudoparticle.
\newblock {\em Phys. Rev.}, D14:3432, 1976.

\bibitem{Banks:1980yr}
T.~Banks and A.~Casher.
\newblock Chiral symmetry breaking in confining theories.
\newblock {\em Nucl. Phys.}, B169:103, 1980.

\bibitem{Coleman:1969sm}
S.~R. Coleman, J.~Wess, and B.~Zumino.
\newblock Structure of phenomenological lagrangians. 1.
\newblock {\em Phys. Rev.}, 177:2239, 1969.

\bibitem{Callan:1969sn}
C.~G. Callan, S.~R. Coleman, J.~Wess, and B.~Zumino.
\newblock Structure of phenomenological lagrangians. 2.
\newblock {\em Phys. Rev.}, 177:2247, 1969.

\bibitem{Weinberg:1979kz}
S.~Weinberg.
\newblock Phenomenological lagrangians.
\newblock {\em Physica}, A96:327, 1979.

\bibitem{Gasser:1984yg}
J.~Gasser and H.~Leutwyler.
\newblock Chiral perturbation theory to one loop.
\newblock {\em Ann. Phys.}, 158:142, 1984.

\bibitem{Gasser:1985gg}
J.~Gasser and H.~Leutwyler.
\newblock Chiral perturbation theory: expansions in the mass of the strange
  quark.
\newblock {\em Nucl. Phys.}, B250:465, 1985.

\bibitem{Gas87}
J.~Gasser and H.~Leutwyler.
\newblock Thermodynamics of chiral symmetry.
\newblock {\em Phys. Lett.}, B188:477, 1987.

\bibitem{Neuberger:1987zz}
H.~Neuberger.
\newblock A better way to measure $f_\pi$ in the linear $\sigma$ model.
\newblock {\em Phys. Rev. Lett.}, 60:889, 1988.

\bibitem{Neuberger:1987fd}
H.~Neuberger.
\newblock Soft pions in large boxes.
\newblock {\em Nucl. Phys.}, B300:180, 1988.

\bibitem{Hasenfratz:1990pk}
P.~Hasenfratz and H.~Leutwyler.
\newblock Goldstone boson related finite size effects in field theory and
  critical phenomena with $o(n)$ symmetry.
\newblock {\em Nucl. Phys.}, B343:241, 1990.

\bibitem{Hansen1990}
F.~C. Hansen.
\newblock Finite size effects in spontaneously broken $su(n)\times su(n)$
  theories.
\newblock {\em Nucl. Phys.}, B345:685, 1990.

\bibitem{Hansen:1990yg}
F.~C. Hansen and H.~Leutwyler.
\newblock Charge correlations and topological susceptibility in qcd.
\newblock {\em Nucl. Phys.}, B350:201, 1991.

\bibitem{Bietenholz:1992ix}
W.~Bietenholz.
\newblock Goldstone bosons in a finite volume: The partition function to three
  loops.
\newblock {\em Helv. Phys. Acta}, 66:633, 1993.

\bibitem{Leut92}
H.~Leutwyler and A.~Smilga.
\newblock Spectrum of dirac operator and role of winding number in qcd.
\newblock {\em Phys. Rev.}, D46:5607, 1992.

\bibitem{quench_success}
T.~Yoshi{\'e}.
\newblock Light hadron spectroscopy.
\newblock {\em Nucl. Phys. Proc. Suppl.}, 63:3, 1998.

\bibitem{Aoki:1999yr}
S.~Aoki and et. al.
\newblock Quenched light hadron spectrum.
\newblock {\em Phys. Rev. Lett.}, 84:238, 2000.

\bibitem{Bernard:1992mk}
C.~W. Bernard and M.~F.~L. Golterman.
\newblock Chiral perturbation theory for the quenched approximation of qcd.
\newblock {\em Phys. Rev.}, D46:853, 1992.

\bibitem{Damgaard:2000gh}
P.~H. Damgaard and K.~Splittorff.
\newblock Partially quenched chiral perturbation theory and the replica method.
\newblock {\em Phys. Rev.}, D62:054509, 2000.

\bibitem{Damgaard2002}
P.~H. Damgaard, P.~Hern{\'a}ndez, K.~Jansen, M.~Laine, and L.~Lellouch.
\newblock Finite-size scaling of vector and axial current correlators.
\newblock {\em Nucl. Phys.}, B656:226, 2003.

\bibitem{Damgaard2001}
P.~H. Damgaard, M.~C. Diamantini, P.~Hern{\'a}ndez, and K.~Jansen.
\newblock Finite-size scaling of meson propagators.
\newblock {\em Nucl. Phys.}, B629:445, 2002.

\bibitem{Giusti:2003iq}
L.~Giusti, P.~Hern{\'a}ndez, M.~Laine, P.~Weisz, and H.~Wittig.
\newblock Low-energy couplings of qcd from topological zero-mode wave
  functions.
\newblock {\em JHEP}, 01:003, 2004.

\bibitem{Damgaard:1998ye}
P.~H. Damgaard and S.~M. Nishigaki.
\newblock Universal spectral correlators and massive dirac operators.
\newblock {\em Nucl. Phys.}, B518:495, 1998.

\bibitem{Damgaard:2000ah}
P.~H. Damgaard and S.~M. Nishigaki.
\newblock Distribution of the k-th smallest dirac operator eigenvalue.
\newblock {\em Phys. Rev.}, D63:045012, 2001.

\bibitem{Wilke:1998gf}
T.~Wilke, T.~Guhr, and T.~Wettig.
\newblock The microscopic spectrum of the {QCD} dirac operator with finite
  quark masses.
\newblock {\em Phys. Rev.}, D57:6486, 1998.

\bibitem{Nishigaki:1998is}
S.~M. Nishigaki, P.~H. Damgaard, and T.~Wettig.
\newblock Smallest dirac eigenvalue distribution from random matrix theory.
\newblock {\em Phys. Rev.}, D58:087704, 1998.

\bibitem{Diakonov:1985nj}
D.~Diakonov and V.~Yu. Petrov.
\newblock Quark propagator and chiral condensate in an instanton vacuum.
\newblock {\em Sov. Phys. JETP}, 62:204, 1985.

\bibitem{Shur93_a}
E.~V. Shuryak and J.~J.~M. Verbaarschot.
\newblock Random matrix theory and spectral sum rules for the dirac operator in
  qcd.
\newblock {\em Nucl. Phys.}, A560:306, 1993.

\bibitem{Shuryak:1982ff}
E.~V. Shuryak.
\newblock The role of instantons in quantum chromodynamics. 1. physical vacuum.
\newblock {\em Nucl. Phys.}, B203:93, 1982.

\bibitem{Shuryak:1982dp}
E.~V. Shuryak.
\newblock The role of instantons in quantum chromodynamics. 2. hadronic
  structure.
\newblock {\em Nucl. Phys.}, B203:116, 1982.

\bibitem{Shuryak:1982hk}
E.~V. Shuryak.
\newblock The role of instantons in quantum chromodynamics. 3. quark - gluon
  plasma.
\newblock {\em Nucl. Phys.}, B203:140, 1982.

\bibitem{SmV}
A.~Smilga and J.~J.~M. Verbaarschot.
\newblock Spectral sum rules and finite volume partition function in gauge
  theories with real and pseudoreal fermions.
\newblock {\em Phys. Rev.}, D51:829, 1995.

\bibitem{Peskin:1980gc}
M.~E. Peskin.
\newblock The alignment of the vacuum in theories of technicolor.
\newblock {\em Nucl. Phys.}, B175:197, 1980.

\bibitem{Dimopoulos:1980sp}
S.~Dimopoulos.
\newblock Technicolored signatures.
\newblock {\em Nucl. Phys.}, B168:69, 1980.

\bibitem{Kogan:1985nb}
Ya.~I. Kogan, M.~A. Shifman, and M.~I. Vysotsky.
\newblock Spontaneous breaking of chiral symmetry for real fermions and n=2
  susy yang-mills theory.
\newblock {\em Sov. J. Nucl. Phys.}, 42:318, 1985.

\bibitem{Altland1}
A.~Altland and M.~R. Zirnbauer.
\newblock Field theory of the quantum kicked rotor.
\newblock {\em Phys. Rev. Lett.}, 77:4536, 1996.

\bibitem{Halasz:1995qb}
M.~A. Halasz and J.~J.~M. Verbaarschot.
\newblock Effective lagrangians and chiral random matrix theory.
\newblock {\em Phys. Rev.}, D52:2563, 1995.

\bibitem{Akemann:1997vr}
G.~Akemann, P.~H. Damgaard, U.~Magnea, and S.~Nishigaki.
\newblock Universality of random matrices in the microscopic limit and the
  dirac operator spectrum.
\newblock {\em Nucl. Phys.}, B487:721, 1997.

\bibitem{Verbaarschot:1993pm}
J.~J.~M. Verbaarschot and I.~Zahed.
\newblock Spectral density of the qcd dirac operator near zero virtuality.
\newblock {\em Phys. Rev. Lett.}, 70:3852, 1993.

\bibitem{Verbaarschot:1994qf}
J.~J.~M. Verbaarschot.
\newblock The spectrum of the qcd dirac operator and chiral random matrix
  theory: The threefold way.
\newblock {\em Phys. Rev. Lett.}, 72:2531, 1994.

\bibitem{Verbaarschot:1994gr}
J.~J.~M. Verbaarschot.
\newblock Spectral sum rules and selberg's integral formula.
\newblock {\em Phys. Lett.}, B329:351, 1994.

\bibitem{Verbaarschot:1994ia}
J.~J.~M. Verbaarschot.
\newblock The spectrum of the dirac operator near zero virtuality for n(c) = 2
  and chiral random matrix theory.
\newblock {\em Nucl. Phys.}, B426:559, 1994.

\bibitem{Verbaarschot:1994ip}
J.~J.~M. Verbaarschot and I.~Zahed.
\newblock Random matrix theory and qcd in three-dimensions.
\newblock {\em Phys. Rev. Lett.}, 73:2288, 1994.

\bibitem{Damgaard:1998xy}
P.~H. Damgaard, J.~C. Osborn, D.~Toublan, and J.~J.~M. Verbaarschot.
\newblock The microscopic spectral density of the {QCD} dirac operator.
\newblock {\em Nucl. Phys.}, B547:305, 1999.

\bibitem{Wigner}
A.~Bohr and B.~Mottelson.
\newblock {\em Nuclear Structure I}.
\newblock Benjamin, New York, 1969.

\bibitem{Wils1974}
K.~G. Wilson.
\newblock Confinement of quarks.
\newblock {\em Phys. Rev.}, D10:2445, 1974.

\bibitem{Rothe}
H.~J. Rothe.
\newblock {\em Lattice Gauge Theories}.
\newblock World Scientific, Singapore, 1992.

\bibitem{Elitzur}
S.~Elitzur.
\newblock Impossibility of spontaneously breaking local symmetries.
\newblock {\em Phys. Rev.}, D12:3978, 1975.

\bibitem{Wil75}
K.~G. Wilson.
\newblock {\em {\it in New Phenomena In Subnuclear Physics}, ed. A.~Zichichi,
  (Plenum Press, New York), Part A (1977) 69}.

\bibitem{Nielsen:1981rz}
H.~B. Nielsen and M.~Ninomiya.
\newblock Absense of neutrinos on a lattice. 1. proof by homotopy theory.
\newblock {\em Nucl. Phys.}, B185:20, 1981.
\newblock Erratum Nucl. Phys. B195 (1982) 541.

\bibitem{Nielsen:1981xu}
H.~B. Nielsen and M.~Ninomiya.
\newblock Absense of neutrinos on a lattice. 2. intuitive topological proof.
\newblock {\em Nucl. Phys.}, B193:173, 1981.

\bibitem{Karsten:1981gd}
L.~H. Karsten.
\newblock Lattice fermions in euclidean space-time.
\newblock {\em Phys. Lett.}, B104:315, 1981.

\bibitem{Kogut:1974ag}
J.~B. Kogut and L.~Susskind.
\newblock Hamiltonian formulation of wilson's lattice gauge theories.
\newblock {\em Phys. Rev.}, D11:395, 1975.

\bibitem{Susskind:1977jm}
L.~Susskind.
\newblock Lattice fermions.
\newblock {\em Phys. Rev.}, D16:3031, 1977.

\bibitem{Kogut:1979wt}
J.~B. Kogut.
\newblock An introduction to lattice gauge theory and spin systems.
\newblock {\em Rev. Mod. Phys.}, 51:659, 1979.

\bibitem{Sharatchandra:1981si}
H.~S. Sharatchandra, H.~J. Thun, and P.~Weisz.
\newblock Susskind fermions on a euclidean lattice.
\newblock {\em Nucl. Phys.}, B192:205, 1981.

\bibitem{Kluberg-Stern:1983dg}
H.~Kluberg-Stern, A.~Morel, O.~Napoly, and B.~Petersson.
\newblock Flavors of lagrangean susskind fermions.
\newblock {\em Nucl. Phys.}, B220:447, 1983.

\bibitem{Golterman:1984cy}
M.~F.~L. Golterman and J.~Smit.
\newblock Selfenergy and flavor interpretation of staggered fermions.
\newblock {\em Nucl. Phys.}, B245:61, 1984.

\bibitem{Bunk:2004kf}
B.~Bunk, M.~Della~Morte, K.~Jansen, and F.~Knechtli.
\newblock The locality problem for two tastes of staggered fermions.
\newblock 2004.

\bibitem{Ginsparg:1982bj}
P.~H. Ginsparg and K.~G. Wilson.
\newblock A remnant of chiral symmetry on the lattice.
\newblock {\em Phys. Rev.}, D25:2649, 1982.

\bibitem{Luscher:1998pq}
M.~L{\"u}scher.
\newblock Exact chiral symmetry on the lattice and the ginsparg- wilson
  relation.
\newblock {\em Phys. Lett.}, B428:342, 1998.

\bibitem{Hasenfratz:1998ri}
P.~Hasenfratz, V.~Laliena, and F.~Niedermayer.
\newblock The index theorem in qcd with a finite cut-off.
\newblock {\em Phys. Lett.}, B427:125, 1998.

\bibitem{Hasenfratz:1998jp}
P.~Hasenfratz.
\newblock Lattice qcd without tuning, mixing and current renormalization.
\newblock {\em Nucl. Phys.}, B525:401, 1998.

\bibitem{Kikukawa:1997qh}
Y.~Kikukawa and H.~Neuberger.
\newblock Overlap in odd dimensions.
\newblock {\em Nucl. Phys.}, B513:735, 1998.

\bibitem{Neuberger:1998fp}
H.~Neuberger.
\newblock Exactly massless quarks on the lattice.
\newblock {\em Phys. Lett.}, B417:141, 1998.

\bibitem{Neuberger:1998wv}
H.~Neuberger.
\newblock More about exactly massless quarks on the lattice.
\newblock {\em Phys. Lett.}, B427:353, 1998.

\bibitem{Bie1998ut}
W.~Bietenholz.
\newblock Solutions of the ginsparg-wilson relation and improved domain wall
  fermions.
\newblock {\em Eur. Phys. J.}, C6:537, 1999.

\bibitem{Bietenholz1998wx}
W.~Bietenholz, N.~Eicker, A.~Frommer, Th. Lippert, B.~Medeke, K.~Schilling, and
  G.~Weuffen.
\newblock Preconditioning of improved and 'perfect' fermion actions.
\newblock {\em Comput. Phys. Commun.}, 119:1, 1999.

\bibitem{WilsKogut1974}
K.~Wilson and J.~Kogut.
\newblock The renormalization group and the $\epsilon$-expansion.
\newblock {\em Phys. Rep.}, C12:75, 1974.

\bibitem{Kadanoff77}
L.~P. Kadanoff.
\newblock Lectures on the application of renormalization group techniques to
  quarks and strings.
\newblock {\em Rev. Mod. Phys.}, 49:267, 1977.

\bibitem{BW96}
W.~Bietenholz and U.~J. Wiese.
\newblock Perfect lattice actions for quarks and gluons.
\newblock {\em Nucl. Phys.}, B464:319, 1996.

\bibitem{Biet94FPA}
W.~Bietenholz.
\newblock Chiral symmetry, renormalization group and fixed points for lattice
  fermions.
\newblock {\em Proc. of XIV Brazilian National Meeting on Particles and
  Fields}, page 360.

\bibitem{BIE96}
W.~Bietenholz, R.~Brower, S.~Chandrasekharan, and U.~J. Wiese.
\newblock Progress on perfect lattice actions for qcd.
\newblock {\em Nucl. Phys. Proc. Suppl.}, 53:921, 1997.

\bibitem{BIE98}
W.~Bietenholz and U.~J. Wiese.
\newblock Perfect actions with chemical potential.
\newblock {\em Phys. Lett.}, B426:114, 1998.

\bibitem{Orginos:1997fh}
K.~Orginos, W.~Bietenholz, R.~Brower, S.~Chandrasekharan, and U.~J. Wiese.
\newblock The perfect quark-gluon vertex function.
\newblock {\em Nucl. Phys. Proc. Suppl.}, 63:904, 1998.

\bibitem{Creutz:1980zw}
M.~Creutz.
\newblock Monte carlo study of quantized $su(2)$ gauge theory.
\newblock {\em Phys. Rev.}, D21:2308, 1980.

\bibitem{metropolis}
N.~Metropolis, A.~W. Rosenbluth, M.~N. Rosenbluth, A.~H. Teller, and E.~Teller.
\newblock {\em J. Chem. Phys.}, 21:1087, 1953.

\bibitem{Cabibbo:1982zn}
N.~Cabibbo and E.~Marinari.
\newblock A new method for updating $su(n)$ matrices in computer simulations of
  gauge theories.
\newblock {\em Phys. Lett.}, B119:387, 1982.

\bibitem{Adler:1981sn}
S.~L. Adler.
\newblock An overrelaxation method for the monte carlo evaluation of the
  partition function for multiquadratic actions.
\newblock {\em Phys. Rev.}, D23:2901, 1981.

\bibitem{Adler:1988ce}
S.~L. Adler.
\newblock Overrelaxation algorithms for lattice field theories.
\newblock {\em Phys. Rev.}, D37:458, 1988.

\bibitem{Brown:1987rr}
F.~R. Brown and T.~J. Woch.
\newblock Overrelaxed heat bath and metropolis algorithms for accelerating pure
  gauge monte carlo calculations.
\newblock {\em Phys. Rev. Lett.}, 58:2394, 1987.

\bibitem{Creutz:1987xi}
M.~Creutz.
\newblock Overrelaxation and monte carlo simulation.
\newblock {\em Phys. Rev.}, D36:515, 1987.

\bibitem{Hernandez:1998et}
P.~Hern{\' a}ndez, K.~Jansen, and M.~L{\"u}scher.
\newblock Locality properties of neuberger's lattice dirac operator.
\newblock {\em Nucl. Phys.}, B552:363, 1999.

\bibitem{bookNR}
W.~H. Press, S.~A. Teukolsky, W.~T. Vetterling, and B.~P. Flannery.
\newblock {\em Numerical Recipes: The Art of Scientific Computing (2nd
  edition)}.
\newblock Cambridge University Press, 1992.

\bibitem{Hernandez:1999cu}
P.~Hern{\' a}ndez, K.~Jansen, and L.~Lellouch.
\newblock Finite-size scaling of the quark condensate in quenched lattice qcd.
\newblock {\em Phys. Lett.}, B469:198, 1999.

\bibitem{Bietenholz:2000iy}
W.~Bietenholz.
\newblock Approximate ginsparg-wilson fermions for qcd.
\newblock {\em Proc. of the International Workshop on Non-Perturbative Methods
  and Lattice QCD Guangzhou (China), X.-Q. Luo and E.B. Gregory (eds.)},
  page~3, 2000.

\bibitem{Jansenprivate}
K.~Jansen.
\newblock Private communication.

\bibitem{Bietenholz:2002ks}
W.~Bietenholz.
\newblock Convergence rate and locality of improved overlap fermions.
\newblock {\em Nucl. Phys.}, B644:223, 2002.

\bibitem{Blum:2000kn}
T.~Blum and et~al.
\newblock Quenched lattice qcd with domain wall fermions and the chiral limit.
\newblock {\em Phys. Rev.}, D69:074502, 2004.

\bibitem{vandenEshof:2002ms}
J.~van~den Eshof, A.~Frommer, Th. Lippert, K.~Schilling, and H.~A. van~der
  Vorst.
\newblock Numerical methods for the qcd overlap operator. i: Sign- function and
  error bounds.
\newblock {\em Comput. Phys. Commun.}, 146:203, 2002.

\bibitem{Giusti:2002sm}
L.~Giusti, C.~Hoelbling, M.~L{\" u}scher, and H.~Wittig.
\newblock Numerical techniques for lattice qcd in the $\epsilon$- regime.
\newblock {\em Comput. Phys. Commun.}, 153:31, 2003.

\bibitem{Arnold:2003sx}
G.~Arnold, N.~Cundy, J.~van~den Eshof, A.~Frommer, S.~Krieg, Th. Lippert, and
  K.~Schafer.
\newblock Numerical methods for the qcd overlap operator. ii: Optimal krylov
  subspace methods.
\newblock 2003.

\bibitem{Cundy:2004pz}
N.~Cundy and et. al.
\newblock Numerical methods for the qcd overlap operator. iii: Nested
  iterations.
\newblock 2004.

\bibitem{Chiarappa:2004ry}
T.~Chiarappa, K.~Jansen, K.-I. Nagai, M.~Papinutto, L.~Scorzato, A.~Shindler,
  C.~Urbach, U.~Wenger, and I.~Wetzorke.
\newblock Comparing iterative methods for overlap and twisted mass fermions.
\newblock 2004.

\bibitem{BietenholzPP}
W.~Bietenholz, S.~Capitani, T.~Chiarappa, N.~Christian, M.~Hasenbusch,
  K.~Jansen, K.-I. Nagai, M.~Papinutto, L.~Scorzato, S.~Shcheredin,
  A.~Shindler, C.~Urbach, Wenger U., and I.~Wetzorke.
\newblock Going chiral: overlap versus twisted mass fermions.
\newblock 2004.

\bibitem{Sommer:1993ce}
R.~Sommer.
\newblock A new way to set the energy scale in lattice gauge theories and its
  applications to the static force and $\alpha_s$ in su(2) yang-mills theory.
\newblock {\em Nucl. Phys.}, B411:839, 1994.

\bibitem{Necco:2001xg}
S.~Necco and R.~Sommer.
\newblock The $n_f = 0$ heavy quark potential from short to intermediate
  distances.
\newblock {\em Nucl. Phys.}, B622:328, 2002.

\bibitem{Guagnelli:1998ud}
M.~Guagnelli, R.~Sommer, and H.~Wittig.
\newblock Precision computation of a low-energy reference scale in quenched
  lattice qcd.
\newblock {\em Nucl. Phys.}, B535:389, 1998.

\bibitem{Damgaard:1999gj}
P.~H. Damgaard, U.~M. Heller, R.~Niclasen, and K.~Rummukainen.
\newblock Looking for effects of topology in the dirac spectrum of staggered
  fermions.
\newblock {\em Nucl. Phys. Proc. Suppl.}, 83:197, 2000.

\bibitem{Damgaard:1999bq}
P.~H. Damgaard, U.~M. Heller, R.~Niclasen, and K.~Rummukainen.
\newblock Staggered fermions and gauge field topology.
\newblock {\em Phys. Rev.}, D61:014501, 2000.

\bibitem{Durr:2003xs}
S.~D{\"u}rr and C.~Hoelbling.
\newblock Staggered versus overlap fermions: A study in the schwinger model
  with $n_f = 0,1,2$.
\newblock {\em Phys. Rev.}, D69:034503, 2004.

\bibitem{Follana:2004sz}
E.~Follana, A.~Hart, and C.~T.~H. Davies.
\newblock The index theorem and universality properties of the low- lying
  eigenvalues of improved staggered quarks.
\newblock 2004.

\bibitem{Wong:2004ai}
K.~Y. Wong and R.~M. Woloshyn.
\newblock Topology and staggered fermion action improvement.
\newblock 2004.

\bibitem{Farchioni:1999yy}
F.~Farchioni, I.~Hip, and C.~B. Lang.
\newblock Effects of topology in the dirac spectrum of staggered fermions.
\newblock {\em Phys. Lett.}, B471:58, 1999.

\bibitem{Farchioni:1998jc}
F.~Farchioni, I.~Hip, C.~B. Lang, and M.~Wohlgenannt.
\newblock Eigenvalue spectrum of massless dirac operators on the lattice.
\newblock {\em Nucl. Phys.}, B549:364, 1999.

\bibitem{Berg:2001nn}
B.~A. Berg, U.~M. Heller, H.~Markum, R.~Pullirsch, and W.~Sakuler.
\newblock Exact zero-modes of the compact qed dirac operator.
\newblock {\em Phys. Lett.}, B514:97, 2001.

\bibitem{Edwards:1999ra}
R.~G. Edwards, U.~M. Heller, J.~E. Kiskis, and R.~Narayanan.
\newblock Quark spectra, topology and random matrix theory.
\newblock {\em Phys. Rev. Lett.}, 82:4188--4191, 1999.

\bibitem{Hasenfratz:2002rp}
P.~Hasenfratz, S.~Hauswirth, T.~J{\"o}rg, F.~Niedermayer, and K.~Holland.
\newblock Testing the fixed-point qcd action and the construction of chiral
  currents.
\newblock {\em Nucl. Phys.}, B643:280--320, 2002.

\bibitem{Bietenholz:2003mi}
W.~Bietenholz, K.~Jansen, and S.~Shcheredin.
\newblock Spectral properties of the overlap dirac operator in qcd.
\newblock {\em JHEP}, 07:033, 2003.

\bibitem{PARPACK1}
C.~Lanczos.
\newblock Solution of systems of linear equations by minimized iteration.
\newblock {\em J. Res. Nat. Bur. Stand.}, 49:33, 1952.

\bibitem{PARPACK2}
D.~C. Sorensen.
\newblock Implicit application of polynomial filters in a k-step arnoldi
  method.
\newblock {\em Siam J. Matrix Anal. Appl.}, 13:357, 1992.

\bibitem{Farchioni:1999se}
F.~Farchioni.
\newblock Leutwyler-smilga sum rules for ginsparg-wilson lattice fermions.
\newblock 1999.

\bibitem{Capitani:1999uz}
S.~Capitani, M.~G{\" o}ckeler, R.~Horsley, P.~E.~L. Rakow, and G.~Schierholz.
\newblock Operator improvement for ginsparg-wilson fermions.
\newblock {\em Phys. Lett.}, B468:150, 1999.

\bibitem{Bietenholz_private}
W.~Bietenholz.
\newblock Private notes.

\bibitem{Damgaard:2001xr}
P.~H. Damgaard.
\newblock Quenched finite volume logarithms.
\newblock {\em Nucl. Phys.}, B608:162, 2001.

\bibitem{Verbaarschot:2000dy}
J.~J.~M. Verbaarschot and T.~Wettig.
\newblock Random matrix theory and chiral symmetry in qcd.
\newblock {\em Ann. Rev. Nucl. Part. Sci.}, 50:343, 2000.

\bibitem{Giusti:2003gf}
L.~Giusti, M.~L{\" u}scher, P.~Weisz, and H.~Wittig.
\newblock Lattice qcd in the $\epsilon$-regime and random matrix theory.
\newblock {\em JHEP}, 11:023, 2003.

\bibitem{Witten:1979vv}
E.~Witten.
\newblock Current algebra theorems for the $u(1)$ 'goldstone boson'.
\newblock {\em Nucl. Phys.}, B156:269, 1979.

\bibitem{Veneziano:1979ec}
G.~Veneziano.
\newblock $u(1)$ without instantons.
\newblock {\em Nucl. Phys.}, B159:213, 1979.

\bibitem{Giusti:2004qd}
L.~Giusti, G.~C. Rossi, and M.~Testa.
\newblock Topological susceptibility in full qcd with ginsparg-wilson fermions.
\newblock {\em Phys. Lett.}, B587:157, 2004.

\bibitem{Luscher:2004fu}
M.~L{\"u}scher.
\newblock Topological effects in qcd and the problem of short-distance
  singularities.
\newblock 2004.

\bibitem{Giusti:2001xh}
L.~Giusti, G.~C. Rossi, M.~Testa, and G.~Veneziano.
\newblock The $u_a(1)$ problem on the lattice with ginsparg-wilson fermions.
\newblock {\em Nucl. Phys.}, B628:234, 2002.

\bibitem{Chandrasekharan:2004cn}
S.~Chandrasekharan and U.~J. Wiese.
\newblock An introduction to chiral symmetry on the lattice.
\newblock {\em Prog. Part. Nucl. Phys.}, 53:373, 2004.

\bibitem{DelDebbio:2003rn}
L.~Del~Debbio and C.~Pica.
\newblock Topological susceptibility from the overlap.
\newblock {\em JHEP}, 02:003, 2004.

\bibitem{DelDebbio:2004ns}
L.~Del~Debbio, L.~Giusti, and C.~Pica.
\newblock Topological susceptibility in the su(3) gauge theory.
\newblock 2004.

\bibitem{Bietenholz:2003bj}
W.~Bietenholz, T.~Chiarappa, K.~Jansen, K.-I. Nagai, and S.~Shcheredin.
\newblock Axial correlation functions in the $\epsilon$-regime: A numerical
  study with overlap fermions.
\newblock {\em JHEP}, 02:023, 2004.

\bibitem{Jansen_notes}
B.~Bunk, K.~Jansen, M.~L\"{u}scher, and H.~Simma.
\newblock Desy report (september 1994).

\bibitem{Kalkreuter:1996mm}
T.~Kalkreuter and H.~Simma.
\newblock An accelerated conjugate gradient algorithm to compute low lying
  eigenvalues: A study for the dirac operator in $su(2)$ lattice qcd.
\newblock {\em Comput. Phys. Commun.}, 93:33, 1996.

\bibitem{Berruto:2003rt}
F.~Berruto, N.~Garron, C.~Hoelbling, L.~Lellouch, C.~Rebbi, and N.~Shoresh.
\newblock Preliminary results from a simulation of quenched qcd with overlap
  fermions on a large lattice.
\newblock {\em Nucl. Phys. Proc. Suppl.}, 129:471, 2004.

\bibitem{Giusti:2004yp}
L.~Giusti, P.~Hernandez, M.~Laine, P.~Weisz, and H.~Wittig.
\newblock Low-energy couplings of qcd from current correlators near the chiral
  limit.
\newblock {\em JHEP}, 04:013, 2004.

\bibitem{Gottlieb:1985rc}
S.~A. Gottlieb, P.~B. MacKenzie, H.~B. Thacker, and D.~Weingarten.
\newblock Hadronic couplings constants in lattice gauge theory.
\newblock {\em Nucl. Phys.}, B263:704, 1986.

\bibitem{Gupta:1987zc}
R.~Gupta, G.~Guralnik, G.~Kilcup, A.~Patel, S.~R. Sharpe, and T.~Warnock.
\newblock The hadron spectrum on a $18^3\times 42$ lattice.
\newblock {\em Phys. Rev.}, D36:2813, 1987.

\bibitem{Neuberger:1999pz}
H.~Neuberger.
\newblock Bounds on the wilson dirac operator.
\newblock {\em Phys. Rev.}, D61:085015, 2000.

\bibitem{Luscher:1998du}
M.~L{\"u}scher.
\newblock Abelian chiral gauge theories on the lattice with exact gauge
  invariance.
\newblock {\em Nucl. Phys.}, B549:295, 1999.

\bibitem{Luscher:1999un}
M.~L{\"u}scher.
\newblock Weyl fermions on the lattice and the non-abelian gauge anomaly.
\newblock {\em Nucl. Phys.}, B568:162, 2000.

\bibitem{Fukaya:2003ph}
H.~Fukaya and T.~Onogi.
\newblock Lattice study of the massive schwinger model with theta term under
  l{\"u}scher's 'admissibility' condition.
\newblock {\em Phys. Rev.}, D68:074503, 2003.

\bibitem{Creutz:2004ir}
M.~Creutz.
\newblock Positivity and topology in lattice gauge theory.
\newblock 2004.

\bibitem{Shcheredin:2004xa}
S.~Shcheredin, W.~Bietenholz, K.~Jansen, K.-I. Nagai, S.~Necco, and
  L.~Scorzato.
\newblock Testing a topology conserving gauge action in qcd.
\newblock 2004.

\bibitem{Ilgenfritz:1985dz}
E.-M. Ilgenfritz, M.~L. Laursen, G.~Schierholz, M.~M{\"u}ller-Preussker, and
  H.~Schiller.
\newblock First evidence for the existence of instantons in the quantized su(2)
  lattice vacuum.
\newblock {\em Nucl. Phys.}, B268:693, 1986.

\bibitem{Ilgenfritz:2002qs}
E.-M. Ilgenfritz, B.~V. Martemyanov, M.~M{\" u}ller-Preussker, S.~Shcheredin,
  and A.~I. Veselov.
\newblock On the topological content of su(2) gauge fields below t(c).
\newblock {\em Phys. Rev.}, D66:074503, 2002.

\bibitem{Nagai:2003gq}
K.-I. Nagai, W.~Bietenholz, T.~Chiarappa, K.~Jansen, and S.~Shcheredin.
\newblock Simulations in the $\epsilon$-regime of chiral perturbation theory.
\newblock {\em Nucl. Phys. Proc. Suppl.}, 129:516, 2004.

\bibitem{Marenzoni:1993im}
P.~Marenzoni, L.~Pugnetti, and P.~Rossi.
\newblock Measure of autocorrelation times of local hybrid monte carlo
  algorithm for lattice qcd.
\newblock {\em Phys. Lett.}, B315:152, 1993.

\bibitem{Scalettar:1986uy}
R.~T. Scalettar, D.~J. Scalapino, and R.~L. Sugar.
\newblock New algorithm for the numerical simulation of fermions.
\newblock {\em Phys. Rev.}, B34:7911, 1986.

\bibitem{Duane:1987de}
S.~Duane, A.~D. Kennedy, B.~J. Pendleton, and D.~Roweth.
\newblock Hybrid monte carlo.
\newblock {\em Phys. Lett.}, B195:216, 1987.

\end{thebibliography}
\bibliographystyle{hunsrt}

%---------------------Anhang-----------------------------
\appendix
% In Latex2e, each appendix is a "chapter"
\chapter{A local Hybrid Monte Carlo algorithm}
In this Appendix we discuss the implementation of the local HMC algorithm~\cite{Marenzoni:1993im} that we used to simulate the L\"uscher gauge action. It is based on the HMC algorithm described in Refs.~\cite{Scalettar:1986uy,Duane:1987de}

Let us consider the Hamiltonian of the $SU(3)$ gauge theory
\begin{equation}
H[\pi, U]=\frac{1}{2}\sum_{x, \mu}\sum_{j=1}^8\pi^2_{x,\mu,j}+ S_{\rm YM}[U]\ .
\end{equation}
We define $\pi_{x, \mu}=\sum_{j=1}^8 \pi_{x,\mu, j}\lambda_j$, where $\lambda_j$ are the Gell-Mann matrices for the $SU(3)$ gauge group. The variables $\pi_{x, \mu}$ are canonical momenta for $U_{x, \mu}$.
The local HMC algorithm for the simulation of the gauge fields $U_{x,\mu}\in SU(3)$ starts off by choosing a random value of the momentum $\pi_{x, \mu}$ according to the Gaussian distribution
\begin{equation}
P_G[\pi_{x,\mu}] \propto \prod_{j=1}^8 \exp{\left\{ -\frac{1}{2}\pi^2_{x,\mu, j} \right\}} \ . 
\end{equation}

Then, using the Hamiltonian $H$ the pair $[\pi_{x,\mu},U_{x,\mu}]$ is changed along the discretized trajectory in phase space up to the end point $[\pi_{x,\mu}',U_{x,\mu}'] \equiv T_H[\pi,U]$, where $T_H$ is the discretized evolution operator corresponding to the Hamiltonian $H$. It is assumed that the discretized classical dynamics equations, which define the operator $T_H$, are such that the mapping $[\pi_{x,\mu}, U_{x,\mu}] \rightarrow [\pi'_{x,\mu}, U'_{x,\mu}]$ is reversible in the sense
\begin{equation}
P_H([\pi'_{x,\mu}, U'_{x,\mu}] \leftarrow [\pi_{x,\mu}, U_{x,\mu}])=P_H([-\pi_{x,\mu}, U_{x,\mu}] \leftarrow [-\pi'_{x,\mu}, U'_{x,\mu}])\ .
\label{reversabilty}
\end{equation} 
$P_H$ is the probability of the configuration transition 
\begin{equation}
P_H([\pi'_{x,\mu}, U'_{x,\mu}] \leftarrow [\pi_{x,\mu}, U_{x,\mu}])=\delta ([\pi'_{x,\mu}, U'_{x, \mu}]-T_H[\pi_{x,\mu}, U_{x, \mu}])\ ,
\end{equation}
which takes this form since the change is deterministic in the case of the classical equations of motion. The reversibility condition holds for instance for the leap frog integration. After the trajectory $[\pi_{x, \mu}, U_{x, \mu}]\rightarrow [\pi'_{x,\mu}, U'_{x,\mu}]$ is determined, the pair $[\pi'_{x,\mu}, U'_{x,\mu}]$ is either accepted or rejected, and the old pair $[\pi_{x,\mu}, U_{x,\mu}]$ is kept. The acceptance is defined by the Metropolis probability
\begin{equation}
P_A([\pi'_{x,\mu}, U'_{x,\mu}] \leftarrow [\pi_{x,\mu}, U_{x,\mu}])={\rm min}\left\{ 1,e^{-\Delta H[\pi'_{x,\mu},U'_{x,\mu}]+\Delta H[\pi_{x,\mu}, U_{x,\mu}]}   \right\}\ ,
\end{equation}
where 
\begin{equation}
\Delta H[\pi_{x,\mu}, U_{x,\mu}]= \sum_{j=1}^8\frac{1}{2} \pi^2_{x,\mu,j}+ \sum_{\mu< \nu}\beta\, S_P(U^{(P)}_{x,\mu \nu})
\end{equation}
is the part of the Hamiltonian $H$ which includes $\pi_{x, \mu}$ and $U_{x,\mu}$.
The total transition probability is
\begin{eqnarray}
P([U'_{x,\mu}\leftarrow U_{x,\mu}])&=&\int [d\pi_{x, \mu} \, d\pi'_{x,\mu}] \, P_A([\pi'_{x,\mu}, U'_{x,\mu}]\leftarrow [\pi_{x,\mu}, U_{x,\mu}])\times \nonumber\\
&&P_H([\pi'_{x,\mu}, U'_{x,\mu}] \leftarrow [\pi_{x,\mu}, U_{x,\mu}]) \, P_G[\pi_{x,\mu}]\ .
\end{eqnarray}
It can be shown that this probability distribution has the canonical distribution proportional to $ \exp{(-S_{\rm YM}[U])}$ as a unique fixed point. It can be proved by using the reversibility condition~(\ref{reversabilty}). Therefore a finite difference approximation to the Hamiltonian equations must also account for it.

Applying now this algorithm to each pair $[\pi_{x,\mu},U_{x,\mu}]$ we perform the update of the gauge field throughout the lattice. 

We close the explanation of the algorithm by giving the description of the finite difference approximation to the Hamiltonian equations, i.e. to the evolution operator $T_H$. It is given by the leapfrog integration, which proceeds as follows: one performs a first half step
\begin{equation}
\pi_{x,\mu ,j}(\frac{\Delta \tau}{2})=\pi_{x,\mu , j}(0)-\frac{\delta \Delta S_{\rm YM}[U_{x,\mu}(0)]}{\delta_j U_{x, \mu}} \frac{\Delta \tau}{2}\ ,
\end{equation}
and then for steps $k=0,1,\dots , n-1$
\begin{equation}
U_{x, \mu}(k\Delta \tau +\Delta \tau)=\exp\left\{ \sum_{j=1}^8 i\lambda_j \pi_{x, \mu, j}(k\Delta \tau +\frac{\Delta \tau}{2} )\Delta \tau  \right\} U_{x, \mu}(k\Delta \tau)\ ,
\end{equation}
and for $k=1,2, \dots , n-1$
\begin{equation}
\pi_{x,\mu ,j}(k\Delta \tau+\frac{\Delta \tau}{2})=\pi_{x,\mu , j}(k\Delta \tau -\frac{\Delta \tau}{2})-\frac{\delta \Delta S_{\rm YM}[U_{x,\mu}(k\Delta \tau)]}{\delta_j U_{x, \mu}} \Delta \tau\ ,
\end{equation}
and finally
\begin{equation}
\pi_{x,\mu ,j}(n\Delta \tau)=\pi_{x,\mu , j}(n\Delta \tau -\frac{\Delta \tau}{2})-\frac{\delta \Delta S_{\rm YM}[U_{x, \mu}(n\Delta \tau)]}{\delta_j U_{x, \mu}} \Delta \tau\ ,
\end{equation}
where $\Delta S_{\rm YM}[U_{x, \mu}]=\sum_{\mu< \nu}\beta\, S_P(U^{(P)}_{x,\mu \nu})$ is the part of the lattice gauge action which contains $U_{x, \mu}$. 
The derivative appearing in these formulae is defined as
\begin{equation}
\frac{\delta f[U_{x, \mu}]}{\delta_j U_{x, \mu}}=\frac{\partial }{\partial \alpha}f\left [ e^{i\alpha \lambda_j}U_{x,\mu} \right ]_{\alpha=0}\ .
\end{equation}

%-------------- Finally, the VITA

\chapter*{Curriculum Vitae}

\begin{tabular}{ll}

Name: & \dcauthorname  \dcauthorsurname \\

1996 - 2002 & Study at Moscow State University \\

 & field Physics\\

02/2002 - 02/2005 & Ph.D. at \\

 & Humboldt University Berlin,\\

 & Chair Prof. M. M\"uller-Preu\ss ker, \\

 & Institute for Physics\\

\end{tabular}

% ------- Declaration

\chapter*{Selbst\"andigkeitserkl\"arung}

\noindent Hiermit erkl\"are ich, die vorliegende Arbeit selbst\"andig ohne fremde Hilfe verfa{\ss}t und nur die angegebene Literatur und Hilfsmittel verwendet zu haben.\\

\vspace{5cm}
\noindent\dcauthorname \dcauthorsurname \\  %%\hspace{-.6cm}
\dcdatesubmitted \\

\end{document}